\documentclass[11pt]{cernrep}
\usepackage{graphicx}
\usepackage{here}
\usepackage{bm}
\usepackage{epsfig}

\newcommand{\la}{\left \langle}
\newcommand{\ra}{\right \rangle}

\newcommand{\tx}{\tilde x}

\newcommand{\2}{\frac{1}{2}}

\newcommand{\nc}{\newcommand}
\nc{\equ}{{\rm eq}}
\nc{\BC}{{_{\rm BC}}}
\nc{\WN}{{_{\rm WN}}}
\nc{\NN}{{_{\rm NN}}}
\nc{\grad}{{\bm{\nabla}}}
\nc{\bb}{{\bm{b}}}
\nc{\be}{{\bm{e}}}
\nc{\bs}{{\bm{s}}}
\nc{\bx}{{\bm{x}}}
\nc{\by}{{\bm{y}}}
\nc{\bu}{{\bm{u}}}
\nc{\bv}{{\bm{v}}}
\nc{\bp}{{\bm{p}}}
\nc{\bq}{{\bm{q}}}
\nc{\br}{{\bm{r}}}
\nc{\bK}{{\bm{K}}}
\nc{\bP}{{\bm{P}}}
\nc{\bKp}{{\bm{K}}_\perp}
\nc{\bvp}{{\bm{v}}_\perp}
\nc{\rp}{{r_\perp}}
\nc{\brp}{{\br_\perp}}
\nc{\bpp}{{\bp_\perp}}
\nc{\mperp}{m_\perp}
\nc{\pperp}{p_\perp}
\nc{\vperp}{v_\perp}
\nc{\pt}{p_\perp}
\nc{\mt}{m_\perp} 
\nc{\Kt}{K_\perp}
\nc{\Mt}{M_\perp}
\nc{\pL}{p_{\rm L}}
\nc{\vL}{v_{\rm L}}
\nc{\etaL}{\eta_{\rm L}}
\nc{\ec}{e_{\rm cr}}
\nc{\Tc}{T_{\rm cr}}
\nc{\scm}{\sqrt{s_{\rm NN}}}
\nc{\Tmunu}{T^{\mu\nu}}
\nc{\gmunu}{g^{\mu\nu}}
\nc{\dmuumu}{\partial_\mu u^\mu}
\nc{\diag}{{\rm diag}}
\nc{\Tdec}{T_{\rm dec}}
\nc{\mT}{m_{\rm T}}
\nc{\pT}{p_{\rm T}}
\nc{\vT}{v_{\rm T}}
\nc{\tauequ}{\tau_{\rm eq}}
\nc{\tauf}{\tau_{\rm f}}
\nc{\eequ}{e_{\rm eq}}
\nc{\nequ}{n_{\rm eq}}
\nc{\sequ}{s_{\rm eq}}
\nc{\Tequ}{T_{\rm eq}}
\nc{\edec}{e_{\rm dec}}
\nc{\eqgp}{e_{\rm QGP}}
\nc{\dec}{{\rm dec}}
\nc{\const}{{\rm const}}

\nc{\Var}{{\rm Var}}
\nc{\fs}{{\rm fs}}
\nc{\kin}{{\rm kin}}
\nc{\form}{{\rm form}}
\nc{\hydro}{{\rm hydro}}
\nc{\ch}{{\rm ch}}
\nc{\sat}{{\rm sat}}
\nc{\half}{\frac{1}{2}}

\nc{\eq}{{\,=\,}}
\nc{\lla}{\la\!\la}
\nc{\rra}{\ra\!\ra}
\nc{\eps}{\epsilon}
\nc{\se}{\section}
\nc{\suse}{\subsection}
\nc{\sususe}{\subsubsection}
\nc{\beq}[1]{\begin{equation}\label{#1}}
\nc{\eeq}{\end{equation}}
\nc{\bea}[1]{\begin{eqnarray}\label{#1}}
\nc{\eea}{\end{eqnarray}}
\nc{\bce}{\begin{center}}
\nc{\ece}{\end{center}}

\renewcommand{\phi}{\varphi}
\renewcommand{\theta}{\vartheta}
\renewcommand{\P}{\partial}
\newcommand{\gapp}{\,{\raisebox{-.2ex}{$\stackrel{>}{_\sim}$}}\,}
\newcommand{\lapp}{\,{\raisebox{-.2ex}{$\stackrel{<}{_\sim}$}}\,}

\begin{document}

\title{CONCEPTS OF HEAVY-ION PHYSICS\thanks{\ These lecture notes are 
an expanded version of the lectures I gave a year earlier  at the
2002 European School of High-Energy Physics in Pylos (Greece) whose 
proceedings were published as a CERN Yellow Report (CERN-2004-001,
N. Ellis and R. Fleischer, eds.). The online version of these lecture 
notes on the arXiv has most graphs presented in color.}}
\author{Ulrich Heinz}
\institute{Department of Physics, The Ohio State University, Columbus, 
           OH 43210, USA}
\maketitle

\begin{abstract}
In these lectures I present the key ideas driving the field of 
relativistic heavy-ion physics and develop some of the theoretical 
tools needed for the description and interpretation of heavy-ion collision
experiments.
\end{abstract}

%%%%%%%% beginning of text %%%%%%%%%%%%%%%

%%%%%%%%%%%%%%%%%%%%%%%%%%%%%%%%%%%%%%%%%%%%%%%%%%%%%%%%%%%%%%%%%%%%%%%%%%%%%
\section{PROLOGUE: THE BIG BANG AND THE EARLY UNIVERSE}
\label{sec1}
%%%%%%%%%%%%%%%%%%%%%%%%%%%%%%%%%%%%%%%%%%%%%%%%%%%%%%%%%%%%%%%%%%%%%%%%%%%%%

Matter as we know it, made up from molecules which consist of atoms
which consist of electrons circling around a nucleus which consists
of protons and neutrons which themselves are bound states of quarks and
gluons, has not existed forever. 
Our universe originated in a ``Big Bang'' from a state of almost
infinite energy density and temperature.
During the first few microseconds of its life the energy density in 
our universe was so high that hadrons (color singlet bound states of 
quarks, antiquarks and gluons), such as the nucleons inside a nucleus, 
could not form. 
Instead, the quarks, antiquarks and gluons were deconfined and permeated 
the entire universe in a thermalized state known as {\bf quark-gluon plasma} 
(QGP).
Only when the energy density of the universe dropped below the critical 
value $\ec\simeq 1$\,GeV/fm$^{ 3}$ and its temperature decreased below 
$\Tc\approx 170$\,MeV, colored degrees of freedom became confined into 
color singlet objects of about 1\,fm diameter: the first hadrons formed.
After the universe hadronized, it took another 200\,s or so until its 
temperature dropped below $\sim100$\,keV such that small atomic nuclei 
could form and survive.
This is known as {\em primordial nucleo\-synthesis}.
At this point (i.e. after ``The First 3 Minutes'') the chemical composition 
of the early universe was fixed ({\bf ``chemical freeze-out''}). 
All unstable hadrons had decayed and all antiparticles had annihilated, 
leaving only a small fraction of excess protons, neutrons and electrons, 
with all surviving neutrons bound inside small atomic nuclei.
The chemical composition of the universe began to change again only
several hundred million years later when the cores of the first stars 
ignited and nuclear fusion processes set in.
After primordial nucleosynthesis the universe was still ionized and
therefore completely opaque to electromagnetic radiation.
About 300\,000 years after the Big Bang, when the temperature had 
reached about 3000\,K, electrons and atomic nuclei were finally able
to combine into electrically neutral atoms, and the universe became 
transparant.
At this point the electromagnetic radiation decoupled, with a perfectly
thermal blackbody spectrum of $T\approx 3000$\,K 
({\bf ``thermal freeze-out''}).
Due to the continuing expansion of our universe this thermal photon 
spectrum has now been redshifted to a temperature of about 2.7\,K
and turned into the ``cosmic microwave background''.
The number of photons in this microwave background is huge (about 250
photons in every cm$^3$ of the universe), and they carry the bulk of
the entropy of the universe.
The entropy-to-baryon ratio of the universe is $S/A\simeq 10^{9\pm 1}$;
its inverse provides a measure for the tiny baryon-antibaryon asymmetry 
of our universe when it hadronized -- a still incompletely understood
small number.
The only other surviving feature of the Big Bang is the ongoing {\bf Hubble
expansion} of our universe, and the structure of its density fluctuations,
amplified over eons by the action of gravity and reflected in today's
distribution of stars, galaxies, galactic and supergalactic clusters, 
and dark matter.
Using these 3 or 4 observational pillars (today's expansion rate or 
``Hubble constant'', the microwave background spectrum and its 
fluctuations, the primordial nuclear abundances and, most recently,
also today's spectrum of density fluctuations) together with the equations 
of motion of general relativity, we have been able to reconstruct the 
cosmological evolution of our universe from its origin in the Big Bang.
However, try as you want, we will never be able to directly see anything 
that happened before 300\,000 years after the Big Bang, due to the opacity 
of the Early Universe. 
In particular, the all-permeating quark-gluon plasma which filled our
universe during the first few microseconds will always remain hidden 
behind the curtain of the cosmic microwave background.
This is were relativistic heavy-ion collisions come in: It turns out
that we can recreate this thermalized QGP matter (or at least some decent 
approximation to it) by colliding large nuclei at high energies.
To elaborate on this is the subject of these lectures.

%%%%%%%%%%%%%%%%%%%%%%%%%%%%%%%%%%%%%%%%%%%%%%%%%%%%%%%%%%%%%%%%%%%%%%%%%%%%%
\section{A FEW IMPORTANT RESULTS FROM LATTICE QCD}
\label{sec2}
%%%%%%%%%%%%%%%%%%%%%%%%%%%%%%%%%%%%%%%%%%%%%%%%%%%%%%%%%%%%%%%%%%%%%%%%%%%%%
\subsection{Lattice QCD in 3 minutes}
\label{sec2a}
%%%%%%%%%%%%%%%%%%%%%%%%%%%%%%%%%%%%%%%%%%%%%%%%%%%%%%%%%%%%%%%%%%%%%%%%%%%%%

We know from lattice QCD that the quark-gluon plasma exists (see 
Ref.~\cite{Karsch01} for a recent review). Lattice QCD is a method
for calculating equilibrium properties of strongly interacting systems
directly from the QCD Lagrangian by numerical evaluation of the 
corresponding path integrals. One starts from the vacuum-to-vacuum
transition amplitude in the Feynman path integral formulation
\bea{Z}
  Z = \int {\cal D}A_\mu^a(x)\,{\cal D}\bar\psi(x)\,
  {\cal D}\psi(x)\,e^{i\int d^4x\,{\cal L}[A_\mu^a,\bar\psi,\psi]}\ ,
\eea
where the phase factor depending on the classical action $\int d^4x\,
{\cal L}[A,\bar\psi,\psi]$ is integrated over all classical field
configurations for the gluon fields, $A_\mu(x)$, and quark and antiquark
fields, $\psi(x)$ and $\bar\psi(x)$. The path integral is dominated
by those field configurations $(A_\mu(x),\bar\psi(x),\psi(x))$ which
minimize the classical action and render the phase factor stationary, 
i.e. which satisfy the classical Euler-Langrange equations of motion.
These classical solutions define the classical chromodynamic field theory.
Dirac and Feynman showed that integrating over {\em all} field 
configurations instead of only the solutions of the classical equations
of motion produces the corresponding quantum field theory. The Pauli
principle for fermions is implemented by postulating that the classical
fermion fields $\psi(x)$ and $\bar\psi(x)$ are Grassmann variables 
satisfying $\psi(x_1)\psi(x_2)+\psi(x_2)\psi(x_1)=0$ etc.

Starting from Eq.~(\ref{Z}), one obtains an expression for the grand 
canonical partition function of an ensemble of quarks, antiquarks and 
gluons in thermal equilibrium by an almost trivial step: 
One replaces time $t$ everywhere
by imaginary time $\tau$, $t\to i\tau$, and restricts the integration range
over $\tau$ in the action to the interval $[0,\beta\eq\frac{1}{T}]$ where
$T$ is the temperature of the system:
\bea{ZE}
  {\cal Z} = \int {\cal D}A_\mu^a(\bx,\tau)\,{\cal D}\bar\psi(\bx,\tau)\,
  {\cal D}\psi(\bx,\tau)\,e^{-\int_0^\beta d\tau \int d^3x\,
  {\cal L}_{\rm E}[A_\mu^a,\bar\psi,\psi]}\ .
\eea
The origin of this replacement is the realization that the partition 
function is defined as ${\cal Z}\eq{\rm tr}\,\hat\rho$ and that the 
density operator $\hat\rho\eq{e}^{-\beta \hat H}$ for the grand canonical 
thermal equilibrium ensemble looks just like the time-evolution 
operator $e^{i\hat Ht}$ in the vacuum theory, with $t$ replaced by 
$i\beta$. The {\em Euclidean} Lagrangian density 
${\cal L}_{\rm E}[A,\bar\psi,\psi]$ arises from the normal QCD Lagrangian
\bea{L}
  {\cal L}_{\rm QCD} = -\frac{1}{4} F^a_{\mu\nu}F_a^{\mu\nu} 
  + i\bar\psi\gamma^\mu \left(\partial_\mu 
  - i g \frac{\lambda_a}{2}A^a_\mu\right)\psi - m\bar\psi\psi,
\eea 
with the non-Abelian gluon field strength tensor
\bea{F}
  F^a_{\mu\nu} = \partial_\mu A^a_\nu - \partial_\nu A^a_\mu
  + g f_{abc} A^b_\mu A^c_\mu,
\eea 
by replacing $\partial_t\eq{-}i\partial_\tau$
as well as $A^a_0\eq{i A^a_4}$ and $j^a_0\eq{i\,j^a_4}$ (where
$j_\mu^a{\eq}g\bar\psi\gamma_\mu\frac{\lambda_a}{2}\psi$ is the
color current vector of the quarks), and summing Lorentz indices over
1 through 4 (instead of 0 through 1) with unit metric tensor. In addition, 
in order to preserve the invariance of 
${\cal Z}[A,\bar\psi,\psi]\eq{\rm tr}\,\hat\rho[\hat A,\hat{\overline\psi},
\hat\psi]$ under cyclic permutations of the field operators under the 
trace, the classical gluon fields in the path integral must be periodic 
in imaginary time, $A^{a}_{\mu}(\bx,\tau)\eq{A}^a_\mu(\bx,\tau{+}\beta)$, 
whereas the Grassmannian fermion fields obey antiperiodic boundary 
conditions, $\psi(\bx,\tau)\eq{-}\psi(\bx,\tau{+}\beta)$ etc.
Note that $\tau$ is not really a time, it only plays a similar formal 
role in the path integral as real time does at zero temperature; a 
system in global thermal equilibrium is completely time independent.

Similar to Eq.~(\ref{ZE}) one can write down a path integral for the 
thermal equilibrium ensemble average $\langle \hat O\rangle\eq{\rm tr}
(\hat\rho\hat O)$ of an arbitrary observable $\hat O[\hat A,
\hat{\overline\psi},\hat\psi]$ which depends on the quark and gluon fields:
\bea{O}
  \langle \hat O\rangle = \frac
  {\int {\cal D}A_\mu^a(\bx,\tau)\,{\cal D}\bar\psi(\bx,\tau)\,
   {\cal D}\psi(\bx,\tau)\,O[A,\bar\psi,\psi]\,
   e^{-\int_0^\beta d\tau \int d^3x\,{\cal L}_{\rm E}[A,\bar\psi,\psi]}}
  {\int {\cal D}A_\mu(\bx,\tau)\,{{\cal D}\bar\psi(\bx,\tau)\,
   \cal D}\psi(\bx,\tau)\,
   e^{-\int_0^\beta d\tau \int d^3x\,{\cal L}_{\rm E}[A,\bar\psi,\psi]}}\ .
\eea
Here $O[A,\bar\psi,\psi]$ is the {\em classical} observable,
expressed through the {\em classical} fields $A^{a}_{\mu}(\bx,\tau)$,
$\psi(\bx,\tau)$ and $\bar\psi(\bx,\tau)$.

So far this expression is exact. Since the QCD Lagrangian is bilinear
in the quark fields, the path integral over the Grassmann field variables
can be done analytically, resulting in an infinite-dimensional determinant
over all space-time points:
\bea{Odet}
  \langle \hat O\rangle = 
  \frac{\int {\cal D}A_\mu(\bx,\tau)\
        \tilde O[A]\ {\rm det}\!\left[i\gamma^\mu\left(\partial_\mu - i g 
        \frac{\lambda_a}{2}A^a_\mu\right)-m\right]\,
        e^{-\int_0^\beta d\tau \int d^3x\,{\cal L}_{\rm E}[A]}}
       {\int {\cal D}A_\mu(\bx,\tau)\
        {\rm det}\!\left[i\gamma^\mu\left(\partial_\mu - i g 
        \frac{\lambda_a}{2}A^a_\mu\right)-m\right]\,
        e^{-\int_0^\beta d\tau \int d^3x\,{\cal L}_{\rm E}[A]}}\,.
\eea
Here ${\cal L}_{\rm E}[A]\eq{-}\frac{1}{4}F^a_{\mu\nu}F_a^{\mu\nu}$ is the
purely gluonic part of the euclidean QCD Lagrangian, and $\tilde O[A]$
arises from $O[A,\bar\psi,\psi]$ when doing the Gaussian integral over 
$\psi$ and $\bar\psi$. 

The words {\em Lattice QCD} stand for an algorithm to numerically 
evaluate this path integral, by discretizing space and time into 
$N_{s}^{3}N_\tau$ space-time lattice points, evaluating the 
corresponding $N_{s}^{3}N_\tau$-dimensional fermion determinant, 
and integrating over the $4{\times}8$ gluon fields $A_{\mu}^{a}$ 
from $-\infty$ to $\infty$ at each of the $N_s^3N_\tau$ 
lattice points. The integrals are performed by Monte Carlo
integration, using the Metropolis method of importance sampling.
This method works well as long as the fermion determinant is positive.
This is indeed the case for the expression given in Eq.~(\ref{Odet}) 
for the grand canonical ensemble at zero chemical potential. It 
describes a quark-gluon plasma with vanishing net baryon density, 
which is a good approximation for the Early Universe. Heavy-ion 
collisions, on the other hand, involve systems with non-zero net baryon
number, brought into the collision by the colliding nuclei. This 
requires introduction of a baryon chemical potential $\mu_B$ which 
enters into the fermion determinant in Eq.~(\ref{Odet}) with a factor 
$i$ and leads to oscillations of the latter. The resulting
``sign problem'' has been a stumbling block for lattice QCD at finite
net baryon density for almost 25 years, and only recently significant 
progress was made, resulting in first lattice QCD results for the 
hadronization phase transition in a quark-gluon plasma with nonzero 
baryon chemical potential \cite{FK02} (although still limited to 
$m_B\lapp3\Tc$ \cite{ST03}).

%%%%%%%%%%%%%%%%%%%%%%%%%%%%%%%%%%%%%%%%%%%%%%%%%%%%%%%%%%%%%%%%%%%%%%%%%%%%%
\subsection{Color deconfinement and chiral symmetry restoration}
\label{sec2b}
%%%%%%%%%%%%%%%%%%%%%%%%%%%%%%%%%%%%%%%%%%%%%%%%%%%%%%%%%%%%%%%%%%%%%%%%%%%%%

\begin{figure}[ht]
\begin{center}
\includegraphics[width=7cm]{polyakov.eps} 
\includegraphics[bb=94 190 467 538,width=7cm]{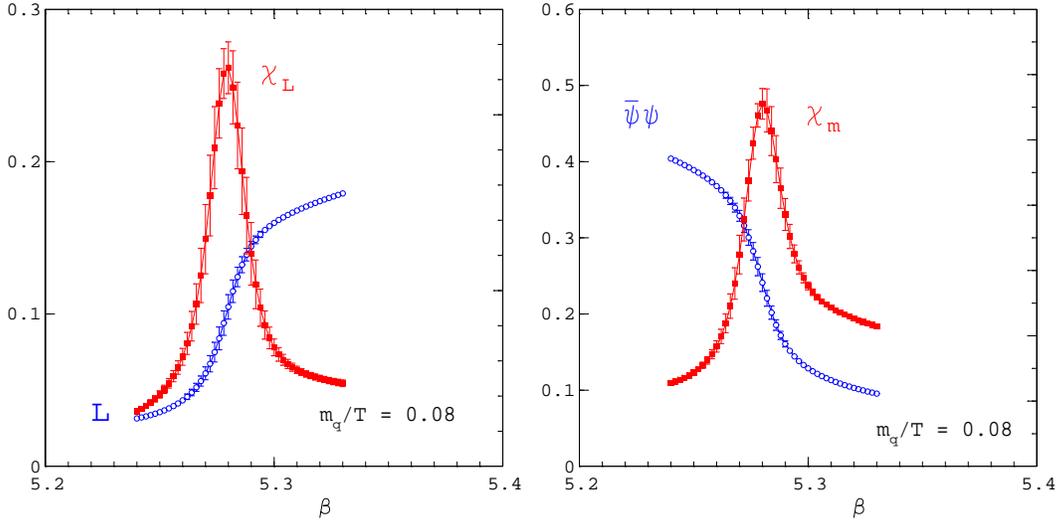} 
\vspace*{-9mm}
\caption{Left: Polyakov loop expectation value $\la L\ra$ and its
temperature derivative (Polyakov loop susceptibility $\chi_L$) as a 
function of the lattice coupling $\beta\eq6/g^2$ which is monotonically 
related to the temperature $T$ (larger $\beta$ correspond to larger $T$).
Right: The chiral condenstate $\la\bar\psi\psi\ra$ and the negative 
of its temperature derivative (chiral susceptibility $\chi_m$) as a
function of temperature. (From Ref.~\protect\cite{KL94}.)
\label{F1}
}
\end{center}
\vspace*{-5mm}
\end{figure}

For our discussion two observables are of particular importance: the 
{\em Polyakov loop operator} 
\bea{Poly}
   L = \frac{1}{3}\,{\rm tr}\left({\cal P}\,e^{ig\int_0^\beta 
   A_4(\bx,\tau)\,d\tau}\right)
\eea
(where $A_4{\eq}A_4^a\frac{\lambda_a}{2}$ is a $3{\times}3$ matrix and
${\cal P}$ stands for path ordering), and the 
scalar quark density $\bar\psi(x)\psi(x)$. Due to translational
invariance of the medium both have $\bx$-independent thermal 
expectation values which, however, show a strong temperature 
dependence. This is shown in Fig.~\ref{F1}. 
%
%%%%%%%%%%%%%%%%%%%%%%% Fig. 1 %%%%%%%%%%%%%%%%%%%%%%%%%%%%%%%%%%%%%%%%%%
%%%%%%%%%%%%%%%%%%%%%%%%%%%%%%%%%%%%%%%%%%%%%%%%%%%%%%%%%%%%%%%%%%%%%%%%
%
The argument of the exponential function in the Polyakov loop operator
$L$ is gauge-dependent although $L$ itself is not (due to the trace and 
the periodicity condition $A_4^a(\bx,\beta)=A_4^a(\bx,0)$); we can thus 
choose a gauge in which its $\tau$-dependence vanishes. It can then be 
interpreted as the interaction energy of an infinitely heavy quark at 
position $\bx$ (whose euclidean color current density four-vector is given 
by $J_\mu^a(y){\eq}ig\frac{\lambda^a}{2}\delta(\by{-}\bx)\left(1,0,0,0\right)$)
with the gluon field $A^\mu_a(y)$:
\bea{Pol}
  e^{ig\int_0^\beta d\tau \, A_4(\bx,\tau)}
  = e^{ig\beta A_4^a(\bx)\frac{\lambda_a}{2}} = e^{-\beta H_{\rm int}}
\eea
where
\bea{Hint}
  H_{\rm int} = - L_{\rm int} = \sum_{\mu=1}^4 \int d^3y \,
  J_\mu^a(\by) A_\mu^a(\by) = \int d^3y \,J^a_4(\by) A_4^a(\by)
  = ig\frac{\lambda^a}{2}A_4^a(\bx).
\eea
A vanishing thermal expectation value $\la L\ra$ of the Polyakov loop
operator thus indicates infinite energy for a free quark, i.e. quark
confinement. The left panel of Fig.~\ref{F1} shows this to be the case 
at small temperatures. However, as the temperature increases, $\la L\ra$ 
increases rapidly to a nonzero value at high temperatures, with a 
relatively sharp peak of its derivative at a critical coupling 
$\beta_{\rm cr}$. This indicates that quark confinement is broken at 
the corresponding critical temperature $\Tc$.

In the right panel of Fig.~\ref{F1} we see that at low temperatures the
scalar quark density has a nonvanishing expectation value (``chiral 
condensate'') which evaporates above a critical coupling. Again the 
corresponding susceptibility shows a relatively sharp peak, at the same 
value $\beta_{\rm cr}$. In the absence of quark masses the QCD Lagrangian
is chirally symmetric, i.e. invariant under separate flavor rotations
of right- and left-handed quarks. Since the up and down quark masses in
${\cal L}_{\rm QCD}$ are very small, neglecting them is a good 
approximation. The nonvanishing chiral condensate at $T\eq0$ breaks this 
chiral symmetry and generates a dynamic mass of order 300\,MeV for the 
quarks; the corresponding ``constituent'' masses in vacuum are
thus about 300\,MeV for the up and down quarks and about 450\,MeV for
the strange quark (whose bare mass in ${\cal L}_{\rm QCD}$ is already 
about 150\,MeV). According to the right panel of Fig.~\ref{F1} the 
dynamically generated mass melts away at $\Tc$, making the quarks
light again above $\Tc$: the approximate chiral symmetry of QCD is 
restored. 

Obviously deconfinement and chiral symmetry restoration happen at the 
same critical temperature $\Tc$. Figure~\ref{F1} shows this for one
specific, temperature-dependent value of the quark mass used in the 
lattice calculation ($m_q\eq0.8\,T$). This value is unrealistically 
large, but calculations with realistic and temperature-independent 
masses are very costly and not yet available. Instead, one repeats 
the calculations for several unrealistically large masses and tries 
to extrapolate to zero mass. The perfect coincidence of the peaks in the
chiral and Polyakov loop susceptibilities is seen for all quark
masses \cite{KL94} and thus expected to survive in the chiral limit.

Both deconfinement and chiral symmetry restoration are phenomenologically
important. Deconfinement leads to the liberation of a large number of
gluons which can produce extra quark-antiquark pairs and drive the
system towards chemical equilibrium among quarks, antiquarks and gluons.
The melting of the dynamical quark masses above $\Tc$ makes the quarks 
lighter and lowers the quark-antiquark pair production threshold. This 
is particularly important for strange quarks whose constituent quark 
mass is much higher than the critical temperature while its current 
mass is comparable to $\Tc$. Above $\Tc$ thermal processes are therefore 
much more likely to equilibrate strange quark and antiquark abundances 
during the relatively short lifetime time of a heavy-ion collision.
 
The dissolution of massive hadrons into almost massles quarks and gluons 
at $\Tc$ leads to a very rapid rise of the energy density near the
deconfinement transition.
%
%%%%%%%%%%%%%%%%%%%%%%% Fig. 2 %%%%%%%%%%%%%%%%%%%%%%%%%%%%%%%%%%%%%%%%%%
\begin{figure}[ht]
\vspace*{-4mm}
\begin{center}
\includegraphics[width=7cm]{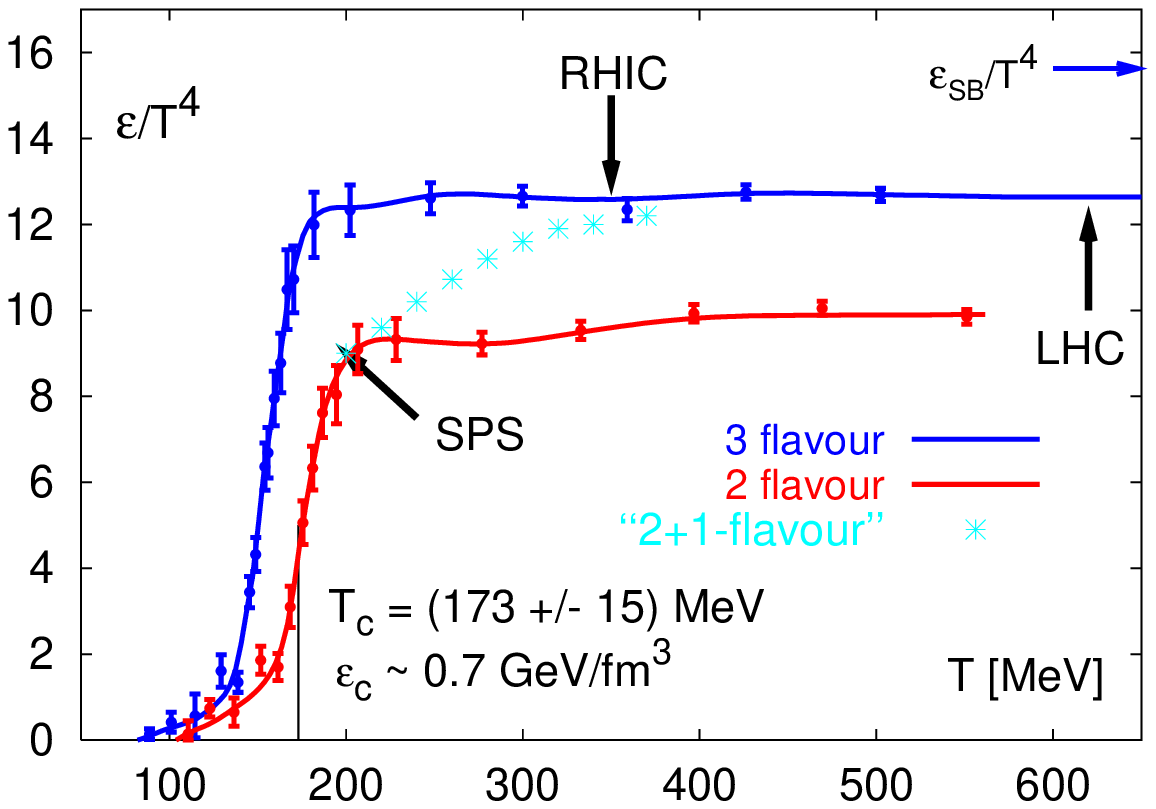} 
\includegraphics[width=8cm,height=6.2cm]{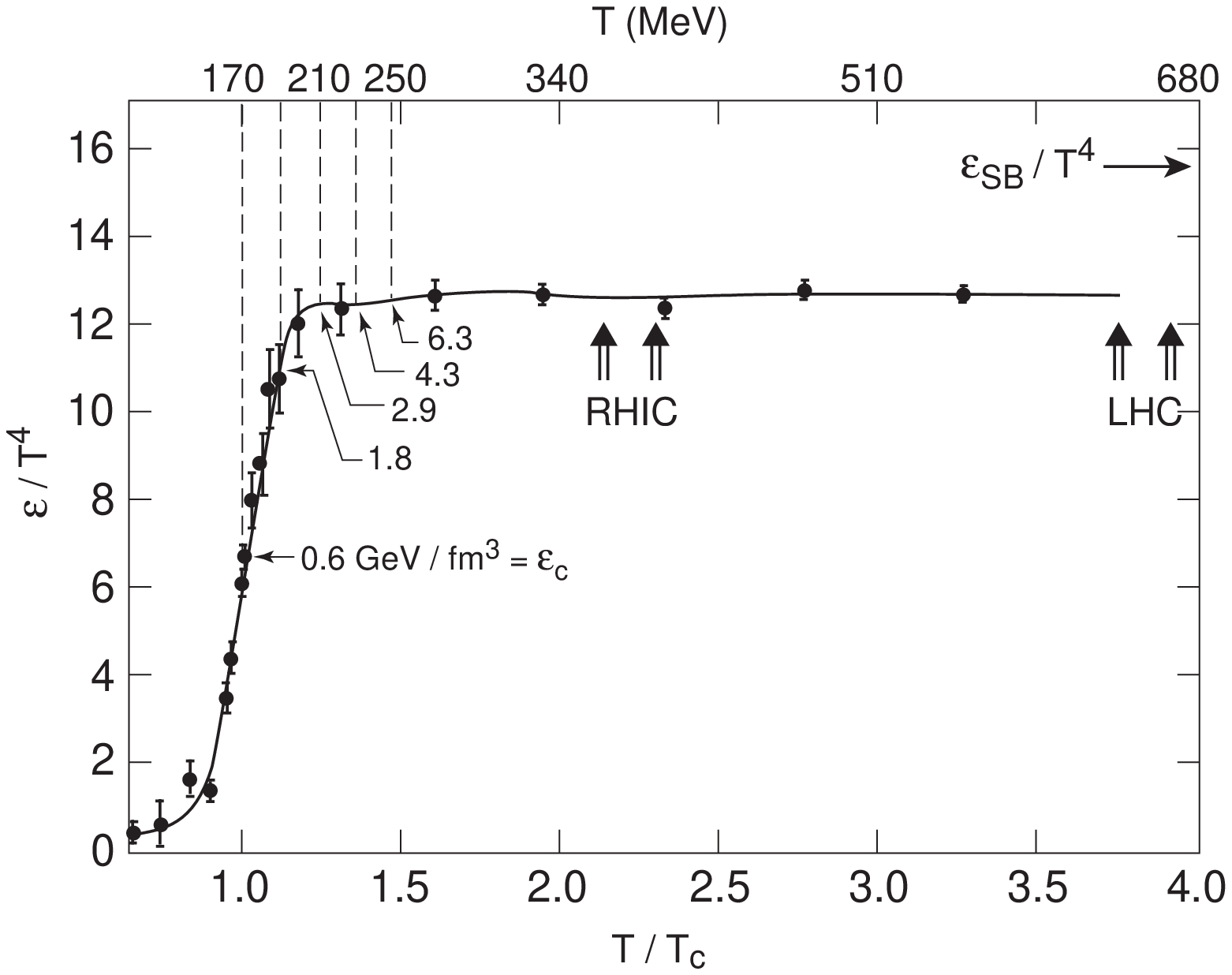} 
\vspace*{-10mm}
\caption{Energy density in units of $T^4$ for QCD with two and three 
dynamical quark flavors~\protect\cite{K00,KL03}. (The symbol $\varepsilon$ 
is here used for the energy density $e$.) The curves labelled ``2 flavour''
and ``3 flavor'' were calculated for two and three light quark flavors of 
mass $\frac{m_q}{T}\eq0.4$, respectively. ``2+1 flavour'' indicates a 
calculation for two light and one heavier strange quark flavor of 
$\frac{m_q}{T}\eq1$. In this case the ratio $\varepsilon/T^4$ interpolates 
between two light flavors at $T\protect\lapp\Tc$ and three light flavors 
at $T\protect\gapp 2\Tc$~\protect\cite{KL03}. For 2 and 2+1 flavors the 
critical temperature is $\Tc\eq173\pm15$\,MeV~\protect\cite{KL03}.
The right figure indicates, for the case of 3 light flavors, absolute 
values for the energy density $\varepsilon$ at several temperatures.
\label{F2}
}
\end{center}
\vspace*{-4mm}
\end{figure}
%%%%%%%%%%%%%%%%%%%%%%%%%%%%%%%%%%%%%%%%%%%%%%%%%%%%%%%%%%%%%%%%%%%%%%%%
% 
This is shown in Fig.~\ref{F2}. For a massless gas of quarks and gluons
the energy density is proportional to $T^4$. The proportionality
constant reflects the number of massless degrees of freedom, multiplied
by $\frac{\pi^2}{30}$ for bosons and by $\frac{7}{8}\frac{\pi^2}{30}$
for fermions. The arrow in the upper right of Fig.~\ref{F2} indicates 
this constant evaluated for 2(helicity)$\times$8(color)=16 gluon degrees 
of freedom plus 
2(spin)$\times$3(color)$\times$3(flavor)$\times2(q{+}\bar q)$=36
massless quark and antiquark degrees of freedom. We see that for $T<4\Tc$
the lattice data remain about 20\% below this Stefan-Boltzmann limit.
Near $\Tc$ the ratio $e/T^4$ drops rapidly by more than a factor 
10. This is due to hadronization. The much heavier hadrons are 
exponentially suppressed below $\Tc$, leading to a much smaller number
of equivalent massless degrees of freedom. (Due to the somewhat too
large quark masses in the simulations presented in Fig.~\ref{F2}
this effect is slightly exaggerated.) According to Fig.~\ref{F2}
the critical energy density for deconfinement is about 0.6--0.7\,GeV/fm$^3$.
The $\pm15$\,MeV uncertainty in $\Tc$ induces a $\pm40\%$ uncertainty
in the critical energy density which could be as large as 1\,GeV/fm$^3$ 
or as small as 500\,MeV/fm$^3$. I'll use $\ec\eq1$\,GeV/fm$^3$ as a 
ball-park number. It is easy to remember because it is the energy density
in the center of a proton and corresponds to one proton mass in a cube
of 1\,fm along each side. 
 
The right panel of Fig.~\ref{F2} also illustrates how painful the factor 
$T^4$ is: if we want to exceed the critical temperature by only 30\% in 
order to reach the upper edge of the transition region, we already need
an energy density $e\simeq 3.5$\,GeV/fm$^3$, five times the critical
value! And to reach $2\Tc$ requires $e\simeq 23$\,GeV/fm$^3$. The former
number is approximately the value one was able to reach at the CERN SPS
in Pb+Pb collisions at $\scm\eq17$\,GeV. The latter value has been 
achieved at the Relativistic Heavy Ion Collider (RHIC) in Au+Au
collisions at $\scm\eq130$\,GeV (see later). An estimate where the 
Large Hadron Collider (LHC) at CERN will take us in 2008 is
indicated near the right edge of the Figure. The corresponding
energy densities are of the order of 500\,GeV/fm$^3$!

%%%%%%%%%%%%%%%%%%%%%%%%%%%%%%%%%%%%%%%%%%%%%%%%%%%%%%%%%%%%%%%%%%%%%%%%%%%%%
\subsection{The QCD phase diagram and how to probe it with heavy-ion 
collisions}
\label{sec2c}
%%%%%%%%%%%%%%%%%%%%%%%%%%%%%%%%%%%%%%%%%%%%%%%%%%%%%%%%%%%%%%%%%%%%%%%%%%%%%

\begin{figure}[ht]
\vspace*{0mm}
\begin{center}
\includegraphics[bb=14 14 601 415,width=13cm]{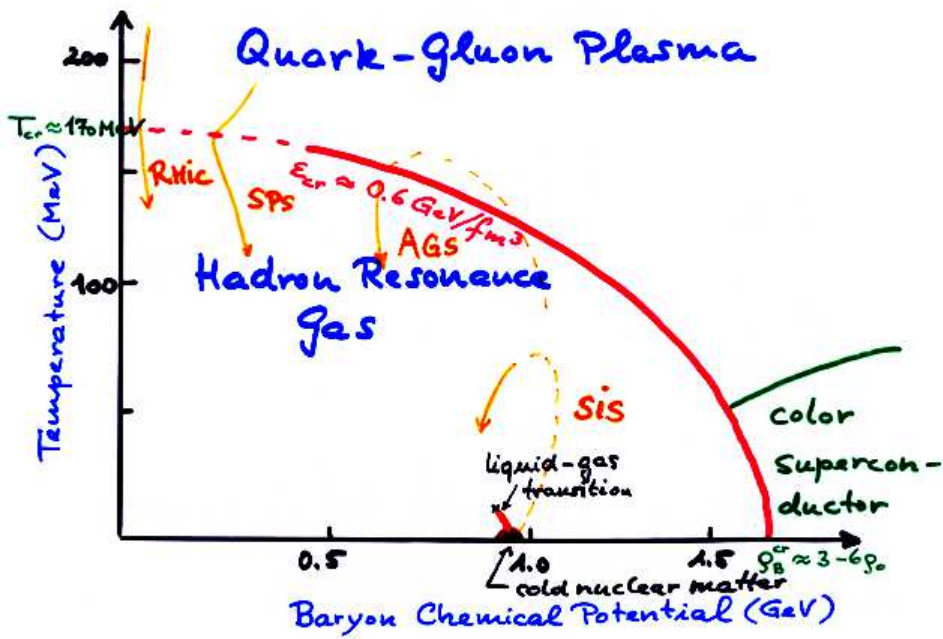} 
\vspace*{-6mm}
\caption{The QCD phase diagram.
\label{F3}
}
\end{center}
\vspace*{-4mm}
\end{figure}

Figure~\ref{F3} shows the phase diagram of strongly interacting matter
in the temperature vs. baryon chemical potential plane $(T,\mu_B)$.
Cold nuclear matter, such as in the interior of, say, a Pb nucleus,
sits a $T\eq0$ and $\mu_B\approx m_N=940$\,MeV. The short line emerging
from this blob indicates the nuclear liquid-gas phase transition,
with a critical endpoint at a temperature of about 7.5\,MeV. At higher
temperatures more and more hadron resonances are excited and we have a
hadron resonance gas. From QCD lattice calculations we know the phase 
structure along the temperature axis, with the deconfinement transition
from a hadron resonance gas to a quark-gluon plasma at $\Tc\approx170$\,MeV.
Lattice QCD tells us that even for realistically small up and down quark
masses the transition at $\mu_B\eq0$ is most likely not a sharp phase
transition but a rapid crossover as shown in Fig.~\ref{F2} \cite{KL03}.
Phenomenological models have long indicated \cite{Heinz:pm} that at 
non-zero $\mu_B$ the QGP and hadron gas are separated by a critical 
line of roughly constant critical energy density $\ec\simeq1$\,GeV/fm$^3$. 
Improved QCD inspired models \cite{R00} and recent lattice calculations 
at moderate non-zero baryon chemical potential \cite{FK02,ST03} indicate 
that the transition becomes first order at non-zero $\mu_B$ although the
precise value where this happens still remains to be determined by
performing calculations with more realistic smaller quark masses.
At low temperatures and asymptotically large baryon densities quarks
are also deconfined, although not in a quark-gluon plasma state but
rather in a color superconductor \cite{R00}. The superconducting state
is separated from the QGP by a first order transition at a critical
temperature estimated to be of order 30-50\,MeV \cite{R00}. Whether
it has a direct transition to normal nuclear matter as indicated in
my sketch or if other phases (e.g. involving pion or kaon conden\-sates) 
intervene is presently not known. There exists a rich spectrum of 
theoretical possibilities \cite{R00,R03}.

%
%%%%%%%%%%%%%%%%%%%%%%% Fig. 3 %%%%%%%%%%%%%%%%%%%%%%%%%%%%%%%%%%%%%%%%%%
%%%%%%%%%%%%%%%%%%%%%%%%%%%%%%%%%%%%%%%%%%%%%%%%%%%%%%%%%%%%%%%%%%%%%%%%
% 
Figure~\ref{F3} also indicates a few typical trajectories through the 
phase diagram which we can follow in relativistic heavy-ion collisions.
The collisions start at the cold nuclear matter point in the phase
diagram, go through an early non-equilibrium stage which can not be
mapped onto the phase diagram (indicated by dotted lines), and then
reappear in the phase diagram after having thermalized at some high
temperature. Unfortunately, one cannot use heavy-ion collisions to 
compress nuclear matter without producing a lot of entropy and therefore 
also heating it; hence it seems impossible to probe with them the
color superconducting phase of strongly interacting matter.
 
At the SIS (heavy-ion synchrotron) at the GSI facility 
in Darmstadt, with beam energies of order 1-2 GeV/nucleon, the center
of mass collision energy is not large enough to reach the deconfined
state of QCD. At the Brookhaven AGS (beam energies of about 
10\,GeV/nucleon) we may have straddled the transition line. First
clear indications that we had crossed into a new state of matter
beyond the hadron resonance gas came from Pb+Pb collisions at 
160\,GeV/nucleon beam energy \cite{HJ00}. Au+Au collisions at RHIC 
thermalize at an initial temperature $T_0\approx 2\,\Tc$ (see later).
By collective expansion the collision fireball cools down and passes
through the hadronization phase transition from above. In this sense
heavy-ion collisions do not probe the {\em de-confinement}, but rather the
{\em confinement} phase transition, just as the early universe. As the
center of mass collision energy increases, the colliding nuclei become
more and more transparent, meaning that a decreasing fraction of the
beam energy and of the incoming baryons get stopped in the center
of mass system. The midrapidity collision fireball therefore contains
fewer and fewer of the incoming net baryons, becoming more and more
baryon-antibaryon symmetric (i.e. $\mu_B$ decreases). At RHIC
the entropy per baryon ratio is between 200 and 300, depending on
collision energy. At the LHC one expects $S/A$ to go up to several 
thousand. While still far from the early universe ratio of $10^9$,
this is, for all practical purposes, ``baryon-free'' ($\mu_B\eq0$) QCD 
matter.  

%%%%%%%%%%%%%%%%%%%%%%%%%%%%%%%%%%%%%%%%%%%%%%%%%%%%%%%%%%%%%%%%%%%%%%%%%%%%%
\section{THE DIFFERENT STAGES OF A HEAVY-ION COLLISION -- THE LITTLE BANG}
\label{sec3}
%%%%%%%%%%%%%%%%%%%%%%%%%%%%%%%%%%%%%%%%%%%%%%%%%%%%%%%%%%%%%%%%%%%%%%%%%%%%%
%%%%%%%%%%%%%%%%%%%%%%%%%%%%%%%%%%%%%%%%%%%%%%%%%%%%%%%%%%%%%%%%%%%%%%%%%%%%%
\subsection{The production of hard probes during the early collision stage}
\label{sec3a}
%%%%%%%%%%%%%%%%%%%%%%%%%%%%%%%%%%%%%%%%%%%%%%%%%%%%%%%%%%%%%%%%%%%%%%%%%%%%%

\begin{figure}[ht]
\begin{center}
\includegraphics[bb=14 34 601 840,width=11cm,height=15cm]{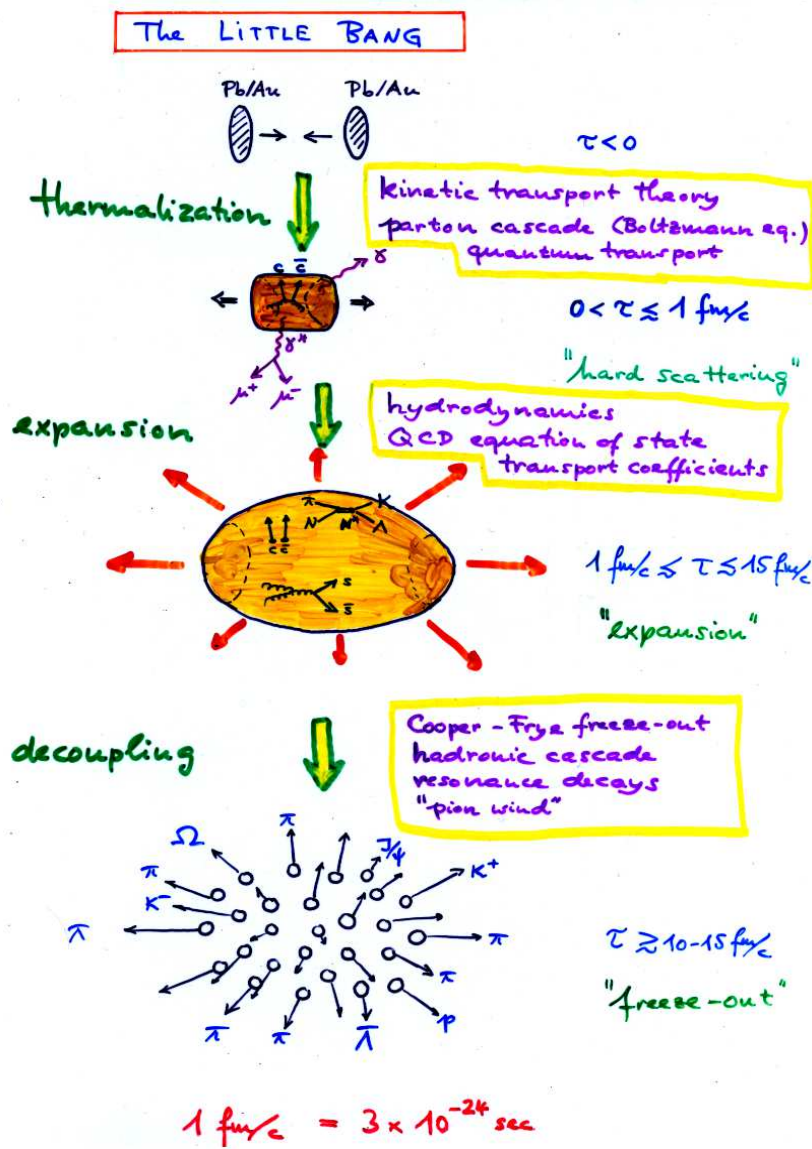} 
\caption{Stages of a relativistic heavy-ion collision and relevant 
theoretical concepts.
\label{F4}
}
\end{center}
\vspace*{-5mm}
\end{figure}

Figure~\ref{F4} summarizes the key stages of relativistic heavy-ion
collisions: thermalization, expansion, and decoupling. In the very early
collision stages, before the bulk of the quanta, which are created from 
the fraction of the beam energy lost in the collision, have time to 
rescatter, ``hard'' particles with either a large mass or large transverse 
momenta $\pt{\,\gg\,}1$\,GeV/$c$ are created. Their creation involves large 
momentum transfers $Q^2 \sim \pt^2 \gg 1$\,GeV$^2$, therefore their
production can be calculated in perturbative QCD, using factorization 
theorems, from the nuclear structure functions. According to the 
uncertainty relation hard particle {\em production} happens on a
time scale $\tau_{\rm form}\simeq 1/\sqrt{Q^2}$; for a 2\,GeV particle 
this means $\tau_{\rm form}\simeq 0.1$\,fm/$c$.

At SPS energies and below, hard particles are essentially only produced 
in the primary collisions between the projectile and target nucleons; the 
bulk of the produced particles have transverse momenta below 2\,GeV/$c$ 
and are too soft to produce hard particles via secondary collisions. 
High-$\pt$ jets from the fragmentation of hard partons have practically
relevant cross sections only at RHIC energies and above. Once produced, 
one can use them to probe the soft matter created by the bulk of soft 
particles \cite{GVWZ03}: In a central collision between two Pb or Au 
nuclei the nuclear reaction zone has a transverse diameter of about 
12\,fm, so a hard particle created near the edge and moving in the 
``wrong'' direction (namely straight inward) needs 12\,fm/$c$ before 
it emerges on the other side. During this time the soft matter
thermalizes, expands, cools down and almost reaches decoupling, all of
which is probed by the fast particle on its way to the other side.
It does so by scattering off the evolving medium and losing energy
which can be measured. The energy loss is proportional to the density
of the medium times the scattering cross section between the probe and
the medium constituents, integrated along the probe's trajectory.

%
%%%%%%%%%%%%%%%%%%%%%%% Fig. 4 %%%%%%%%%%%%%%%%%%%%%%%%%%%%%%%%%%%%%%%%%%
%%%%%%%%%%%%%%%%%%%%%%%%%%%%%%%%%%%%%%%%%%%%%%%%%%%%%%%%%%%%%%%%%%%%%%%%
% 

At SPS energies the ``hardest'' particles which can be produced and well
identified by their distinctive decay pattern are $c\bar c$ pairs. If
produced in a proton-proton collision where either no hot soft matter
is created or the charmed quarks leave the region in which soft particles
are formed before that happens, these charmed quarks and antiquarks form 
either a bound charmonium state ($J/\psi$, $\psi'$, or $\chi$, 
``hidden charm'' production) or they find light quark partners
to hadronize into ``open charm'' states ($D$ and $\bar D$ mesons or 
charmed baryons). The corresponding branching ratios are well-known --
the hidden charm states form a very small fraction (less than 1\%) of 
all final states. If the $c\bar c$ pair is created in a heavy-ion 
collision, something similar as with the jets discussed above happens:
the two heavy quarks have to travel through a dense medium of soft
particles which interferes with their intention to hadronize and 
modifies their branching ratios into open and hidden charm states.
In particular, if the soft medium thermalizes into a quark-gluon plasma,
the color interaction between the $c$ and $\bar c$ is Debye screened
by the colored quarks and gluons in the plasma, thereby prohibiting 
their normal binding into one of the charmonium states. This should 
cause ``$J/\psi$ suppression'' \cite{MS86}. 

Other probes of the early collision stage are direct photons \cite{R92},
either real or virtual, in which case they materialize a lepton-antilepton
($e^+e^-$ or $\mu^+\mu^-$) pairs, generally known as ``dileptons''. 
Such photons are produced by the electric charges in the medium (i.e. by
quarks and antiquarks during the early collision stage). Their
production cross section is proportional to the square of the Sommerfeld
fine structure constant $\alpha\eq\frac{1}{137}$ and thus small, but since
they also reinteract only with this small electromagnetic cross section
their mean free path even in a very dense quark-gluon plasma is of
the order of 50\,000\,fm and thus much larger than any conceivable 
heavy-ion fireball. In contrast to all hadronic probes, they thus 
escape from the collision zone without reinteraction and carry pristine 
information about the momentum distributions of their parent quarks 
and antiquarks into the detector. Real and virtual photons are emitted 
throughout the collision and expansion stage, and their measured 
spectrum thus integrates over the expansion history of the collision 
fireball. However, their production rate from a thermal system scales 
with a high power of the temperature (${\sim\,}T^4$ in a static heat bath 
but ${\,\sim\,}T^6$ once the effects of expansion are taken into account 
\cite{R92}) and is thus strongly biased towards the early
collision stages. Unfortunately, the directly emitted photons and 
dileptons must be dug out from a huge background of indirect photons
and and a large combinatorical background of uncorrelated lepton pairs 
resulting from electromagnetic and weak decays of hadrons after 
hadronic freeze-out. This renders the measurement of these clean 
electromagnetic signals difficult.

%%%%%%%%%%%%%%%%%%%%%%%%%%%%%%%%%%%%%%%%%%%%%%%%%%%%%%%%%%%%%%%%%%%%%%%%%%%%%
\subsection{Thermalization and expansion}
\label{sec3b}
%%%%%%%%%%%%%%%%%%%%%%%%%%%%%%%%%%%%%%%%%%%%%%%%%%%%%%%%%%%%%%%%%%%%%%%%%%%%%

The key difference between elementary particle and nucleus-nucleus 
collisions is that the quanta created in the primary collisions between
the incoming nucleons can't right away escape into the surrounding vacuum, 
but rescatter of each other. In this way they create a form of dense,
strongly interacting matter which, when it thermalizes quickly enough
and at sufficiently large energy density, is a quark-gluon plasma. This
is why with heavy-ion collisions we have a chance to recreate the matter
in the very early universe whereas with high energy collisions between
leptons or single hadrons we don't.

The produced partons rescatter both elastically and inelastically. Both
types of collisions lead to equipartitioning of the deposited energy,
but only the inelastic collisions change the relative abundances of
gluons, light and strange quarks. (To also change the abundance of the 
heavier charm quarks ($m_c\simeq 1200$\,MeV) requires secondary collisions
of sufficient energy which do not happen at the SPS but may play a role
at RHIC and above.) From the phenomenology of $pp$ collisions it is
known \cite{Hag65,BH97} that the produced hadron abundances are 
distributed statistically, but that strange hadrons are systematically 
suppressed (probably because strange quarks are not present in the
initial state and their large constituent mass of ${\,\simeq\,}450$\,MeV
makes them hard to create from the vacuum as the produced partons
hadronize). In a heavy-ion collision, if the reaction zone thermalizes
at energy density ${\,>\,}\ec$ such that gluons are deconfined and
chiral symmetry is restored, strange quarks are much lighter 
($m_s{\,\simeq\,}150$\,MeV) and can be relatively easily created
by secondary collisions among the many gluons, leading to chemical
equilibration between light and strange quarks \cite{RM82}. The observed
strangeness suppression in $pp$ collisions should thus be reduced or
absent in relativistic heavy-ion collisions \cite{R82}.

A thermalized system has thermal pressure which, when acting against 
the surrounding vacuum, leads to collective (hydrodynamic) expansion of
the collision fireball. As a consequence, the fireball cools and
its energy density decreases. When the latter reaches 
$\ec{\,\simeq\,}1$\,GeV/fm$^3$, the partons convert to hadrons. During
this phase transition the entropy density drops steeply over a small
temperature interval (similar to the energy density shown in Fig.~\ref{F2}). 
Since the total entropy cannot decrease this implies that the fireball
volume must increase by a large factor while the temperature remains
approximately constant. The growth of the fireball volume takes time, so
the fireball ends up spending significant time near $\Tc$. Furthermore,
while the matter hadronizes its speed of sound $c_s\eq\sqrt{\P p/\P e}$
is small \cite{KL03}, causing inefficient acceleration so that the 
collective flow does not increase during this period. This may be
visible in the direct photon spectrum: an inverse Laplace transform 
should show a particularly strong weight at the transition temperature, 
blueshifted by the prevailing radial flow as the system crosses $\Tc$
\cite{Sch94}. The softness of the equation of state near the phase
transition should also manifest itself in the center-of-mass energy
dependence of the collective flow, as discussed later.

%%%%%%%%%%%%%%%%%%%%%%%%%%%%%%%%%%%%%%%%%%%%%%%%%%%%%%%%%%%%%%%%%%%%%%%%%%%%%
\subsection{Hadronic freeze-out and post-freeze-out decays}
\label{sec3c}
%%%%%%%%%%%%%%%%%%%%%%%%%%%%%%%%%%%%%%%%%%%%%%%%%%%%%%%%%%%%%%%%%%%%%%%%%%%%%

After hadronization of the fireball, the hadrons keep rescattering
with each other for a while, continuing to build up expansion flow,
until the matter becomes so dilute that the average distance between 
hadrons exceeds the range of the strong interactions. At this points all
scattering stops and the hadrons decouple (``freeze out''). Actually,
their {\em abundances} already freeze out earlier when the rates for inelastic
processes, in which the hadrons change their identity, become too small
to keep up with the expansion. Since the corresponding inelastic cross
sections are only a small fraction of the total cross section, inelastic
processes stop long before the elastic ones, leading to earlier freeze-out
for the hadron abundances than for their momenta: {\em chemical
freeze-out} precedes {\em thermal} or {\em kinetic freeze-out}. What I 
call ``elastic'' includes resonant processes such as 
$\pi{+}N\to\Delta\to\pi{+}N$ where two hadrons form a short-lived 
resonance which subsequently decays back into the same particles 
(possible with different electric charge assignments). Such processes 
don't change the finally observed chemical composition, but they 
contribute to the thermalization of the momenta and have large ``resonant''
cross sections. Since most of the hadrons in a relativistic heavy-ion
fireball are pions (since they are so light), resonances with pions
are very efficient in keeping the system in thermal equilibrium (even 
after chemical equilibrium has been broken). For scattering among pions
the $\rho$ resonance plays a large role while kaons and (anti)baryons
couple to the pion fluid via the $K^*$, $\Delta$ and $Y^*$ resonances.
Due to the particularly large $\Delta$ and $Y^*$ resonance cross sections,
even at RHIC and LHC (where the net baryon density is small) 
baryon-antibaryon pairs play an important role as part of the ``glue'' 
that keeps the expanding pion fluid thermalized well below the chemical
freeze-out point.  

At kinetic freeze-out all hadrons, including the then present unstable
resonances, have an approximately exponential transverse momentum 
spectrum reflecting the temperature of the fireball at that point, 
blueshifted by the average transverse collective flow. The unstable
resonances decay, however, producing daughter particles with, on average, 
smaller transverse momenta. The experimentally measured spectra off stable
hadrons cannot be understood without adding these decay products to the 
originally emitted spectra. Since most resonances decay by emitting a pion,
this effect is particularly important for the pion spectrum which at
low $\pt$ are completely dominated by decay products. This seriously 
affects the slope of their spectrum out to $\pt{\,\simeq\,}700$\,MeV 
\cite{SKH91,SSH93}, making it steeper than the blueshifted thermal
spectrum of the directly emitted hadrons.

%%%%%%%%%%%%%%%%%%%%%%%%%%%%%%%%%%%%%%%%%%%%%%%%%%%%%%%%%%%%%%%%%%%%%%%%%%%%%
\subsection{Theoretical tools}
\label{sec3d}
%%%%%%%%%%%%%%%%%%%%%%%%%%%%%%%%%%%%%%%%%%%%%%%%%%%%%%%%%%%%%%%%%%%%%%%%%%%%%

The idea that the parton production process factorizes into a perturbatively
calculable ``hard'' QCD cross section and a non-perturbative, experimentally
determined nuclear parton structure function (or parton distribution function)
can be used to describe the initial production of hard partons, with
$p_{\rm T}{\,>\,}p_0$ where $p_0\gapp1{-}2$\,GeV describes the lower 
applicability 
limit of this type of approach \cite{EKL88,EKVT99}. The production of soft 
partons is usually non-perturbative and requires phenomenological models, 
such as string models \cite{String}, for their description. At very high
collision energies particle production at midrapidity (i.e. particles
with small longitudinal momenta in the center-of-momentum frame) probes
the nuclear structure functions at small $x$ where $x$ is the fraction of
the beam momentum carried by the partons whose collision produces the
secondary partons. At small $x$ the gluon distribution function becomes
very big and gluons begin to fill the transverse area of the colliding 
nuclei densely \cite{GLR83,MQ86}. The gluons start to recombine, leading
to gluon saturation, and low $\pT$ gluon modes are occupied by a 
macroscopic number of gluons $\sim 1/\alpha_s$ where the effective 
strong coupling $\alpha_s$ is small due to the high density of gluons
\cite{MV93}. As a result, the initial gluons can be effectively 
described by a classical gluon field \cite{MV93,KM98} in which the 
coupling is weak but nonlinear density effects are important (this
state has become known as the ``color glass condensate'' \cite{ILM02}).
The production of soft secondary gluons can be understood as the 
liberation of these gluons by breaking the coherence of their 
multiparticle wavefunction \cite{KM98,Kov00}. The collision energies
where these modern ideas become applicable lie probably beyond the
RHIC range, but this is an exciting and active field of ongoing research
which may come to fruition at the LHC.

After the initial parton production or ``liberation'' process one must
describe the rescattering and thermalization of the produced quanta.
For not too dense systems this can be done with classical kinetic 
transport theory (relativistic Boltzmann equation), also known
as parton cascade \cite{G94,MG00}. In the early collision stages at
RHIC and LHC energies the densities are probably too high for this
to remain a reliable approach, and one must switch to quantum transport
theory \cite{EH89}. This formalism has not yet been developed very
well for practical applications, and much interesting work is presently
going on in this direction (e.g. \cite{BMSS00,MS02}). None of these
approaches, which usually invoke perturbative QCD arguments to describe
the microscopic scattering processes, has so far been able to produce
parton thermalization time scales at RHIC which are shorter than about
5\,fm/$c$ \cite{BMSS00}. This is too long for heavy-ion collisions
whose typical expansion time scale (Hubble time) is only a couple
of fm/$c$. As I will show later there is strong phenomenological
evidence that thermalization must happen much more quickly. This
presents an interesting challenge for theory and indicates the importance
of strong non-perturbative effects in the early collision stages and
the QGP. 

Once local thermal equilibrium has been reached, the further evolution can 
be described hydrodynamically. The simplest version of such an approach
is ideal fluid dynamics which will be discussed in more detail later.
The fireball can be described as an ideal fluid if the microscopic
scattering time scale is much shorter than any macroscopic time scale
associated with the fireball evolution. If this is not the case, one
should include non-ideal effects such as shear, diffusion and heat
conduction. This requires the calculation of the corresponding transport
coefficients in a partonic system close to thermal equilibrium. Much
work is going on in this direction (see e.g. \cite{J95,AMY03,AM03}), but
systematic approaches based on perturbatively resummed thermal QCD 
\cite{J95,AMY03,AM03} give phenomenologically unacceptable large values
for these transport coefficients, again indicating the importance of
non-perturbative effects.

The hydrodynamic equations require knowledge of the equation of state,
i.e. the relationship $p(e,n_B)$ between the pressure, energy density and 
baryon density. For small $n_B$ this is known from lattice QCD; for large
$n_B$ one extrapolates the lattice results with phenomenological models 
(see \cite{R03} for a recent review). Hydrodynamics is the ideal 
language for relating observed collective flow phenomena to the
equation of state. Since it only requires the equation of state but no 
detailed knowledge of the microscopic collision dynamics, it allows
for an easy description of the hadronization phase transition without
any need for a microscopic understanding of how hadrons form from
quarks and gluons. Of course, the underlying assumption is that all
these microscopic processes happen so fast that the system never strays
appreciably from a local thermal equilibrium.

After hadronization the system continues to expand and dilute until
the average distance between hadrons becomes larger than the range of
the strong interaction. At this point the hadrons decouple, i.e. their
momenta stop changing until they are recorded by the detector. One
can implement this decoupling or ``kinetic freeze-out'' in different ways,
either by truncating the hydrodynamic phase abruptly with the 
Cooper-Frye algorithm (see below) or by switching back from hydrodynamics
to a (this time hadronic) cascade \cite{BDBBZSG99,TLS01,TLS02} in which
decoupling happens automatically and selfconsistently. In the Cooper-Frye
algorithm one also must take into account that in this approach all
kinds of hadrons decouple simultaneously, and that unstable hadron 
resonances decay subsequently by strong (and in some situations also by
weak) interactions before their stable daughter products reach the 
detector. To compute the measured spectra one must therefore fold the
initially emitted Cooper-Frye hadron spectra with these decays 
\cite{SKH91}. Some studies, such as two-particle momentum correlations,
also require the computation of long-range final state interaction
effects, such as the Coulomb repulsion/attraction between charged 
hadrons which continues long after their strong interactions with 
each other have ceased.

%%%%%%%%%%%%%%%%%%%%%%%%%%%%%%%%%%%%%%%%%%%%%%%%%%%%%%%%%%%%%%%%%%%%%%%%%%%%%
\subsection{Strategies for reconstructing the Little Bang}
\label{sec3e}
%%%%%%%%%%%%%%%%%%%%%%%%%%%%%%%%%%%%%%%%%%%%%%%%%%%%%%%%%%%%%%%%%%%%%%%%%%%%%

The bulk (over 99\%) of the particles produced in heavy-ion collisions are
hadrons. These are strongly interacting particles which cannot decouple
from the fireball before the system is so dilute that strong interactions
cease. Their observed momenta thus provide a snapshot of the kinetic
decoupling stage (``thermal freeze-out''). In this sense hadrons are the
Little Bang analogue of the cosmic microwave background in the Big Bang. 
As discussed in Sec.~\ref{sec3c}, hadron abundances freeze-out earlier. As
we will see later, this ``chemical freeze-out'' happens right after
hadronization, i.e. the finally observed hadron abundances are generated
by the hadronization process itself. This {\em primordial hadrosynthesis}
is the Little Bang analogue of primordial nucleosynthesis in the Big Bang.
As the hadrons decouple, they carry not only thermal information
about the prevalent temperature at chemical resp. thermal freeze-out,
but in the momentum spectra this information is folded with (i.e. 
blueshifted by) the collective expansion flow, just as the temperature 
of the cosmic microwave radiation is redshifted by the cosmological 
expansion. The hydrodynamic expansion flow is the Little Bang analogue
of the cosmic Hubble expansion in the Big Bang. (Of course, the
origin of the expansion is entirely different in the two cases: in
heavy-ion collisions it is generated hydrodynamically by pressure gradients
whereas in the Big Bang it reflects an initial condition, modified 
over billions of years by the effects of the gravitational interaction.
But the velocity {\em profiles} turn out to be surprisingly similar, as 
we will see!)

We reconstruct the Little Bang from these ``late'' hadronic observables
very much like we reconstructed the Big Bang from the 3 pillars of
cosmology, Hubble expansion, CMB, and primordial nuclear abundances.
The hadronic observables are abundant and can be measured with high
statistical accuracy. As I will show, their theoretical analysis allows
to separate the thermal from the collective motion. The collective flow
provides a memory of the pressure and other thermodynamic conditions
during the earlier collision stages. In fact, we will see that certain
anisotropies in the flow patterns seen in non-central heavy-ion collisions
(``elliptic flow'') provide a unique window into the very early collision
stages and are no longer changed after about 5\,fm/$c$ after initial
impact. As you will see, a (in my opinion) watertight proof for 
thermalization during this early stage, at a time 
$\tau_{\rm therm}{\,<\,}1$\,fm/$c$ and at prevalent energy densities
which exceed the critical value $\ec$ for deconfinement by at least an 
order of magnitude, can be based on the accurately measured elliptic 
flow of the final state hadrons. (This is a bit analogous to the 
indelible imprint that cosmic inflation has left on the density 
fluctuations in the universe, which can be accurately measured through 
the anisotropy of the cosmic microwave radiation which only decoupled 
300\,000 years later.) 

The reconstruction of the global space-time evolution of the reaction zone
from the finally observed soft hadrons is the cornerstone of the program.
The reconstructed dynamical picture will be the basis on which other
rarer observables, in particular the ``deep'' or ``hard'' probes, will 
be interpreted. For example, jet quenching and heavy quarkonium suppression 
cannot be quantitatively interpreted without knowledge of the fireball 
density and its space-time evolution, and direct photon and dilepton 
spectra cannot be properly understood without a relatively accurate 
idea about the transverse flow patterns at the time of photon emission. 
Still, these probes will in the end be the only {\em direct access} we 
will ever have to the temperature and energy density at the beginning 
of the expansion when the system was a quark-gluon plasma, and although
they cannot be fully interpreted and exploited without the later emitted 
soft particles they are still an indispensable part of the picture. This 
illustrates the network-like interdependence between soft and hard 
observables in their role for elucidating heavy-ion collision and 
quark-gluon plasma dynamics.

%%%%%%%%%%%%%%%%%%%%%%%%%%%%%%%%%%%%%%%%%%%%%%%%%%%%%%%%%%%%%%%%%%%%%%%%%%%%%%
\section{THE LITTLE BANG -- COLLECTIVE EXPLOSION OF A THERMALIZED SYSTEM}
\label{sec4}
%%%%%%%%%%%%%%%%%%%%%%%%%%%%%%%%%%%%%%%%%%%%%%%%%%%%%%%%%%%%%%%%%%%%%%%%%%%%%%

An unavoidable consequence of quark-gluon plasma formation in heavy-ion
collisions is collective flow. Since a quark-gluon plasma is by definition
an (approximately) thermalized system of quarks and gluons, it has  
thermal pressure, and the pressure gradients with respect to the 
surrounding vacuum cause the quark-gluon plasma to explode. Absence
of collective flow would indicate absence of pressure and imply absence
of a hot thermalized system and, {\em a fortiori}, of a quark-gluon plasma.

In this chapter we'll therefore do two things: (i) learn how to analyze 
the measured particle spectra for the presence of collective flow and how
to separate it from random thermal motion, and (ii) compute the expected
collective flow patterns from reasonable initial conditions using a
hydrodynamic model. By comparing with experiments we can fine-tune the
initial conditions, learn about the equation of state of the hot expanding
matter and, most importantly, find out about when and at which energy 
densities the thermal pressure builds up and begins to drive the collective
expansion.  

%%%%%%%%%%%%%%%%%%%%%%%%%%%%%%%%%%%%%%%%%%%%%%%%%%%%%%%%%%%%%%%%%%%%%%%%%%%%%
\subsection{Radial flow}
\label{sec4a}
%%%%%%%%%%%%%%%%%%%%%%%%%%%%%%%%%%%%%%%%%%%%%%%%%%%%%%%%%%%%%%%%%%%%%%%%%%%%%%
%%%%%%%%%%%%%%%%%%%%%%%%%%%%%%%%%%%%%%%%%%%%%%%%%%%%%%%%%%%%%%%%%%%%%%%%%%%%
\subsubsection{Flow defined}
\label{sec4a1}
%%%%%%%%%%%%%%%%%%%%%%%%%%%%%%%%%%%%%%%%%%%%%%%%%%%%%%%%%%%%%%%%%%%%%%%%%%%%%

Consider a nuclear fireball undergoing collective expansion. Collective
flow is defined by the following operational procedure: at any space-time
point $x$ in the fireball, we consider an infinitesimal volume element
centered at that point and add up all the 4-momenta of the quanta in
it. The total 3-momentum $\bP$ obtained in this way, divided by the 
associated total energy $P^0$, defines the average ``flow'' velocity 
$\bv(x)$ of the matter at point $x$ through the relation 
$\bP{/}P^0\eq\bv$. Collective flow thus describes a correlation 
between the average momentum of the particles with their space-time 
position, a so-called {\em $x$-$p$-correlation}. With $\bv(x)$ we can 
associate a normalized 4-velocity $u^\mu\eq\gamma(1,\bv)$ 
where $\gamma(x)\eq1/\sqrt{1{-}\bv^2(x)}$ is the corresponding Lorentz 
dilation factor and $u{\cdot}u{\eq}u^\mu u_\mu\eq1$ (in units where
the speed of light $c\eq1$).

I separate the flow velocity $\bv(x)$ into its components along the beam
direction (``longitudinal flow'' $\vL$) and in the plane perpendicular to
the beam (``transverse plane'') which I call ``transverse flow'' $\bvp$.
The magnitude $\vperp$ may depend on the azimuthal angle around the beam 
direction, i.e. on the angle between $\bvp$ and the impact parameter 
$\bb$ of the collision. In this case we call the transverse flow 
``anisotropic''. Its azimuthal average we call {\em radial flow}.

%%%%%%%%%%%%%%%%%%%%%%%%%%%%%%%%%%%%%%%%%%%%%%%%%%%%%%%%%%%%%%%%%%%%%%%%%%%%%
\subsubsection{Local thermodynamic equilibrium}
\label{sec4a2}
%%%%%%%%%%%%%%%%%%%%%%%%%%%%%%%%%%%%%%%%%%%%%%%%%%%%%%%%%%%%%%%%%%%%%%%%%%%%%

If the fireball is in local thermodynamic equilibrium, we can not only
define a local flow 4-velocity $u^\mu(x)$, but also a local temperature
$T(x)$ and, for each particle species $i$, a chemical potential $\mu_i(x)$
which controls its particle density at point $x$. In this case the
phase-space distribution of particles of type $i$ is given by the 
Lorentz covariant local equilibrium distribution 
\bea{eq}
   f_{i,\equ}(x,p) = \frac{g_i}{e^{[p\cdot u(x)-\mu_i(x)]/T(x)}\pm1}
   = g_i \sum_{n=1}^\infty (\mp)^{n+1} \,e^{-n[p\cdot u(x)-\mu_i(x)]/T(x)}\,.
\eea
Here $g_i$ is a spin-isospin-color-flavor-etc. degeneracy factor which
counts all particles with the same mass $m_i$ and chemical potential
$\mu_i$. The factor $p\cdot u(x)$ in the exponent is the energy of the
particle in the local rest frame (local heat bath frame), boosted to
the observer frame by the flow 4-velocity $u^\mu(x)$ of the fluid cell
at point $x$ ($p\cdot u{\,\to\,}p^0\eq{E}$ when $u^\mu{\,\to\,}(1,\bm{0})$). 
The $\pm1$ in the denominator accounts for the proper quantum statistics
of particle species $i$ [upper (lower) sign for fermions (bosons)].
The Boltzmann approximation corresponds to keeping only the first term 
in the sum over $n$ in the last expression. In our applications this is
an excellent analytical approximation for all hadrons except for the pion.
Pions and quarks and gluons are too light, $m_i\lapp{T}$, and one must
use the proper quantum statistical distributions. 

%%%%%%%%%%%%%%%%%%%%%%%%%%%%%%%%%%%%%%%%%%%%%%%%%%%%%%%%%%%%%%%%%%%%%%%%%%%%%
\subsubsection{Rapidity coordinates}
\label{sec4a3}
%%%%%%%%%%%%%%%%%%%%%%%%%%%%%%%%%%%%%%%%%%%%%%%%%%%%%%%%%%%%%%%%%%%%%%%%%%%%%

At relativistic energies it is convenient to parametrize the longitudinal 
flow velocities and momenta in terms of rapidities (for any velocity $v$ 
the associated rapidity is $\eta\eq\half\ln\frac{1{+}v}{1{-}v}$ or
$v\eq\tanh\eta$):
\bea{rap}
   \eta_{\rm L} = \half\ln\frac{1{+}\vL}{1{-}\vL} \, ,\qquad\qquad
   y = \half\ln\frac{1{+}\frac{\pL}{E}}{1{-}\frac{\pL}{E}}
     = \half\ln\frac{E{+}\pL}{E{-}\pL} \,.
\eea
As $\vL{\,\to\,}1$, $\etaL{\,\to\,}\infty$. Rapidities have the advantage 
over longitudinal velocities that they are additive under longitudinal
boosts: a fluid cell with flow rapidity $\eta_{\rm L}$ in a given inertial
frame has rapidity $\eta'_{\rm L}\eq\etaL+\Delta\eta$ in another inertial 
frame which moves relative to the first frame with rapidity $\Delta\eta$ 
in the $-z$ direction. The flow 4-vector 
$(u^0,\bu)\eq(u^0,\bu_\perp,u_{\rm L})$ is then parametrized as
\bea{flow}
   u^\mu=\gamma_\perp\Bigl(\cosh\etaL,\,v_x,\,v_y,\,\sinh\etaL\Bigr)
    \qquad {\rm with}\qquad
    \gamma_\perp=\frac{1}{\sqrt{1{-}\vperp^2}}
                =\frac{1}{\sqrt{1{-}v_x^2{-}v_y^2}}
\eea
where $\bvp\eq(v_x,v_y)$ is the transverse flow velocity.
Similarly the 4-momentum $p^\mu\eq(E,\bpp,\pL)$ is written as
\bea{p}
    p^\mu = \Bigl(\mperp\cosh y,\,p_x,\,p_y,\,\mperp\sinh y\Bigr)
    \qquad {\rm with}\qquad
    \mperp=\sqrt{m^2{+}p_\perp^2}=\sqrt{m^2{+}p_x^2{+}p_y^2}
\eea
which obviously satisfies the mass-shell constraint 
$p^2\eq{p^\mu}p_\mu\eq{m^2}$.
[We use the notations $\pperp{\eq}p_{\rm T}$ interchangeably for the
transverse momentum, and similarly $m_{\rm T}{\eq}m_\perp$ for the 
``transverse mass'' -- in the literature one often also finds the 
notations $p_{\rm t}$ and $m_{\rm t}$.] The scalar product $p\cdot u(x)$
in the exponent of the Boltzmann factor then becomes
\bea{pdotu}
   p\cdot u = \gamma_\perp \Bigl(\mperp\cosh(y{-}\etaL)-\bvp\cdot\bpp\Bigr).
\eea

%%%%%%%%%%%%%%%%%%%%%%%%%%%%%%%%%%%%%%%%%%%%%%%%%%%%%%%%%%%%%%%%%%%%%%%%%%%%%
\subsubsection{Longitudinal boost-invariance and Bjorken scaling}
\label{sec4a4}
%%%%%%%%%%%%%%%%%%%%%%%%%%%%%%%%%%%%%%%%%%%%%%%%%%%%%%%%%%%%%%%%%%%%%%%%%%%%%

So far no approximations have been made as long as we allow $\bvp$ and 
$\etaL$ to be arbitrary functions of $x^\mu\eq(t,\br_\perp,z)$ where
$\br_\perp\eq(x,y)$ denotes the transverse coordinates. Things simplify,
however, enormously if one assumes longitudinal boost-invariance.
Bjorken \cite{B83} argued that at asymptotically high energies the 
physics of secondary particle production should be independent of the 
longitudinal reference frame. This condition can be easily expressed
if one puts the nuclear collision at longitudinal position $z\eq0$
and introduces, instead of $z$ and $t$, the following ``space-time 
rapidity'' and ``longitudinal proper time'' coordinates to describe 
the forward light cone emanating from the collision point:
\bea{LC}
   \eta=\half\ln\frac{t{+}z}{t{-}z} 
       =\half\ln\frac{1{+}\frac{z}{t}}{1{-}\frac{z}{t}}\,, \qquad
   \tau = \sqrt{t^2{-}z^2}\,.
\eea
In these coordinates $x^\mu\eq(t,\brp,z)$ reads
$x^\mu\eq(\tau\cosh\eta,\brp,\tau\sinh\eta)$, and the space-time 
integration measure takes the form $d^4x\eq\tau\,d\tau\,d\eta\,d^2\rp$.
In a $t$-$z$-diagram, lines of constant space-time rapidity are rays 
through the origin with slope $1/\tanh\eta$. Longitudinal boost-invariance
of particle production then implies that the initial conditions for 
local observables (such as particle and energy densities) are only 
functions of $\tau$ and $\brp$, but independent of $\eta$ \cite{B83}.
Furthermore, the boost-invariance of these initial conditions is preserved
in longitudinal proper time if the system expands collectively along
the longitudinal direction with a very specific ``scaling'' velocity 
profile $v_{\rm L}\eq\frac{z}{t}$ \cite{B83}. Inserting this into the 
definition (\ref{rap}) for the longitudinal fluid rapidity $\eta_{\rm L}$ 
and comparing with (\ref{LC}) we see that longitudinal boost-invariance 
implies the identity 
\bea{Bj}
   \eta_{\rm L} = \eta \qquad {\rm (Bjorken\ scaling)}
\eea
of the longitudinal fluid rapidity with the space-time rapidity. 
Since $\eta_{\rm L}(x)$ characterizes the average longitudinal momentum of 
the produced particles at point $x$ whereas $\eta$ characterizes the 
coordinate $x$ itself, this identity implies a very strong correlation
between the average longitudinal momentum and the longitudinal position.

The Bjorken scaling approximation is expected to be good at high 
energies and not too close to the beam and target rapidities,
i.e. in safe distance from the longitudinal kinematic limits. We will 
use it to describe particle production near midrapidity ($y\approx 0$
in the c.m. frame). As a result of Bjorken scaling, the Boltzmann 
exponent reduces to
\bea{p.u}
   p\cdot u(x) = \gamma_\perp(\brp,\tau) 
  \Bigl(\mperp\cosh(y{-}\eta)-\bpp\cdot\bvp(\brp,\tau)\Bigr).
\eea
Due to the additivity of rapidities, this is manifestly invariant under
longitudinal boosts since it only involves rapidity differences. The 
transverse flow velocity, as well as the temperature $T$ and chemical 
potential $\mu_i$, are independent of $\eta$ and depend only on the 
transverse position $\brp$ and longitudinal proper time $\tau$.

%%%%%%%%%%%%%%%%%%%%%%%%%%%%%%%%%%%%%%%%%%%%%%%%%%%%%%%%%%%%%%%%%%%%%%%%%%%%%
\subsubsection{The Cooper-Frye formula}
\label{sec4a5}
%%%%%%%%%%%%%%%%%%%%%%%%%%%%%%%%%%%%%%%%%%%%%%%%%%%%%%%%%%%%%%%%%%%%%%%%%%%%%

Suppose we want to count the total number of particles of species $i$
after produced in the collision. Since this number does not depend on
the reference frame of the observer, we must be able to express it in
a Lorentz-invariant way. We define a three-dimensional hypersurface 
$\Sigma(x)$ in 4-dimensional space-time along which we perform the 
counting. The simplest case would be a measurement in all space at a 
fixed global time $t$. In a space-time diagram this would correspond
to a horizontal line at fixed $t$. But this is not how a real detector
works. In an ideal detector, we surround the collision region hermetically
by detector elements which sit stationary at a fixed distance from the 
collision point, and we wait until the particles pass through these 
detector elements (which particles with different velocities will do
at different times). Assuming, for example, a spherical detector of radius 
$R$, the corresponding detection hypersurface would in a $(t,r)$ 
space-time diagram be represented by a vertical line at fixed $r\eq{R}$,
extending from $t\eq{-}\infty$ to $t\eq\infty$. In this case the 
hypersurface is a 2-dimensional sphere extending over all time, i.e.
it is again 3-dimensional.

You see that different choices for the 3-dimensional hypersurface $\Sigma$ 
are possible as long as it completely closes off the future light cone 
emerging from the collision point. We count particles crossing the 
surface by subdividing it into infinitesimal elements $d^3\sigma$,
defining an outward-pointing 4-vector $d^{3}\sigma_\mu(x)$ perpendicular 
to $\Sigma(x)$ at point $x$ with the magnitude $d^3\sigma$, computing 
the scalar product of the 4-vector $j_i^\mu(x)$ describing the current
of particles $i$ through point $x$, and summing over all such infinitesimal
hypersurface elements:
\bea{N}
   N_i = \int_\Sigma d^3\sigma_\mu(x)\,j_i^\mu(x)
       = \int_\Sigma d^3\sigma_\mu(x)\left(\frac{1}{(2\pi)^3}
         \int \frac{d^3p}{E}\,p^\mu\,f_i(x,p)\right).
\eea
The particle number current density $j_i^\mu(x)$ is given in terms
of the Lorentz-invariant phase-space distribution (giving the probability
of finding a particle with momentum $p$ at point $x$) by multiplying
it with the velocity $\frac{p^\mu}{E}$ and integrating over all momenta 
with measure $\frac{d^3p}{h^3}\eq\frac{d^3p}{(2\pi\hbar)^3}$ where 
$h\eq2\pi\hbar\eq2\pi$ is Planck's quantum of action and we are using units
where $\hbar\eq1$.

Dividing by the Lorentz-invariant momentum-space measure $\frac{d^3p}{E}$
(which in rapidity coordinates reads 
$dy\,\pperp d\pperp\,d\phi_p\eq{dy}\,\mperp d\mperp\,d\phi_p$ where
$\phi_p$ is the azimuthal angle of $\bpp$) we obtain the invariant 
momentum distribution for particle species $i$:
\bea{CF}
   E \frac{dN_i}{d^3p} = \frac{dN_i}{dy\,\pperp d\pperp\,d\phi_p}
   =  \frac{dN_i}{dy\,\mperp d\mperp\,d\phi_p}
   =  \frac{1}{(2\pi)^3} \int_\Sigma p\cdot d^3\sigma_\mu(x)\,f_i(x,p).
\eea 
This is the {\bf Cooper-Frye formula} \cite{CF74}. One can show that
two different surfaces $\Sigma_1$ and $\Sigma_2$ give the {\em same particle
number} $N_i$ if between $\Sigma_1$ and $\Sigma_2$ the distribution function 
$f_i(x,p)$ evolves via a Boltzmann equation with a collision kernel which 
preserves the number of particles $i$, and that we obtain the {\em same
momentum spectrum} if and only if $f_i(x,p)$ evolves from $\Sigma_1$ to
$\Sigma_2$ by free-streaming, i.e. if $f_i$ is a solution of the collisionless
Boltzmann equation. To compute the measured momentum spectrum we can 
therefore replace the surface $\Sigma$ corresponding to the detector 
by shrinking it to the smallest and earliest surface that still 
encloses all scattering processes. We call this the ``surface of last 
scattering'' or ``freeze-out surface'' $\Sigma_{\rm f}$.

Any 3-dimensional space can be parametrized by a set of three locally
orthogonal coordinates $u,v,w$. The points on the surface $\Sigma(x)$
then have coordinates $\Sigma^\mu(u,v,w)$, $\mu\eq0,1,2,3$.
The normal vector $d^3\sigma_\mu(x)$ is computed through the formula
\bea{dsigma}
  d^3\sigma_\mu = - \epsilon_{\mu\nu\lambda\rho}
  \frac{\P\sigma^\nu}{\P u}\frac{\P\sigma^\lambda}{\P v}
  \frac{\P\sigma^\rho}{\P w} du\, dv\, dw
\eea
where $\epsilon_{\mu\nu\lambda\rho}$ is the completely antisymmetric 
Levi-Civita symbol in four dimensions, with 
$\epsilon^{0123}=-\epsilon_{0123}=1$. If we assume longitudinal 
boost-invariance, the freeze-out surface $\Sigma_{\rm f}$ can be 
characterized by a longitudinal proper time $\tau_{\rm f}(\brp)$ such that
\bea{sigma}
   \Sigma^\mu_{\rm f}(\brp,\eta) = (t_{\rm f},x_{\rm f},y_{\rm f},z_{\rm f})
   = \Bigl(\tauf(\brp)\cosh\eta,\brp,\tauf(\brp)\sinh\eta\Bigr).
\eea
In this case
\bea{dsig}
   d^3\sigma_\mu  = \Bigl(\cosh\eta,-\frac{\P\tauf}{\P x},
   -\frac{\P\tauf}{\P y},-\sinh\eta\Bigr)\, \tauf(\brp)\,d^2r_\perp\,d\eta,
\eea
and the integration measure in the Cooper-Frye formula (\ref{CF}) 
becomes (with $(\P_x,\P_y)\eq\grad_{\!\perp}$)
\bea{pdsig}
   p\cdot d^3\sigma(x) = \Bigl( \mperp\cosh(y{-}\eta) -
   \bpp\cdot\grad_{\!\perp}\tauf(\brp)\Bigr) \,\tauf(\brp)\,d^2r_\perp\,d\eta.
\eea

%%%%%%%%%%%%%%%%%%%%%%%%%%%%%%%%%%%%%%%%%%%%%%%%%%%%%%%%%%%%%%%%%%%%%%%%%%%%%
\subsubsection{Thermal spectra from an exploding source}
\label{sec4a6}
%%%%%%%%%%%%%%%%%%%%%%%%%%%%%%%%%%%%%%%%%%%%%%%%%%%%%%%%%%%%%%%%%%%%%%%%%%%%%

The local thermal equilibrium form (\ref{eq}) for the phase-space 
distribution is a good approximation as long as the particles rescatter
intensely, i.e. as long as the relaxation time for returning to equilibrium 
is shorter than any macroscopic time scale which changes the parameters 
$T(x)$, $\mu_i(x)$ and $u^\mu(x)$. Use of Eq.~(\ref{CF}) to compute the 
measured
momentum spectrum requires knowledge of the phase-space distribution on
the surface of last scattering, i.e. when all scattering has ceased. How 
can we relate the two? If the system expands very fast, its density
decreases rapidly and the mean free path of the particles growth quickly.
The transition from strong coupling to free-steaming thus happens in a 
short time interval. During this short time it is unlikely that the 
phase-space distribution undergoes qualitative changes, and we may 
approximate $f_i(x,p)$ on the last scattering surface by its thermal 
equilibrium form that it still had just a little earlier. This results 
in the famous Cooper-Frye freeze-out algorithm: determine the surface 
where local equilibrium {\em begins} to break down and use the local 
equilibrium distribution (\ref{eq}), with parameters $T(x)$, $\mu_i(x)$
and $u^\mu(x)$ determined on this surface, in Eq.~(\ref{CF}) to compute
the final momentum distribution. The result is \cite{R87,HLS90} (note that
$\tau_{\rm f}$, $T$, $\mu_i$, $\bvp$ and $\gamma_\perp$ are all functions
of $\brp$)
\bea{flowspec}
   &&\frac{dN_i}{dy\,\mperp d\mperp\,d\phi_p} =
   \frac{g_i}{(2\pi)^3} \sum_{n=1}^\infty (\mp)^{n+1}
   \int d^2r_\perp\,\tauf\, e^{n\mu_i/T}\,
   e^{n\gamma_\perp\bvp\cdot\bpp}
\nonumber\\
   &&\hspace*{3cm}
   \times\int_{-\infty}^\infty d\eta\, 
   \Bigl(\mperp\cosh(y{-}\eta) - \bpp{\cdot}\grad_{\!\perp}\tauf\Bigr)\,
   e^{-n\gamma_\perp\mperp\cosh(y{-}\eta)/T}
\\\nonumber 
   &&\qquad\quad =
   \frac{2g_i}{(2\pi)^3} \sum_{n=1}^\infty (\mp)^{n+1}
   \int d^2r_\perp\,\tauf\, e^{n\mu_i/T}\,
   e^{n\gamma_\perp\bvp\cdot\bpp}
   \Bigl(\mperp {\rm K}_1(n\beta_\perp) -\bpp{\cdot}\grad_{\!\perp}\tauf\, 
   {\rm K}_0(n\beta_\perp)\Bigr),
\eea
where $\beta_\perp(\brp)\equiv\mperp\frac{\gamma_\perp(\brp)}{T(\brp)}$. 
Due to the assumed boost-invariance, which allowed for an easy integration 
over $\eta$, this result is independent of rapidity $y$.

For central collisions ($b\eq0$) the fireball is azimuthally symmetric,
so $\bvp=v_\perp\be_r$ and (in polar coordinates $\brp\eq(r_\perp,\phi_s)$)
the functions $\tau_{\rm f}$, $T$, $\mu_i$, $v_\perp$, $\gamma_\perp$ are 
all independent of $\phi_s$. This also implies 
$\grad_{\!\perp}\tauf\eq(\P\tauf/\P\rp)\be_\rp$. The only angular 
dependences then arise from the scalar products $\bvp{\cdot}\bpp$ and 
$\bpp{\cdot}\grad_{\!\perp}\tauf$ which both involve the relative angle 
between $\bpp$ and $\brp$, i.e. $\cos(\phi_s{-}\phi_p)$. The azimuthal
integral can thus be done analytically, producing another set of modified
Bessel functions:
\bea{flowspec1}
   \frac{dN_i}{dy\,\mperp d\mperp} &=&
   \frac{g_i}{\pi^2} \sum_{n=1}^\infty (\mp)^{n+1}
   \int_0^\infty r_\perp dr_\perp\,\tauf\, e^{n\mu_i/T}\,
\nonumber\\
   && \times
   \Bigl(\mperp {\rm K}_1(n\beta_\perp)\,{\rm I}_0(n\alpha_\perp)
    -\pperp\frac{\P\tauf}{\P\rp}
   {\rm K}_0(n\beta_\perp)\,{\rm I}_1(n\alpha_\perp)\Bigr),
\eea
where 
$\alpha_\perp\eq\frac{\gamma_\perp v_\perp\pperp}{T}\eq\alpha_\perp(\rp)$. 
For all hadrons except pions this can be used in the Boltzmann approximation, 
by keeping only the term $n\eq1$. The factor
$\tau_{\rm f}(\rp)\,e^{\mu_i(\rp)/T(\rp)}{\equiv}n_i(\rp)$ can be 
interpreted as the (unnormalized) radial density profile of the particles 
$i$. Introducing the {\em radial flow rapdity} $\rho$ via 
$\vperp\eq\tanh\rho$, which allows to write 
$\beta_\perp\eq\frac{\mperp\cosh\rho}{T}$ and 
$\alpha_\perp\eq\frac{\pperp\sinh\rho}{T}$, we then obtain the
following ``flow spectrum'' \cite{HLS90,LHS90}:
\bea{flowspec2}
   \frac{dN_i}{dy\,\mperp d\mperp} &=&
   \frac{g_i}{\pi^2} \int_0^\infty \rp d\rp\,n_i(\rp)\,\biggl[
   \mperp {\rm K}_1\Bigl(\frac{\mperp\cosh\rho(\rp)}{T(\rp)}\Bigr)
          {\rm I}_0\Bigl(\frac{\pperp\sinh\rho(\rp)}{T(\rp)}\Bigr)
\\\nonumber
   &&\hspace*{3.2cm}
   -\pperp\frac{\P\tauf}{\P\rp}
   {\rm K}_0\Bigl(\frac{\mperp\cosh\rho(\rp)}{T(\rp)}\Bigr)
   {\rm I}_1\Bigl(\frac{\pperp\sinh\rho(\rp)}{T(\rp)}\Bigr)\biggr].
\eea
This formula is useful because it allows to easily perform systematic
studies of the influence of the radial profiles of temperature, density
and transverse flow on the transverse momentum spectrum, in order to 
better understand which features of a real dynamical calculation of 
these profiles control the shape of the observed spectra. There are 
many such case studies documented in the literature (see, e.g., 
\cite{LHS90,Peitz02}), and I will here discuss only the most important
and generic characteristics. 

%%%%%%%%%%%%%%%%%%%%%%%%%%%%%%%%%%%%%%%%%%%%%%%%%%%%%%%%%%%%%%%%%%%%%%%%%%%%%
\subsubsection{How radial flow affects single-particle transverse 
momentum spectra}
\label{sec4a7}
%%%%%%%%%%%%%%%%%%%%%%%%%%%%%%%%%%%%%%%%%%%%%%%%%%%%%%%%%%%%%%%%%%%%%%%%%%%%%

Since for all hadrons $\mperp/T{\,>\,}1$, the modified Bessel functions 
K$_\nu$ can be approximated by exponentials 
${\,\sim\,}e^{-\mperp\cosh\rho/T}$. The temperature on the freeze-out 
hypersurface is approximately constant \cite{IlCiocco,MH97} since 
freeze-out is controlled by the mean free path which is inversely 
proportional to the density, which itself is a steep function of 
temperature \cite{SH94}. Nevertheless, the flow spectra are 
characteristically curved, due to two effects: the influence of the 
I$_\nu$ Bessel functions at low $\pperp$ and the integration over 
the radial flow profile $\rho(\rp)$. 

\begin{figure}[ht]
\begin{center}
\includegraphics[bb=215 452 565 775,width=5.5cm]{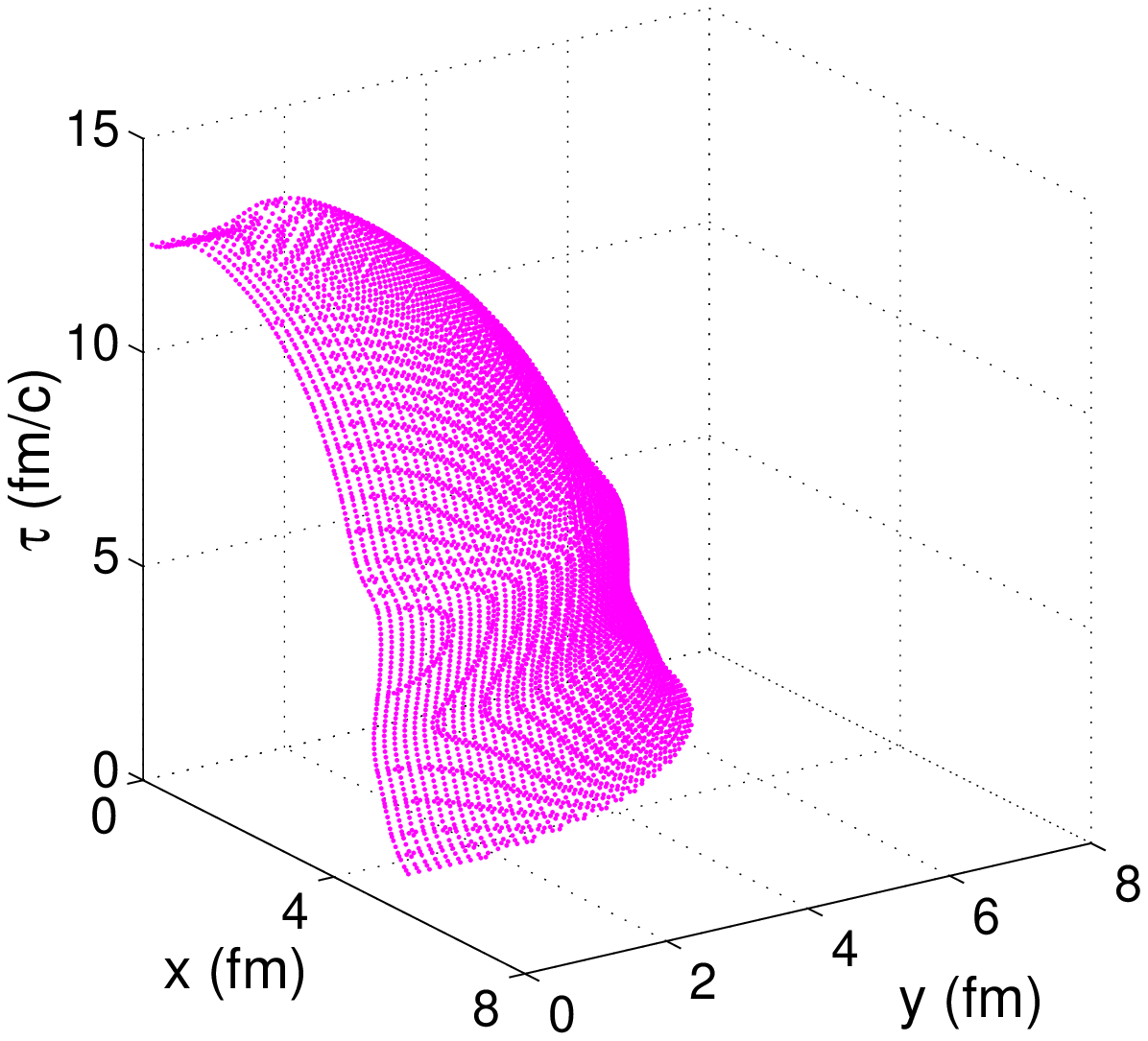} 
\includegraphics[bb=215 452 565 775,width=5.5cm]{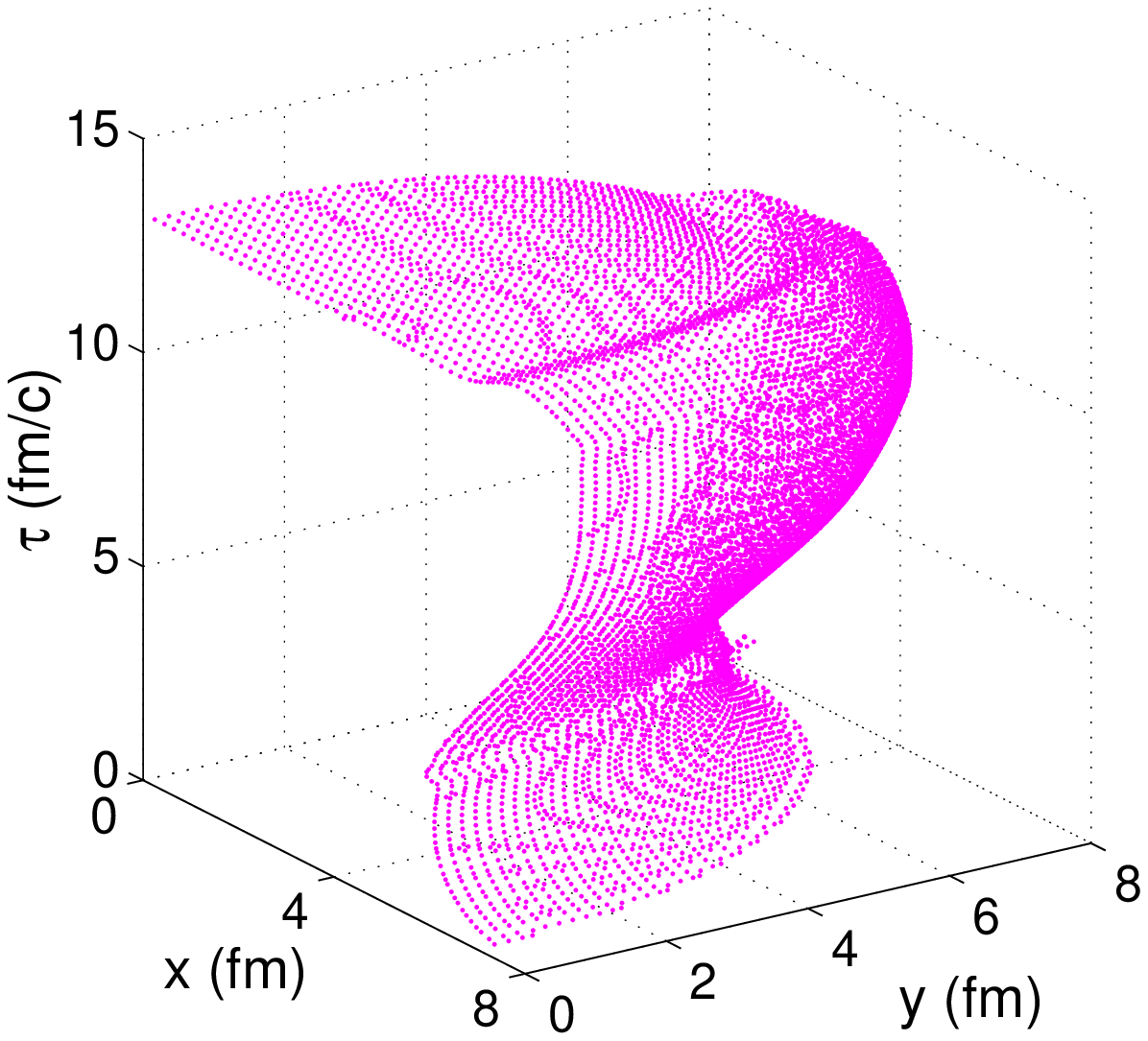} 
\includegraphics[width=4.8cm]{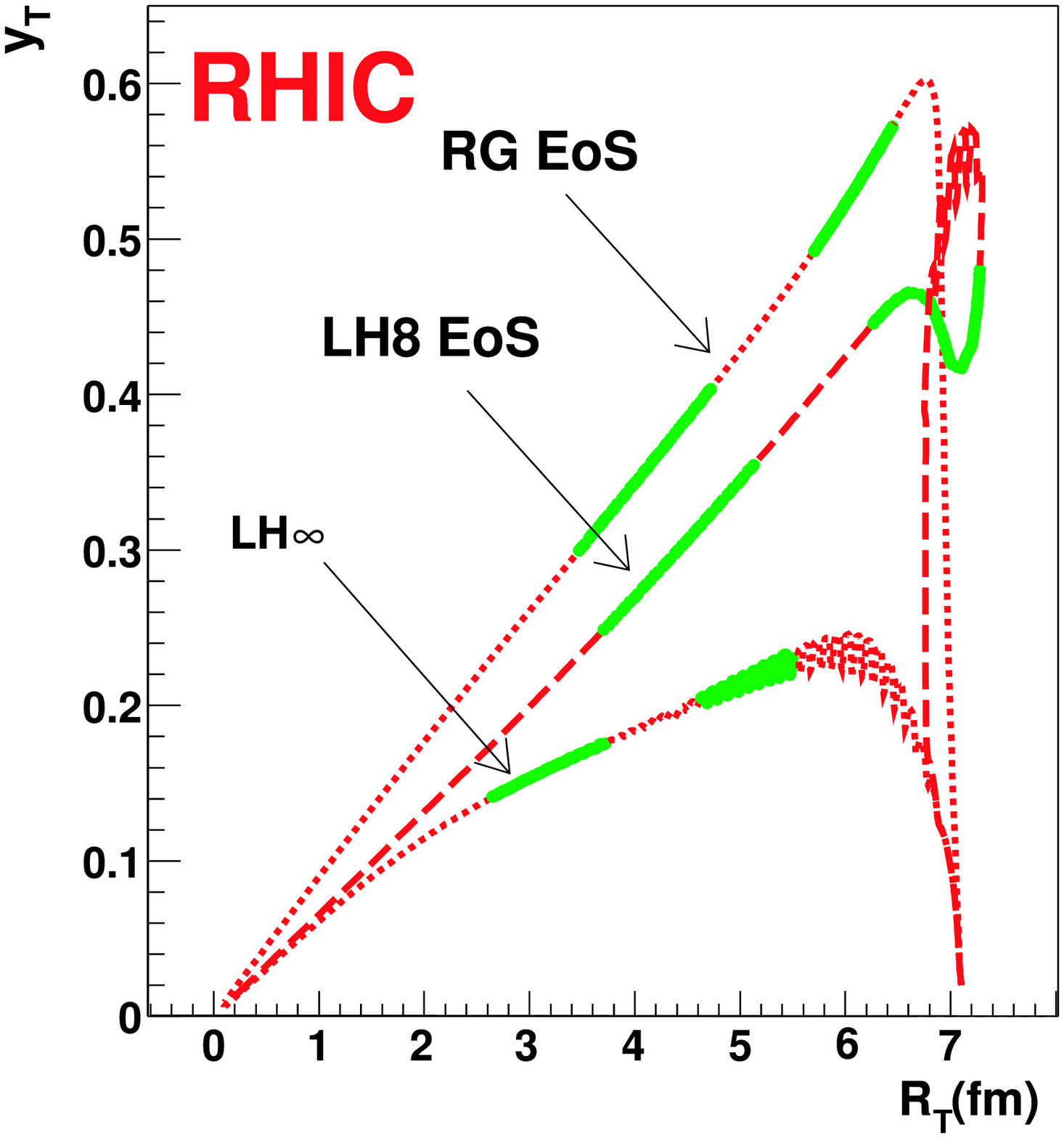} 
\caption{Left panels: Hydrodynamical freeze-out surface $\tauf(\brp)$ 
for non-central Pb+Pb collisions at $b\eq8$\,fm ($\bb$ points in $x$ 
direction) with two different initial energy densities, one corresponding
to SPS energies (left), the other corresponding to LHC energies (middle)
\protect\cite{KSH99}. Right panel: Radial flow rapidity profile
$\rho(\rp){\,\equiv\,}y_{\rm T}(\rp)$ for central Au+Au collisions at 
RHIC, from hydrodynamic calculations employing three different
equations of state (for details see Ref.~\cite{TLS01} from where this
figure was taken).
\label{F5}
}
\end{center}
\vspace*{-5mm}
\end{figure}

Let us first see what kind of flow profiles we should consider. At 
$\rp\eq0$ the radial flow velocity must vanish by symmetry; as you 
follow the freeze-out surface out to larger $\rp$, $\vperp$ typically 
rises linearly with $\rp$ \cite{IlCiocco,MH97,TLS01}. As shown in the 
right panel of Fig.~\ref{F5}, it eventually reaches a maximum value
%
%%%%%%%%%%%%%%%%%%%%%%% Fig. 5 %%%%%%%%%%%%%%%%%%%%%%%%%%%%%%%%%%%%%%%%%%
%%%%%%%%%%%%%%%%%%%%%%%%%%%%%%%%%%%%%%%%%%%%%%%%%%%%%%%%%%%%%%%%%%%%%%%%
% 
and drops again to zero since the dilute tail of the initial 
density distribution freezes out early before radial flow develops.
The left two panels of Fig.~\ref{F5} show typical freeze-out surfaces 
$\tauf(\brp)$ from hydrodynamic calculations (see later) for SPS 
(left) and LHC energies (middle) \cite{KSH99}. One sees that at SPS 
energies the freeze-out surface moves from the edge inward since the 
fireball matter cools and freezes out faster than the developing radial 
flow can push it out. At LHC energies things begin similarly, but then 
the much stronger radial flow generated by the much higher internal 
pressure makes the fireball grow considerably before suddenly freezing 
out after about 13\,fm/$c$. 
The freeze-out surface for RHIC energies lies in between \cite{HK02osci}, 
with transverse flow almost exactly balancing freeze-out and leading to 
an almost constant freeze-out radius (also seen in the right panel of 
Fig.~\ref{F5}) until after about 12\,fm/$c$ it rather suddenly shrinks 
to zero, again indicating sudden bulk freeze-out. 

So how does the radial flow affect the spectra? Looking at 
Eq.~(ref{flowspec2}) we see that in the absence of flow 
($\rho\eq0$) the second term vanishes (since I$_1(0)\eq0$) and the
first term simply becomes (for constant decoupling temperature)
\bea{noflow} 
   \frac{dN_i}{dy\,\mperp d\mperp} \sim 
   \mperp {\rm K}_1\Bigl(\frac{\mperp}{T}\Bigr).
\eea
This depends only $\mperp$ (rather than on both $\mperp$ and 
$\pperp\eq\sqrt{\mperp^2{-}m_i^2}$) and, except for the fact that
$\mperp$ cannot be less than the rest mass, the spectrum is therefore 
identical for all hadrons! This is known as {\em ``$\mperp$-scaling''}: 
In a static fireball all hadron spectra follow the same exponential 
$dN_i/(dy\,\mperp d\mperp) \sim \mperp^{1/2}\,e^{-\mperp/T}$, and the
fireball temperature can be immediately extracted from their slope.

\begin{figure}[ht]
\begin{center}
  \includegraphics[bb=122.0 237.0 490.5 605.5,width=7cm]{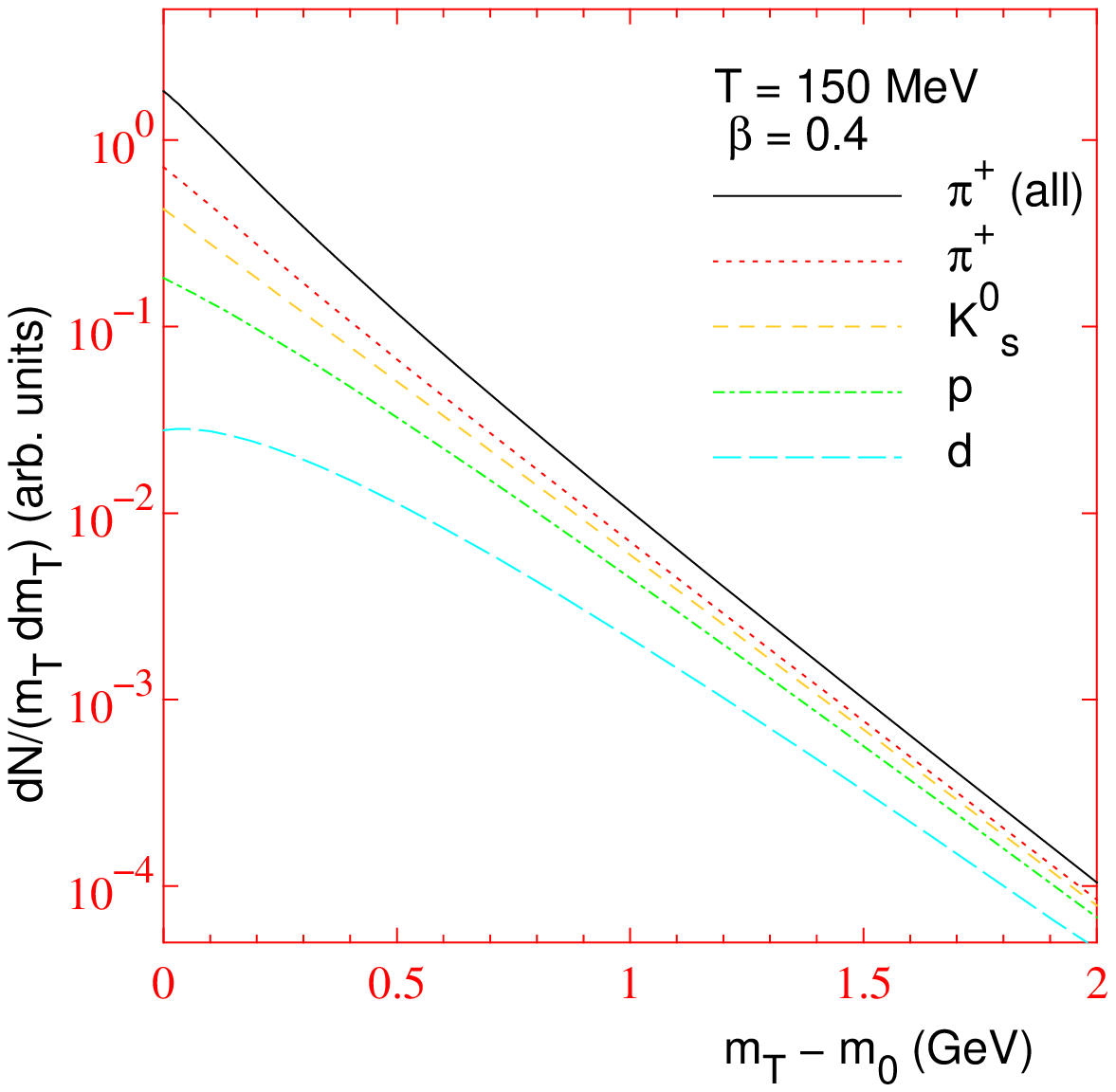} 
  \includegraphics[bb=122.0 237.0 490.5 605.5,width=7cm]{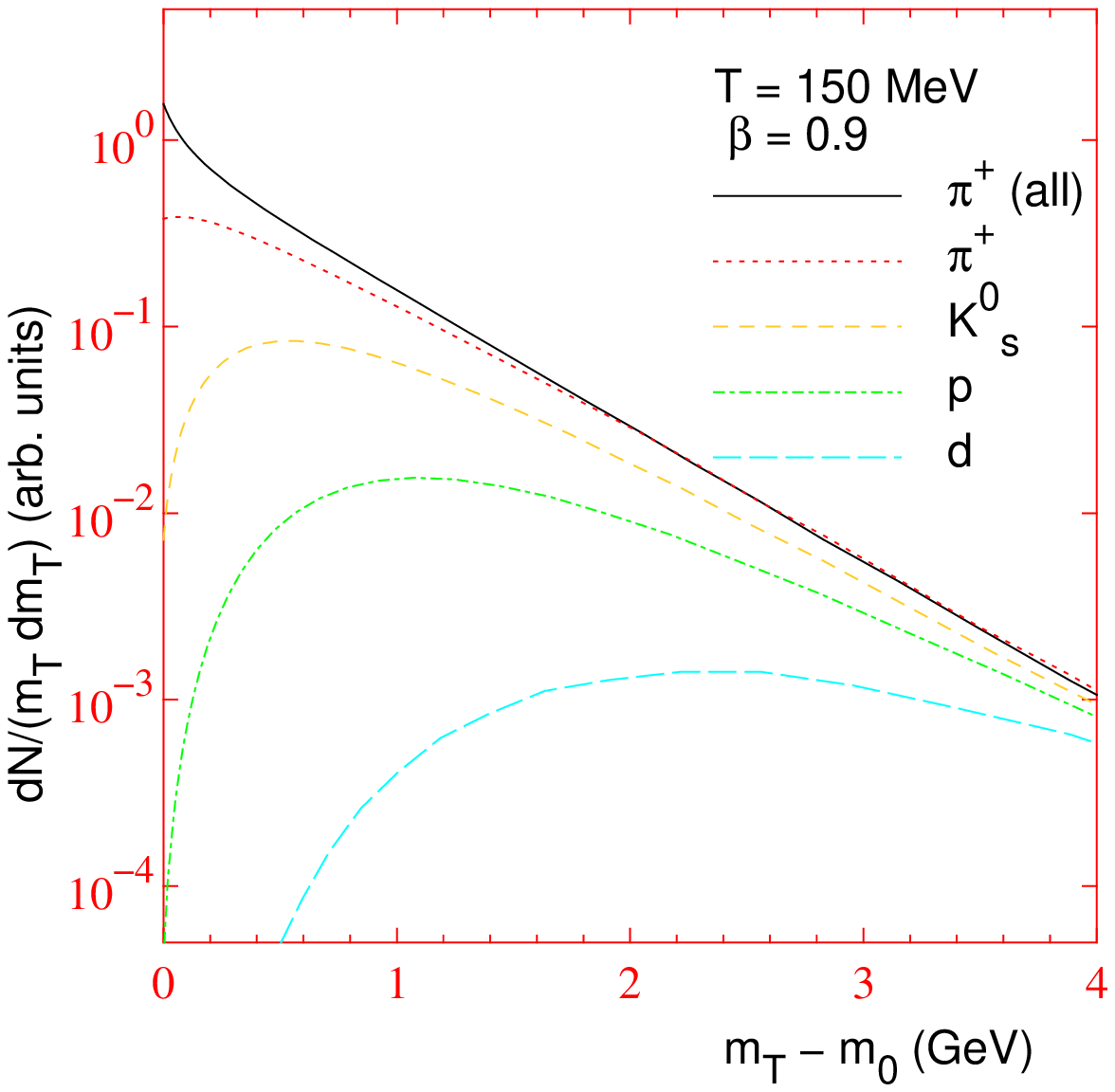} 
\caption{Flow spectra for various hadrons as a function of 
$\mperp{-}m_0$ where $m_0$ is their rest mass. The calculation 
assumes an infinitesimally thin shell of temperature $T\eq150$\,MeV 
expanding with $\vperp\eq0.4$ (left) and $\vperp\eq0.9$ (right). The
curve labelled ``$\pi^+$ (all)'' includes pions from resonance decays
in addition to the thermally emitted pions. 
\label{F6}
}
\end{center}
\vspace*{-5mm}
\end{figure}

Collective flow breaks $\mperp$-scaling. This is shown in Fig.~\ref{F6}.
%
%%%%%%%%%%%%%%%%%%%%%%% Fig. 6 %%%%%%%%%%%%%%%%%%%%%%%%%%%%%%%%%%%%%%%%%%
%%%%%%%%%%%%%%%%%%%%%%%%%%%%%%%%%%%%%%%%%%%%%%%%%%%%%%%%%%%%%%%%%%%%%%%%
% 
The breaking of $\mperp$-scaling does not occur at large $\mperp$
where again all spectra approach exponentials with identical slope.
However, the inverse slope of these spectra now reflects a blueshifted
freezeout temperature: Approximating $\pperp{\,\approx\,}\mperp{\,\gg\,}m_i$ 
and combining the K$_0$, K$_1$ and I$_1$, I$_0$ Bessel functions gives 
$dN_i/(dy\,\mperp d\mperp){\,\sim\,}$\break $\exp\bigl[-\frac{\mperp}{T}
 (\cosh\rho{-}\sinh\rho)\bigr]\eq\exp(-\mperp/T_{\rm slope})$
with the inverse slope parameter $T_{\rm slope}\eq{T}\sqrt{\frac{1{+}\vperp}
{1{-}\vperp}}$. This is a standard relativistic blueshift factor
reflecting the boost of the thermal radiation {\em towards} the detector
with radial flow velocity $\vperp$.

Transverse flow breaks $\mperp$-scaling at small transverse 
momenta $\pperp\lapp{m_i}$ where momenta and velocities can be added
non-relativistically. The exact form of the of this breaking of 
$\mperp$-scaling depends on the density and velocity profiles.
It is most extreme for a thin shell expanding with fixed velocity
(``blast wave''), shown in Fig.~\ref{F6}, in which case for sufficiently 
large hadron mass and flow velocity the spectrum develops a ``blast 
wave peak'' at nonzero transverse momentum \cite{SR79}. More
realistic calculations take into account the integration over a
velocity profile in which the hole at low $\pperp$ in the blast wave
spectrum is filled in by contributions with smaller radial boost 
velocities from the fireball interior \cite{LHS90}. For a Gaussian
density profile combined with a non-relativistic linear transverse
velocity profile the integrated spectrum can be calculated analytically 
\cite{Scheibl}, and one finds again an exponential $\mperp$-spectrum,
but this time with inverse slope $T_{i,{\rm slope}}\eq{T_{\rm f}}+
\half m_i\langle\vperp\rangle^2$. We summarize these two important limits:
\bea{slopes}
   &\bullet& {\rm Non-relativistic},\ \pperp\ll m_i:
   \qquad  T_{i,{\rm slope}}\approx{T_{\rm f}}+
           \half m_i\langle\vperp\rangle^2
\\\label{slopes1}
   &\bullet& {\rm Relativistic},\ \pperp\gg m_i:
   \hspace*{1.85cm}
   T_{\rm slope}\ \approx T_{\rm f}\,\sqrt{\frac{1{+}\vperp}{1{-}\vperp}} 
   \qquad ({\rm for\ all\ } m_i)
\eea
The slope systematics for hadrons with different masses in the 
low-$\pperp$ region thus allow to separate thermal from collective 
flow motion. Note that Eq.~(\ref{slopes}) can never be applied to 
pions since in the region $\pperp\ll{m}_\pi$ their slope is affected
by Bose statistics and by the contamination from resonance decays 
(see Fig.~\ref{F6}) neither of which is accounted for by 
Eq.~(\ref{slopes}).

%%%%%%%%%%%%%%%%%%%%%%%%%%%%%%%%%%%%%%%%%%%%%%%%%%%%%%%%%%%%%%%%%%%%%%%%%%%%%
\subsubsection{Extracting the freeze-out temperature and flow from measured
transverse momentum spectra}
\label{sec4a8}
%%%%%%%%%%%%%%%%%%%%%%%%%%%%%%%%%%%%%%%%%%%%%%%%%%%%%%%%%%%%%%%%%%%%%%%%%%%%%

Equations (\ref{slopes}) and (\ref{slopes1}) show that, as long as 
$m_i\langle\vperp\rangle\gapp2T_{\rm f}$, the spectra are steeper
at high $\pperp$ and bend over becoming flatter at low $\pperp$. 
It is therefore difficult to characterize them by a single slope,
especially when the detector measures different hadrons in different 
$\pperp$-windows, or when two different experiments measure the same 
hadron in different $\pperp$-windows. To extract the flow velocity
%
%%%%%%%%%%%%%%%%%%%%%%% Fig. 7 %%%%%%%%%%%%%%%%%%%%%%%%%%%%%%%%%%%%%%%%%%
\begin{figure}[ht]
\begin{center}
\includegraphics[width=15cm]{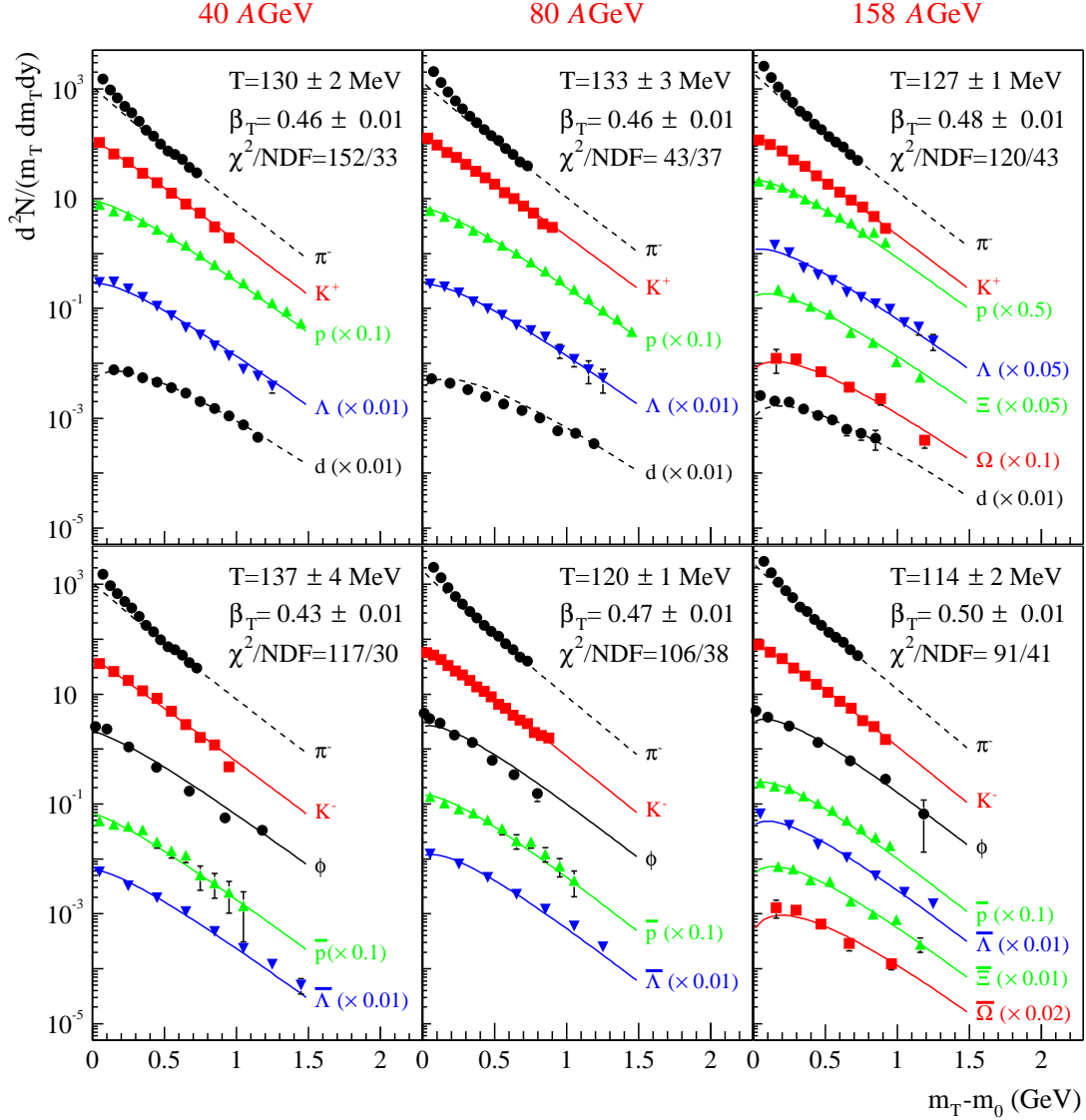} 
\caption{Positively and negatively charged hadron spectra from
central Pb+Pb collisions at 40, 80 and 160\,$A$\,GeV beam energy
at the SPS (left to right), measured by the NA49 Collaboration
\protect\cite{vL02}. Also shown are 2-parameter fits with 
Eq.~(\ref{flowspec3}), assuming a sharp transverse flow velocity 
$\beta_{\rm T}$ (i.e. $w(\rho)\eq\delta(\rho{-}\tanh^{-1}\beta_{\rm T})$)
and constant $\tauf$. The resulting fit values for $T$ and $\beta_{\rm T}$ 
are given in the figures. Positive and negativ hadrons were fitted 
independently; dashed lines indicate hadrons which were not included 
in the fit. The dashed pion curve does not include resonance decay 
contributions and Bose statistics.
\label{F7}
}
\end{center}
\vspace*{-3mm}
\end{figure}
%%%%%%%%%%%%%%%%%%%%%%%%%%%%%%%%%%%%%%%%%%%%%%%%%%%%%%%%%%%%%%%%%%%%%%%%
% 
using Eq.~(\ref{slopes}) requires measuring all hadrons in a common 
interval of nonrelativistic transverse kinetic energy satisfying
$\mperp{-}m_i\lapp{m_i}$. Since such a procedure throws away
information outside the common window, it is not very efficient.
A much preferred method is to use the entire experimentally available 
information on the spectra by performing a simultaneous fit to 
all hadrons over all $\mperp$ using Eq.~(\ref{flowspec2}). 
Assuming a constant freeze-out temperature, a common shape for the 
density profiles $n_i(\rp)$, and a specific shape for $\tauf(\brp)$, 
we can rewrite Eq.~(\ref{flowspec2}) as an integral over flow 
rapidities \cite{Peitz02},
\bea{flowspec3}
   \frac{dN_i}{dy\,\mperp d\mperp} &=&
   N_i \int_0^\infty d\rho\, w(\rho)\,\biggl[
   \mperp {\rm K}_1\Bigl(\frac{\mperp\cosh\rho}{T}\Bigr)
          {\rm I}_0\Bigl(\frac{\pperp\sinh\rho}{T}\Bigr)
\\\nonumber
   &&\hspace*{2.3cm}
   -\pperp\frac{\P\tauf}{\P\rp}
   {\rm K}_0\Bigl(\frac{\mperp\cosh\rho}{T}\Bigr)
   {\rm I}_1\Bigl(\frac{\pperp\sinh\rho}{T}\Bigr)\biggr],
\eea
with the flow rapidity distribution 
\bea{}
   N_i\,w(\rho) = \frac{g_i}{\pi^2}\, \rp(\rho)\, n_i(\rp(\rho))\,
   \frac{d\rp}{d\rho}
\eea
where $\rp(\rho)$ is the inverse of the velocity profile $\rho(\rp)$.
We see that for the shape of the spectrum the density and velocity 
profiles $n_i(\rp)$ and $\vperp(\rp)$ have no independent relevance --
only the flow rapidity distribution $w(\rho)$ matters. If we fix a 
reasonable shape for $w(\rho)$ (see \cite{Peitz02}), leaving its
mean $\la\rho\ra$ free, we end up with a 2-parameter fit in terms of
the freeze-out temperature $T\eq{T_{\rm f}}$ and average transverse 
flow velocity $\la\vperp\ra\eq\tanh^{-1}\la\rho\ra$.

Figure~\ref{F7} shows such a fit to hadronic $\mperp$-spectra measured
by the NA49 Collaboration \cite{vL02} in Pb+Pb collisions at the SPS
at three different beam energies. Even disregarding the fitted lines 
the flow-typical hierarchy of slopes in the low-$\mperp$ region is 
obvious. The fit was done with a thin-shell model with fixed radial 
flow velocity. As discussed above this tends to exaggerate the bending 
of the spectra at low $\pperp$, especially for the heavier particles
$\Xi$, $\Omega$ and the deuteron. This tendency can be seen in 
Fig.~\ref{F7}. Otherwise the quality of the fit, its ability to 
describe all spectra simultaneously, and the consistency of the 
fit parameters extracted from positively and negatively charged 
hadrons is impressive. 

%
%%%%%%%%%%%%%%%%%%%%%%%%%%  Fig. 7a %%%%%%%%%%%%%%%%%%%%%%%%%%%%%%%%%%%%%%%%%
\begin{figure}[htb] 
\begin{center}
\begin{minipage}[h]{11cm}
\includegraphics[width=11cm,height=6cm]{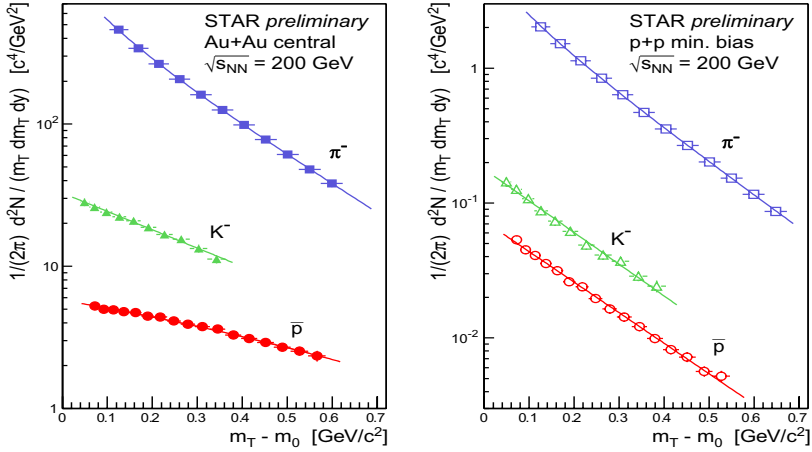}
\end{minipage}
\hspace*{3mm}
\begin{minipage}[h]{4.2cm}
\vspace*{-6mm}
\caption{%
Pion, kaon and antiproton spectra from 200\,$A$\,GeV central Au+Au (left)
and minimum bias p+p collisions (right), measured by the STAR experiment
\cite{ppAAspec}. Note the similar slopes for kaons and antiprotons
in p+p collisions and their dramatically different slopes at low transverse
kinetic energy in central Au+Au collisions.
\label{F7a}
}
\end{minipage}
\end{center} 
\vspace*{-6mm}
\end{figure} 
%%%%%%%%%%%%%%%%%%%%%%%%%%%%%%%%%%%%%%%%%%%%%%%%%%%%%%%%%%%%%%%%%%%%%%%
%

The flattening of the spectra at low transverse kinetic energy
$\mperp{-}m_0$ by transverse collective flow is even more dramatic 
at RHIC. Figure~\ref{F7a} shows a direct comparison of the negative
pion, kaon and antiproton spectra in central Au+Au and minimum bias 
proton-proton collisions, at the same center of mass energy.
Clearly, in p+p collisions the kaon and antiproton spectra have the same
slope, indicating the absence of transverse collective flow. That
the pion spectra are steeper than both kaons and antiprotons can be 
attributed to the contribution of resonance decay pions which accumulate
at low transverse momenta. However, the pion spectra in Au+Au are 
obviously flatter in Au+Au than in p+p, and this is even more true for
kaons and antiprotons, with a large difference in slope between those
last two. Even without a quantitative fit this is a clear manifestation 
of strong radial flow.

%
%%%%%%%%%%%%%%%%%%%%%%%%%%  Fig.8 %%%%%%%%%%%%%%%%%%%%%%%%%%%%%%%%%%%%%%%%%
\begin{figure}[htb] 
\vspace*{-2mm}
\begin{center}
\begin{minipage}[h]{8cm}
\includegraphics[width=8cm,height=9cm]{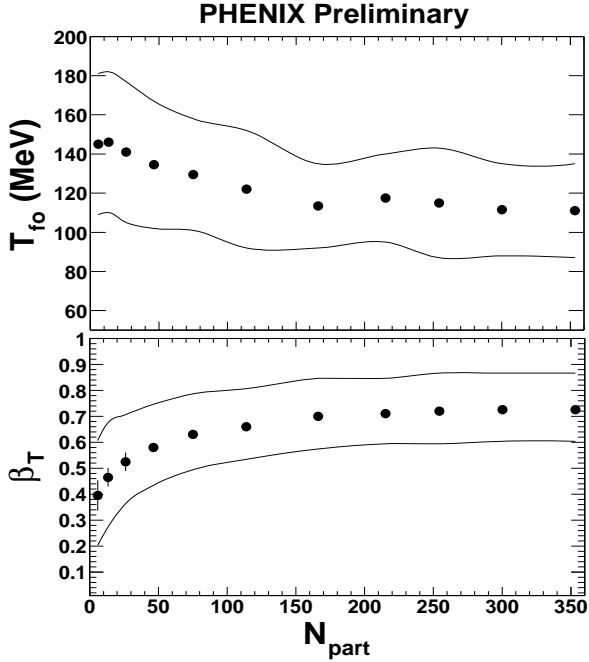}
\end{minipage}
\hspace*{5mm}
\begin{minipage}[h]{6cm}
\vspace*{-6mm}
\caption{%
Kinetic freeze-out temperature $T_{\rm f}$ and transverse flow velocity 
$\beta_{\rm T}$ at the fireball edge, extracted from a simultaneous fit of a 
flow spectrum parametrization to $\pi^{\pm}$, $K^{\pm}$, $p$ and $\bar p$ 
spectra from 200\,$A$\,GeV Au+Au collisions over the entire range of 
collision centralities.\protect\cite{JBH03} More peripheral collisions
(small numbers $N_{\rm part}$ of participating nucleons) are seen
to decouple earlier, at higher freeze-out temperature and with less
transverse flow.
\label{F8}
}
\end{minipage}
\end{center} 
\vspace*{-4mm}
\end{figure} 
%%%%%%%%%%%%%%%%%%%%%%%%%%%%%%%%%%%%%%%%%%%%%%%%%%%%%%%%%%%%%%%%%%%%%%%
%

% 
A two-parameter flow fit on RHIC data from 200\,$A$,GeV Au+Au collisions 
was performed by J. Burward-Hoy \cite{JBH03}, see Fig.~\ref{F8}.
Here a box-profile for the transverse density $n_i(\rp)$ and a linear 
transverse velocity profile were used where the number $\beta_{\rm T}$
given in Fig.~\ref{F8} is the surface velocity (the average
transverse velocity is $\la\vperp\ra\eq\frac{2}{3}\beta_{\rm T}$). Again
in the most central collisions freeze-out temperatures of about 120\,MeV
and average radial flow velocities $\la\vperp\ra{,\simeq\,}0.45$ are
found. One should note, however, that in this case the pions were 
included in the fit, without correction for resonance decay feeddown.
Resonance decay contributions steepen the pion spectrum, biasing the 
fit towards lower temperatures and/or flow velocities. The bands in the
Figure attempt to provide an estimate of the systematic uncertainty
associated with details of the model assumptions on the density and 
velocity profile and of the $\pperp$ range of the pion spectrum used 
in the fit \cite{JBH03}. 

An interesting aspect of Fig.~\ref{F8} is the centrality dependence of 
the fit parameters: The fits were performed over a very wide range of 
collision centralities, and one observes that more peripheral collisions
tend to develop less radial flow and freeze out at a higher temperature.
This is consistent with the expectation that the collision fireballs
in peripheral collisions don't live as long and have less time to build
up radial flow. However, even though their average flow is smaller,
the {\em flow gradients} are bigger (due to the smaller size), leading 
to decoupling at higher particle density and temperature. In $pp$ collisions
one expects no or very little collectivity; Fig.~\ref{F8} indicates that
the transition to this limit in the most peripheral Au+Au collisions is 
very steep, and that already a small number of nucleons participating in 
the collision generates significant collectivity and sizeable radial flow.

Since the effects of the freeze-out temperature and flow on the spectral 
slope are strongly anticorrelated \cite{SSH93}, flow fits from 
single-particle spectra tend to produce narrow $\chi^2$ valleys in the
$T$-$\la\vperp\ra$ plane, oriented from the upper left to the lower 
right. A direct comparison of the spectra from Pb+Pb at the SPS and 
from Au+Au at RHIC shows that the RHIC spectra are flatter, especially 
for the heavier hadrons ((anti)protons, (anti)Lambdas, etc.) 
\cite{Peitz02,BF02}. So, when fitted with exactly the same procedure, 
RHIC data give $\chi^2$ valleys which in the $T$-$\la\vperp\ra$ plane 
lie to the upper right of those obtained from the SPS data 
\cite{Peitz02,BF02}. At the same $T_{\rm f}$ they produce a somewhat 
larger flow whereas with the same flow RHIC data yield somewhat higher 
freeze-out temperatures \cite{Peitz02}. In either case the effect is 
about 15\%.  

This flow analysis of the single-hadron spectra leads us to our 
{\bf First Lesson:} When the collision is over and the hadron 
momentum spectra decouple, the fireball has a ``thermal freeze-out 
temperature'' $T_{\rm f}{\;\simeq\,}110-130$\,MeV and is in a state of
rapid transverse collective expansion with 
$\la\vperp\ra{\,\simeq\;}0.45-0.55\,c$ (``The Little Bang''). 
These parameters have a systematic uncertainty ${\,\sim\,}\pm10\%$
and are strongly anticorrelated. In the $T$-$\la\vperp\ra$ plane
the $\chi^2$ valleys for the RHIC data are shifted systematically 
about 15\% towards the upper right compared to the SPS data.

%%%%%%%%%%%%%%%%%%%%%%%%%%%%%%%%%%%%%%%%%%%%%%%%%%%%%%%%%%%%%%%%%%%%%%%%%%%%%
\subsubsection{Hydrodynamic calculation of radial flow}
\label{sec4a9}
%%%%%%%%%%%%%%%%%%%%%%%%%%%%%%%%%%%%%%%%%%%%%%%%%%%%%%%%%%%%%%%%%%%%%%%%%%%%%

Of course, a flow fit to the spectra is just a fit, and their is no 
{\em a priori} guarantee that the extracted fit parameters make physical
sense. In order to assess whether the extracted freeze-out temperatures
and radial flow velocities are physically meaningful we must check that
they can be the result of reasonable models for the dynamical evolution
of the fireball. Is it possible to generate that much collective transverse
flow during the short lifetime of the collision zone before everything
fizzles?

The natural language for studying flow phenomena is hydrodynamics. The 
hydrodynamic equations of motion are obtained from the local conservation 
laws for energy, momentum and baryon number,
\beq{equ:dmuTmunu}
  \P_\mu \, T^{\mu \nu}(x)=0 \quad (\nu=0,\dots, 3)\qquad {\rm and}
  \quad \P_\mu j_B^\mu (x) = 0,
\end{equation}
by inserting the {\em ideal fluid decompositions} for the energy-momentum 
tensor and baryon number current:
\bea{equ:Tmunu}
  \Tmunu(x) &=& \Bigl(e(x)+p(x)\Bigr) u^\mu(x) u^\nu(x) - p(x) \, \gmunu\,,
\qquad
  j_B^\mu(x) = n_B(x) u^\mu(x)\,.
\eea  
These are 5 partial differential equations for 6 fields ($p,e,n_B$ and
the 3 independent components of $u^\mu$), and the system must be closed
by providing an equation of state $p(e,n_B)$ for the fluid matter. We use
an equation of state which mimicks the lattice QCD results from 
Fig.~\ref{F2} by matching a hadron resonance gas below $\Tc$ to an ideal
gas of massless quarks and gluons above $\Tc$. For details I refer to the
literature \cite{KH03rev,KSH00}. The numerical solution is considerably
simplified by assuming Bjorken scaling flow \cite{B83} along the beam
direction \cite{O92} (see Sec.~\ref{sec4a4}). The form of these simplified
equations can be found e.g. in \cite{KH03rev,KSH00}. They are solved for the 
transverse flow and hadron spectra near midrapidity where, at sufficiently 
high collision energies, the assumption of longitudinal boost-invariance 
should yield a good approxiumation. To obtain the entire rapidity 
distribution, on the other hand, requires a solution of the full 
(3+1)-dimensional set of equations (see, e.g., Refs.~\cite{HMMN02,MMNH02})
since boost-invariance breaks down near the target and projectile 
rapidities. 
 
As discussed in Sec.~\ref{sec3d}, the applicability of fluid dynamics 
requires mean free paths which are short on any relevant macroscopic 
length scale. It can therefore not be used for the initial 
pre-equilibrium stage of the collision just 
after nuclear impact nor for the very last stage when the matter becomes
so dilute that particles start to decouple. In addition to an equation of 
state (see Sec.~\ref{sec2b}), hydrodynamics thus needs {\em initial 
conditions} for the time at which it starts to be a good 
approximation and for the energy density, baryon number density 
and flow velocity profiles at that time, and a {\em freeze-out 
algorithm} which tells us where to stop the hydrodynamic evolution and 
how to translate the hydrodynamic output (energy and baryon densities 
and flow at freeze-out) into hadron spectra. For the freeze-out
algorithm we use the Cooper-Frye formula discussed in Sec.~\ref{sec4a5}
with a constant freeze-out temperature. The initial time, the initial
entropy (or energy) and baryon density in the fireball, and the freeze-out 
temperature (or equivalently the decoupling energy density) are treated 
as parameters to be adjusted to experimental observables. The shape of 
the initial density profiles is calculated from the overlap geometry of 
the two nuclei (for details see \cite{KHHET01}). For the initial flow 
we assume a Bjorken scaling profile along the beam direction 
(see Sec.~\ref{sec4a4}) and zero initial transverse flow. 

We thus need to fix four parameters, $\tauequ$, $\sequ$, $n_{B,{\rm eq}}$ 
and $\edec$. The ratio of the net baryon baryon density $n_{B,{\rm eq}}$ 
to the entropy density $\sequ$ is fixed by the measured proton/pion ratio.
By entropy conservation, the final total charged multiplicity 
$dN_{\rm ch}/dy$ fixes the initial product 
$(s\cdot\tau)_\equ$ \cite{KSH99,B83,O92}.
The value of $\tauequ$ controls how much transverse flow can be 
generated until freeze-out.
It can thus be determined by trying to fit the final pion and proton 
spectra simultaneously whose different slopes at low $\pperp$ are
sensitive to the radial flow at freeze-out.
The associated freeze-out temperature, also implicit in these spectra,
then fixes the decoupling energy density $e_\dec$.
%

%%%%%%%%%%%%%%%%%%%%% Fig. 9 %%%%%%%%%%%%%%%%%%%%%%%%%%%%%%%%%%%%%%%%%%%
\begin{figure}[htb]
\begin{center}
\epsfig{file=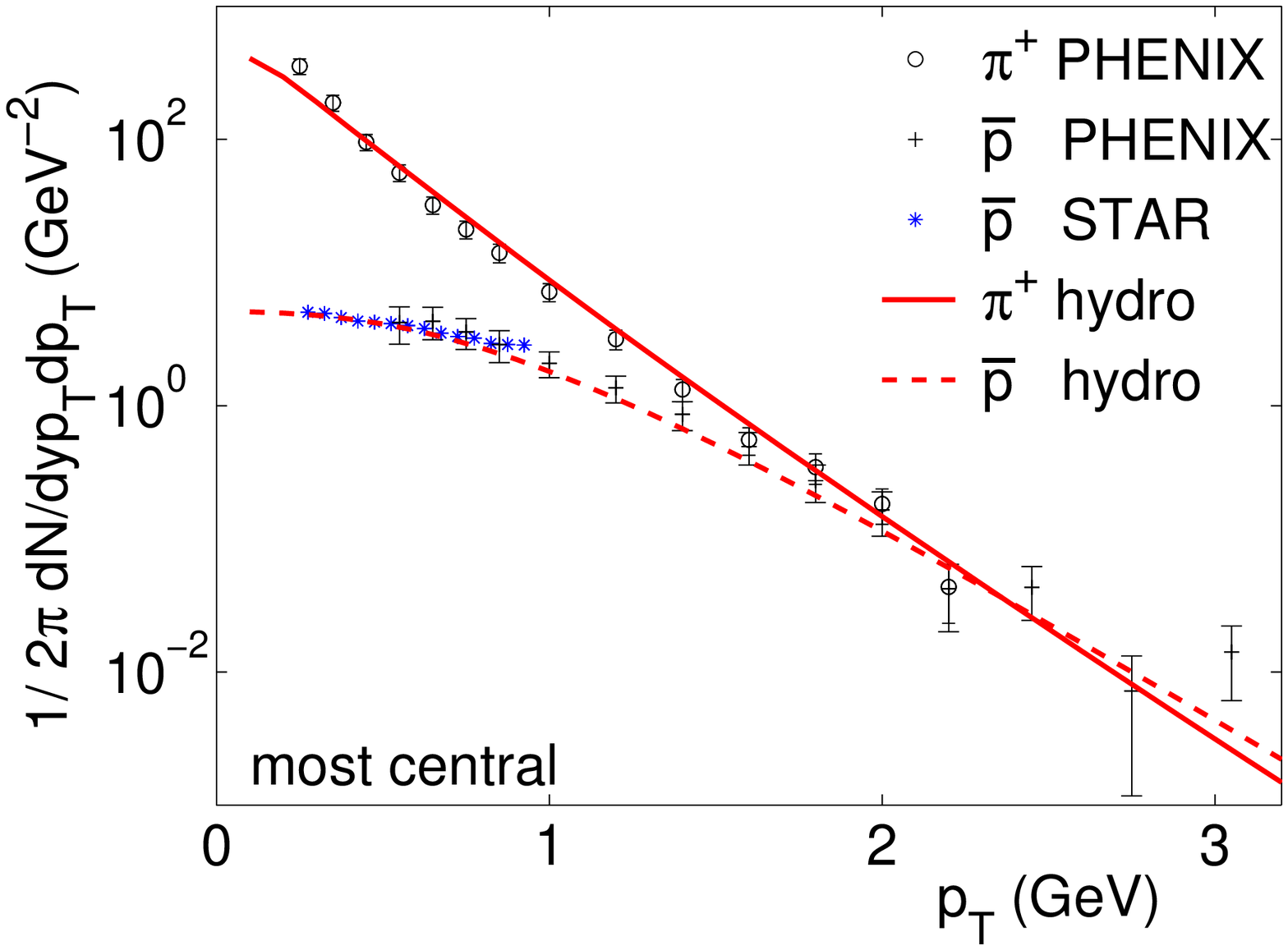,bb=42 205 571 594,width=56mm,height=50mm}
\epsfig{file=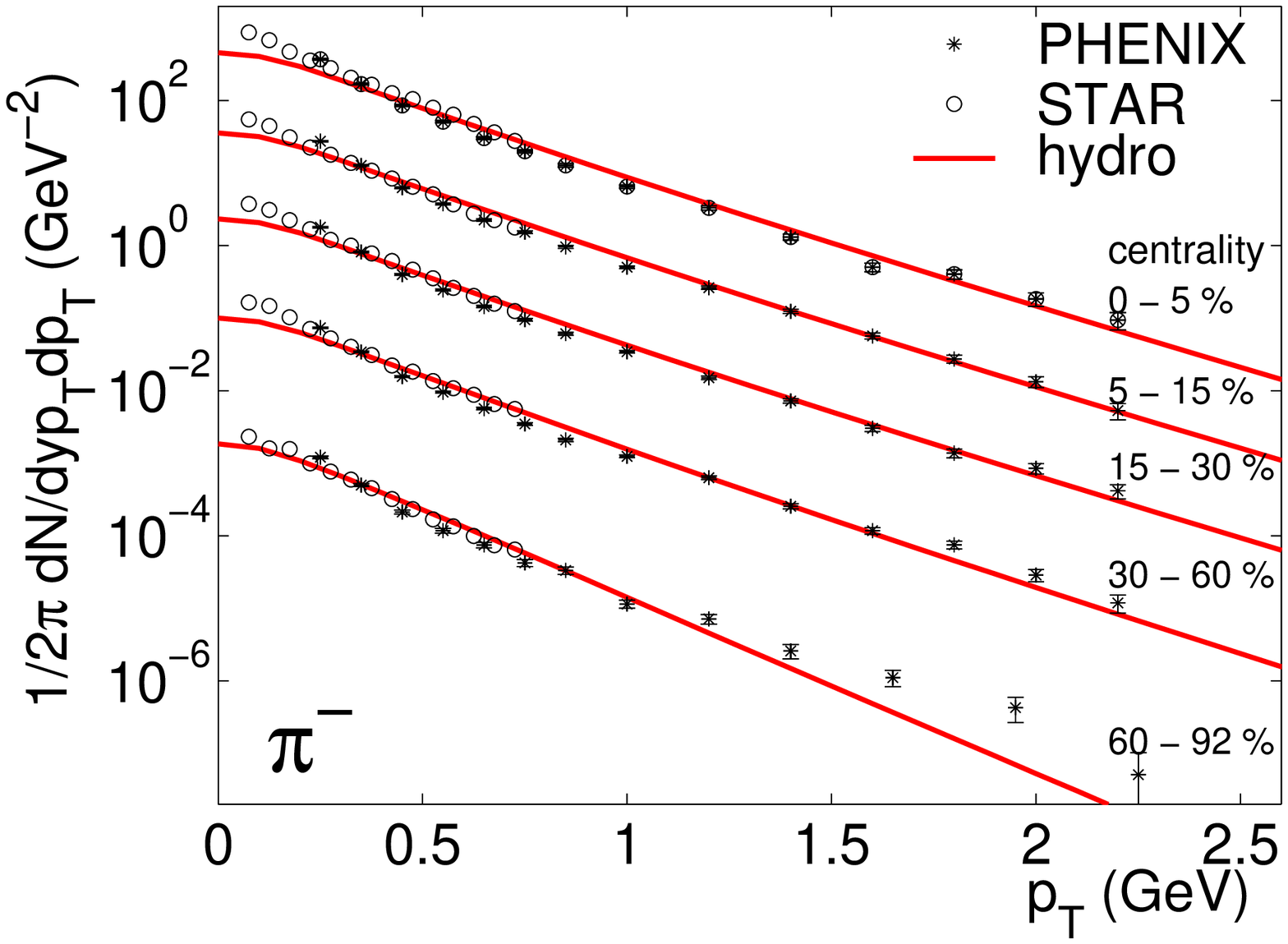,bb=42 207 568 594,width=56mm,height=50mm}
\\
\epsfig{file=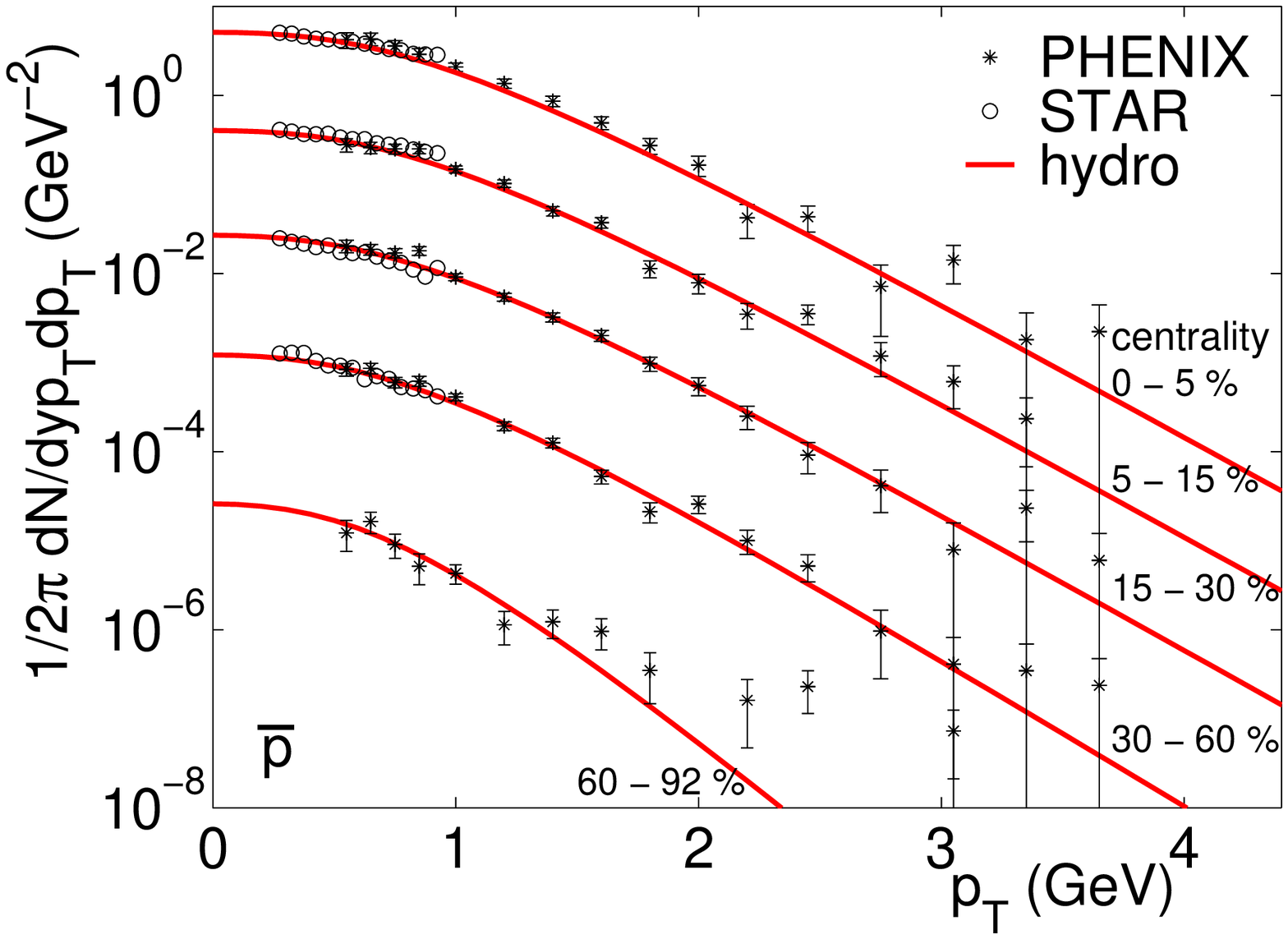,bb=44 211 568 594,width=56mm,height=50mm}
\epsfig{file=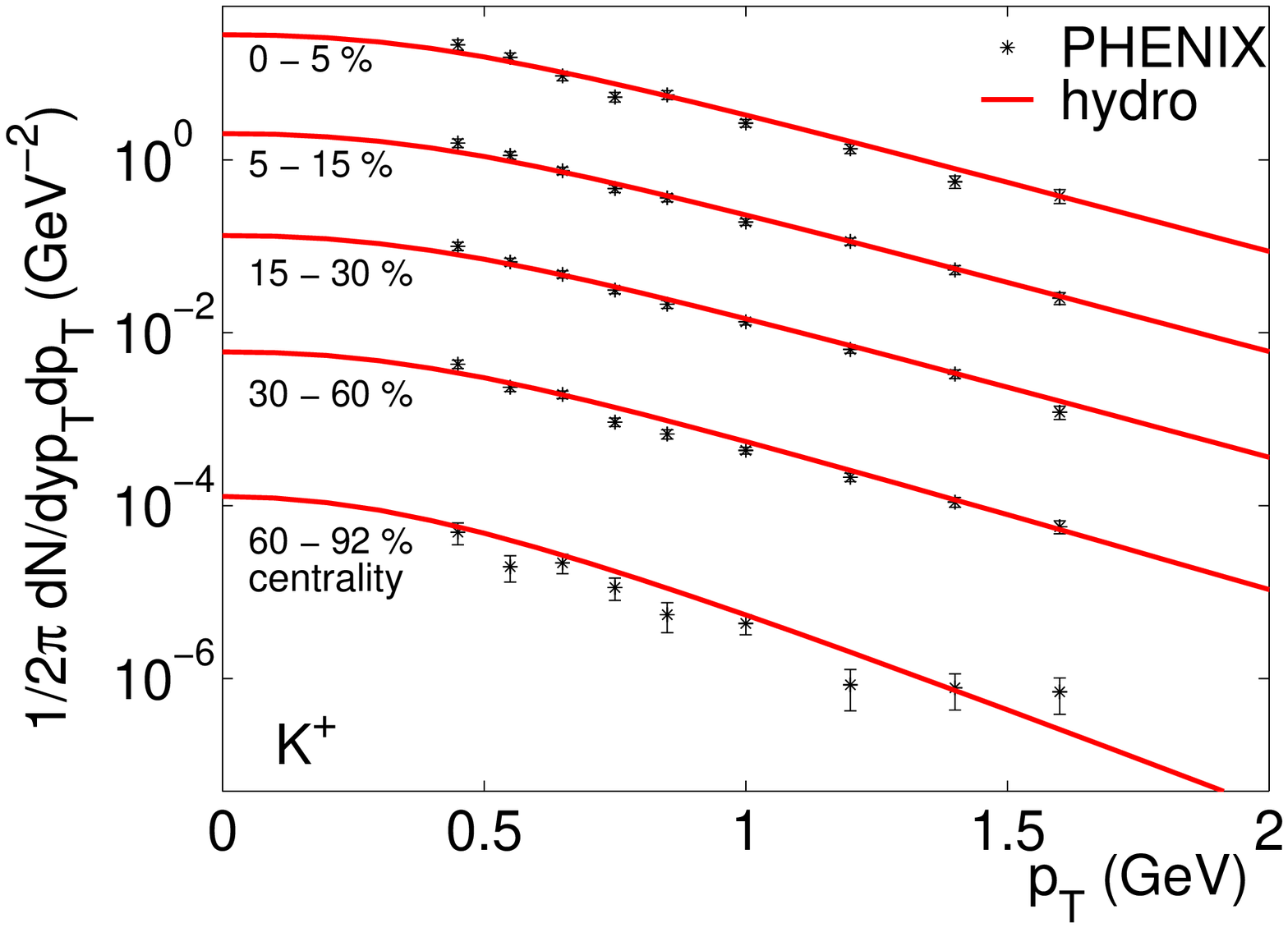,bb=38 208 574 594,width=56mm,height=50mm}
\end{center}
\vspace*{-7mm}
\caption{\label{fig:spectra130}
         Identified pion, antiproton and kaon spectra for 
         $\sqrt{s_{\rm NN}}=130$~GeV from the 
         PHENIX~\protect\cite{PHENIX01spec,PHENIX02spec} and 
         STAR~\protect\cite{STAR01spec,Sanchez02} collaborations
         in comparison with results from a hydrodynamic 
         calculation \protect\cite{HK02WWND}.
         The top left panel shows pion and (anti-)\-proton spectra from
         central collisions. Shown in the other panels are spectra
         of five different centralities: from most central (top) to 
         the most peripheral (bottom). The spectra are successively
         scaled by a factor 0.1 for clarity.
}
\vspace*{-2mm}
\end{figure}
%%%%%%%%%%%%%%%%%%%%%%%%%%%%%%%%%%%%%%%%%%%%%%%%%%%%%%%%%%%%%%%%%%%%%%%%%%%%%

%
The top left panel of Fig.~\ref{fig:spectra130} shows the hydrodynamic 
fit \cite{HK02WWND} to the transverse momentum spectra of positive pions 
and antiprotons, as measured by the PHENIX and STAR collaborations in 
central ($b\eq0$) Au+Au collisions at $\sqrt{s}\eq130\,A$\,GeV 
\cite{PHENIX01spec,STAR01spec,Sanchez02}.
The fit yields an initial central entropy density $s_\equ\eq95$~fm$^{-3}$ 
at an equilibration time $\tau_\equ\eq0.6$~fm.
This corresponds to an initial temperature of $T_\equ\eq340$~MeV and
an initial energy density $e\eq25$\,GeV/fm$^3$ in the fireball center.
(Note that these parameters satisfy the ``uncertainty relation''
$\tau_\equ\cdot T_\equ \approx 1$.)
Freeze-out was implemented on a hypersurface of constant energy density 
with $e_\dec\eq0.075$~GeV/fm$^3$.
Table~1 summarizes the initial conditions applied in hydrodynamic 
studies at SPS and RHIC energies (see the recent review \cite{KH03rev} 
for references).
%

%%%%%%%%%%%%%%%%%%%%%%%  Table 1 %%%%%%%%%%%%%%%%%%%%%%%%%%%%%%%%%%%%%%%%%%%%
\begin{table}[htdp]
\begin{center}
\begin{tabular}{|c|c|c|c|}
\hline
                       &  SPS   & RHIC\,1     & RHIC\,2    \\
$\scm$ (GeV)           &  17    &   130     &  200      \\
\hline
$s_\equ$ (fm$^{-3}$)   &  43    &    95     &  110      \\
$T_\equ$ (MeV)         &  257   &   340     &  360      \\
$\tau_\equ$ (fm/$c$)   &  0.8   &   0.6     &  0.6      \\
\hline
\end{tabular}
\end{center}
{\footnotesize
         Table 1. Initial conditions for SPS and RHIC energies used to 
         fit the particle spectra from central Pb+Pb or Au+Au collisions.
         $s_\equ$ and $T_\equ$ refer to the maximum values at $\tau_\equ$ 
         in the fireball center. 
\label{T1}
}
\vspace*{-2mm}
\end{table}
%%%%%%%%%%%%%%%%%%%%%%%%%%%%%%%%%%%%%%%%%%%%%%%%%%%%%%%%%%%%%%%%%%%%%%%%%

%
The fit in the top left panel of Fig.~\ref{fig:spectra130} was 
performed with a chemical equilibrium equation of state.
Use of such an equation of state implicitly assumes that even below 
the hadronization temperature $\Tc$ chemical equilibrium among the 
different hadron species can be maintained all the way down to
kinetic freeze-out.
With such an equation of state the decoupling energy 
$e_\dec\eq0.075$~GeV/fm$^3$ translates into a kinetic freeze-out 
temperature of $T_\dec{\,\approx\,}130$\,MeV.
The data, on the other hand, show \cite{BMMRS01} that the hadron 
abundances freeze out at $T_{\rm chem}{\,\approx\,}\Tc$, i.e. already when 
hadrons first coalesce from the expanding quark-gluon soup the 
inelastic processes which could transform different hadron species 
into each other are too slow to keep up with the expansion.
The measured $\bar p/\pi$ ratio thus does not agree with the one 
computed from the chemical equilibrium equation of state at the 
kinetic freeze-out temperature $T_\dec=130$\,MeV, and the latter
must be rescaled by hand if one wants to reproduce not only the shape,
but also the correct normalization of the measured spectra in
Fig.~\ref{fig:spectra130}.
A better procedure would be to use a chemical non-equilibrium
equation of state for the hadronic phase \cite{Teaney02,Rapp02,HT02} 
in which for temperatures $T$ below $T_{\rm chem}$ the chemical 
potentials for each hadronic species are readjusted in such a way
that their total abundances (after decay of unstable resonances) are 
kept constant at the observed values.
This approach has recently been applied \cite{KR03} to newer RHIC data 
at $\sqrt{s}\eq200\,A$\,GeV.
Once the parameters have been fixed in central collisions, spectra at
other centralities and for different hadron species can be predicted
without introducing additional parameters. The remaining three panels 
of Fig.~\ref{fig:spectra130} show the transverse momentum spectra of 
pions, kaons and antiprotons in five different centrality bins as 
observed by the PHENIX \cite{PHENIX01spec,PHENIX02spec} and 
STAR \cite{STAR01spec,Sanchez02} collaborations. 
For all centrality classes, except the most peripheral one, the hydrodynamic
predictions (solid lines) agree pretty well with the data.
The kaon spectra are reproduced almost perfectly, but for pions the 
model consistently underpredicts the data at low $\pperp$.
This has now been understood to be largely an artifact of having employed 
in these calculations a chemical equilibrium equation of state all the
way down to kinetic freeze-out.
More recent calculations \cite{KR03} with a chemical non-equilibrium 
equation of state show that, as the system cools below the chemical 
freeze-out point $T_{\rm chem}{\,\approx\,}\Tc$, a significant positive 
pion chemical potential builds up, emphasizing the concave curvature of 
the spectrum from Bose effects and increasing the feeddown corrections
from heavier resonances at low $\pperp$.
Significant discrepancies are also seen at large impact parameters
and large transverse momenta $\pperp\,\gapp\,2.5$\,GeV/$c$.
This is not surprising since high-$\pperp$ particles require more 
rescatterings to thermalize and escape from the fireball before doing so. 
This is in particular true in more peripheral collisions where the 
reaction zone is smaller.
%

%
%%%%%%%%%%%%%%%%%%% Fig. 11 %%%%%%%%%%%%%%%%%%%%%%%%%%%%%%%%%%%%%%%%%%%%
\vspace*{2mm}
\begin{figure}[htb]
\begin{center}
\begin{minipage}[h]{7cm}
\includegraphics[bb=57 203 569 594,width=7cm,height=6cm]{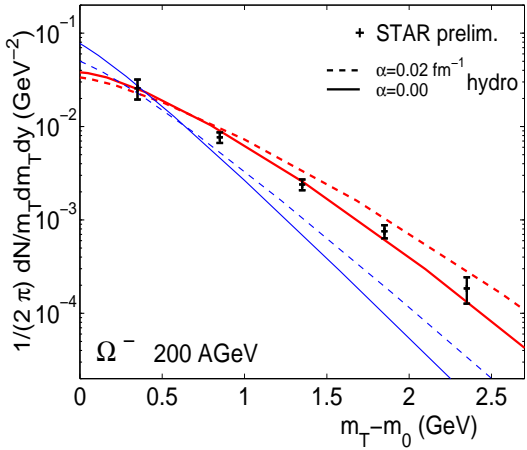}
\end{minipage}
\hspace*{5mm}
\begin{minipage}[h]{8cm}
\vspace*{-6mm}
\caption{
\label{fig:omegaspec}
   Transverse mass spectrum of $\Omega$ hyperons from central 
   200\,$A$\,GeV Au+Au collisions at RHIC \protect\cite{STAR03omega}.
   The curves are hydrodynamic calculations  \protect\cite{KR03} with 
   different initial and freeze-out conditions: Solid lines correspond 
   to the default
   of no initial transverse flow at $\tau_\equ$, dashed lines assume
   a small but non-zero radial flow, $\vperp\eq\tanh(\alpha\rp)$ with 
   $\alpha\eq0.02$~fm$^{-1}$, already at $\tau_\equ$. The lower (thin) 
   set of curves assumes $\Omega$-decoupling at $\Tc\eq164$\,MeV,
   the upper (thick) set of curves decouples the $\Omega$ together
   with the pions and protons at $\edec\eq0.075$\,GeV/fm$^3$.
   (In the chemical non-equilibrium hadronic equation of state 
   employed here this value of $\edec$ corresponds to a freeze-out 
   temperature $T_\dec{\,\approx\,}100$\,MeV -- see 
   Ref.~\protect\cite{KR03} for details.)
}
\end{minipage}
\end{center} 
\end{figure}
\vspace*{-5mm}
%%%%%%%%%%%%%%%%%%%%%%%%%%%%%%%%%%%%%%%%%%%%%%%%%%%%%%%%%%%%%%%%%%%%%%%%%%%%%%%
%

%
Figure~\ref{fig:omegaspec} compares preliminary spectra of $\,\Omega$ 
hyperons \cite{STAR03omega} with hydrodynamic predictions 
\cite{KR03,HKHRV01}. 
For this comparison the original calculations for 130\,$A$\,GeV Au+Au 
collisions \cite{HKHRV01} were repeated with RHIC2 initial conditions 
and a chemical non-equilibrium equation of state in the hadronic 
phase \cite{KR03}. 
Following a suggestion that $\Omega$ hyperons, being heavy and not 
having any known strong coupling resonances with pions, should not be 
able to participate in any increase of the radial flow during the 
hadronic phase and thus decouple early \cite{HSN98}, 
Fig.~\ref{fig:omegaspec} shows two solid lines, the steeper one 
corresponding to decoupling at $\edec\eq0.45$\,GeV/fm$^3$, i.e. directly 
after hadronization at $\Tc$, whereas the flatter one assumes decoupling 
together with pions and other hadrons at $\edec\eq0.075$\,GeV/fm$^3$.
The data clearly favor the flatter curve, suggesting intense rescattering
of the $\Omega$'s in the hadronic phase.
The microscopic mechanism for this rescattering is still unclear.
However, without hadronic rescattering the hydrodynamic model, in spite
of its perfect local thermalization during the early expansion stages, 
is unable to generate enough transverse flow to flatten the $\Omega$
spectra as much as required by the data. 
Partonic hydrodynamic flow alone can not explain the $\Omega$ spectrum.
Similar comments apply to the $\Omega$ and $\bar\Omega$ spectra from 
Pb+Pb collisions at the SPS shown in Fig.~\ref{F7}.
We close this subsection by concluding that the freeze-out parameters
extracted from flow fits to the measured spectra are {\em dynamically
consistent}, i.e. they can be reproduced in hydrodynamic calculations
of the fireball evolution using realistic initial conditions. The
successful hydrodynamic calculations require fast thermalization
at $\tauequ{\,\lapp\,}1$\,fm/$c$ and high initial energy density,
$e_\equ\gapp20$\,GeV/fm$^3$. This conclusion will be solidified by the
study of elliptic flow in the next subsection. The hydrodynamically 
generated strong transverse flow naturally explains the much flatter 
proton than pion spectra at low $\pperp$ (see Fig.~\ref{fig:spectra130}).
Together with the approximate baryon/antibaryon symmetry at RHIC
($\bar p/p{\,\approx\,}0.7$ in Au+Au collisions) this provides a 
simple understanding for the initially puzzling observation that 
in Au+Au collisions at RHIC for $\pperp\gapp2$\,GeV/$c$ 
{\em antiprotons are as abundant as negative pions} (upper left panel
in Fig.~\ref{fig:spectra130})! This is completely unexpected from standard
string and jet fragmentation phenomenology in $pp$ and $e^+e^-$ collisions 
where the $\bar p/\pi^-$ ratio never exceeds 10-15\%, but a simple (albeit
dramatic) consequence of the collective transverse flow in heavy-ion 
collisions.

%%%%%%%%%%%%%%%%%%%%%%%%%%%%%%%%%%%%%%%%%%%%%%%%%%%%%%%%%%%%%%%%%%%%%%%%%%%%%
\subsection{Elliptic flow}
\label{sec4b}
%%%%%%%%%%%%%%%%%%%%%%%%%%%%%%%%%%%%%%%%%%%%%%%%%%%%%%%%%%%%%%%%%%%%%%%%%%%%%%
%%%%%%%%%%%%%%%%%%%%%%%%%%%%%%%%%%%%%%%%%%%%%%%%%%%%%%%%%%%%%%%%%%%%%%%%%%%%
\subsubsection{Anisotropic flow in non-central collisions}
\label{sec4b1}
%%%%%%%%%%%%%%%%%%%%%%%%%%%%%%%%%%%%%%%%%%%%%%%%%%%%%%%%%%%%%%%%%%%%%%%%%%%%%

\begin{figure}[ht]
\begin{center}
\begin{minipage}[h]{4cm}
\vspace*{-5mm}
\epsfig{file=radial_flow.eps,width=40mm}
\end{minipage}
\hspace*{5mm} 
\begin{minipage}[h]{5cm}
\vspace*{4mm}
\epsfig{file=aniso_flow.eps,width=45mm,angle=-90}
\end{minipage}
\begin{minipage}[h]{6cm}
\vspace*{1mm}
\epsfig{file=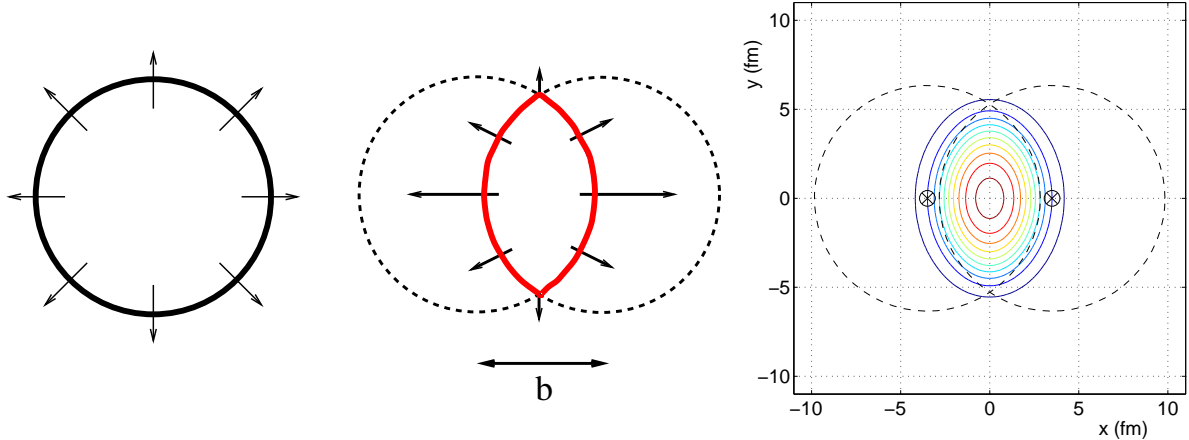,width=60mm}
\end{minipage}
\caption{Left and center panels: Schematic picture of radial and elliptic 
flow. Right panel: Initial energy density contours in the transverse plane
for a Au+Au collision at impact parameter $b\eq7$\,fm. The dashed circles
indicate transverse projections of the colliding nuclei. 
\label{F11}
\vspace*{-7mm}
}
\end{center}
\end{figure}

For central collisions ($b\eq0$) between equal spherical nuclei, radial 
flow (left panel of Fig.~\ref{F11}) is the only possible type of 
transverse flow allowed by symmetry. 
%
%%%%%%%%%%%%%%%%%%%%%%% Fig. 11 %%%%%%%%%%%%%%%%%%%%%%%%%%%%%%%%%%%%%%%%%%
%%%%%%%%%%%%%%%%%%%%%%%%%%%%%%%%%%%%%%%%%%%%%%%%%%%%%%%%%%%%%%%%%%%%%%%%
%
In non-central ($b{\,\ne\,}0$) collisions between spherical nuclei (right
panels of Fig.~\ref{F11}), or in $b\eq0$ collisions between suitably 
aligned deformed nuclei (such as $^{238}$U), this azimuthal symmetry 
is broken and anisotropic transverse flow patterns can develop. The 
overlap region of the two colliding nuclei is then spatially deformed
in the transverse plane, i.e. the {\em spatial eccentricity} of the 
reaction zone
\beq{equ:epsilonx}
  \varepsilon_x(b) = \frac{\la y^2 - x^2 \ra}{\la y^2+x^2 \ra} 
\end{equation}
(where the average is taken with the energy density distribution) is 
initially non-zero and positive. Rescattering processes among the produced
particles transfer this spatial deformation onto momentum space, i.e.
the initially locally isotropic transverse momentum distribution of the
produced matter begins to become anisotropic. The driving force for
this momentum anisotropy is the spatial eccentricity; the momentum 
anisotropy can only grow as long as $\varepsilon_x{\,>\,}0$. As a function 
of time $\varepsilon_x$ decreases, either spontaneously due to free-streaming
radial expansion (if no rescattering happens) or somewhat more quickly
due to the development of elliptic flow (if rescattering occurs) which 
makes the system expand faster into the reaction plane than perpendicular 
to it (see middle panel in Fig.~\ref{F11}). Once the spatial eccentricity 
has disappeared, the momentum anisotropy saturates \cite{Sorge97}. Since 
this happens quite early in the collision (as we will see shortly), the 
finally observed momentum anisotropy opens a window onto the early stage 
of the fireball expansion.

%
%%%%%%%%%%%%%%%%%%%%%%% Fig. 12 %%%%%%%%%%%%%%%%%%%%%%%%%%%%%%%%%%%%%%%%%%
\begin{figure}[ht]
\begin{center}
\epsfig{file=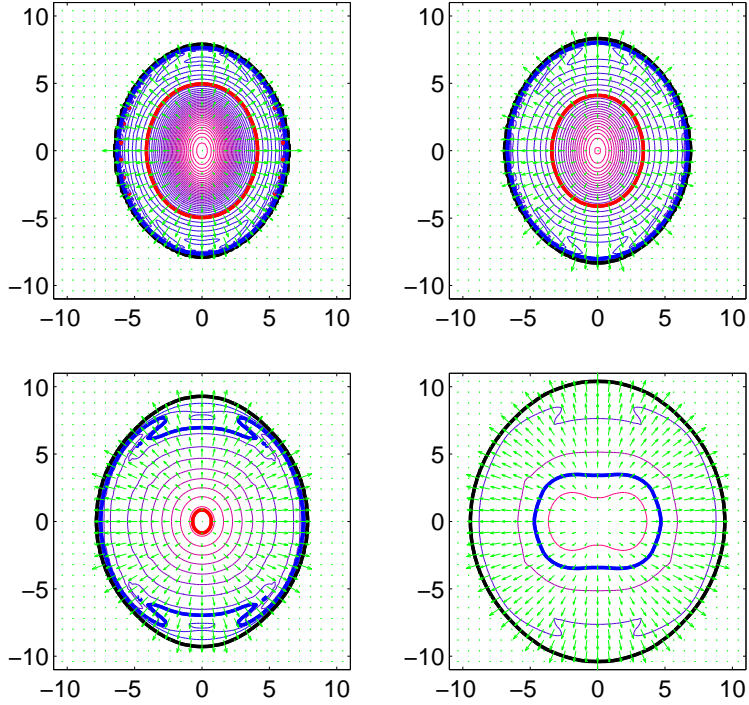,bb=73 202 511 616,width=100mm}
\end{center}
\vspace*{-5mm}
\caption{Snapshots of the energy density contours (lines) and velocity
profiles (small arrows) taken at $\tau{-}\tauequ\eq3.2$, 4.0, 5.6 and
8.0\,fm/$c$ after initialization, for Pb+Pb collisions at $b\eq7$\,fm 
for an initial central temperature $T_\equ\eq500$\,MeV (from 
Ref.~\cite{KSH00}). The outer thick (black) line indicates the freeze-out 
surface at $\edec$, the next thick (blue) line separates the hadron 
resonance gas from the enclosed mixed phase, and the innermost thick (red)
line separates the mixed phase from the quark-gluon plasma in the center.
The quark-gluon plasma phase disappears shortly after the third 
snapshot at $\tau{-}\tauequ\eq5.6$\,fm/$c$ (lower left panel). The
horizontal and vertical axes are $x$ and $y$ in fm, respectively (as in 
Fig.~\protect\ref{F11}).
%\vspace*{-3mm}
\label{F12}
}
\end{figure}
%%%%%%%%%%%%%%%%%%%%%%%%%%%%%%%%%%%%%%%%%%%%%%%%%%%%%%%%%%%%%%%%%%%%%%%%
%
The mechanism for the creation of elliptic flow is most easily understood
in the hydrodynamic limit: After thermalization there is high pressure in
the interior of the reaction zone which falls off to zero outside. 
Obviously the pressure gradient is steeper in the short direction, 
leading to stronger hydrodynamic acceleration into the reaction plane. 
This is seen in Fig.~\ref{F12} where I show hydrodynamically computed 
energy density contours and flow vector fields for non-central Pb+Pb 
collisions for 4 different times after the onset of hydrodynamic expansion.
One sees that the reaction zone expands faster into the reaction plane
and that the spatial eccentricity is almost gone after about 8\,fm/$c$.

\begin{figure}[ht] 
\begin{center}
\begin{minipage}[h]{8.5cm}
  \epsfig{file=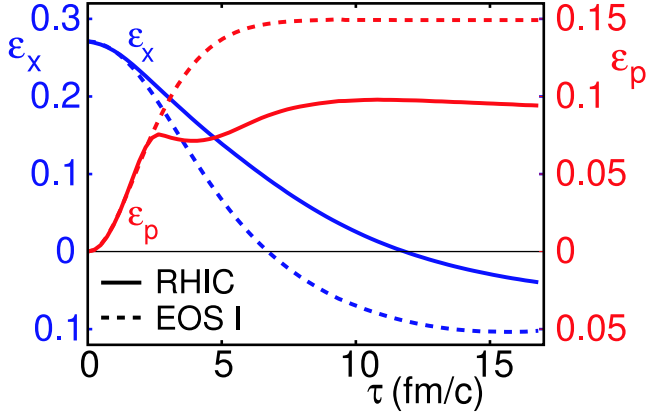,width=8.5cm}  
\end{minipage}
\hspace*{5mm}
\begin{minipage}[h]{6cm}
\vspace*{-6mm}
\caption{Time evolution of the spatial eccentricity $\varepsilon_x$ 
         and the momentum anisotropy $\varepsilon_p$ for Au+Au collisions 
         with $b\eq7$\,fm at RHIC and at much higher collision energies
         (from Ref.~\protect\cite{KH03}). 
\label{F13} 
} 
\end{minipage}
\end{center} 
\vspace*{-6mm}
\end{figure}

Figure~\ref{F13} shows the time evolution of the spatial eccentricity
$\varepsilon_x$ and of the momentum anisotropy
\beq{equ:epsilonpdef}
     \varepsilon_p(\tau) = \frac{\int dx dy \, (T^{xx}-T^{yy})}
                        {\int dx dy \, (T^{xx}+T^{yy})}\;.
\end{equation}
Note that with these sign conventions, the spatial eccentricity is 
positive for out-of-plane elongation (as is the case initially)
whereas the momentum anisotropy is positive if the preferred flow 
direction is {\em into} the reaction plane. The calculations were
done for Au+Au collisions at impact parameter $b\eq7$\,fm, for RHIC 
initial conditions with a realistic equation of state (EOS~Q, solid 
lines) and for a much higher initial energy density (initial 
temperature at the fireball center =\,2\,GeV) with a massless 
ideal gas equation of state (EOS~I, dashed lines), relevant for 
future experiments at the LHC \cite{KH03}. 

%
%%%%%%%%%%%%%%%%%%%%%%%%% Fig. 13 %%%%%%%%%%%%%%%%%%%%%%%%%%%%%%%%%%%%%%%%%
%%%%%%%%%%%%%%%%%%%%%%%%%%%%%%%%%%%%%%%%%%%%%%%%%%%%%%%%%%%%%%%%%%%%%%%
%
The initial spatial asymmetry at this impact parameter is 
$\varepsilon_x(\tau_{\rm equ})\eq0.27$, and clearly
$\varepsilon_p(\tau_{\rm equ})\eq0$ since the fluid is initially at 
rest in the transverse plane.
The spatial eccentricity is seen to disappear before the fireball 
matter freezes out, in particular for the case with the very high
initial temperature (dashed lines) where the source is seen to
switch orientation after about 6\,fm/$c$ and becomes in-plane-elongated
at late times~\cite{HK02osci}.
One also sees that the momentum anisotropy $\varepsilon_p$ saturates 
at about the same time when the spatial eccentricity $\varepsilon_x$ 
vanishes; all of the momentum anisotropy is built up during the 
first 5\,fm/$c$.

At RHIC the saturation happens even earlier because part of the matter 
begins to hadronize even before the spatial eccentricity has fully 
disappeared. 
Near the phase transition (in particular if as in our case it is modelled 
as a first order transition) the equation of state becomes very soft, and 
this inhibits the generation of transverse flow.
This also affects the generation of transverse flow {\em anisotropies} 
as seen from the solid curves in Figure~\ref{F13}:
The rapid initial rise of $\varepsilon_p$ suddenly stops as a significant
fraction of the fireball matter enters the mixed phase. 
It then even decreases somewhat as the system expands radially without 
further acceleration, thereby becoming more isotropic in both coordinate
and momentum space. 
Only after the phase transition is complete and pressure gradients 
reappear, the system reacts to the remaining spatial eccentricity
by a slight further increase of the momentum anisotropy.
The softness of the equation of state near the phase transition 
thus focusses the generation of anisotropic flow to even earlier
times, when the system is still entirely partonic and has not even
begun to hadronize.
At RHIC energies this means that almost all of the finally observed
elliptic flow is created during the first 3-4 fm/$c$ of the collision
and reflects the hard QGP equation of state of an ideal gas of
massless particles ($c_s^2\eq\frac{1}{3}$) \cite{KSH00}. 
Microscopic kinetic studies of the evolution of elliptic flow lead to 
similar estimates for this time scale \cite{Sorge97,Sorge99,ZGK99,MG02}.
% 

%%%%%%%%%%%%%%%%%%%%%%%%%%%%%%%%%%%%%%%%%%%%%%%%%%%%%%%%%%%%%%%%%%%%%%%%%%%%
\subsubsection{The elliptic flow coefficient $v_2$}
\label{sec4b2}
%%%%%%%%%%%%%%%%%%%%%%%%%%%%%%%%%%%%%%%%%%%%%%%%%%%%%%%%%%%%%%%%%%%%%%%%%%%%%

The momentum anisotropy $\varepsilon_p$ manifests itself as an azimuthal
anisotropy of the measured hadron spectra. One quantifies this
anisotropy in terms of the azimuthal Fourier coefficients of the
transverse momentum spectrum:
\beq{equ:fourierexpansion}
    \frac{dN_i}{dy\,\pperp d\pperp\,d\phi_p}(b) 
  = \frac{1}{2 \pi}\frac{dN_i}{dy\,\pperp d\pperp}(b) 
    \Bigl( 1 + 2 \,v_2^i(\pperp,b) \cos(2\phi_p) + \dots\Bigr) \,.
\end{equation}
These coefficients depend on the impact parameter $b$, the transverse 
momentum $\pperp$ and the particle species $i$ (through their rest mass
$m_i$). We have suppressed the dependence on rapidity $y$ since the 
calculation assumes longitudinal boost-invariance and thus makes meaningful
statements only near $y\eq0$. At midrapidity $v_2$ is (for collisions 
between equal nuclei) the lowest nonvanishing Fourier coefficient. 
Reflection symmetry with respect to the reaction plane forbids the 
appearance of sine terms in the expansion. 
Equation~(\ref{equ:fourierexpansion}) implies that the elliptic flow 
coefficient can be calculated as 
\beq{v2}
  v_2^i(\pperp,b) = \la\cos(2\phi_p)\ra_{\pperp,b}^i
\eeq
were the average is performed with the transverse momentum spectrum of 
particle species $i$ at fixed impact parameter and $\pperp$. If the 
average is performed with the $\pperp$-integrated spectrum one obtains
the ``$\pperp$-integrated elliptic flow'' $v_2^i(b)$. 

%
%%%%%%%%%%%%%%%%%%%%%%%%% Fig. 14 %%%%%%%%%%%%%%%%%%%%%%%%%%%%%%%%%%%%%%%%%
\begin{figure}[ht] 
\begin{center}
  \epsfig{file=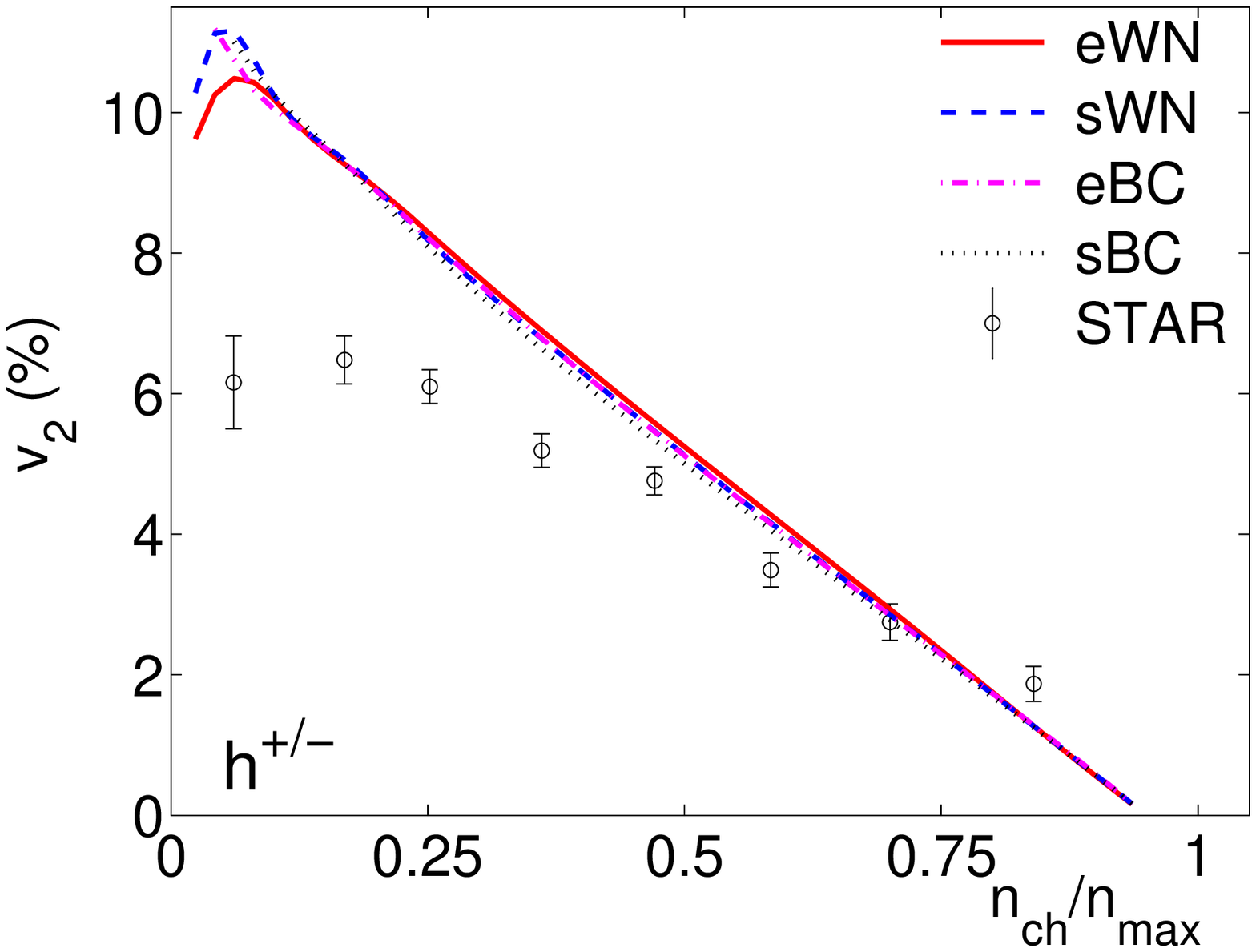,bb=61 210 568 594,width=7cm}  
  \epsfig{file=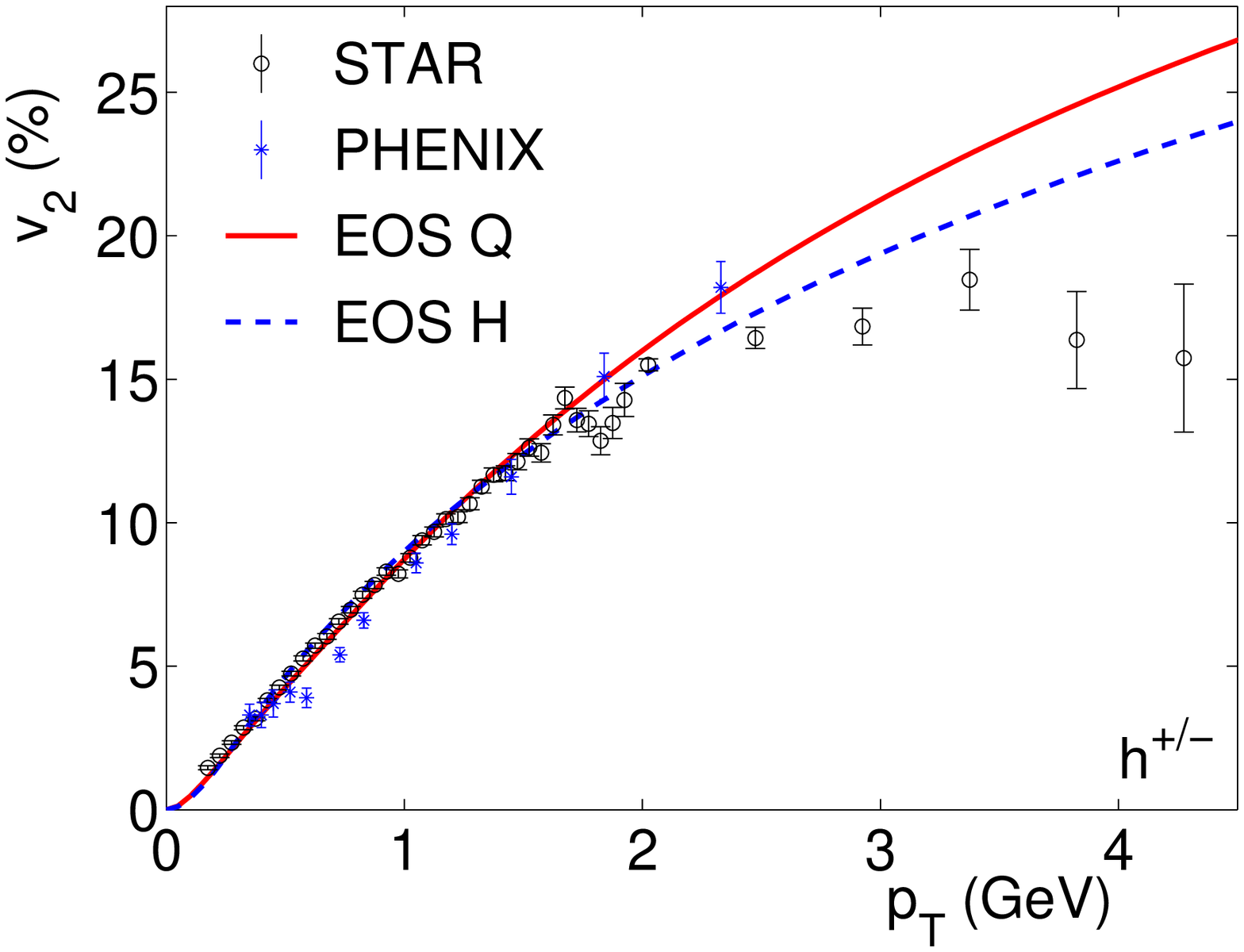,bb=62 207 568 594,width=7cm}  
\end{center}
\vspace*{-2mm}
\caption{Elliptic flow of all charged hadrons in 130\,$A$\,GeV Au+Au 
         collisions at RHIC. Left: $\pperp$-integrated elliptic flow
         vs. collision centrality (measured by the normalized total
         charged particle multiplicity $n_{\rm ch}$). Right: 
         differential elliptic flow $v_2(\pperp)$ for minimum bias
         collisions (i.e. inegrated over all impact parameters).
         The curves are hydrodynamic calculations for various 
         parametrizations of the initial transverse density profile
         (for details see \protect\cite{KHHET01}). Data are from the
         STAR \protect\cite{STAR01v2} and PHENIX 
         \protect\cite{PHENIX02v2overeps} Collaborations.
\label{F14} 
} 
\vspace*{-2mm}
\end{figure} 
%%%%%%%%%%%%%%%%%%%%%%%%%%%%%%%%%%%%%%%%%%%%%%%%%%%%%%%%%%%%%%%%%%%%%%%
%

Figure~\ref{F14} shows a comparison of elliptic flow data for charged
hadrons from Au+Au collisions at RHIC with hydrodynamic calculations.
The hydrodynamic results are genuine {\em predictions} since all model
parameters have been fixed in central ($b\eq0$) collisions where $v_2$
vanishes. The impact parameter dependence of $v_2(\pperp)$ then follows 
from the overlap geometry and its implications for the initial 
transverse density profile and its spatial eccentricity \cite{KHHET01},
without additional parameters. The data follow the hydrodynamic predictions
up to impact parameters $b\gapp7$\,fm ($n_{\rm ch}/n_{\rm max}\lapp0.5$
in Fig.~\ref{F14}a) and up to transverse momenta $\pperp\lapp2$\,GeV/$c$
(Fig.~\ref{F14}b). For $\pperp\gapp2$\,GeV/$c$ (note that this concerns
less than 0.5\% of all produced hadrons!) the measured elliptic flow 
saturates and falls behind the hydrodynamically predicted continuous
rise with $\pperp$. This indicates a gradual breakdown of local 
thermal equilibrium for high-$\pperp$ particles -- not a big surprise.
Similarly, the elliptic flow in peripheral collisions is also smaller
than predicted -- again presumably a consequence of incomplete local
thermalization in very peripheral collisions where the nuclear overlap
zone becomes quite small.  

Hydrodynamics predicts a clear mass-ordering of elliptic flow 
\cite{HKHRV01}. As the collective radial motion boosts particles to 
higher average velocities, heavier particles gain more momentum than 
lighter ones, leading to the previously discussed flattening of their 
spectra at low transverse kinetic energies.
When plotted against $\pperp$ this effect is further enhanced by a 
kinematic factor arising from the transformation from $\mperp$ to 
$\pperp$ (see Fig.~\ref{fig:spectra130}).
This flattening reduces the momentum anisotropy coefficient $v_2$ 
at low $\pperp$ \cite{HKHRV01}, and the heavier the particle the more
the rise of $v_2(\pperp)$ is shifted towards larger $\pperp$.
This effect, which is a consequence of both the thermal shape of the 
single-particle spectra at low $\pperp$ and the superimposed collective 
radial flow, has been nicely confirmed by the experiments:
%
%
%%%%%%%%%%%%%%%%%%%%%%%%% Fig. 15 %%%%%%%%%%%%%%%%%%%%%%%%%%%%%%%%%%%%%%%%%
\begin{figure}[ht] 
\begin{center}
  \epsfig{file=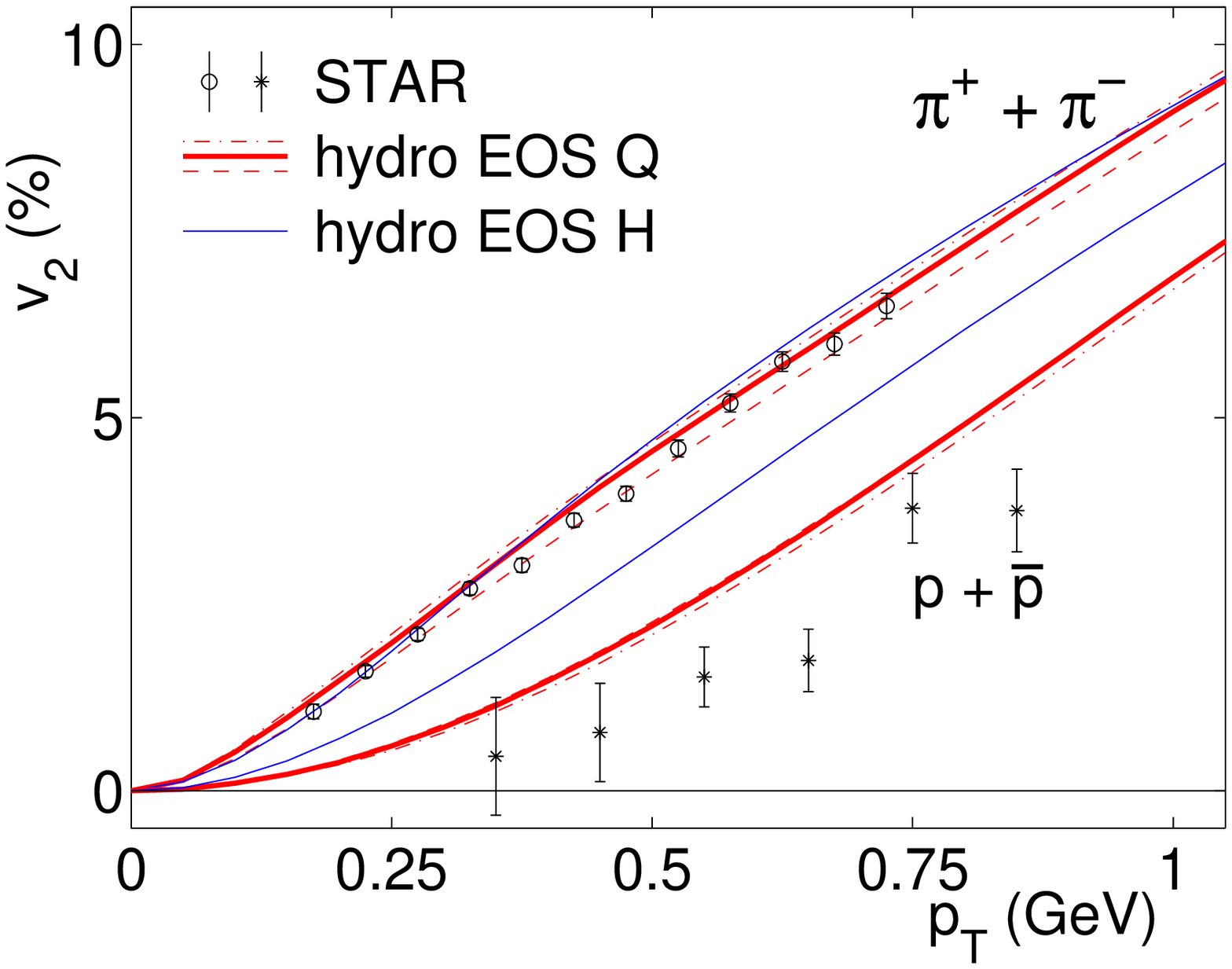,bb=77 209 568 594,width=6.5cm,height=5.4cm}  
  \epsfig{file=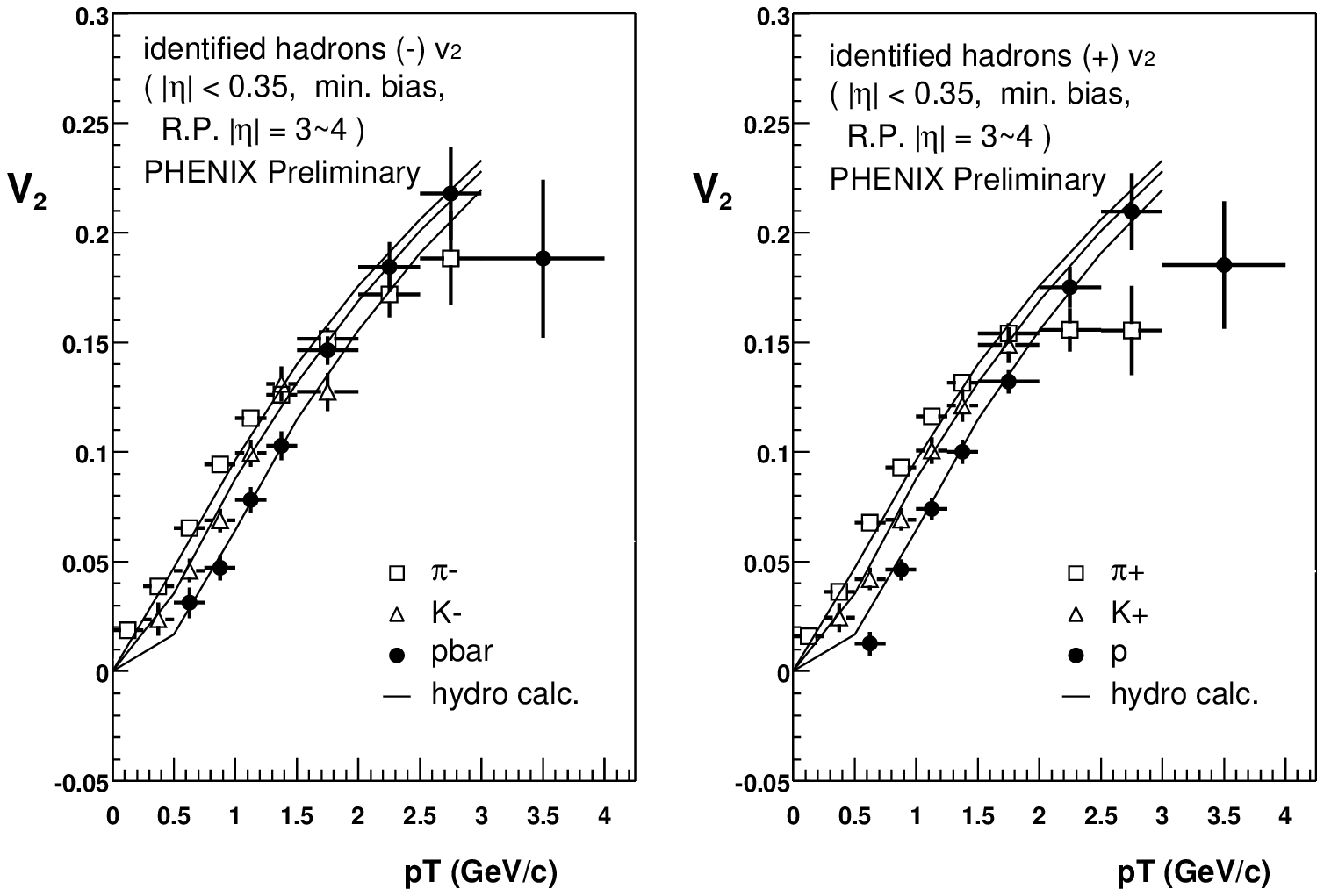,width=9cm,height=6cm}  
\end{center}
\vspace*{-2mm}
\caption{Differential elliptic flow for identified pions, kaons and 
         (anti)protons from Au+Au collisions at $\scm\eq130$\,GeV 
         \protect\cite{STAR01v2piKp} (left) and at $\scm\eq130$\,GeV 
         \protect\cite{PHENIX02v2id200} (right), together with 
         hydrodynamic calculations \protect\cite{HK02WWND}.
\label{F15} 
} 
\vspace*{-2mm}
\end{figure} 
%%%%%%%%%%%%%%%%%%%%%%%%%%%%%%%%%%%%%%%%%%%%%%%%%%%%%%%%%%%%%%%%%%%%%%%
%
Figure~\ref{F15} shows that the data \cite{STAR01v2piKp,PHENIX02v2id200} 
follow the predicted mass ordering out to transverse momenta of about 
1.5\,GeV/$c$.
For $K_s^0$ and $\Lambda+\bar\Lambda$ very accurate data have recently 
become available from 200\,$A$\,GeV Au+Au collisions \cite{Sorensen}, 
again in quantitative agreement with hydrodynamic calculations up to 
$\pperp{\,\simeq\,}1.5$\,GeV/$c$ for kaons and up to 
$\pperp{\,\simeq\,}2.5$\,GeV/$c$ for $\Lambda+\bar\Lambda$.
The inversion of the mass-ordering in the data at large $\pperp$ is caused 
by the mesons whose $v_2(\pperp)$ breaks away from the hydrodynamic rise 
and begins to saturate at $\pperp\gapp1.5$\,GeV/$c$.
In contrast, baryons appear to behave hydrodynamically to 
$\pperp{\,\simeq\,}2.5$\,GeV/$c$, breaking away from the flow prediction
and saturating at significantly larger $\pperp$ than the mesons.
This is consistent with the idea that the hadronic elleptic flow really 
reflects a partonic elliptic flow already established before hadronization, 
that the latter exhibits a hydrodynamic rise at low $\pperp$ followed 
by saturation above $\pperp{\,\simeq\,}750-800$\,MeV/$c$,
and that these features are transferred to the observed hadrons by quark 
coalescence, manifesting themselves there at twice resp. three times
larger $\pperp$-values \cite{MV03}.

The comparison of the $v_2$ data with the hydrodynamic model makes the 
implicit assumption that all of the measured $v_2$ is collective, i.e.
that non-flow contributions to the angular average 
$v_2\eq\la\cos(2\phi_p)\ra$ can be eliminated. Various experimental 
cuts \cite{STAR01v2} ensure that the influence of non-collective
2-particle correlation effects (jet correlations \cite{KT02}, resonance 
decays, HBT effects) on $v_2$ are minimized. An alternative method to 
extract $v_2$ from higher-order cumulants of the azimuthal particle 
distribution \cite{BDO00a,BDO01} has recently shown \cite{STARPRC02v2}
that, in the interesting range $\pperp\lapp2$\,GeV/$c$ where the emitted
hadrons behave hydrodynamically, non-flow corrections do not exceed
15\% of the value extracted from the prescription 
$v_2\eq\la\cos(2\phi_p)\ra$. (This may be different at much higher 
$\pperp$ where $v_2$ from 2-particle angular correlations seems to 
approach a $\pperp$-independent constant value \cite{STAR03v2highpt} 
but statistics is not good enough for a 4-particle cumulant analysis 
which could exclude that at high $\pperp$ an increasing fraction of 
this $v_2$ stems from angular 2-particle jet correlations.) Further
support of the collective nature of the measured $v_2$ comes from the
observation that the event plane angles reconstructed from $v_2$ at 
forward and backward rapidities are, within the statistical error, 
perfectly correlated. We can therefore assume that at least 85\% of
the measured $v_2$ values shown in Figs.~\ref{F14}--\ref{F16} are
due to collective flow.

%%%%%%%%%%%%%%%%%%%%%%%%%%%%%%%%%%%%%%%%%%%%%%%%%%%%%%%%%%%%%%%%%%%%%%%%%%%%
\subsubsection{Rapid thermalization and early pressure -- the creation of 
the Quark-Gluon Plasma}
\label{sec4b3}
%%%%%%%%%%%%%%%%%%%%%%%%%%%%%%%%%%%%%%%%%%%%%%%%%%%%%%%%%%%%%%%%%%%%%%%%%%%%%

I already mentioned that momentum anisotropies, in particular elliptic flow,
can only build as long as the source is spatially deformed. Even if the
created particles don't rescatter but simply fly off ballistically 
with their isotropically distributed initial momenta, the spatial 
eccentricity decreases as a function of time as \cite{KSH00}
\beq{equ:epsilondepletion}
  \frac{\varepsilon_x(\tau_0{+}\Delta\tau)}{\varepsilon_x(\tau_0)}
  = \left[ 
  1+\frac{(c\,\Delta\tau)^2}{\la\br^2\ra_{\tau_0}} \right]^{\!-1}\,,
\end{equation}
where $\tau_0$ is the time when the particles were created and 
$\la\br^2\ra_{\tau_0}$ is the azimuthally averaged initial
transverse radius squared of the reaction zone. So if thermalization is
delayed by a time $\Delta\tau$, any elliptic flow would have to build on
a reduced spatial deformation and would come out smaller.

\begin{figure}[ht] 
\begin{center}
\begin{minipage}[h]{8.5cm}
  \epsfig{file=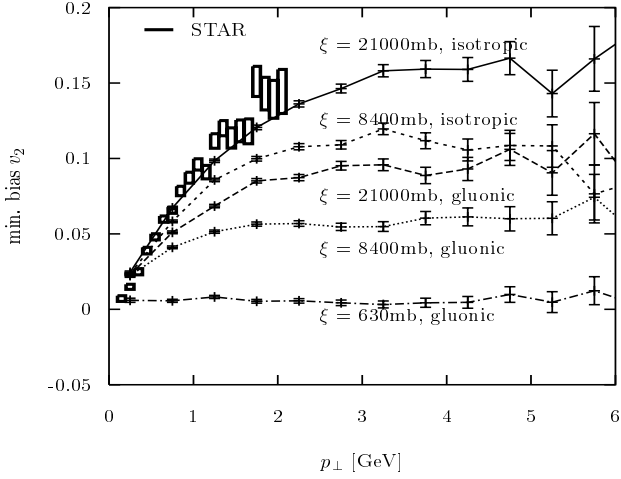,width=8.5cm,height=6.5cm}
\end{minipage}
\hspace*{5mm}
\begin{minipage}[h]{6cm}
\vspace*{-10mm}
\caption{Impact parameter averaged elliptic flow as a function of 
         transverse momentum. The experimental data points from 
         STAR\protect\cite{STAR01v2} are compared with parton cascade  
         calculations \protect\cite{MG02} with varying transport 
         opacities $\xi$ (from \protect\cite{MG02}).
\label{F16} 
} 
\end{minipage}
\end{center} 
\vspace*{-7mm}
\end{figure}

It is also known from microscopic kinetic studies \cite{ZGK99,MG02} that,
for a given collision geometry, i.e. a given initial spatial eccentricity
$\varepsilon_x$, the magnitude of the generated elliptic flow is a monotonic 
function of the mean free path (or the product of density and scattering 
cross section) in the fireball. This is shown in Fig.~\ref{F16}. 
%
%%%%%%%%%%%%%%%%%%%%%%%%% Fig. 16 %%%%%%%%%%%%%%%%%%%%%%%%%%%%%%%%%%%%%%%%%
%%%%%%%%%%%%%%%%%%%%%%%%%%%%%%%%%%%%%%%%%%%%%%%%%%%%%%%%%%%%%%%%%%%%%%%
%
Without reinteractions (i.e. for infinite mean free path) no momentum 
anisotropies develop at all \cite{ZGK99}; the maximal value for $v_2$ 
is reached in the hydrodynamic limit of zero mean free path \cite{HK02}. 

This hydrodynamic upper limit for the charged particle elliptic flow 
was shown to be insensitive to shape details of the initial transverse
density distribution \cite{KHHET01} and, as long as the shape and 
normalization of the single particle spectra from central collisions 
were held fixed by proper retuning of the initial conditions, any 
sensitivity to details of the QGP equation of state remained below 25\%
\cite{HKHRV01}. (Note that this statement refers to the $v_2$ of all 
charged particles which are strongly dominated by pions and kaons; 
Fig.~\ref{F15}a shows that the elliptic flow of heavier particles such 
as protons, which are more sensitive to radial flow, shows indeed 
significant dependence on the equation of state, and that the data 
disfavor, for example, a pure hadron resonance gas over an equation 
of state with a deconfining phase transition and a hard QGP equation 
of state above $\Tc$.)  

The observation that for $b\lapp8$\,fm and $\pperp\lapp1.5-2$\,GeV/$c$
the measured $v_2$ exhausts at least 85\% of the hydrodynamically
predicted value thus puts severe constraints on the dynamical evolution
and on our global picture of the collision event. It requires early 
thermalization (such that the spatial eccentricity does not decay before 
elliptic flow starts to build) and almost complete local equilibrium 
throughout the crucial first 5\,fm/$c$ of the expansion when the 
momentum anisotropies are generated. The hydrodynamic model 
calculations for RHIC were done with a thermalization time 
$\tauequ\eq0.6$\,fm/$c$; increasing $\tauequ$ beyond 1\,fm/$c$ begins 
to not only seriously degrade the description of the angle-averaged 
single-particle spectra, but also cuts down $v_2$ below the 
experimentally measured level. To obtain the full measured amount of
elliptic flow requires almost complete thermalization in less than
1\,fm/$c$.

The sensitivity of $v_2$ to any kind of deviation from local thermal 
equilibrium seems to be particularly strong \cite{Teaney03,HW02}. 
(Of course, we are talking here about the momentum distribution for 
particles with average momenta $\pperp{\,\sim\,}3\,T$; no claim for
thermalization can be made for particles with much larger than average 
$\pperp$ because they are too rare to affect bulk properties such as 
the pressure and the hydrodynamic expansion.) Turning this observation 
around and remembering that most of the momentum anisotropy is generated 
at energy densities well above the critical value for hadronization, the 
good agreement of the data with ideal fluid dynamics points to a very 
small viscosity of the quark-gluon plasma. (Note my use of the word
{\em quark-gluon plasma} for the matter in the early expansion stage!
What other name could I use for a thermalized system of strongly 
interacting matter at energy densities ${\,>\,}10\,e_{\rm cr}$?) 
The QGP does not behave as a weakly interacting quark-gluon gas, as 
suggested by naive perturbation theory, nor does it behave like 
viscous honey, as suggested by the value of the shear viscosity 
extracted from state-of-the-art resummed thermal field theory at 
leading order in the strong coupling constant \cite{AMY03}. The 
fact that the QGP behaves like an ideal fluid implies {\em strong 
non-perturbative interactions} in the quark-gluon plasma 
phase.

So here is our {\bf Second Lesson}: The strong measured elliptic flow 
at RHIC requires rapid thermalization at $\tauequ\lapp1$\,fm/$c$. At this
time the average energy density in the thermalized fireball is 
$e\gapp10-15\ec$, so thermalization leads into a quark-gluon plasma 
state. In Au+Au collisions at RHIC this QGP lives for about 5--7\,fm/$c$
before hadronizing.

So what about the SPS? Have we made quark-gluon plasma there, too
\cite{HJ00}? The energy densities $e\gapp3.5$\,GeV/fm$^3$ were sufficient
and, on the 30\% accuracy level, hydrodynamics did a good job for the
angle-averaged single particle spectra from S+S and Pb+Pb collisions
at the SPS, too \cite{SSH93,Schlei}. 
However, the measured elliptic flow at the SPS is only
about 50\% of the value measured at RHIC \cite{NA49-03v2,CERES02v2}
whereas hydrodynamics predicts it to be even slightly larger than at
RHIC \cite{KSH00}.\footnote{At the SPS most of the elliptic flow is 
generated in the hadronic phase, and the softening effects of the 
phase transition on the development of flow are not as important as at 
RHIC \cite{KSH00}.} So thermalization seems to be less efficient 
at the SPS than at RHIC. This may not be unreasonable, given the 
significantly lower (by about a factor 3 \cite{KSH00}) initial density 
and the rapid conversion of the matter to a less strongly coupled 
hadron resonance gas before the momentum anisotropy has been able to 
reach saturation. It would be good to test this with central U+U collisions 
at the lower end of the collision energy range of RHIC \cite{KSH00}. 
Such collisions provide similar spatial eccentricities $\varepsilon_x$ as
semiperipheral Au+Au collisions, but for a fireball which covers
a much larger transverse area. This should improve thermalization also
at lower energies, making hydrodynamics a better approximation, and help 
to test our understanding of the dynamical origin of elliptic flow. -- I 
should also mention that at RHIC hydrodynamics fails to reproduce the 
rapidity-dependence of elliptic flow, predicting a wide rapidity plateau 
for $v_{2}$ \cite{HT02} whereas the measured elliptic flow rapidly
decreases as one moves away from midrapidity \cite{PHOBOS02v2eta}.
The origin of this discrepancy is not yet clear, but I suspect that it
again indicates a breakdown of local equilibrium away from midrapidity
\cite{Hirano01v2etaRHIC}, for reasons which are related to the 
analogous breakdown at midrapidity at lower collision energies
\cite{HK04}.

%%%%%%%%%%%%%%%%%%%%%%%%%%%%%%%%%%%%%%%%%%%%%%%%%%%%%%%%%%%%%%%%%%%%%%%%%%%%%%
\section{TWO-PARTICLE BOSE-EINSTEIN CORRELATIONS AND HBT INTERFEROMETRY}
\label{secV}
%%%%%%%%%%%%%%%%%%%%%%%%%%%%%%%%%%%%%%%%%%%%%%%%%%%%%%%%%%%%%%%%%%%%%%%%%%%%%%

Single-particle momentum spectra, including their anisotropies, provide 
no direct information on the space-time structure of the source. It is 
therefore in principle possible to describe these spectra with models
that have the same momentum structure but completely different 
freeze-out distributions in space and time. Two-particle correlations,
on the other hand, depend on the average separation of the particles
at decoupling and therefore provide valuable spatial and temporal 
information. This is most easily seen for correlations generated by 
final state Coulomb interactions between charged hadrons which influence
their distribution of {\em relative} momenta depending on how close they
were at time of emission. The measured relative momentum distribution
of particle pairs thus allows to constrain their average space-time
distance at decoupling. Conceptually more involved, but theoretically 
easier to analyze are two-particle momentum correlations between pairs
of identical particles (``Bose-Einstein correlations'' in the case of 
identical bosons) caused by quantum statistical (wave function 
symmetrization) effects. Again, these affect the relative momentum 
distribution at low relative momenta, and the relative momentum range
of these quantum statistical correlations yields information
about the average space-time separation of the emission points of the 
two particles, and thereby also about the overall size and emission 
duration of the source. There exist a number recent reviews of these 
methods to which I refer the reader for more details 
\cite{HJ99,WH99,W99,TW02}.

%
%%%%%%%%%%%%%%%%%%%%%%%%% Fig. 16a %%%%%%%%%%%%%%%%%%%%%%%%%%%%%%%%%%%%%%%%%
\begin{figure}[ht] 
\begin{center}
  \epsfig{file=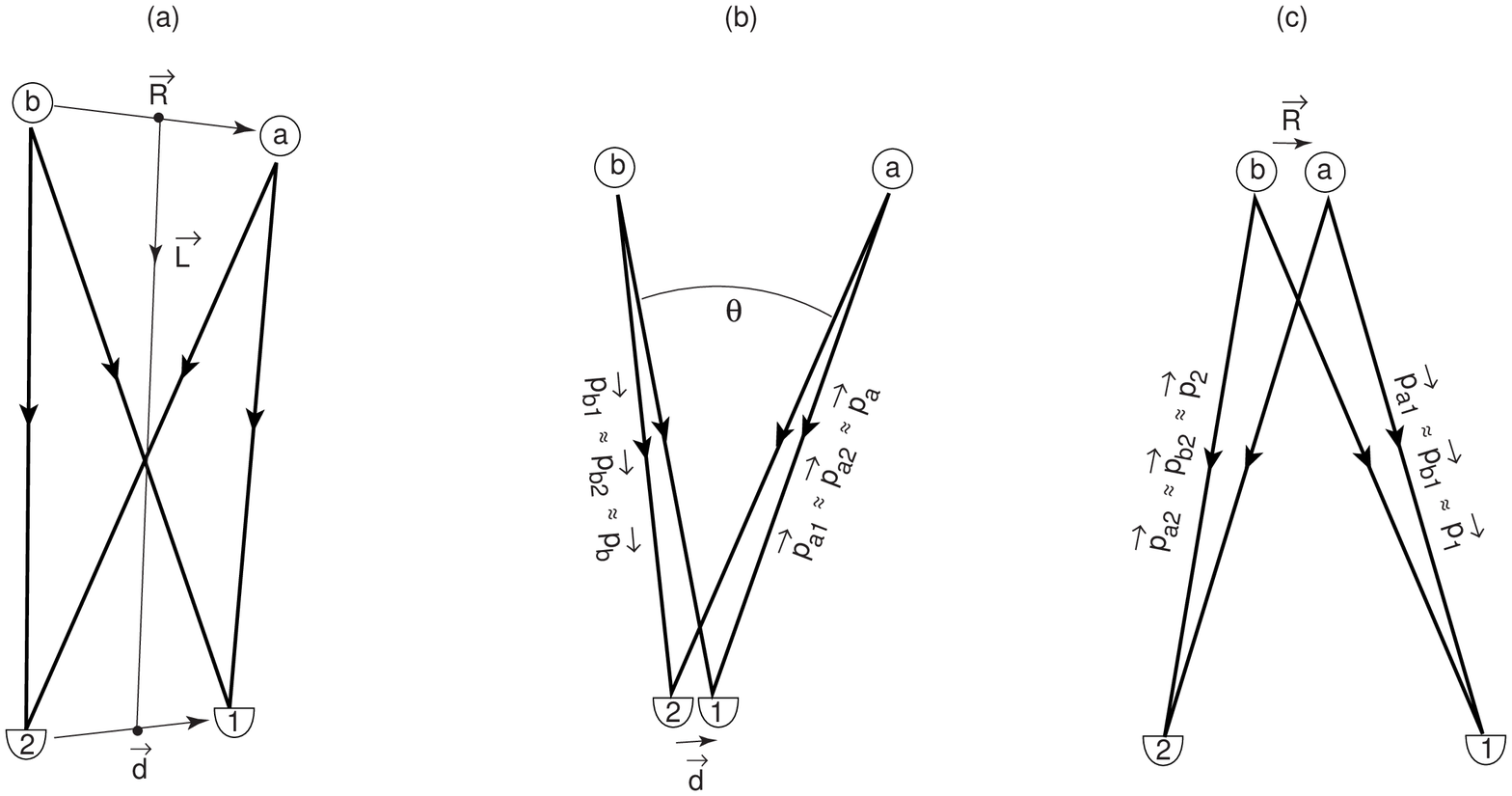,bb=470 278 648 600,width=6.5cm,angle=90,clip}
\end{center}
\vspace*{-7mm}
\caption{The principle of two-particle interferometry: a pair of 
indistinguishable particles is emitted from points $a$ and $b$ and
detected by detectors 1 and 2 which measure their momenta $\bp_1$ and
$\bp_2$. The total pair yield is obtained by integrating over all 
possible positions $(a,b)$ in the source. The measurement is performed
as a function of the average momentum $\bK=(\bp_1+\bp_2)/2$ of the
pair and its relative momentum $\bq=\bp_1-\bp_2$. The $\bq$-dependence
yields information about the average separation 
$\langle R^2 \rangle^{1/2}$ of the emitters $a$ and $b$ which contribute
pairs of given pair momentum $\bK$.
\label{F16a} 
} 
\vspace*{-2mm}
\end{figure} 
%%%%%%%%%%%%%%%%%%%%%%%%%%%%%%%%%%%%%%%%%%%%%%%%%%%%%%%%%%%%%%%%%%%%%%%
%

For pairs of identical bosons (e.g. pions), this technique is known 
as Hanbury Brown-Twiss (HBT) interferometry \cite{HBT54}. The basic 
principle, as applied in particle and nuclear physics, is illustrated 
in Figure~\ref{F16a}. I will now shortly outline its theoretical
treatment, referring to the literature \cite{HJ99,WH99,W99,TW02}
for more details.

%%%%%%%%%%%%%%%%%%%%%%%%%%%%%%%%%%%%%%%%%%%%%%%%%%%%%%%%%%%%%%%%%%%%%%%%%%%%%%
\subsection{Single-particle spectra and two-particle correlations}
\label{secVa}
%%%%%%%%%%%%%%%%%%%%%%%%%%%%%%%%%%%%%%%%%%%%%%%%%%%%%%%%%%%%%%%%%%%%%%%%%%%%%%

The covariant single- and two-particle distributions are defined by
 \begin{eqnarray}
   P_1(\bp) 
  & = & E\, \frac{dN}{d^3p} 
        = E \, \langle\hat{a}^+_{\bp} \hat{a}_{\bp}\rangle \, ,
 \label{1} \\
   P_2(\bp_1,\bp_2) 
  & = & E_1\, E_2\, \frac{dN}{d^3p_1 d^3p_2}
        = E_1 \, E_2\, 
          \langle\hat{a}^+_{\bp_1} \hat{a}^+_{\bp_2}
                 \hat{a}_{\bp_2} \hat{a}_{\bp_1} \rangle \, ,
 \label{2}
 \end{eqnarray}
where $\hat{a}^+_{\bp}$ ($\hat{a}_{\bp}$) creates (destroys) a
particle with momentum $\bp$. The angular brackets denote an 
ensemble average $\langle \hat O \rangle = {\rm tr}\, (\hat \rho \hat O)$
where $\hat \rho$ is the density operator associated with the 
ensemble. (When talking about an ensemble we may think of either a 
single large, thermalized source, or a large number of similar, but 
not necessarily thermalized collision events.) The single-particle 
spectrum is normalized to the average number of particles, $\langle N 
\rangle$, per collision, 
 \begin{equation}
 \label{norm1}
   \int {d^3p \over E}\, P_1(\bp) = \langle N \rangle \, ,
 \end{equation}
while the two-particle distribution is normalized to the number 
of particles in pairs, $\langle N (N-1) \rangle$, per event:
 \begin{equation}
 \label{norm2}
   \int {d^3p_1 \over E_1}\,{d^3p_2 \over E_2}\, P_2(\bp_1,\bp_2) 
   = \langle N (N-1) \rangle \, .
 \end{equation}
The two-particle correlation function is defined as 
 \begin{equation}
 \label{3}
   C(\bp_1,\bp_2)
   = \frac{P_2(\bp_1,\bp_2)}{P_1(\bp_1)P_1(\bp_2)} \, .
 \end{equation}
If the two particles are emitted independently and final state 
interactions are neglected one can prove a generalized Wick 
theorem \cite{He96,HSZ01}
 \begin{equation}
 \label{corr}
  C(\bp_1, \bp_2) = 1 \pm 
  {\vert \langle \hat a^+_{\bp_1} \hat a_{\bp_2} \rangle \vert^2
   \over
   \langle \hat a^+_{\bp_1} \hat a_{\bp_1} \rangle 
   \langle \hat a^+_{\bp_2} \hat a_{\bp_2} \rangle } \, .
 \end{equation}
This assumption is often called ``chaoticity assumption'', having in 
mind a completely chaotic (uncorrelated) emitter, in contrast to 
processes where the phases of the wave functions of the two emitted 
particles are to some extent correlated with each other. The degree 
of ``chaoticity'' or ``phase correlation'' in the source can be studied
by comparing three-particle correlations with two-particle ones 
\cite{Heinz:1997mr}. This was recently done by the STAR Collaboration 
\cite{STAR3pi} who established that in central Au+Au collisions at 
$\scm\eq130$\,GeV the pion emitting source is indeed completely chaotic.

From now on I will assume that the emitted particles are bosons, 
e.g. pions, and use the $+$ sign in Eq.~(\ref{corr}). We next discuss 
the numerator of the Bose-Einstein correlation term, 
$\vert \langle \hat a^+_{\bp_1} \hat a_{\bp_2} \rangle \vert^2$.

%%%%%%%%%%%%%%%%%%%%%%%%%%%%%%%%%%%%%%%%%%%%%%%%%%%%%%%%%%%%%%%%%%%%%%
\subsection{Source Wigner function and spectra} 
\label{secVb}
%%%%%%%%%%%%%%%%%%%%%%%%%%%%%%%%%%%%%%%%%%%%%%%%%%%%%%%%%%%%%%%%%%%%%%

As reviewed in Refs.~\cite{WH99,He96,HSZ01}, the pion emitting source 
can be described in terms of classical currents $J(x)$ which act as 
classical sources of freely propagating pions. These currents represent 
a parametrization of the last collision from which the free outgoing 
pion emerges. Using this language in one form or another 
\cite{WH99,He96,HSZ01}, one can express the ``exchange amplitude''
$\langle \hat a^+_{\bp_1} \hat a_{\bp_2} \rangle$ as the Fourier
transform of the single-particle Wigner density $S(x,K)$ which is
the quantum mechanical analogue of the classical phase-space density
at pion freeze-out, describing the probability of emitting a pion with 
momentum $K$ from space-time point $x$:
 \begin{equation}
 \label{SxK}
   \langle \hat a^+_{\bp_1} \hat a_{\bp_2} \rangle = \int d^4x\,
   S(x,K)\,e^{iq\cdot x}.
 \end{equation}
$S(x,K)$ is also called ``emission function''. 
$K={1\over 2}(E_1{+}E_2,\bp_1{+}\bp_2)$ is the average momentum of the 
pair while $q=(E_1{-}E_2,\bp_1{-}\bp_2)$ is the relative momentum between
the two pions. Since the pion momenta $\bp_1,\bp_2$ are on-shell, 
$p^0_i = E_i = (m^2 + \bp_i^2)^{1/2}$, the 4-momenta $q$ and $K$ are in 
general off-shell. They satisfy the orthogonality relation (``on-shell
constraint'')
 \begin{equation}
 \label{ortho}
   q \cdot K = 0\quad\Longrightarrow\quad q^0=\bm{\beta}\cdot\bq\quad {\rm
   with}\quad \bm{\beta}={\bK\over K^0} \approx {\bK\over E_K}.
 \end{equation}
$\bm{\beta}$ is the velocity of the pair; the last, so-called ``on-shell'' 
approximation holds for small relative momenta $\bq$ which is where the 
correlation term is non-zero:
 \begin{equation}
 \label{Konshell}
   K^0 = E_K \, \left( 1 + {\bq^2 \over 8 E_K^2} + 
   {\cal O}\left({\bq^4 \over E_K^4}\right) \right) 
   \approx E_K \, .
 \end{equation}
Since the relevant range of $q$ is given by the inverse size of
the source (more properly: the inverse size of the regions of 
homogeneity in the source -- see below), the validity of this 
approximation is ensured in practice as long as the Compton wavelength 
of the particles is small compared to this ``source size". For the 
case of pion, kaon, or proton interferometry for heavy-ion collisions 
this is true automatically due to the rest mass of the particles: even 
for pions at rest, the Compton wavelength of 1.4 fm is comfortably 
smaller than any typical nuclear source size. This is of enormous 
practical importance because it allows you essentially to replace the 
source Wigner density by a classical phase-space distribution function
for on-shell particles. This provides a necessary theoretical 
foundation for the calculation of HBT correlations from classical 
hydrodynamic or kinetic (e.g. cascade) simulations of the collision.  

With the definition (\ref{SxK}) we can express the single-particle 
spectra (\ref{1}) and correlation function (\ref{corr}) through
the emission function $S(x,K)$:
 \begin{eqnarray}
 \label{spectrum} 
   E_p {dN \over d^3p} &=& \int d^4x\, S(x,p) \, ,
 \\
 \label{corrapp}
  C(\bq,\bK) &=& 1 + \frac{\left(P_1(\bK)\right)^2}
                          {P_1(\bK{+}\half\bq)P_1(\bK{-}\half\bq)}
                 \left\vert {\int d^4x\, e^{iq{\cdot}x}\, S(x,K) 
                             \over
                             \int d^4x\, S(x,K)} \right\vert^2
\nonumber\\
  &\approx& 1 + 
  \left\vert {\int d^4x\, e^{iq{\cdot}x}\, S(x,K) 
              \over
              \int d^4x\, S(x,K)} 
  \right\vert^2
  \equiv 1 + \left\vert \langle e^{iq{\cdot}x} \rangle \right\vert^2
  \, .
 \end{eqnarray}
The last line uses the ``smoothness approximation'' 
$P_1(\bK{+}\half\bq)P_1(\bK{-}\half\bq)\approx |P_1(\bK)|^2$ which can
be justified for sufficiently large sources; it is not really 
necessary because the single-particle spectra are measured and the 
ratio $|P_1(\bK)|^2/P_1(\bK{+}\half\bq)P_1(\bK{-}\half\bq)$ can thus
be divided out from the correlator $C(\bq,\bK)-1$ before analyzing
it theoretically, but we use it here for simplicity.

The last equality in Eq.~(\ref{corrapp}) introduces the notation 
$\langle f(x)\rangle$ for the average of a space-time observable 
$f(x)$ with the emission function $S(x,K)$. Note that, due to the
$K$-dependence of the emission function, this average depends on
the pair momentum $K$.

The fundamental relations (\ref{spectrum}) and (\ref{corrapp}) show 
that {\em both the single-particle spectrum and the two-particle 
correlation function can be expressed as simple integrals over the 
emission function}. Whereas the single-particle momentum spectrum 
(\ref{spectrum}) integrates over the spatial structure of the 
emission function and thus provides no spatial information about the 
source, Eq.~(\ref{corrapp}) shows that the $\bq$-dependence of the 
correlation function $C(\bq,\bK)$ provides access to the space-time 
structure of the source. However, it depends only on the modulus 
square of the spatial Fourier transform of the source, so the phase
information is missing, and furthermore, due to the ``on-shell constraint''
(\ref{ortho}), not all four components of $q$ can be varied independently.
The Fourier transform in Eq.~(\ref{corrapp}) thus mixes space and 
time in a particular way, and the emission function $S(x,K)$ cannot
be fully reconstructed from the measured correlation function $C(\bq,\bK)$
without taking recourse to model assumptions about the source. Still,
as I will now show, the correlator $C(\bq,\bK)$ provides very important 
constraints for any dynamical model for the heavy-ion collision.

%%%%%%%%%%%%%%%%%%%%%%%%%%%%%%%%%%%%%%%%%%%%%%%%%%%%%%%%%%%%%%%%%%%%
\subsection{HBT radii and source sizes}
\label{secVc}
%%%%%%%%%%%%%%%%%%%%%%%%%%%%%%%%%%%%%%%%%%%%%%%%%%%%%%%%%%%%%%%%%%%%

The most interesting feature of the two-particle correlation function
is its half-width as a function of $q$. Actually, since the relative 
momentum $\bq = \bp_1{-}\bp_2$ has three Cartesian components, the 
fall-off of the correlator for increasing $q$ is not described by a 
single half-width, but rather by a (symmetric) 3$\times$3 tensor. 
Through the Fourier transform in Eq.~(\ref{corrapp}), these half-widths
of the correlator (at a given pair momentum $K$) can be related to the 
widths of the emission function $S(x,K)$ in space and time (at the same 
$K$). This relation becomes exact if particle emission is not complicated
by processes involving multiple time scales. Multiple time scales,
with corresponding dynamically induced multiple length scales, are
an issue in pion emission from heavy-ion collisions: a large fraction 
of the observed pions is not emitted from the source directly, but 
appears later from the decay of unstable resonances which were
originally emitted directly from the source but than travelled a 
certain distance without interactions before decaying. Resonance
decay effects on the HBT correlation function have been extensively
discussed in the literature (see e.g. \cite{CLZ96,WH96}). The results
can be shortly summarized: Most resonances are so short-lived that
they don't travel far before decaying, leading to a pion emission 
function of almost the same size and shape as for directly emitted 
pions. Some resonances are very long-lived and decay far outside the
original source; the corresponding temporal and spatial tails in the
emission function affect the correlator at such small values of $q$ 
that they can not be experimentally resolved. They effectively reduce
the visible Bose-Einstein correlation strength and can be absorbed by 
reducing the second term in Eq.~(\ref{corrapp}) by a $\bK$-dependent
``intercept parameter'' $\lambda(\bK)$: $C(\bq,\bK)=1+\lambda(\bK)
|\langle e^{iq{\cdot}x}\rangle_{\rm dir}|^2(\bK)$ where
$\langle\dots\rangle_{\rm dir}$ denotes the average with the emission
function for (almost) directly emitted pions. The only resonance that
can neither be classified as short-lived nor as long-lived in this sense
is the $\omega$ meson with its lifetime of about 20\,fm/$c$. I will
assume here that its abundance is small enough to be negligible (this
depends on the freeze-out temperature), but this is mostly for didactical
purposes and needs to be reassessed in the actual comparison between 
theory and data.

\begin{figure}[ht] 
\begin{center}
  \epsfig{file=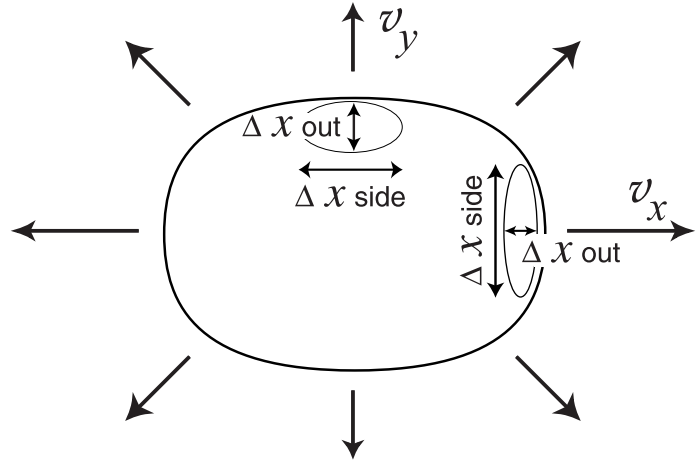,width=10cm}
\end{center}
\vspace*{-7mm}
\caption{Illustration of the concept of homogeneity regions
in an expanding source (see text).
\label{F16b} 
} 
\vspace*{-2mm}
\end{figure}

With these assumptions the ``direct'' emission function can be fully
characterized by its ($K$-dependent) ``widths''
\beq{Smunu}
    S_{\mu\nu}(\bK) = \left[ \langle x_\mu x_\nu\rangle_{\rm dir}
    - \langle x_\mu\rangle_{\rm dir}\langle x_\nu\rangle_{\rm dir}\rangle
                      \right](\bK)
    \equiv \langle \tilde x_\mu \tilde x_\nu\rangle_{\rm dir}
\eeq
which describe the widths of the effective emission region inside the 
source (``homogeneity region'') which effectively contributes to the 
emission of particles with momentum $\bK$. Note that for an expanding 
source these effective emission regions do not coincide with the entire 
source: As illustrated in Fig.~\ref{F16b}, pions whose momentum $\bK$ is 
large and points in the positive $x$ direction will predominantly 
originate from the elliptic region near the right edge of the figure,
since this region anyway moves collectively in the $+x$ direction. On 
the other hand, the elliptic region near the top of the figure moves in 
the $+y$ direction and thus mostly contributes pions with (large) positive
$y$-momentum $K_y$. Eq.~(\ref{corrapp}) shows that the correlation function
$C(\bq,\bK)$ of pairs with pair momentum $\bK$ probes only these effective
emission regions for particles with fixed $\bK$, not the entire source.
Only for static fireball, where each volume element radiates the same 
momentum spectrum, does the HBT correlation function measure the entire 
source. (The notation $x_{\rm side}$ and $x_{\rm out}$ used in 
Fig.~\ref{F16a} will be explained below.) 
%
%%%%%%%%%%%%%%%%%%%%%%%%% Fig. 16b %%%%%%%%%%%%%%%%%%%%%%%%%%%%%%%%%%%%
%%%%%%%%%%%%%%%%%%%%%%%%%%%%%%%%%%%%%%%%%%%%%%%%%%%%%%%%%%%%%%%%%%%%%%%
%

If, as assumed, the emission function $S(x,K)$ can be fully characterized
by its ``spatial correlation tensor'' $S_{\mu\nu}(\bK)$ (\ref{Smunu})
which describes the size and shape of the effective emission regions
for particles of momentum $\bK$, we can make the following Gaussian
ansatz for $S(x,K)$:
 \begin{equation}
 \label{7}
   S_{\rm dir}(x,K) \sim 
   e^{ - \half \tilde x^\mu(\bK)\, (S^{-1})_{\mu\nu}(\bK)\,\tilde x^\nu(\bK)}.
 \end{equation} 
The normalization is fixed by the single-particle spectrum \cite{WH99}
but drops out from the correlation function. Note the dependence only on
the distance $\tilde x_\mu(\bK) = x_\mu -\langle x_\mu\rangle(\bK)$ to
the point of maximum emissivity at momentum $\bK$ -- the latter itself 
can not be measured, i.e. the absolute position of the regions of 
homogeneity is experimentally inaccessible. Reconstructing the entire 
source as a sum of effective emission regions for fixed $\bK$ is 
therefore always a model-dependent theoretical enterprise.

The Fourier transform of the above Gaussian parametrization is easily 
evaluated. Eliminating $q^0$ with the on-shell constraint (\ref{ortho})
and introducing the intercept parameter $\lambda(\bK)$ as described 
above to account for the unresolvable tail in the emission function
from long-lived resonances gives for the correlation function the 
Gaussian form
 \begin{equation}
 \label{corrgauss}
   C(\bq,\bK) = 1 + \lambda(\bK)\,\exp\left[
   - \!\!\!\!\! \sum_{i,j=x,y,z} q_i\, q_j\, R_{ij}^2(\bK)\right],
 \end{equation}
with
 \beq{Rij}
    R_{ij} = S_{ij} - \beta_i S_{it} - \beta_j S_{jt} + 
             \beta_i \beta_j S_{tt}, \qquad i,j=x,y,z.
 \eeq 
The $R_ij(\bK)$ are called ``HBT radius parameters''; note that the 
off-diagonal terms with $i\ne j$ need not be positive. The last 
equation shows explicitly the mixing between spatial and temporal 
source information in the experimentally measured HBT radius parameters, 
caused by the on-shell constraint (\ref{ortho}). 

Note that this mixing involves the components of the pair velocity 
$\bm{\beta}$; it can be minimized by rotating the Cartesian coordinate 
system $(x,y,z)$ such that $\bK$ and $\bm{\beta}$ point in $+x'$ 
direction \cite{PCZ90}. In
relativistic heavy-ion collisions it is, however, advantageous to
keep the $z$-axis aligned with the beam direction, and therefore one
rotates only the tranverse $(x,y)$ axes such that the transverse
pair momentum $\bK_\perp$ points in $x'$ direction. This eliminates
the $y'$ components of $\bK_\perp$ and $\bm{\beta}$. Traditionally
\cite{PCZ90}, the rotated $x'$ axis is called the ``outward'' direction
(denoted as $x_{\rm out}$ or $x_o$), the $y'$ axis is called the 
``sideward'' direction ($x_{\rm side}$ or $x_s$), while the beam 
direction $z$ is called ``longitudinal'' (denoted by $x_{\rm long}$ or 
$x_l$). In this frame $\bm{\beta}=(\beta_\perp,0,\beta_L)$ and 
$\bK=(K_\perp,0,K_L)$. Note, however, that in non-central heavy-ion 
collisions the direction of the impact parameter $\bm{b}$ defines a 
distinguished direction in the transverse plane, and that therefore 
even in the out-side-long system the HBT radius parameters $R_{ij}(\bK)$ 
still depend on both magnitude and direction of $\bK_\perp$.  

%%%%%%%%%%%%%%%%%%%%%%%%%%%%%%%%%%%%%%%%%%%%%%%%%%%%%%%%%%%%%%%%%%%%
\subsection{Central ($\bm{b{=}0}$) heavy-ion collisions}
\label{secVd}
%%%%%%%%%%%%%%%%%%%%%%%%%%%%%%%%%%%%%%%%%%%%%%%%%%%%%%%%%%%%%%%%%%%%
\subsubsection{Formalism}
\label{secVd1}
%%%%%%%%%%%%%%%%%%%%%%%%%%%%%%%%%%%%%%%%%%%%%%%%%%%%%%%%%%%%%%%%%%%%

In central collisions between spherical nuclei, the entire source is 
azimuthally symmetric around the beam axis. After fixing the transverse
coordinate system such that the ``out''-axis points in the emission
direction $\bK_\perp$ of the pair, the HBT radii $R_{ij}$ can only
depend on the magnitude, but not the direction of $\bK_\perp$. 
Furthermore, the source is clearly symmetric under reflections 
with respect to the plane defined by the beam direction and $\bK_\perp$,
so all components of the spatial correlation tensor 
$S_{\mu\nu}\eq\langle \tilde x_\mu \tilde x_\nu\rangle$ which are linear in 
$\tilde x_{\rm side}$ must vanish. As a result, only {\em six} nonvanishing
components of $S_{\mu\nu}(K_\perp,Y)$ survive. (I here introduce the
pair rapidity $Y\eq\tanh^{-1}\beta_L$ instead of the longitudinal pair 
momentum $K_L$.) On the other hand, the same reflection symmetry also
eliminates terms from the exponent in Eq.~(\ref{corrgauss}) that are 
linear in $q_s\eq{q}_{\rm side}$ such that only {\em four} HBT radius
parameters are experimentally accessible:
 \begin{equation}
 \label{azsym}
   C(\bq,\bK) = 1 + \lambda(\bK)\,\exp\left[- q_s^2 R_s^2 - q_o^2 R_o^2
   - q_l^2 R_l^2 - 2 q_o q_l R_{ol}^2 \right].
 \end{equation}
The HBT radii are again functions of two variables only, 
$R_{ij}^2(K_\perp,Y)$. Still, there are two more functions 
$S_{\mu\nu}(K_\perp,Y)$ needed to fully characterize the source than 
can be measured, due to the on-shell constraint. 

In symmetric $A+A$ collisions at midrapidity $Y\eq0$, things simplify
further: the cross-term $R_{ol}^2$ vanishes \cite{CNH95}, leaving
only three functions $R_s(K_\perp),\,R_o(K_\perp),$ and 
$R_l(K_\perp)$ to be determined. They are related to the source
widths $S_{\mu\nu}\eq\langle\tilde x_\mu \tilde x_\nu\rangle$ as follows:
\bea{Rs2}
  R_s^2 &=& \langle \tilde x_s^2\rangle,
\\
\label{Ro2}
  R_o^2 &=& \langle \tilde x_o^2\rangle 
  - 2 \beta_\perp \langle \tilde x_o\tilde t\rangle
  + \beta_\perp^2 \langle \tilde t^2 \rangle,
\\
\label{Rl2}
  R_l^2 &=& \langle \tilde z^2\rangle,
\eea  
where I used that $\beta_L\eq0$ at $Y\eq0$. I will concentrate on this
simple limit since RHIC HBT data have so far only been obtained near
$Y\eq0$.
 
The HBT radii (\ref{Rs2}-\ref{Rl2}) depend on $K_\perp$ both
{\em explicitly} (through the $\beta_\perp$ factors in (\ref{Ro2})) and
{\em implicitly} (since $\la\dots\ra$ denotes an average with
the $K$-dependent emission function). If $\bm{K}_\perp$ is the 
only vector in the problem, there is no distinction between
the outward and sideward directions in the limit $K_\perp{\,\to\,}0$,
so $R_o$ and $R_s$ must be equal to each other in that limit:
\bea{RoeqRs}
  \lim_{K_\perp\to0}\left(R_o^2{-}R_s^2\right) =
  \lim_{K_\perp\to0}\left(\langle\tx_o^2\rangle-\langle\tx_s^2\rangle
                    \right) = 0.
\eea  
This theorem is not valid for non-central collisions, where the impact
parameter $\bm{b}$ introduces a preferred direction, and it may 
also break down for ``opaque'' (surface emitting) sources where, at 
least in principle, the outward normal vector can play a similar role
\cite{Heiselberg:1997bt,Tomasik:1998qt}.  

The geometric contributions $\langle \tilde x_s^2\rangle,\,
\langle \tilde x_o^2\rangle$, and $\langle \tilde z^2\rangle$ are
controlled by an interplay between thermal motion and collective 
expansion: stronger expansion, resulting in larger flow velocity 
gradients, tends to reduce these variances whereas higher freeze-out 
temperatures, resulting in broader momentum spectra, tend to increase 
the size of the homogeneity regions. This size increase due to thermal 
smearing is easily understood by considering the limit $T\to 0$: in this 
limit particles have no thermal motion, and particles of momentum 
$\bK$ can therefore only come from individual points in the source
whose collective flow velocity agrees exactly with the pair velocity
$\bm{\beta}$. The corresponding homogeneity regions are pointlike, i.e.
have zero size. Thermal particle motion can smear out the collective 
flow velocity gradients such that particles of given $\bK$ can now come
from a larger homogeneity region. As $T\to\infty$, the homegeneity regions
approach the size of the entire fireball, irrespective of its dynamical \
state.  

Making these features explicit requires a parametrization of the emission
function, for example the extensively studied model function 
\cite{HJ99,WH99,CNH95,CL95}
 \begin{equation}
 \label{3.15}
    S(x,K) = {M_\perp \cosh(\eta{-}Y) \over 8 \pi^4 \Delta \tau}
    \exp\left[- {K{\cdot}u(x) \over T(x)}
              - {(\tau-\tau_0)^2 \over 2(\Delta \tau)^2}
              - {r^2 \over 2 R^2} 
              - {(\eta- \eta_0)^2 \over 2 (\Delta \eta)^2}
           \right].
 \end{equation}
Here the transverse radius $r^2{\,=\,}x^2{+}y^2$, the spacetime rapidity 
$\eta = {1 \over 2} \ln[(t{+}z)/(t{-}z)]$, and the longitudinal proper 
time $\tau= \sqrt{t^2{-}z^2}$ parametrize the spacetime coordinates 
$x^\mu$, with measure $d^4x = \tau\, d\tau\, d\eta\, r\, dr\, d\phi$. 
$Y = \half \ln[(E_K{+}K_L)/(E_K{-}K_L)]$ and 
$M_\perp = \sqrt{m^2{+}K_\perp^2}$ 
parametrize the longitudinal and transverse components of the pair 
momentum $\bK$. $\sqrt{2} R$ is the transverse geometric
(Gaussian) radius of the source, $\tau_0$ its average freeze-out
proper time, $\Delta \tau$ the mean proper time duration of particle
emission, and $\Delta \eta$ parametrizes the finite longitudinal
extension of the source. $T(x)$ is the freeze-out temperature; if you
don't like the idea of invoking thermal equilibrium during the decoupling 
stage, you can think of $T$ as a parameter that describes the random 
distribution of the particle momenta at each space-time point around 
their average value. The latter is parametrized by a collective flow 
velocity $u^\mu(x)$ in the form 
 \begin{equation}
 \label{26}
   u^\mu(x) = \left( \cosh \eta \cosh \eta_t(r), \,
                     \sinh \eta_t(r)\, \be_r,  \,
                     \sinh \eta \cosh \eta_t(r) \right) ,
 \end{equation}
with a boost-invariant longitudinal flow rapidity $\eta_l{\,=\,}\eta$ 
($v_l{\,=\,}z/t$) and a linear transverse flow rapidity profile 
 \begin{equation}
 \label{27}
  \eta_t(r) = \eta_f \left( {r \over R} \right)\, .
 \end{equation} 
$\eta_f$ scales the strength of the transverse flow. The exponent of 
the Boltzmann factor in (\ref{3.15}) can then be written as
 \begin{equation}
 \label{26a}
  K\cdot u(x) = M_\perp \cosh(Y-\eta) \cosh\eta_t(r) - 
                \bK_\perp{\cdot}\be_r \sinh\eta_t(r)\, .
 \end{equation}
For vanishing transverse flow ($\eta_f=0$) the source depends only 
on $M_\perp$ and remains azimuthally symmetric for all $K_\perp$.
In this case the sideward and longitudinal HBT radii (\ref{Rs2})
and (\ref{Rl2}) exhibit perfect $M_\perp$-scaling, i.e. when 
plotted as functions of $M_\perp$ they coincide for pion and kaon 
pairs \cite{WH99}. [In fact, $R_s$ does not depend on $M_\perp$ at all 
in this limit.] For the outward radius $M_\perp$-scaling is broken by 
the $\beta_\perp$-dependent terms in Eq.~ (\ref{Ro2}). For non-zero 
transverse flow, $R_s$ does depend on $M_\perp$, at a level that 
directly reflects the strength of the transverse flow (see below).
In this case the $M_\perp$-scaling of $R_s$ and $R_l$ is broken by 
the $K_\perp$-dependent second term in the thermal exponent (\ref{26a}),
albeit only weakly \cite{WH99}.

Using saddle-point integration, the HBT radii (\ref{Rs2}-\ref{Rl2}) 
for this source can be approximately evaluated analytically \cite{CNH95}.
For pairs with $Y{\,=\,}0$ one finds the simple pocket formulae
 \begin{eqnarray}
 \label{Rs}
    R_s^2 &=& R_*^2 \, ,
 \\
 \label{R0}
    R_o^2 &=& R_*^2 + \beta_\perp^2 (\Delta t_*)^2\, ,
 \\
 \label{Rl}
    R_l^2 &=& L_*^2 \, ,
 \end{eqnarray}
with
 \begin{eqnarray}
 \label{Rstar}
   {1\over R_*^2} &=& {1\over R^2} + {1\over R_{\rm flow}^2}\, , 
 \\
 \label{tstar}
   (\Delta t_*)^2 &=& (\Delta\tau)^2 + 
   2 \left( \sqrt{\tau_0^2 + L_*^2} - \tau_0 \right)^2 \, , 
 \\
 \label{Lstar}
   {1\over L_*^2} &=& {1\over (\tau_0\Delta\eta)^2} 
   + {1\over L_{\rm flow}^2}\, . 
 \end{eqnarray}
Here $R_{\rm flow}$ and $L_{\rm flow}$ are the transverse and 
longitudinal ``dynamical lengths of homogeneity'' generated by the 
expansion velocity gradients:
 \begin{eqnarray}
 \label{RH}
   R_{\rm flow}(M_\perp) &=& {R\over \eta_f}\, \sqrt{{T\over M_\perp}}
   = {1\over \partial \eta_t(r)/\partial r} \, \sqrt{{T\over M_\perp}}\, ,
 \\
 \label{LH}
   L_{\rm flow}(M_\perp) &=& \tau_0\, \sqrt{{T\over M_\perp}}
   = {1\over \partial{\cdot}u_l} \, \sqrt{{T\over M_\perp}}\, ,
 \end{eqnarray}
where $u_l$ is the longitudinal 4-velocity. These expressions show 
explicitly the competition between flow gradients and thermal 
smearing (reflected in the factor $\sqrt{T/M_\perp}$) which I 
mentioned above. They also show the competition between the overall 
geometric size of the source and the ``dynamical homogeneity lengths'',
with the smaller of the two controlling the HBT radii. The second term
in the expression (\ref{tstar}) for the emission duration reflects
the fact that the emission duration is measured in terms of the 
{\em coordinate} time in the fixed center-of-mass reference frame, whereas
freeze-out occurs at constant {\em proper} time $\tau_0$ which, over the 
longitudinal range $L_*$ probed by the longitudinal HBT radius, explores 
a finite range of coordinate times even if the proper time spread 
$\Delta\tau$ is set to zero. For freeze-out along a hyperbola of fixed
proper time, it is therefore impossible to get a vanishing emission 
duration $\Delta t_*$.

For a longitudinally boost-invariant source ($\Delta\eta\to\infty$) and
weak transverse expansion the source function integral in Eq.~(\ref{Rl2})
can actually be done exactly \cite{HB95}, and one finds instead of 
(\ref{Rl},\ref{LH}) 
 \begin{eqnarray}
    R_l^2 &=& 
    \tau_0^2\, {T\over M_\perp}\,
    \frac{K_2(M_\perp/T)}{K_1(M_\perp/T)}\,.
 \end{eqnarray}
For small $M_\perp\simeq T$, this differs by up to a factor of 2 from 
Eq.~(\ref{LH}).

%%%%%%%%%%%%%%%%%%%%%%%%%%%%%%%%%%%%%%%%%%%%%%%%%%%%%%%%%%%%%%%%%%%%
\subsubsection{HBT radii from Au+Au collisions at RHIC --- 
 the ``HBT Puzzle''}
\label{secVd2}
%%%%%%%%%%%%%%%%%%%%%%%%%%%%%%%%%%%%%%%%%%%%%%%%%%%%%%%%%%%%%%%%%%%%

Let us now study the HBT radii predicted by the hydrodynamic model
introduced in Section \ref{sec4a9}, which was very successful in 
reproducing the single particle momentum spectra, and compare them
with data from RHIC.  

The hydrodynamic emission function, which implements sudden decoupling 
on a sharp freeze-out surface $\Sigma$ as described in Section \ref{sec4a5},
takes the form \cite{SOPW92,CH94}
\beq{equ:sourcefunction2}
  S_i(x,K) = \frac{g_i}{(2 \pi)^3} \int_\Sigma 
           \frac{K{\cdot}d^3\sigma(x')\, \delta^4 (x-x')}
                {\exp\{[K{\cdot}u(x')-\mu_i(x')]/\Tdec(x')\}\pm 1}\,.
\end{equation}
Phenomenological fits to spectra and HBT data often use a generalization
of this form which replaces the $\delta$-function by allowing for a spread 
of emission times (``fuzzy freeze-out'') \cite{CNH95,CL95,TWH99}. 
For a longitudinally boost invariant source the freeze-out hypersurface
can be parametrized in terms of the freeze-out eigentime as a 
function of the transverse coordinates, $\tau_f(x,y)$. 
The normal vector $d^3\sigma_\mu$ on such a surface is
given by
\beq{equ:d3sigma}
  d^3\sigma = \Bigl(\cosh\eta,\,\grad_{\!\perp}\tau_f(x,y),\, 
                    \sinh\eta \Bigr)
              \tau_f(x,y)\,dx\, dy\, d\eta \,.
\end{equation}
With the four momentum $K^\mu=(\Mt \cosh Y, \bKp, \Mt \sinh Y)$, 
where $Y$ and $\Mt$ are the rapidity and transverse mass associated with 
$K$, Eq.~(\ref{equ:sourcefunction2}) becomes
\bea{equ:sourcefunction3}
  S_i(x,K) =
   \frac{g_i}{(2 \pi)^3} \int_{-\infty}^\infty d\eta\,dx\,dy
   \bigl[\Mt\cosh(Y-\eta){-}\bKp \cdot \grad_{\!\perp}\tau_f(x,y) 
   \bigr] \nonumber
\\
   \times f\bigl(K{\cdot}u(x),x\bigr) \,
   \delta\bigl(\tau{-}\tau_f(x,y)\bigr) \hspace*{1cm}
\end{eqnarray}
with the flow-boosted local equilibrium distribution $f(K{\cdot}u(x),x)$ 
from (\ref{eq}). 
With this expression we can now study the emissivity of the source 
as a function of mass and momentum of the particles.
For the purpose of presentation we integrate the emission function over  
two of the four space-time coordinates and discuss the contours of
equal emission density in the remaining two coordinates.
We begin with calculations describing central Au+Au collisions at 
$\scm\eq130$~GeV and focus on directly emitted pions, neglecting pions 
from unstable resonance decays.
Resonance decay pions are known to produce non-Gaussian tails in the
spatial emission distribution, increasing its width, but these tails are 
not efficiently picked up by a Gaussian fit to the width of the measured 
two-particle momentum correlation function \cite{WH96,LKS02}. 
A comparison of the experimental HBT size parameters extracted from 
such fits with the spatial widths of the emission function is thus best 
performed by plotting the latter without resonance decay contributions.
% 

%
%%%%%%%%%%%%%%%%%%%%%%%%%%% Fig. 16c %%%%%%%%%%%%%%%%%%%%%%%%%%%%%%%%%%%%%%%%
\begin{figure}[htb] 
\vspace*{-2mm}
\centerline{
            \epsfig{file=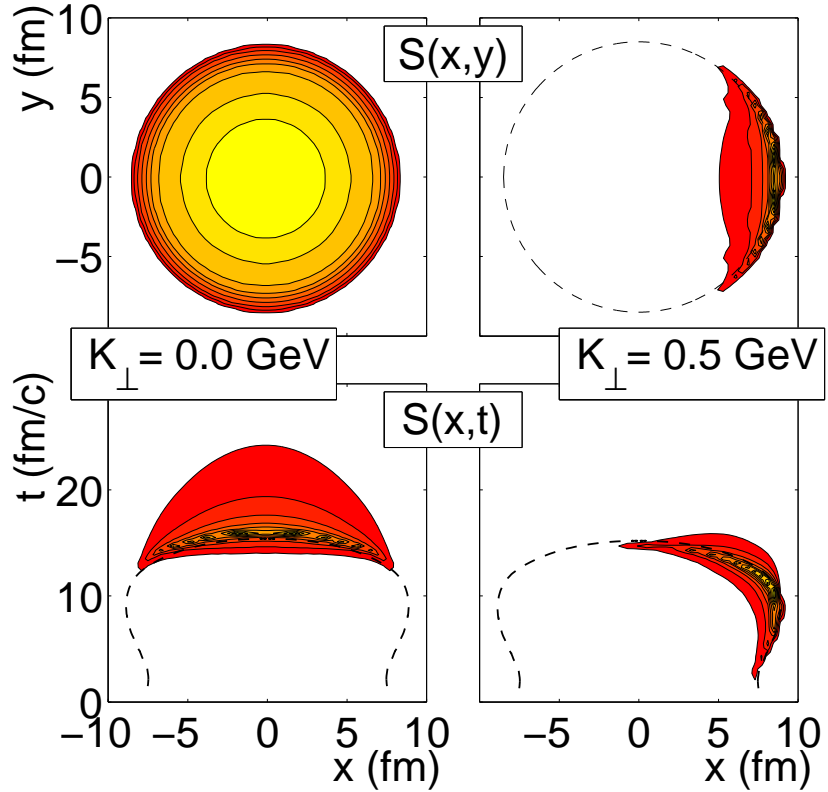,bb=81 183 530 616,width=11cm}
            }  
\caption{Pion source function $S(x,K)$ for central Au+Au collisions 
         at $\scm\eq130$\,GeV. The upper row shows the source after
         integrating out the longitudinal and temporal coordinate, 
         in the lower row the source is integrated over the longitudinal 
         and one transverse coordinate ($y$). In the left column we 
         investigate the case $\bK\eq0$, in the right column 
         the pions have rapidity $Y\eq0$ and transverse momentum 
         $K_\perp\eq0.5$\,GeV in $x$ direction \protect\cite{HK02}.         
\label{fig:sourcexyxt} 
} 
\vspace*{-3mm}
\end{figure} 
%%%%%%%%%%%%%%%%%%%%%%%%%%%%%%%%%%%%%%%%%%%%%%%%%%%%%%%%%%%%%%%%%%%%%%%
%
Figure~\ref{fig:sourcexyxt} shows equal density contours at 
10,\,20,\,\dots,\,90\% of the maximum in a transverse cut integrated
over time and $\eta$ (top row) and as a function of radius and time 
integrated over $\eta$ and the second transverse coordinate (bottom row).
The dashed circle in the top row indicates the largest freeze-out
radius reached during the expansion, the dashed line in the bottom
row gives the freeze-out surface $\tau_f(x,y{=}0)=t_f(x,y{=}z{=}0)$.
Pions with vanishing transverse momentum (left column) are seen to 
come from a broad region symmetric around the center and are emitted
rather late.
Pions with $K_\perp\eq0.5$\,GeV pointing in $x$-direction, on the 
other hand, are emitted on average somewhat earlier and only from a 
rather thin, crescent shaped sliver along the surface of the fireball 
at its point of largest transverse extension.
The reason for this apparent ``opacity'' (surface dominated emission)
of high-$\pt$ particles is that they profit most from the radial 
collective flow which is largest near the fireball surface.
Low-$\pt$ pions don't need the collective flow boost and are preferably
emitted from smaller radii (where the flow velocity is smaller) when the
freeze-out surface eventually reaches these points during the final stage 
of the decoupling process.
%

%
%%%%%%%%%%%%%%%%%%%%%%%%%%%%% Fig. 16d %%%%%%%%%%%%%%%%%%%%%%%%%%%%%%%%%
\begin{figure}[htb]
\vspace*{-2mm}
\centerline{
            \epsfig{file=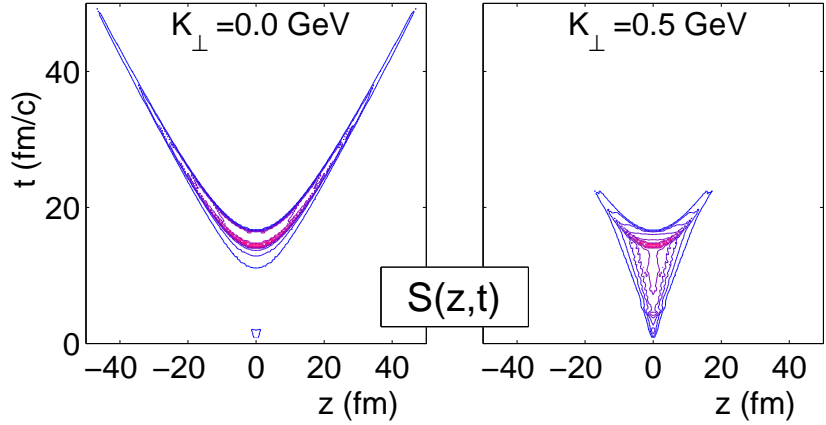,,bb=96 390 516 607,width=11cm}
            }  
\caption{Longitudinal cuts through the pion emission function $S(x,K)$, 
         integrated over transverse coordinates, for $Y\eq0$ pions with
         two different values of $K_\perp$ as indicated \protect\cite{KH03rev}.
\label{fig:sourcezt} 
} 
\vspace*{-3mm}
\end{figure} 
%%%%%%%%%%%%%%%%%%%%%%%%%%%%%%%%%%%%%%%%%%%%%%%%%%%%%%%%%%%%%%%%%%%%%%%
%
%
Fig.~\ref{fig:sourcezt} displays the time structure of the emission 
process along the beam direction. 
Especially for low tansverse momenta one clearly sees a very long
emission duration, as measured in the laboratory (center-of-momentum
frame). 
The reason is that, according to the assumed longitudinal boost 
invariance, freeze-out happens at constant proper time 
$\tau\eq{t^2{-}z^2}$ and extends over a significant range in
longitudinal position $z$.
This range is controlled by the competition between the longitudinal
expansion velocity gradient (which makes emission of $Y\eq0$ pions
from points at large $z$ values unlikely) and the thermal velocity 
smearing. 
If freeze-out happens late (large $\tau$), the longitudinal velocity 
gradient $\sim 1/\tau$ is small and pions with zero longitudinal 
momentum are emitted with significant probability even from large
values of $|z|$, i.e. very late in coordinate time $t$.
Note that this is also visible in the lower left panel of 
Fig.~\ref{fig:sourcexyxt} where significant particle emission 
still happens at times where the matter at $z\eq0$ has already 
fully decoupled.
We will see shortly that this poses a problem when compared with the 
data. 
The long tails at large values of $|z|$ and $t$ can only be avoided
by reducing $\tau_f$ (thereby increasing the longitudinal velocity 
gradient and reducing the $z$-range which contributes $Y\eq0$ pions)
and/or by additionally breaking longitudinal boost-invariance by 
reducing the particle density or postulating earlier freeze-out  
at larger space-time rapidities $|\eta|$ \cite{HT02}.
From earlier hydrodynamic calculations \cite{RG96} it was expected that
a fireball evolving through the quark-hadron phase transition would
emit pions over a long time period, resulting in a large contribution
$\beta_\perp^2 \langle \tilde t^2\rangle$ to the outward HBT radius and a 
large ratio $R_o/R_s$.
This should be a clear signal of the time-delay induced by the phase 
transition.
It was therefore a big surprise when the first RHIC HBT 
data \cite{STAR01HBT,PHENIX02HBT} yielded $R_o/R_s{\,\approx\,}1$ 
in the entire accessible $\Kt$ region (up to 0.7\,GeV/$c$).
In the meantime this finding has been shown to hold true out to
$K_\perp\gapp1.2$~GeV/$c$ \cite{PHENIXHBT200}. 
%

%
%%%%%%%%%%%%%%%%%%%%%%% Fig. 16e %%%%%%%%%%%%%%%%%%%%%%%%%%%%%%%%%%%%%%%%%%
\begin{figure} 
\vspace*{-1mm}
\centerline{
            \epsfig{file=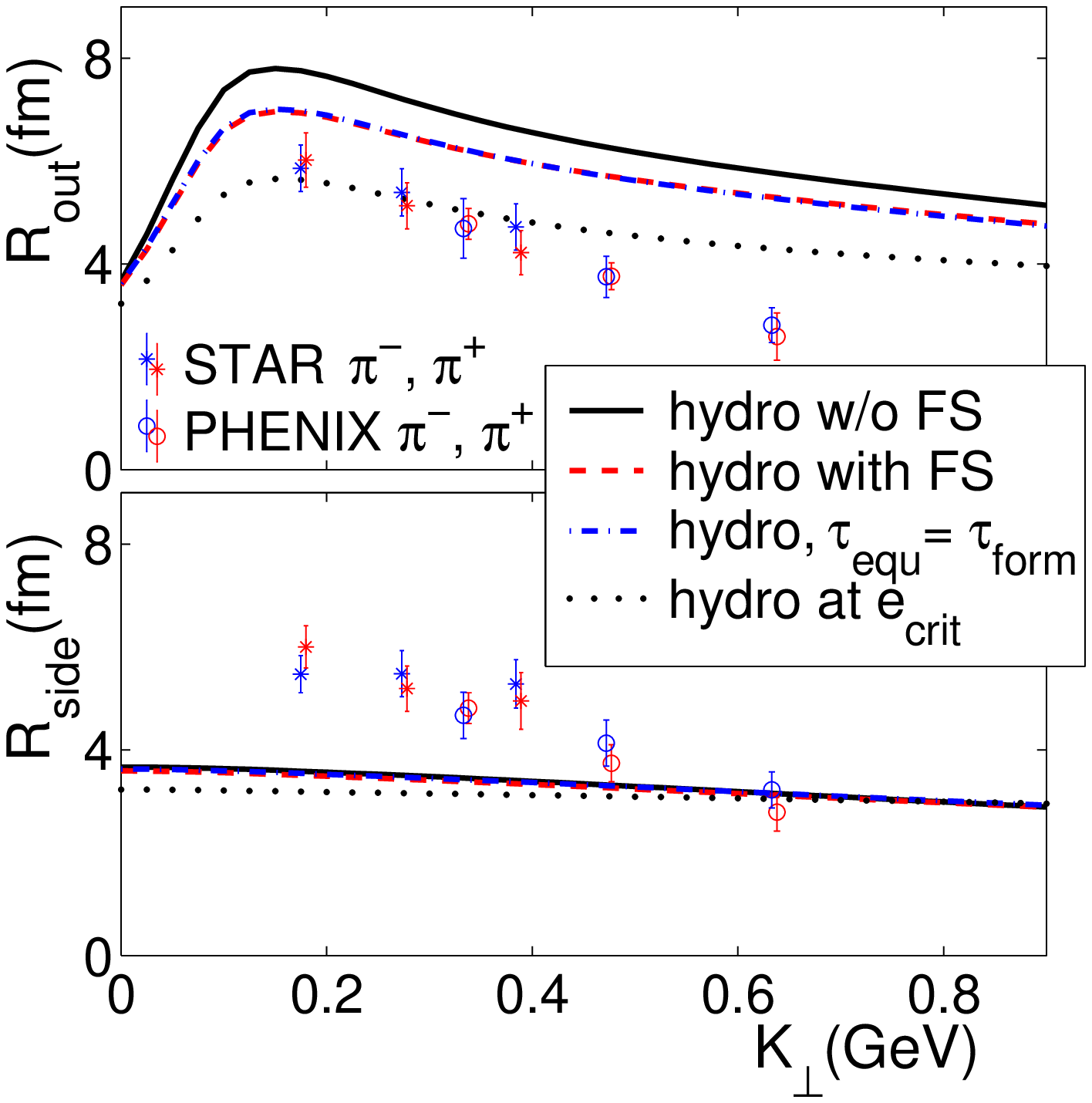,bb=55 180 467 594,width=7cm}
            \epsfig{file=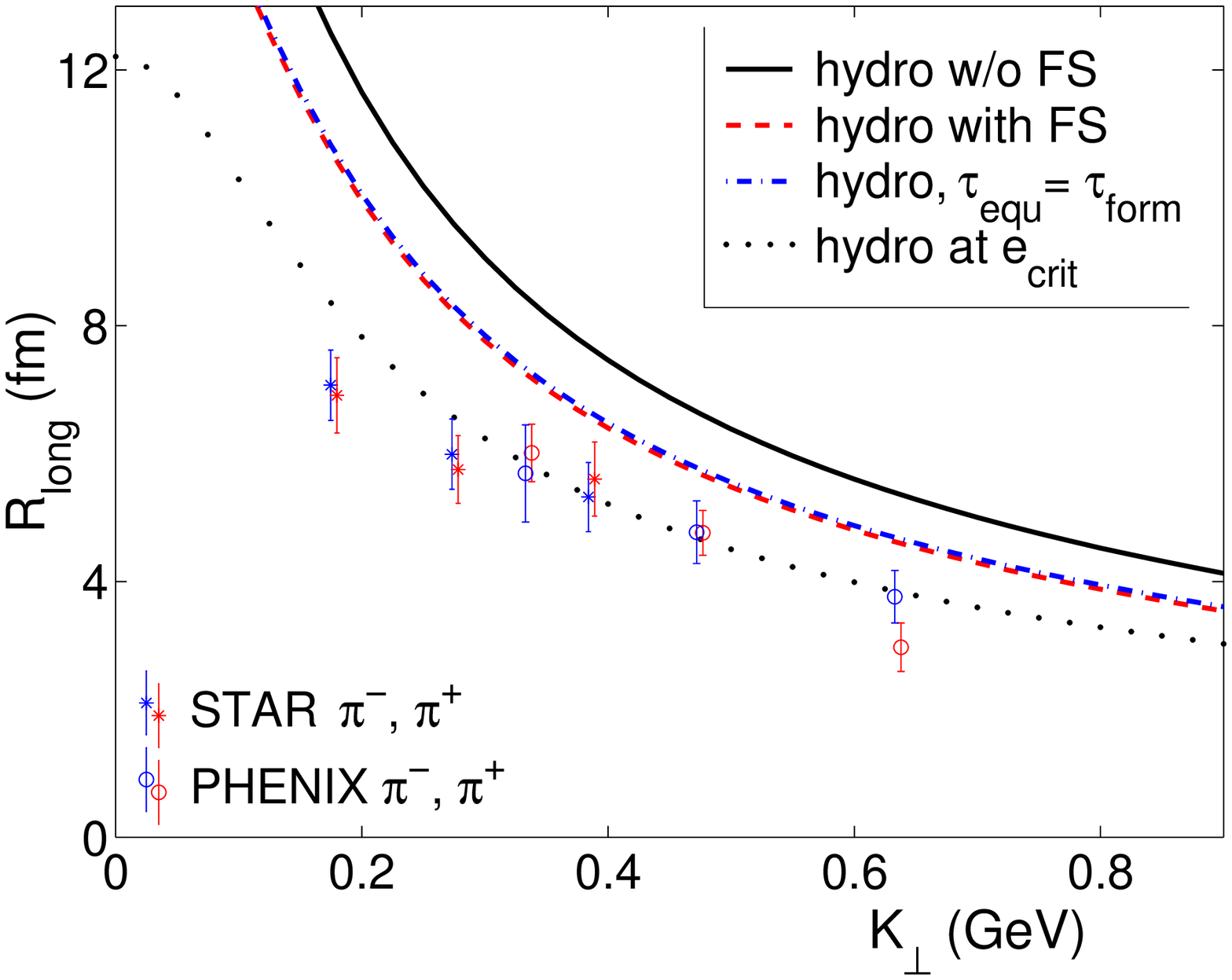,bb=52 192 571 594,width=7cm,height=7.05cm}
            }  
\caption{HBT radii from hydrodynamic calculations \protect\cite{HK02WWND}  
         (solid lines) together with data from STAR \protect\cite{STAR01HBT}
         and PHENIX \protect\cite{PHENIX02HBT}.         
         The dotted lines give hydrodynamic radii calculated directly
         after hadronization whereas the other two lines refer
         to different assumptions about the initial conditions
         (see text).
\label{fig:HBTcentral} 
} 
\vspace*{-2mm}
\end{figure} 
%%%%%%%%%%%%%%%%%%%%%%%%%%%%%%%%%%%%%%%%%%%%%%%%%%%%%%%%%%%%%%%%%%%%%%%
%
Figure~\ref{fig:HBTcentral} shows a comparison of the experimental 
data with results from hydrodynamic calculations \cite{HK02WWND}.
Clearly, a purely hydrodynamic description with default initial 
conditions (solid lines in Fig.~\ref{fig:HBTcentral}) fails to
describe the measured HBT radii.
The longitudinal and outward radii $R_l$ and $R_o$ are too
large, $R_s$ is too small, and the $\Kt$-dependence of both $R_o$ and
$R_s$ are too weak in the model.
Since for longitudinally boost-invariant sources $R_l$ is entirely 
controlled by the longitudinal velocity gradient at freeze-out which 
decreases as $1/\tau_f$, making $R_l$ smaller within the hydrodynamic
approach requires letting the fireball decouple earlier \cite{HK02WWND} 
or breaking the boost invariance \cite{HT02}.
As noted above, a smaller $R_l$ would also reduce the emission duration, 
i.e. $\la \tilde t^2\ra$, and thus help to bring down $R_o$, especially
if freeze-out at nonzero rapidities $|\eta|{\,>\,}0$ happens earlier
than at midrapidity.
The fireball can be forced to freeze out earlier by changing the
freeze-out condition (e.g. by imposing freeze-out directly at hadronization, 
see dotted line in Fig.~\ref{fig:HBTcentral}).
This generates, however, serious conflicts with the single-particle 
spectra (see Sec.~\ref{sec4a9}).
Alternatively, one can allow transverse flow to build up sooner, either 
by seeding it with a non-zero value already at $\tau_\equ$ (short dashed 
line labelled ``hydro with FS'' \cite{HK02WWND,KR03}) or by letting the 
hydrodynamic stage begin even earlier (e.g. at $\tau_{\rm 
form}\eq0.2$\,fm/$c$, long-dashed line). 
The last two options produce similar results, but do not fully 
resolve the problems with the magnitudes of $R_l$ and $R_o$.
The situation may improve by taking also the breaking of longitudinal 
boost-invariance into account \cite{HT02}, but a fully consistent
hydrodynamic description has not yet been found.
In particular, the sideward radius $R_s$ is still too small and the 
$\Kt$-depenences of both $R_s$ and $R_o$ are still to 
weak \cite{HK02WWND,HT02}.
It is hard to see how to increase $R_s$ without also increasing
$R_o$ and $R_l$ which are already too large.
Hybrid calculations \cite{SBD01} in which the hydrodynamic 
Cooper-Frye freeze-out is replaced by transition to a hadronic
cascade at $\Tc$, followed by self-consistent kinetic freeze-out,
tend to increase $R_s$ by making freeze-out more ``fuzzy'', but at
the expense of also increasing $R_o$ and $R_l$ in a disproportionate
manner, mainly due to an increase in the emission duration. 
This makes the problems with the $R_o/R_s$ ratio even worse.
In the past, a strong $\Kt$-dependence of $R_s$ has been associated
with strong transverse flow \cite{WH99,HJ99}.
It is therefore surprising that even the hydrodynamic model with its 
strong radial flow cannot reproduce the strong $\Kt$-dependence of
$R_s$ measured at RHIC.
Also, according to Eqs.~(\ref{Rs2},\ref{Ro2}) the difference between 
$R_o^2$ and $R_s^2$ can be reduced, especially at large $\Kt$ \cite{TWH99},
if the positive contribution from the emission duration $\la \tilde t^2\ra$ 
is compensated by ``source opacity'', i.e. by a strongly surface-dominated 
emission process \cite{Heiselberg:1997bt,Tomasik:1998qt,MP02}.
In this case the geometric contribution $\la \tilde x_o^2\ra$ to
$R_o^2$ is much smaller than $R_s^2\eq\la \tilde x_s^2\ra$.
Again, the source produced by the hydrodynamic model (see top right
panel in Fig.~\ref{fig:sourcexyxt}) is about as ``opaque'' as one
can imagine \cite{HK02osci}, and it will be difficult to
further increase the difference $\la \tilde x_s^2{-}x_o^2\ra$ \cite{MP02}.
This leaves almost only one way out of the ``HBT puzzle'', namely
the space-time correlation term $-2\beta_\perp \la\tilde x_o\tilde t\ra$
in expression (\ref{Ro2}) for $R_o^2$.
It correlates the freeze-out position along the outward direction with
the freeze-out time.
The hydrodynamic model has the generic feature that, in the region where
most particles are emitted (see Fig.~\ref{fig:sourcexyxt}), these
two quantities are negatively correlated, because the freeze-out
surface moves from the outside towards the center rather than the 
other way around. 
Hence the term $-2\beta_\perp \la\tilde x_o\tilde t\ra$ is positive
and tends to make $R_o^2$ larger than $R_s^2$.
The small measured ratio $R_o/R_s\lapp 1$ may instead call for strong 
{\em positive} $x_o{-}t$ correlations, implying that particles emitted 
from {\em larger} $x_o$ values decouple {\em later}.
Hydrodynamics can not produce such a positive $x_o{-}t$ correlation
(at least not at RHIC energies).
On the other hand, there are indications that microscopic models, such 
as the AMPT \cite{LKS02} and MPC \cite{MG02a} models, may produce them,
for reasons which are not yet completely understood.
One should also not forget that Fig.~\ref{fig:HBTcentral} really
compares two different things: 
The data are extracted from the width of the 2-particle correlator
in momentum space while in theory one calculates the same quantities
from the source width parameters in coordinate space.
The two produce identical results only for Gaussian sources.
The hydrodynamic source function shown in Figs.~\ref{fig:sourcexyxt} 
and \ref{fig:sourcezt} are not very good Gaussians and show a lot of
additional structure.
We have checked, however, by explicit computation of the momentum-space 
correlation function using Eq.~(\ref{corrapp}) that the non-Gaussian
effects are small.%
\footnote{They would have been larger if we had included resonance decay 
pions in the emission function, as found by Lin {\it et al.} \cite{LKS02},
which was our main reason for not doing so.}
The largest non-Gaussian effects are seen in the longitudinal radius 
$R_l$ \cite{WH96}, but although the corresponding corrections go in the 
right direction by making the $R_l$ extracted from the momentum-space 
correlator smaller, the effect is only a fraction of 1\,fm and not 
large enough to bridge the discrepancy with the data.
It was recently suggested \cite{Teaney03,Dumitru02} that
neglecting dissipative effects might be at the origin of the discrepancy 
between the purely hydrodynamic calculations and the data.
A calculation of first-order dissipative corrections to the spectra and
HBT radii at freeze-out \cite{Teaney03}, with a ``reasonable'' value
for the viscosity, yielded a significant decrease of $R_l$ along with 
a corresponding strong reduction of the emission duration contribution 
to $R_o$, both as desired by the data.
There was no effect on the $x_o{-}t$ correlations, however, and only
a weak effect on $R_s$ which went in the wrong direction, making
it even flatter as a function of $\Kt$.
The rather steep $\Kt$-dependence of the data for both $R_s$ and $R_o$
and the larger than predicted size of $R_s$ at low $\Kt$ are therefore 
not explained by this mechanism \cite{Teaney03,MG02a}.
Furthermore, the elliptic flow $v_2$ has been shown to be very sensitive
to viscosity \cite{Teaney03,HW02}, and the viscosity values needed to 
produce the desired reduction in $R_l$ turned out \cite{Teaney03} to 
reduce $v_2$ almost by a factor 2, incompatible with the data.
The ``RHIC HBT puzzle'' thus still awaits its resolution.

%%%%%%%%%%%%%%%%%%%%%%%%%%%%%%%%%%%%%%%%%%%%%%%%%%%%%%%%%%%%%%%%%%%%
\subsection{Non-central ($\bm{b{\ne}0}$) collisions --- azimuthally 
sensitive HBT interferometry (asHBT)}
\label{secVe}
%%%%%%%%%%%%%%%%%%%%%%%%%%%%%%%%%%%%%%%%%%%%%%%%%%%%%%%%%%%%%%%%%%%%

In non-central collisions the azimuthal symmetry of the source is 
broken by the impact parameter vector $\bb$ which, together with 
the beam direction, defines the reaction plane. Pairs emitted
at different azimuthal angles $\Phi\eq\angle(\bb,\bKp)$ 
relative to the reaction plane therefore originate in general from 
effective emission regions (homogeneity regions) of different sizes 
and shapes. 

As a result, the components of the spatial correlation tensor 
$S_{\mu\nu}\eq\langle \tilde x_\mu \tilde x_\nu\rangle$ now depend
not only on the magnitude of the transverse pair momentum $K_\perp$,
but also on its angle $\Phi$ relative to the reactuion plane. Furthermore,
the source is no longer reflection symmetric with respect to the plane
defined by the beam axis and $\bKp$ (i.e. symmetric under
$x_s{\,\to\,}{-}x_s$), so $R_{os}$ and $R_{ls}$ no longer vanish.
This means that for non-central heavy-ion collisions we have to
measure 6 (not 4) HBT radii, and they are functions not only of $Y$ and
$K_\perp$, but also of the emission angle $\Phi$ relative to the reaction
plane. A complete HBT analysis of non-central heavy-ion collisions
thus requires the event-wise determination of the reaction plane 
(through a Fourier analysis of the single particle spectrum as,
for example, used for extracting the elliptic flow) and the investigation
of the dependence of the HBT radii on the angle $\Phi$ relative to 
that plane.

%%%%%%%%%%%%%%%%%%%%%%%%%%%%%%%%%%%%%%%%%%%%%%%%%%%%%%%%%%%%%%%%%%%%
\subsubsection{asHBT formalism}
\label{secVe1}
%%%%%%%%%%%%%%%%%%%%%%%%%%%%%%%%%%%%%%%%%%%%%%%%%%%%%%%%%%%%%%%%%%%%

Experimentally it is still useful to characterize the source in terms
of HBT radii $R_o$ and $R_s$ parallel and perpendicular to the transverse
emission direction defined by the pair momentum $\bKp$. The emitting 
source $S(x,K)$, however, is most conveniently written down in 
transverse coordinates $x,y$ aligned with the reaction plane, since 
it typically (at least for collisions between spherical nuclei) is 
symmetric under reflection with respect to that plane, 
$y{\,\to\,}{-}y$ combined with $\Phi{\,\to\,}{-}\Phi$. The analogue
relations to Eqs.~(\ref{Rs2}-\ref{Rl2}), which express the HBT radii
through the components $S_{\mu\nu}\eq\langle\tx_\mu\tx_\nu\rangle$,
therefore involve an explicit rotation by the angle $\Phi$ between
the $(x,y)$ axes characterizing the source $S_{\mu\nu}$ and the
$(x_o,x_s)$ axes defined by the emission direction $\bKp$
\cite{Wiedemann:1997cr,Lisa:2000ip,Heinz:2002au}:
  \begin{eqnarray}
  \label{azim}
    R_s^2 &=& \textstyle{\2}(S_{xx}{+}S_{yy}) 
            - \textstyle{\2}(S_{xx}{-}S_{yy})\cos(2\Phi)
            - S_{xy} \sin(2\Phi)
  \nonumber\\
    R_o^2 &=& \textstyle{\2}(S_{xx}{+}S_{yy}) 
            + \textstyle{\2}(S_{xx}{-}S_{yy})\cos(2\Phi)
            + S_{xy} \sin(2\Phi)
  \nonumber\\
          &&- 2\beta_\perp (S_{tx} \cos\Phi{+}S_{ty} \sin\Phi)
             + \beta_\perp^2 S_{tt}, 
  \nonumber\\
     R_{os}^2 &=& S_{xy} \cos(2\Phi) 
        - \textstyle{\2} \left(S_{xx}{-}S_{yy}\right)\sin(2\Phi)
        + \beta_\perp (S_{tx} \sin\Phi{-}S_{ty} \cos\Phi), 
  \nonumber\\
    R_{l}^2 &=& S_{zz} -2 \beta_L S_{tz} + \beta_L^2 S_{tt}, 
  \nonumber\\
    R_{ol}^2 &=& \left( S_{xz}{-}\beta_L S_{tx}\right) \cos\Phi
               + \left( S_{yz}{-}\beta_L S_{ty}\right) \sin\Phi
               - \beta_\perp S_{tz} 
               + \beta_L\beta_\perp S_{tt},
  \nonumber\\
    R_{sl}^2 &=& \left(S_{yz}{-}\beta_L S_{ty}\right) \cos\Phi
               - \left(S_{xz}{-}\beta_L S_{tx}\right) \sin\Phi .
  \end{eqnarray}
These equations exhibit the {\em explicit} $\Phi$-dependence resulting from
the rotation between the $(x,y)$ and $(x_o,x_s)$ axes, but not the
{\em implicit} one reflecting the $\Phi$ dependence of the spatial
correlation tensor $S_{\mu\nu}(Y,K_\perp,\Phi)$. The total emission 
angle dependence of the HBT radii results from the combination of both.

The implicit $\Phi$-dependence of the spatial correlation tensor is
restricted by symmetries of the source \cite{Heinz:2002au}. It is a 
relativistic effect associated with an azimuthal spatial source 
deformation superimposed by strong transverse collective flow
\cite{Wiedemann:1997cr,H02} which vanishes with the $4^{\rm th}$ 
power of the transverse flow velocity $v_\perp/c$ for weak or no 
collective expansion \cite{Lisa:2000ip,H02}. 

A full analysis of the symmetry constraints on $S_{\mu\nu}(Y,\Kt,\Phi)$
for symmetric collisions between spherical nuclei and for pairs detected
in a symmetric rapidity window around $Y\eq0$ was performed in 
Ref.~\cite{Heinz:2002au}. One finds the following {\em most general} form
for the azimuthal oscillations of the HBT radii:
\begin{eqnarray}
 \label{24}
   R_s^2 &=& R_{s,0}^2 + 2\sum_{n=2,4,6,\dots} R_{s,n}^2\cos(n\Phi),
 \nonumber\\
   R_{os}^2 &=& \hspace*{11.5mm} 2\sum_{n=2,4,6,\dots} R_{os,n}^2\sin(n\Phi),
 \nonumber\\
   R_o^2 &=& R_{o,0}^2 + 2\sum_{n=2,4,6,\dots} R_{o,n}^2\cos(n\Phi),
 \nonumber\\
   R_{ol}^2 &=& \hspace*{11.5mm} 2\sum_{n=1,3,5,\dots} R_{ol,n}^2\cos(n\Phi),
 \nonumber\\
   R_l^2 &=& R_{l,0}^2 + 2\sum_{n=2,4,6,\dots}
   R_{l,n}^2\cos(n\Phi),
 \nonumber\\
   R_{sl}^2 &=& \hspace*{11.5mm} 2\sum_{n=1,3,5,\dots} R_{sl,n}^2\sin(n\Phi).\ 
\end{eqnarray}
We see that only even {\em or} odd sine {\em or} cosine terms occur, but
no mixtures of such terms. Statistical errors in the resolution of the 
reaction plane angle as well as finite angular bin sizes in $\Phi$ tend 
to reduce the actually measured oscillation amplitudes; fortunately, 
these dilution effects can be fully corrected by a model-independent 
correction algorithm \cite{Heinz:2002au,Borghini:2004ra}.
A Gaussian fit to the thus corrected correlation function, binned in $Y$,
$K_\perp$ and emission angle $\Phi$, then yields the ``true'' HBT radius
parameters $R_{ij}^2(Y,K_\perp,\Phi)$ from which the $n^{\rm th}$ order 
azimuthal oscillation amplitudes are extracted via
\begin{equation}
  R_{ij,n}^2(Y,K_\perp) = \frac{1}{n_{\rm bin}} \sum_{j=1}^{n_{\rm bin}} 
          R_{ij}^2(Y,K_\perp,\Phi_j) {\rm osc}(n \Phi_j).
\end{equation}
Here $n_{\rm bin}$ indicates the number of (equally spaced) $\Phi$ bins 
in the data and ${\rm osc}(n\Phi_j)$ stands for $\sin(n\Phi_j)$ or 
$\cos(n\Phi_j)$ as appropriate, see Eqs.\,(\ref{24}). (Note that 
Nyquist's theorem limits the number of harmonics that can be extracted 
to $n\leq n_{\rm bin}$.) 

We would like to relate the azimuthal oscillation amplitudes of the 
6 HBT radius parameters to the geometric and dynamical anisotropies 
of the source, as reflected in the azimuthal oscillations of the 10 
independent components of the spatial correlation tensor. Their allowed
oscillation patterns at midrapidity $Y\eq0$ are given by 
\cite{Heinz:2002au}
\begin{eqnarray}
\label{A-J}
  A(\Phi) \equiv \textstyle{\2}\langle\tilde x^2{+}\tilde y^2\rangle
  &=& \textstyle{A_0{+}2\sum_{n\geq2,{\rm even}} A_n\cos(n\Phi)},
\nonumber\\   
  B(\Phi) \equiv \textstyle{\2}\langle\tilde x^2{-}\tilde y^2\rangle
  &=& \textstyle{B_0{+}2\sum_{n\geq2,{\rm even}} B_n\cos(n\Phi)},
\nonumber\\  
 C(\Phi) \equiv \langle\tilde x\tilde y\rangle
  &=& \textstyle{\phantom{A_0{+}}2\sum_{n\geq2,{\rm even}} C_n\sin(n\Phi)},
\nonumber\\  
  D(\Phi) \equiv \langle\tilde t^2\rangle
  &=& \textstyle{D_0{+}2\sum_{n\geq2,{\rm even}} D_n\cos(n\Phi)},
\nonumber\\  
  E(\Phi) \equiv \langle\tilde t\tilde x\rangle
  &=& \textstyle{\phantom{A_0{+}}2\sum_{n\geq1,{\rm odd}} E_n\cos(n\Phi)},
\nonumber\\  
  F(\Phi) \equiv \langle\tilde t\tilde y\rangle
  &=& \textstyle{\phantom{A_0{+}}2\sum_{n\geq1,{\rm odd}} F_n\sin(n\Phi)},
\nonumber\\  
  G(\Phi) \equiv \langle\tilde t\tilde z\rangle
  &=& \textstyle{\phantom{A_0{+}}2\sum_{n\geq1,{\rm odd}} G_n\cos(n\Phi)},
\nonumber\\  
  H(\Phi) \equiv \langle\tilde x\tilde z\rangle
  &=& \textstyle{H_0{+}2\sum_{n\geq2,{\rm even}}\!H_n\cos(n\Phi)},
\nonumber\\  
  I(\Phi) \equiv \langle\tilde y\tilde z\rangle
  &=& \textstyle{\phantom{A_0{+}}2\sum_{n\geq2,{\rm even}} I_n\cos(n\Phi)},
\nonumber\\  
  J(\Phi) \equiv \langle\tilde z^2\rangle
  &=& \textstyle{J_0\,{+}2\sum_{n\geq2,{\rm even}} J_n\cos(n\Phi)}.
\end{eqnarray}
The missing terms in the sums over $n$ have amplitudes which are odd
functions of $Y$ and vanish at midrapidity. They do, however, contribute 
to the HBT radii if the data are averaged over a finite, symmetric 
rapidity window around $Y\eq0$ \cite{Heinz:2002au}. Their contributions 
can be eliminated by varying the width $\Delta Y$ of this rapidity window 
and extrapolating quadratically to $\Delta Y\to0$. Note that 
$C_0\eq{E}_0\eq{F}_0\eq{G}_0\eq{I_0}\eq0$ by symmetry, i.e. the 
corresponding components of $S_{\mu\nu}$ oscillate around zero.

The oscillation amplitudes of the HBT radii relate to the oscillation
amplitudes of the source parameters as follows 
\cite{Heinz:2002au,Heinz:2003zt}: For the odd harmonics $n\eq1,3,5,\dots$ 
we have
\begin{eqnarray}
  R_{ol,n}^2 &=& {\textstyle\2}
 \langle H_{n{-}1}{+}H_{n{+}1}{-}I_{n{-}1}{+}I_{n{+}1}
 -\beta_L(E_{n{-}1}{+}E_{n{+}1}{-}F_{n{-}1}{+}F_{n{+}1})\rangle
\nonumber\\
  && -\langle  \beta_\perp G_n -\beta_L D_n\rangle,
\\\nonumber
  R_{sl,n}^2 &=& {\textstyle\2}
 \langle{-}H_{n{-}1}{+}H_{n{+}1}{+}I_{n{-}1}{+}I_{n{+}1}
 -\beta_L({-}E_{n{-}1}{+}E_{n{+}1}{+}F_{n{-}1}{+}F_{n{+}1})\rangle,
\end{eqnarray}
whereas the even harmonics $n=0,2,4,\dots$ satisfy
\begin{eqnarray}
  R_{s,n}^2 &=& \langle A_n\rangle 
  +{\textstyle\2}\langle {-}B_{n{-}2}{-}B_{n{+}2}{+}C_{n{-}2}{-}C_{n{+}2}
  \rangle,
\nonumber\\
  R_{o,n}^2 &=& \langle A_n\rangle 
 +{\textstyle\2} \langle B_{n{-}2}{+}B_{n{+}2}{-}C_{n{-}2}{+}C_{n{+}2}\rangle
\nonumber\\
 && - \beta_\perp\langle E_{n{-}1}{+}E_{n{+}1}{-}F_{n{-}1}{+}F_{n{+}1}\rangle
                + \beta_\perp^2 \langle D_n\rangle,
\nonumber\\
  R_{os,n}^2 &=& {\textstyle\2}\langle{-}B_{n{-}2}{+}B_{n{+}2}
   {+}C_{n{-}2}{+}C_{n{+}2}
  + \beta_\perp(E_{n{-}1}{-}E_{n{+}1}{-}F_{n{-}1}{-}F_{n{+}1})\rangle,
\nonumber\\
  R_{l,n}^2 &=& \langle J_n\rangle -2\langle\beta_L G_n\rangle
   +\langle\beta_L^2 D_n\rangle.
\end{eqnarray}
In these relations it is understood that all negative harmonic coefficients
$n<0$ as well as $C_0,E_0,F_0,G_0$ and $I_0$ are zero. The angular brackets 
$\langle\dots\rangle$ indicate an average over a finite, symmetric rapidity 
window around $Y\eq0$. The terms involving the longitudinal pair velocity 
$\beta_L$ vanish quadratically as the width $\Delta Y$ of that window 
shrinks to zero. 

Even after extrapolating to $Y\eq0$ in this way, we have still many
more source parameters than measurable HBT amplitudes. One counts easily 
that up to $n\eq2$ there are 9 measurable Fourier coefficients which 
(at $Y\eq0$) depend on 19 source amplitudes. From there on, increasing 
$n$ by 2 yields 6 additional measured amplitudes which depend on 10 
additional source amplitudes. This lack of analysis power is an 
intrinsic weakness of the HBT microscope and due to the fundamental
restrictions arising from the mass-shell constraint 
$q^0\eq\bm{\beta}\cdot\bq$. The reconstruction of the source
thus must necessarily rely on additional assumptions. 

One such assumption which may not be too unreasonable is that the 
emission duration $D\eq\langle\tilde t^2\rangle$ is approximately
independent of emission angle and that the source is sufficiently
smooth that higher order harmonics $n\geq3$ of $S_{\mu\nu}$ can be
neglected. Such source properties would result in the ``Wiedemann sum
rule'' \cite{Wiedemann:1997cr} 
\begin{equation}
  R_{o,2}^2-R_{s,2}^2 + 2 R_{os,2}^2 = 0
\end{equation}
which can be experimentally tested. If verified for all $\Kt$ it would
provide strong support for the underlying assumptions on the source. 
In this case we can measure 3 azimuthally averaged HBT radii and
5 independent oscillation amplitudes with $n\leq2$, depending
on 14 source parameters of which 5 can be eliminated by going to 
$\Kt\eq\beta_\perp\eq0$ (see \cite{Heinz:2002au} for explicit 
expressions). This makes the geometry of the effective source for 
particles with $\Kt\eq0$ ``almost solvable''\footnote{Note that the 
  limit $\Kt\to0$ eliminates all influence from the 
  temporal structure of the source; the emission duration and 
  correlations between position and time at freeze-out
  must be extracted from correlation data at non-zero $\Kt$ where,
  however, the explicit $\beta_\perp$-dependence associated with 
  factors $t$ in the variances $\langle\tilde x_\mu\tilde x_\nu\rangle$
  interferes with the implicit $\Kt$-dependence in the spatial
  variances which result from collective expansion flow \cite{WH99}.
  Models for disentangling these different contributions to the 
  $\Kt$-dependence have been extensively studied for azimuthally
  symmetric sources \cite{WH99}, but must be generalized for
  azimuthally deformed sources \cite{Retiere:2003kf}.}, as confirmed 
by hydrodynamical calculations \cite{HK02osci} which show that at 
$\Kt\eq0$ the effective emission region closely tracks the overall 
geometry of the source even if it is strongly and anisotropically 
expanding. 

The source geometry can be completely reconstructed from HBT data if 
transverse flow is so weak that all implicit $\Phi$-dependence (i.e. 
all higher harmonics $n\geq1$) of $S_{\mu\nu}$ can be neglected. In 
this case one obtains at $Y\eq0$ the ``geometric relations'' 
\cite{Lisa:2000ip} 
\begin{eqnarray}
  &&R_{s,0}^2 = A_0 = \textstyle{\2}\langle\tilde x^2{+}\tilde y^2\rangle_0,
\nonumber\\
  &&R_{o,0}^2-R_{s,0}^2 = \beta_\perp^2 D_0 
                        = \beta_\perp^2 \langle\tilde t^2\rangle_0,
\nonumber\\
  &&R_{l,0}^2 = J_0 = \langle\tilde z^2\rangle_0,
\nonumber\\
  &&R_{ol,1}^2 = - R_{sl,1}^2 = \textstyle{\2} H_0
                  = \textstyle{\2}\langle\tilde x\tilde z\rangle_0,
\nonumber\\
  &&R_{o,2}^2 = - R_{s,2}^2 = - R_{os,2}^2 = \textstyle{\2} B_0
    = \textstyle{\frac{1}{4}}\langle\tilde x^2{-}\tilde y^2\rangle_0.
\end{eqnarray}
$A_0$ describes the average transverse size and $B_0$ (which generates 
a second-order harmic in the transverse HBT radii) the transverse 
deformation of the source. $H_0$ generates a first-order harmonic 
in the $ol$ and $sl$ cross terms and describes a longitudinal tilt of 
the source away from the beam direction \cite{Lisa:2000ip}. Such a tilt 
was found in Au+Au collisions at the AGS \cite{E895}. Its sign yielded 
important information on the kinetic pion production mechanism 
\cite{Lisa:2000ip,E895}.

%%%%%%%%%%%%%%%%%%%%%%%%%%%%%%%%%%%%%%%%%%%%%%%%%%%%%%%%%%%%%%%%%%%%
\subsubsection{Azimuthal oscillations of HBT radii in hydrodynamics 
and RHIC data}
\label{secVe2}
%%%%%%%%%%%%%%%%%%%%%%%%%%%%%%%%%%%%%%%%%%%%%%%%%%%%%%%%%%%%%%%%%%%%

We can use the hydrodynamic model introduced in Section \ref{sec4a9} to
describe the momentum spectra and explore what it would predict for the
azimuthal oscillations of the HBT radii in non-central Au+Au collisions 
at RHIC \cite{HK02osci}. The most interesting question, perhaps, is 
whether we can reconstruct the orientation and magnitude of the 
{\em spatial} deformation of the source at freeze-out from the HBT
analysis, thereby complementing our knowledge of the orientation
and magnitude of the {\em momentum} anisotropy from the elliptic
flow analysis.

To see how different orientations of the spatial deformation 
(in-plane elongated vs. out-of-plane elongated) manifest themselves
in the azimuthal oscillations of the HBT radii, one can explicitly 
construct sources of either orientation and perform an HBT analysis on 
them. Since the elliptic flow makes the fireball expand faster into
the reaction plane than perpendiculer to it, we can generate an 
in-plane elongated source (IPES) at freeze-out from the initial 
out-of-plane oriented overlap region simply by making sure that the 
elliptic flow has enough time to act before the system decouples.
By increasing the initial energy density or temperature, we can make
the total time until decoupling arbitrarily large. Let us therefore 
study hydrodynamic sources evolving from different sets of initial 
conditions \cite{HK02osci}.

The first set (labelled ``RHIC1'') corresponds to optimized values for 
central Au+Au collisions at $\scm\eq130$\,GeV which were used
earlier in Section \ref{sec4a9} (see Table~1). The second set of 
initial conditions (labelled ``IPES'') uses a much higher initial 
temperature $T_{\rm eq}\eq2$\,GeV (at $b\eq0$), and the hydrodynamic 
evolution is started at $\tau_{\rm eq}\eq0.1$\,fm/$c$ and 
stopped when a freeze-out temperature $T_{\rm dec}\eq100$\,MeV has 
been reached. Such a high initial temperature can probably not even 
be achieved at the LHC, and the strong transverse flow generated in 
this case probably causes the system to decouple already much closer 
to the hadronization temperature $\Tc\eq164$\,MeV. But if we force 
the system to start at such a high temperature and freeze out so low, 
we give it enough time to convert the initial out-of-plane deformation 
into a final in-plane deformation (which is the phenomenon we want to 
study). The following analysis refers to semi-peripheral Au+Au collisions 
with impact parameter $b\eq7$\,fm.

%%%%%%%%%%%%%%%%%%%%%%%%%%% Fig. 16f %%%%%%%%%%%%%%%%%%%%%%%%%%%%%%%%%%%%%%%%%
\begin{figure}[htb]
\begin{center}
\epsfig{file=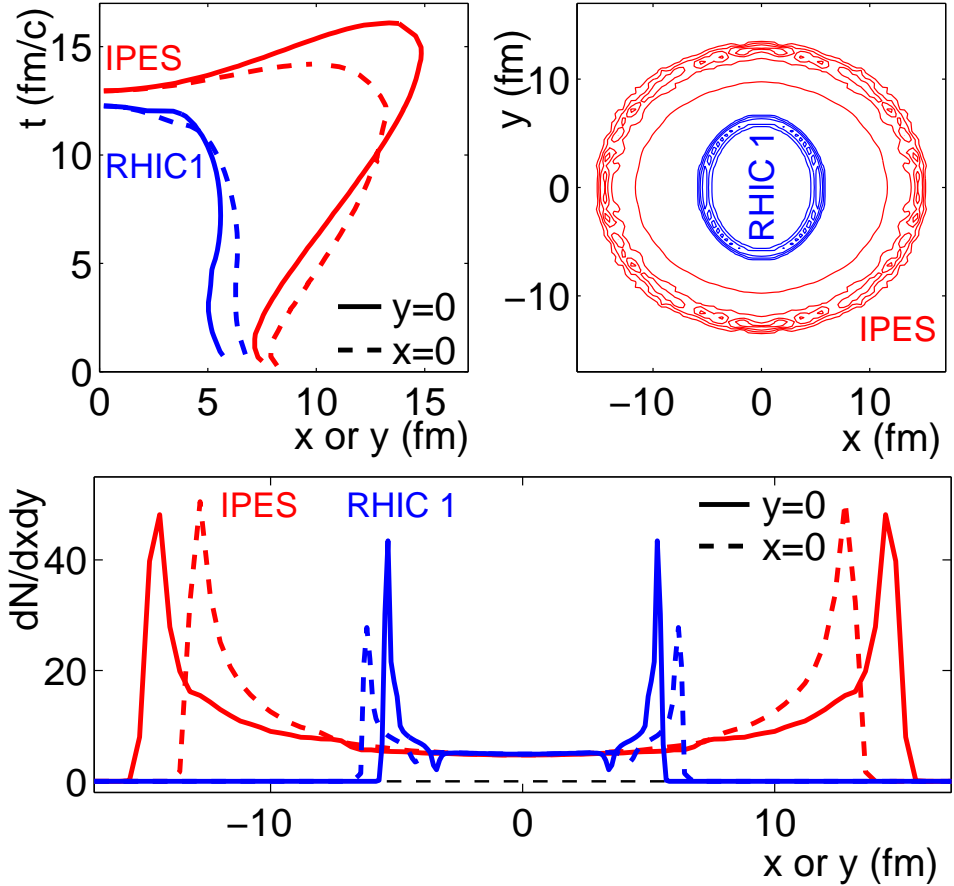,bb=117 228 512 594,width=0.8\linewidth}
\caption{(a) Cuts through the freeze-out surface $t_{\rm f}(\br)$ 
         at $z\eq0$ along and perpendicular to the reaction plane.
         (b) Contour plots in the transverse plane of the time-, $z$-, 
         and momentum-integrated emission function $d^2N/d^2r$ for 
         RHIC1 and IPES initial conditions (see text). 
         (c) Cuts through diagram (b) along and perpendicular to the 
         reaction plane, for RHIC1 and IPES initial conditions. In all
         cases $b\eq7$\,fm.
         }  
\label{F16f} 
\end{center}
\vspace*{-5mm}
\end{figure}
%%%%%%%%%%%%%%%%%%%%%%%%%%%%%%%%%%%%%%%%%%%%%%%%%%%%%%%%%%%%%%%%%%%%%%%%%%%%%%%

%%%%%%%%%%%%%%%%%%%%%%%%%%% Fig. 16g %%%%%%%%%%%%%%%%%%%%%%%%%%%%%%%%%%%%%%%%%%
\begin{figure}[htb]
\begin{center}
\epsfig{file=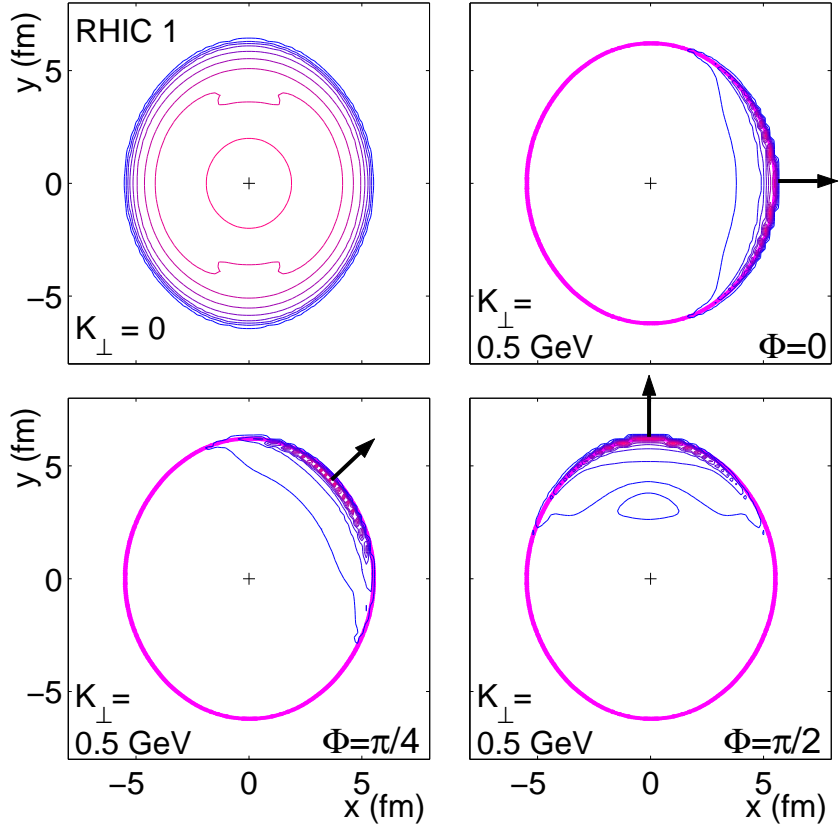,bb=108 190 531 607,width=0.7\linewidth}
\caption{Contours of constant emission density in the transverse plane
         for RHIC1 initial conditions (see text). The thick line 
         indicates the largest transverse extension of the freeze-out 
         hypersurface (see Figs.~\ref{F16f}a,b). The four panels show 
         emission regions for midrapidity pions ($Y\eq0$) with 
         $\Kt\eq0$ and, for three emission angles indicated by arrows, 
         with $\Kt\eq0.5$\,GeV.
}  
\label{F16g} 
\end{center}
\vspace*{-5mm}
\end{figure}
%%%%%%%%%%%%%%%%%%%%%%%%%%%%%%%%%%%%%%%%%%%%%%%%%%%%%%%%%%%%%%%%%%%%%%%%%%%%%%%

Figure~\ref{F16f}a shows the freeze-out surface $t_{\rm dec}(\br)$
in cuts along ($y\eq0$) and perpendicular to the reaction plane 
($x\eq0$). Initially the source is extended out-of-plane (larger in 
$y$ than in $x$ direction), but it then expands more rapidly into the 
$x$-direction, becoming in-plane elongated at later times. For RHIC1 
initial conditions this only happens after most of the matter has 
already decoupled; as a consequence the time-integrated source, shown 
in Fig.~\ref{F16f}b, is still longer in $y$ than in $x$ direction in 
the RHIC1 case. For IPES initial conditions the deformation changes 
sign before most particles decouple, and the time-integrated source 
appears in-plane-extended (see again Fig.~\ref{F16f}b). Also, it is 
much larger due to the much higher initial energy density and longer 
lifetime. 

Figure~\ref{F16f}c shows cuts along and perpendicular to the reaction 
plane through the density contour plots 
$d^2N/d^2r\eq\int (d^3K/E_K) dz\, dt\,S(x,K)$ of Fig.~\ref{F16f}b.
Pion emission is seen to be strongly surface peaked, in particular
at RHIC1 where the freeze-out radius is almost constant for a long
time. This ``opacity'' is weaker both at lower collision energies 
(where the freeze-out surface shrinks to zero continuously 
\cite{KSH99}) and at higher energies, due to larger temporal 
variations of the freeze-out radius.

\begin{figure}[htb]
\begin{center}
\epsfig{file=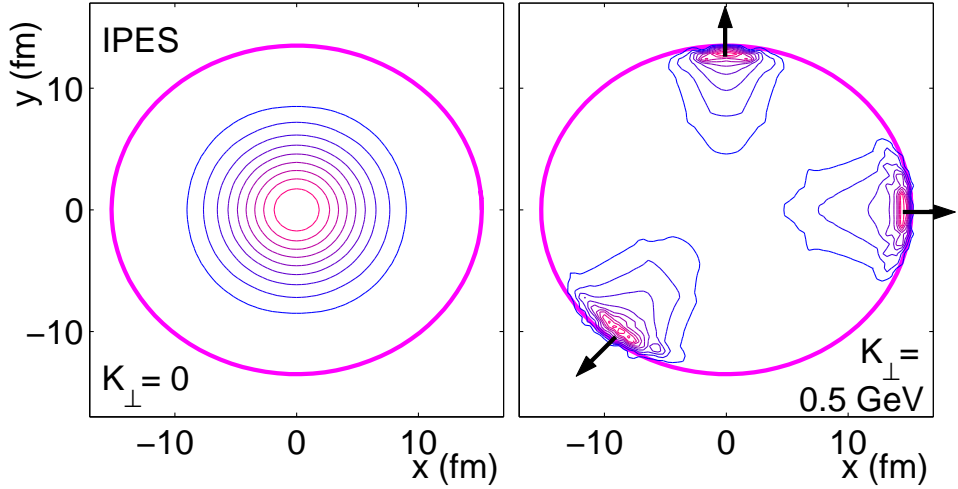,bb=95 363 547 594,width=0.8\linewidth}
\caption{Same as Fig.~\ref{F16g}, but for IPES initial conditions.
}  
\label{F16h} 
\end{center}
\vspace*{-5mm}
\end{figure}

Figures \ref{F16g} and \ref{F16h} show the spatial distributions of 
pions emitted with fixed momentum. Shown are density contours of 
$\int dz\,dt\,S(x,K)$ in the transverse plane $(x,y)$ for pions
%
%%%%%%%%%%%%%%%%%%%%%%%%%%% Fig. 16h %%%%%%%%%%%%%%%%%%%%%%%%%%%%%%%%%%%%%%%%%%
%%%%%%%%%%%%%%%%%%%%%%%%%%%%%%%%%%%%%%%%%%%%%%%%%%%%%%%%%%%%%%%%%%%%%%%%%%%%%%%
%
with rapidity $Y\eq0$ and fixed $\Kt$, for three emission angles 
$\Phi\eq0,\,45^\circ$ and $90^\circ$ relative to the reaction 
plane. (In Fig.~\ref{F16h} we replaced $\Phi\eq45^\circ$ for clarity
by the equivalent angle $\Phi\eq225^\circ$.) Particles with vanishing 
transverse momentum $\Kt$ are seen to be emitted from almost the 
entire interior of the ``bathtub'' shown in Figs.~\ref{F16f}b,c; for 
RHIC1 (IPES) initial conditions this region is elongated out-of-plane 
(in-plane). For slow pions the source thus looks transparent. Pions 
with sufficiently large transverse momenta are emitted from relatively 
thin regions close to the rim of the ``bathtub'' where the flow 
velocity is largest and points into the direction of the emitted 
pions. For fast pions the source thus looks opaque, more so at RHIC than 
at higher energies. Their emission regions rotate with the emission 
angle, constrained by the shape of the ``bathtub''.

\begin{figure}[htb]
\begin{center}
\epsfig{file=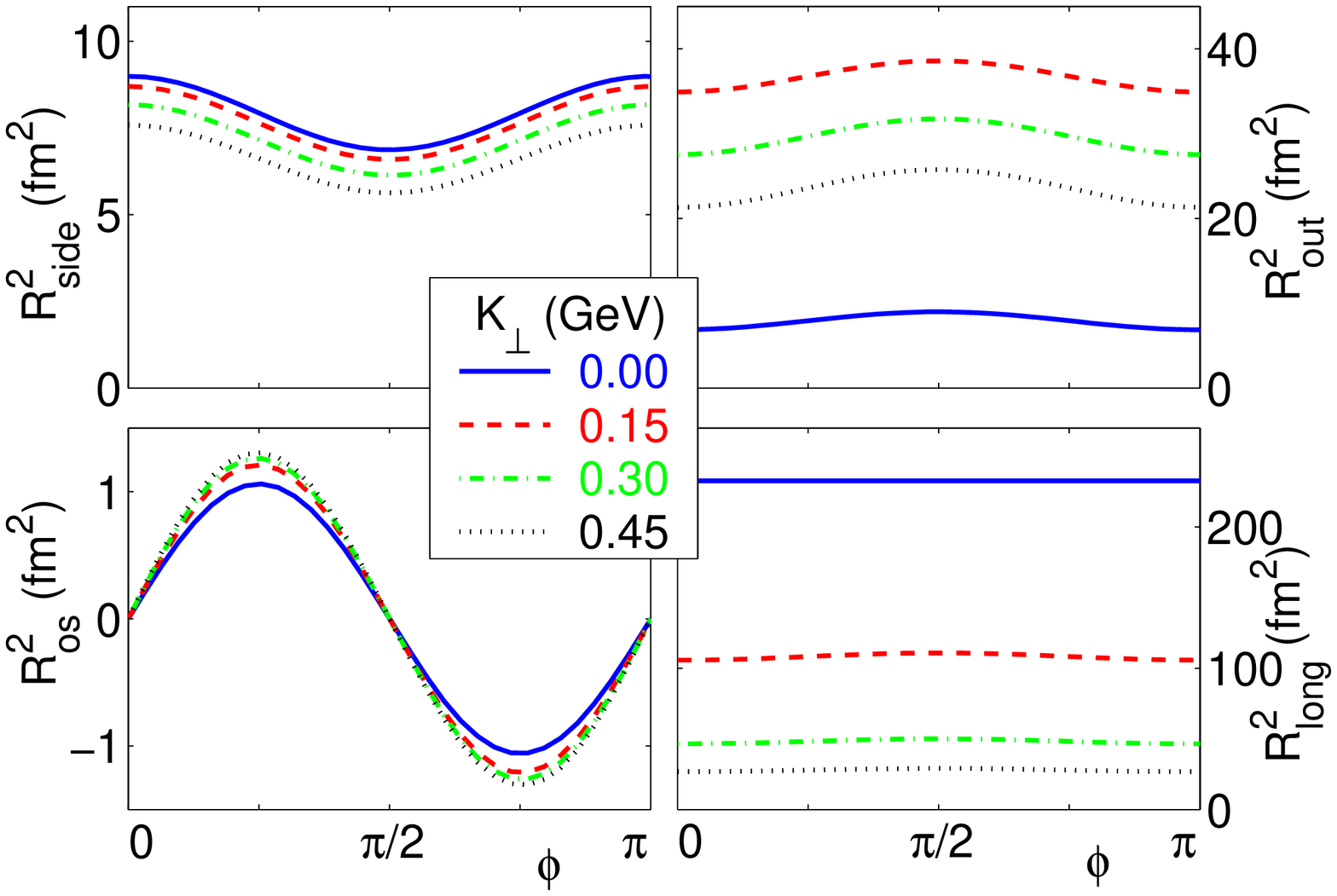,bb=19 208 597 594,width=0.495\linewidth}
\epsfig{file=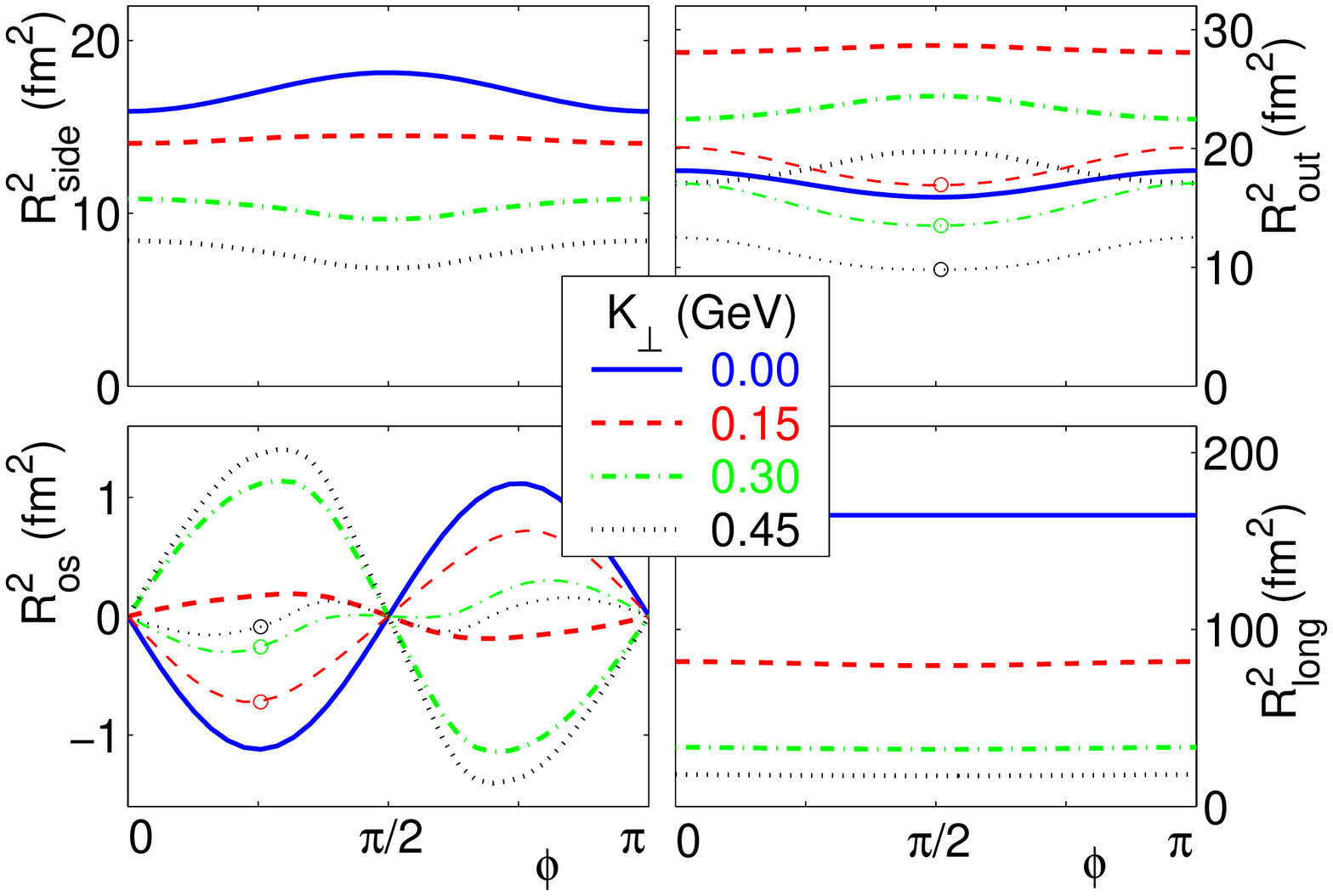,bb=19 208 597 594,width=0.495\linewidth}
\caption{Left: Azimuthal oscillations of the HBT radii at $Y\eq0$ for 
         $b\eq7$\,fm Au+Au collisions at $\scm\eq130$\,GeV (RHIC1),
         for four values of the transverse momentum $\Kt$ 
         as indicated. Right: The same for IPES initial conditions.
         For $R_o^2$ and $R_{os}^2$ the geometric contributions 
         are shown separately as thin circled lines.}  
\label{F16i} 
\end{center}
\vspace*{-5mm}
\end{figure}

The HBT radii, calculated from the widths of the $K$-dependent 
emission regions according to Eqs.~(\ref{azim}), are shown in
Fig.~\ref{F16i}, as functions of the azimuthal emission angle
$\Phi$. (Since $Y\eq0$, all terms $\sim\beta_L$ vanish.) Note that 
$R_o^2$ and $R_{os}^2$ receive purely geometric and mixed space-time 
correlation contributions; the latter are proportional to the pair 
velocity $\beta_\perp$. Since we want to use pion interferometry to 
obtain information on the geometric deformation of the source at 
freeze-out, these should be analyzed separately. In the right panel 
of Fig.~\ref{F16i} we therefore show the geometric contributions 
separately as thin circled lines. For a detailed discussion of all 
the curves in Fig.~\ref{F16i} I refer the reader to Ref.~\cite{HK02osci};
here I will only discuss the most important point.

%
%%%%%%%%%%%%%%%%%%%%%%%%%%% Fig. 16i %%%%%%%%%%%%%%%%%%%%%%%%%%%%%%%%%%%%%%%%%%
%%%%%%%%%%%%%%%%%%%%%%%%%%%%%%%%%%%%%%%%%%%%%%%%%%%%%%%%%%%%%%%%%%%%%%%%%%%%%%%
%

Let us first look at the geometric contributions. From Figs.~\ref{F16g} 
and \ref{F16h} we already know that pions with $\Kt\eq0$ are emitted 
from almost the entire fireball and thus probe the different sign of
the spatial deformation of the {\em total} (momentum-integrated) RHIC1 
and IPES sources shown in Fig.~\ref{F16f}. This is reflected by the 
opposite sign of the oscillation amplitudes of $R_s^2$ and of the 
geometric contribution to $R_o^2$ at $\Kt\eq0$ in the left and right 
panels of Fig.~\ref{F16i}. At higher $\Kt$-values the oscillations 
for $R_s^2$ in the right panel Fig.~\ref{F16i} change sign, but those 
of the geometric contribution to $R_o^2$ do not. This reflects an 
intricate interplay between geometric and flow effects \cite{HK02osci}. 
However, the opposite signs of the oscillation amplitudes near $\Kt\eq0$
of $R_s^2$, $R_o^2$, and $R{_os}^2$ for the RHIC1 and IPES sources
are clear and unique indicators of the opposite directions of the spatial
deformation for these two sources and can thus be used to diagnose
the latter. In particular, the ratio of the second harmonic oscillation 
amplitude $R_{s,2}^2$ of the sideward radius to its $\Phi$-averaged
value $R_{s,0}^2$ is seen to be almost independent of $\Kt$ 
\cite{HK02osci,Retiere:2003kf}. This ratio can therefore be used to 
extract the spatial deformation of the {\em entire} source even at 
non-zero (albeit not too large) $\Kt$, even though the homogeneity 
regions at non-zero $\Kt$ do not probe the entire source
\cite{Retiere:2003kf,STARasHBT}.

%
%%%%%%%%%%%%%%%%%%%%%%%%%%% Fig. 16k %%%%%%%%%%%%%%%%%%%%%%%%%%%%%%%%%%%%%%%%%%
\begin{figure}[htb]
\begin{center}
\epsfig{file=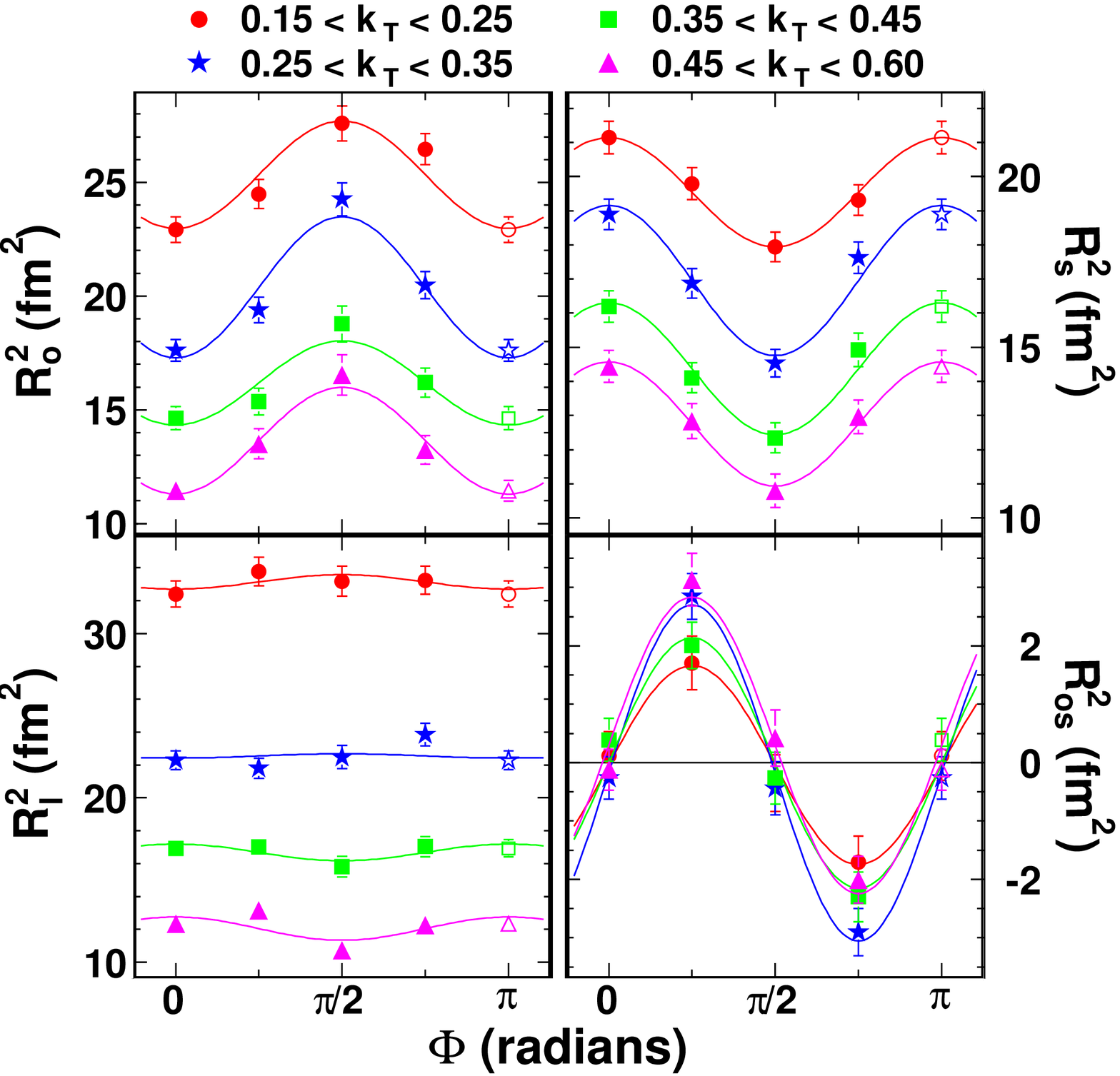,bb=0 15 567 549,width=0.5\linewidth,height=9cm}
\epsfig{file=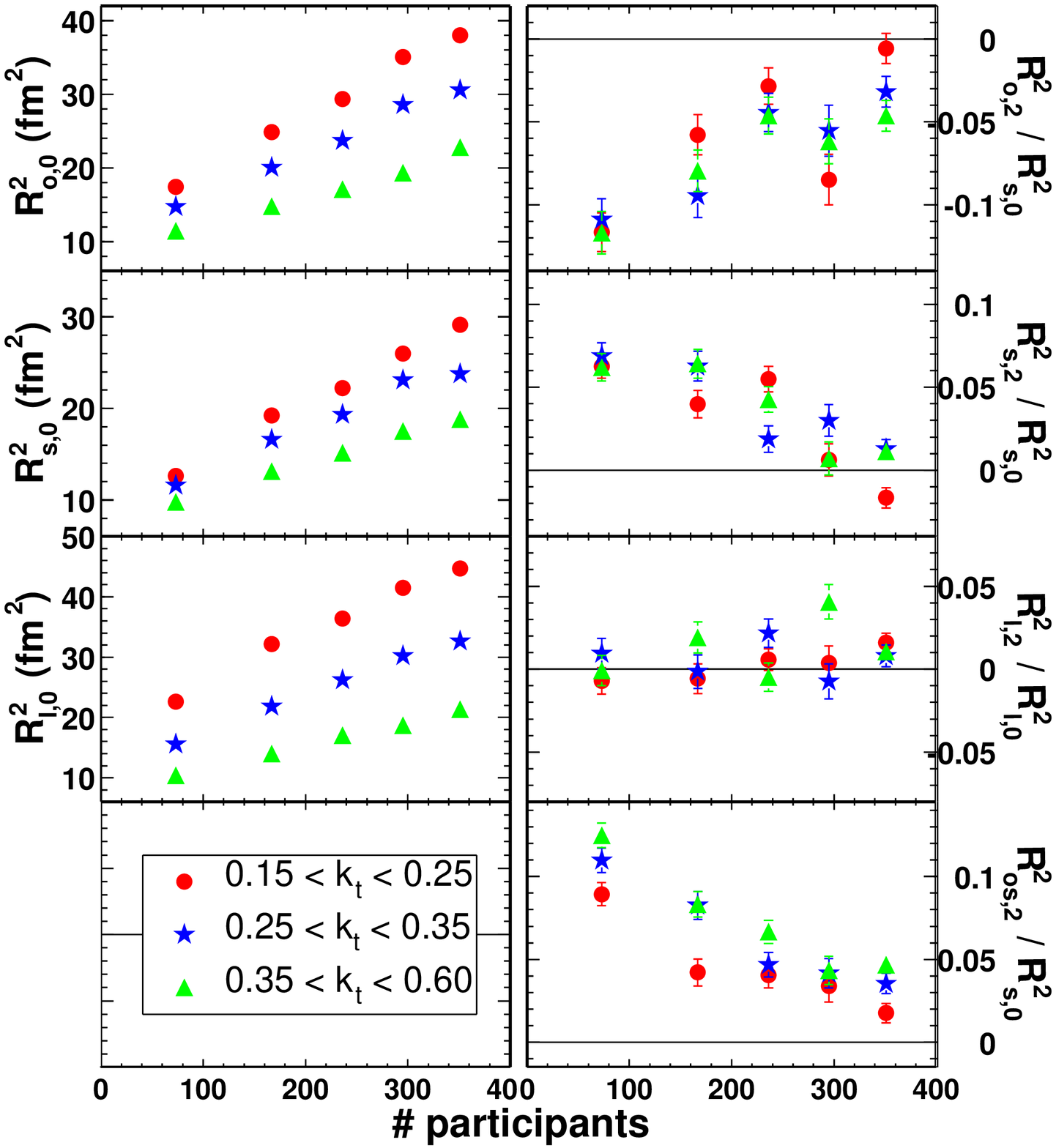,bb=0 0 567 611,width=0.48\linewidth}
\caption{Left: Azimuthal oscillations of the HBT radii near $Y\eq0$ from 
         Au+Au collisions at $\scm\eq200$\,GeV, for 20{\%}-30{\%} centrality
         events and four bins of the transverse momentum $\Kt$ (GeV/$c$)
         \protect\cite{STARasHBT}. The lines show fits with Eqs.~(\ref{24}).
         Right: Zeroth order (left column) and normalized second order 
         (right column) harmonic coefficients of the HBT radii, for 
         three $\Kt$ bins (stars, circles, and triangles), as a function
         of collision centrality, from the same STAR experiment 
         \cite{STARasHBT}. Larger participant numbers correspond to more
         central collisions.} 
\label{F16k} 
\end{center}
\vspace*{-5mm}
\end{figure}
%%%%%%%%%%%%%%%%%%%%%%%%%%%%%%%%%%%%%%%%%%%%%%%%%%%%%%%%%%%%%%%%%%%%%%%%%%%%%%%
%

Recently, the STAR Collaboration published the first azimuthally
sensitive HBT analysis of Au+Au collisions at RHIC \cite{STARasHBT}. 
As you can see in Fig.~\ref{F16k}, the quality and level of detail
of the data is impressive.
The analysis is fully 3-dimensional in the relative momentum $\bq$ and 
sufficiently finely binned in $\Kt$, $\Phi$, and collision centrality
to obtain a reasonably complete HBT snapshot of the final state of
the collision fireball created at RHIC. (Of course, theorists will 
always keep asking for more detail, but even that is coming.) It is 
clear that with these asHBT data a new dimension has opened, with 
unprecedented constraining power on theories!

Comparing the left panels in Figs.~\ref{F16i} and \ref{F16k} one sees 
very strong similarity. The absolute scales on the vertical axes don't 
match, but this just reflects the ``RHIC HBT puzzle'' from the previous 
subsection, namely that the hydrodynamic model does not reproduce the 
angle-averaged HBT radii and overpredicts $R_l$ and $R_o$ while 
underpredicting $R_s$. The signs of the oscillation amplitudes match,
however, clearly indicating that the measured source is still 
somewhat out-of-plane elongated at freeze-out, as predicted by the
hydrodynamic model (see Figs.~\ref{F16f} and \ref{F16g}). The 
centrality systematics shown in the right panel of Fig.~\ref{F16k}
shows that the HBT radii grow as the collisions become more central
(left column -- no surprise there), but that the growth rate decreases,
together with the radii themselves, as the pair momentum $\Kt$ increases.
This is qualitatively consistent with strong longitudinal and radial flow, 
as discussed in the previous section, even though the measured 
$\Kt$-dependence of $R_{s,0}^2$ and $R_{o,0}^2$ is stronger than predicted.
The experimental data show that the strong $\Kt$-dependence of $R_{s,0}^2$
in particular is not only a feature of central collisions, but seems to
persist towards peripheral collisions.

The rightmost column of Fig.~\ref{F16k} shows that the oscillation 
amplitudes are are largest in peripheral collisions and decrease to zero
in central collisions. Again this is no surprise since the 
oscillations reflect the geometric deformation of the source and 
peripheral collisions produce, at least initially, the most deformed 
fireballs. Note the very weak $\Kt$-dependence of the {\em relative}
oscillation amplitudes which is, at least qualitatively, consistent 
with hydrodynamic predictions. A more quantitative analysis of the
%
%%%%%%%%%%%%%%%%%%%%%%%%%%% Fig. 16l %%%%%%%%%%%%%%%%%%%%%%%%%%%%%%%%%%%%%%%%%%
\begin{figure}[htb]
\begin{center}
\epsfig{file=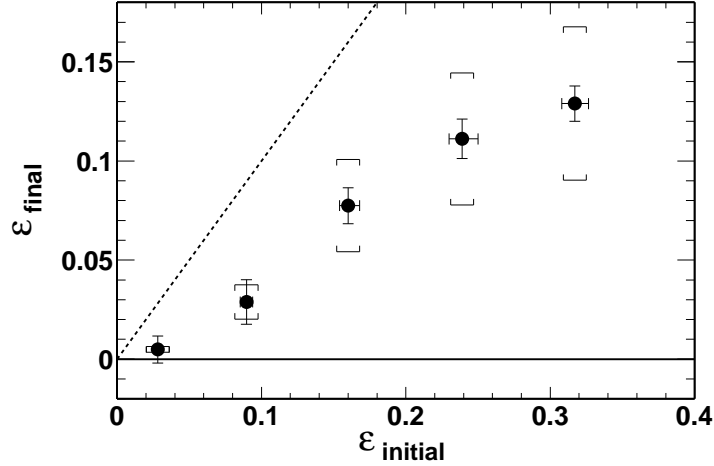,width=0.6\linewidth}
\caption{Final source eccentricity at freeze-out, $\varepsilon_{\rm final}$,
         extracted from the normalized second order harmonic coefficient 
         of the sideward HBT radius, $R_{s,2}^2/R_{s,0}^2$
         \cite{Retiere:2003kf} (see text), as a function of the initial 
         eccentricity, $\varepsilon_{\rm initial}$, from a Glauber model.
         Smaller eccentricities correspond to more central collisions.
         The dotted line indicates 
         $\varepsilon_{\rm initial}\eq\varepsilon_{\rm final}$. The data
         are from Au+Au collisions at $\scm\eq200$\,GeV meassured by
         STAR \cite{STARasHBT}.} 
\label{F16l} 
\end{center}
\vspace*{-5mm}
\end{figure}
%%%%%%%%%%%%%%%%%%%%%%%%%%%%%%%%%%%%%%%%%%%%%%%%%%%%%%%%%%%%%%%%%%%%%%%%%%%%%%%
%
source deformation as a function of centrality is presented in 
Fig.~\ref{F16l} from the same publication \cite{STARasHBT}, where
the final spatial eccentricity $\varepsilon_{\rm final}$ at decoupling,
extracted from the normalized oscillation amplitude $R_{s,2}^2/R_{s,0}^2$
with the help of a blast-wave model parametrization of the source
\cite{Retiere:2003kf}, is contrasted with the initial spatial 
deformation $\varepsilon_{\rm initial}$ of the original nuclear 
overlap zone, calculated from the Glauber model. At all impact 
parameters, the final deformation is smaller than the initial one,
showing the consequences of elliptic flow. Relatively speaking, the
reduction of the eccentricity is stronger in almost central than
in peripheral collisions, but at no centrality does the freeze-out
configuration exhibit a negative value for $\varepsilon_{\rm final}$,
which would indicate an in-plane-elongated source. For mid-peripheral
($b\sim7$\,fm) collisions (second point from the right), the extracted 
value $\varepsilon_{\rm final}\eq0.11\pm0.035$ agrees (within the 
relatively large systematic error of about 30\%) with the eccentricity 
of the hydrodynamic source shown in Figs.~\ref{F16f},\ref{F16g}
which has $\varepsilon_x\simeq0.14$. This provides further support for 
the hydrodynamic model, since it indicates an intrinsic dynamical 
consistency between the evolution of the eccentricity from its initial 
to its final value and the hydrodynamic expansion time scales and flow 
profiles. A stronger and more confident statement requires, of course, 
the resolution of the ``HBT puzzle'', i.e. a quantitative theoretical 
understanding of the angle-averaged HBT radii which is so far missing.   

%%%%%%%%%%%%%%%%%%%%%%%%%%%%%%%%%%%%%%%%%%%%%%%%%%%%%%%%%%%%%%%%%%%%%%%%%%%%%%
\subsection{What has HBT interferometry of heavy-ion collisions taught us?}
\label{secVf}
%%%%%%%%%%%%%%%%%%%%%%%%%%%%%%%%%%%%%%%%%%%%%%%%%%%%%%%%%%%%%%%%%%%%%%%%%%%%%%

Let me quickly summarize this long chapter by stating the following as 
our {\bf Third Lesson}: Two-particle momentum correlations, in 
particular Bose-Einstein correlations between pairs of identical bosons,
are a powerful tool to constrain the {\em space-time structure and evolution} 
of the heavy-ion fireball, thereby yielding complementary information
to its {\em momentum-space structure} extracted from the single-particle 
spectra. Emission-angle dependent HBT interferometry for non-central
collisions (or central collisions between deformed nuclei) allows to 
extract the shape and orientation of the deformed source at freeze-out.
By comparing it with the initial spatial deformation of the nuclear
overlap region when the nuclei first hit each other, we get an idea
how the system evolved and can constrain the time between impact and
decoupling. Azimuthally sensitive HBT interferometry (asHBT) is a 
newly developing field, but first results indicate consistency of the 
extracted final deformation with hydrodynamic evolution models. On the 
other hand, the absolute magnitudes of the emission-angle averaged HBT 
radii and their dependence on the transverse momentum $\Kt$ of the 
emitted pairs is still poorly understood: Almost all dynamical models 
overpredict $R_l$ and $R_o$, underpredict $R_s$ and are unable to 
reproduce the measured strong $\Kt$-dependence of the transverse radii 
$R_s$ and $R_o$. This {\bf ``RHIC HBT Puzzle''} indicates that our 
understanding of the kinetics of the freeze-out process is still 
insufficient and constitutes, in my opinion, the most important open 
problem in our present global picture of heavy-ion collision dynamics.

%%%%%%%%%%%%%%%%%%%%%%%%%%%%%%%%%%%%%%%%%%%%%%%%%%%%%%%%%%%%%%%%%%%%%%%%%%%%%%
\section{STATISTICAL HADRONIZATION AND PRIMORDIAL HADROSYNTHESIS}
\label{sec5}
%%%%%%%%%%%%%%%%%%%%%%%%%%%%%%%%%%%%%%%%%%%%%%%%%%%%%%%%%%%%%%%%%%%%%%%%%%%%%%

In this section we'll discuss what kind of information can be extracted
from the measured hadron abundances (i.e. from the {\em normalization}
of the hadron spectra, in contrast to their {\em shape} which we discussed
in the previous section). Up to this point we did not discuss the 
normalization of any of the spectra, nor did we care about it. All we 
needed for local thermal equilibrium and the build-up of thermal pressure 
and collective flow was rapid momentum equilibration, and this can in 
principle be achieved entirely by elastic collisions (including resonant 
elastic collisions in the hadronic phase such as $\pi{+}N\to\Delta\to\pi{+}N$)
which do not change any of the particle abundances. Whether or not the 
microscopic interactions alter the chemical composition or leave it 
unchanged is, for this part of the discussion, entirely irrelevant.

This analysis of the spectral shapes has already taught us that hadrons
are not formed right away but appear with some delay via the hadronization
of a dense, thermalized, collectively expanding quark-gluon system. Let us 
now consider this hadronization process and ask how it affects the 
hadron abundances, i.e. the {\em chemical composition} of the hadronic
system emerging from this process.

%%%%%%%%%%%%%%%%%%%%%%%%%%%%%%%%%%%%%%%%%%%%%%%%%%%%%%%%%%%%%%%%%%%%%%%%%%%%
\subsection{Statistical hadronization and the Maximum Entropy Principle}
\label{sec5a}
%%%%%%%%%%%%%%%%%%%%%%%%%%%%%%%%%%%%%%%%%%%%%%%%%%%%%%%%%%%%%%%%%%%%%%%%%%%%%

The microscopic description of the process in which thousand of quarks and 
gluons combine to form thousands of final state hadrons is clearly an
impractical problem. But it is also not needed. A microscopic approach 
is asked for if we want to understand the (relatively rare) processes 
by which, say, a high-$\pperp$ hadron of a given flavor is created. Such
particles probe the microscopic dynamics of high-$\pperp$ parton 
production, which is controlled by QCD, and although they may rescatter
in the dense fireball medium they never completely lose their memory
of the primary production process. We can exploit this and use them 
as ``deep probes'' for the fireball matter, but extracting the 
information on their production and subsequent medium modification 
then clearly requires a detailed microscopic treatment, including the 
hadronization of such hard partons. This is completely different for 
soft partons and hadrons: in their case the rescattering
by the medium wipes out all memories about their initial production 
process, and there is no way their final momenta could ever be traced
back to their initial momenta. This is in particular true for the 
hadronization process which is probably the most non-perturbative process
imaginable: there are literally innumerable different microscopic channels
which might lead to the production of, say, a negative pion with 
a soft momentum of order 300\,MeV/$c$ in the laboratory.    

The proper approach to such a non-perturbative, multi-channel multi-particle
problem is (as already recognized by Fermi, Landau and Hagedorn more than
half a century ago) a statistical one. From a microscopic point of view,
there is really very little information we can extract from the yields
and momentum spectra of soft single hadrons. Due to the strong interactions
among the quarks and gluons as they are forming hadrons, anything that is 
not explicitly forbidden by the conservation laws for energy, momentum, 
baryon number, electric charge, and net strangeness will in fact happen.
(Net strangeness can only be changed by weak interactions, on a time 
scale which is much too long compared to a heavy-ion collision.)

Note that such a statistical approach has, of course, its limitations: 
when we start to investigate correlations between pairs, triplets, 
quadruplets etc. of particles, at some level they will begin to realize
that they all belong to a single heavy-ion collision event and are {\em not}
produced entirely inependently, but correlated in a non-statistical way. 
(For example, the momenta of all 2479 particles emerging from the 
collision have to add up to the total momentum of the initially 
colliding nuclei, and this is likely to generate non-statistical 
momentum correlations among considerably smaller subclusters of 
particles.) The statistical approach is certainly best justified for
single-particle observables, such as hadron spectra and abundances.

These words are translated into equations via the Maximum Entropy 
Principle \cite{MEP}: Let us consider the entropy ${\cal S}$ of a 
small fluid cell of size $\Delta V$ in its own rest system,
\beq{entropie}
  {\cal S} = -\sum_k \int_{\Delta V}\int\frac{d^3x\,d^3p}{(2\pi)^3}
  \Bigl[ f_i\,\ln f_i +\theta_i\big(1-\theta_i f_i\big)\,
      \ln \big(1-\theta_i f_i\big) \Bigr],
\eeq
where the sum over $i$ includes all particle species in the cell 
and $\theta_i\eq+1 (-1)$ for fermions (bosons). We assume that soft 
hadronization happens through many different channels which are 
constrained only by the local (in $\Delta V$) conservation laws for 
energy, baryon number and strangeness ($\int d\omega = \int_{\Delta V}
\int\frac{d^3x\,d^3p}{(2\pi)^3}$):
\beq{rand}
   \la E\ra = \int d\omega \sum_i E_i f_i\,,\;\;\;
   \la B\ra = \int d\omega \sum_i B_i f_i\,,\;\;\;
   \la S\ra = \int d\omega \sum_i S_i f_i = 0\,.
\eeq
Here $B_i$ and $S_i$ are the baryon number and strangeness carried by
each particle of species $i$, and we used that the net strangeness of 
the two colliding nuclei is zero and that strong interactions keep it 
that way {\em locally}.
Maximizing ${\cal S}$ with these constraints gives for the most likely 
distribution for particle species $i$ \cite{SH95}
\beq{mep}
   f_i(E_i,\Delta V) = \frac{1}{e^{\beta(\Delta V)[E_i-\mu_i(\Delta V)]}
   +\theta_i}
\quad {\rm with}\quad
   \mu_i(\Delta V) = B_i\mu_B(\Delta V) + S_i\mu_S(\Delta V),
\eeq
where $\beta$, $\mu_B$ and $\mu_S$ are Lagrange multipliers related to
the values of the constraints (\ref{rand}) in $\Delta V$. If the 
hadronization process does not change the number of $s\bar s$ pairs 
(for example, because strange valence quarks are too heavy), we should
implement an additional constraint on the total number of strange 
quarks and antiquarks which fixes that number to the value before 
hadronization begins. In this case the expression for $\mu_i$ in 
(\ref{mep}) generalizes to 
$\mu_i\eq{B}_i\mu_B{+}S_i\mu_S{+}|s_i|\tilde\mu_s$ where $|s_i|$ is
the total number of strange quarks and antiquarks in hadron $i$ and
$\tilde\mu_s$ is a further Lagrange multiplier related to this 
additional constraint. The associated {\em fugacity}
$\gamma_s\eq{e}^{\beta\tilde\mu_s}$ is known as the 
{\em ``strangeness saturation factor''} \cite{Let92}. An 
under- resp. oversaturation of the strange particle phase space relative to
its chemical equilibrium value corresponds to $\gamma_s{\,<\,}1$ resp. 
$\gamma_s{\,>\,}1$.

Equation (\ref{mep}) is a local thermal and chemical equilibrium 
distribution function. If the system could be kept at constant volume,
any type of strong interaction among the hadrons would leave
this distribution unchanged since such microscopic processes 
again conserve energy, baryon number and strangeness. However, the
apparent ``equilibrium'' expressed by Eq.~(\ref{mep}) is {\em not
achieved kinetically} (i.e. as a result of hadronic rescattering)
but {\em statistically} (by interference of many different hadron 
production channels). It does not require a thermalized prehadronic 
state (although {\em very} strong deviations from equilibrium, 
e.g. at high $\pperp$, in the prehadronic state may survive the 
hadronization process and invalidate the Maximum Entropy Principle).

It is very important to realize that the same local equilibrium 
distribution (\ref{mep}) can be the result of two conceptually 
entirely different types of processes: we can either take system
of hadrons with an arbitrary (except for the constraint
on total energy, baryon number and strangeness) initial phase-space 
distribution and let it evolve for a sufficiently long time to obtain
Eq.~(\ref{mep}) as a result of the action of elastic and inelastic 
processes among the hadrons. This is {\it kinetic equilibration}.
Or we produce the system of hadrons of given energy, baryon number an
strangeness from some non-hadronic state by a statistical process
which fills hadronic phase-space in the statistically most probable
configuration. This is {\em statistical equilibrium}. Both processes
share the property that they lead to a state of Maximum Entropy. However,
statistical hadronization can produce a Maximum Entropy distribution
through non-hadronic processes which occur much faster than any inelastic 
scattering among hadrons at the given energy and baryon number density.

If you try to create, via statistical hadronization, a ``pre-established
equilibrium'' distribution (\ref{mep}) of hadrons at energy density
$e{\,>\,}\ec$ ($T{\,>\,}\Tc$), lattice QCD tells us that such a 
``superheated'' hadronic equilibrium configuration is unstable against 
deconfinement. If you leave this system to itself, the hadrons will 
dissolve again into quarks and gluons. Statistical hadronization thus
can only proceed once the energy density has dropped to 
$e_{\rm had}\eq\ec{\,\simeq\,}1$\,GeV/fm$^3$. Therefore the hadrons,
when first formed from the hadronizing quark-gluon state, will have
apparent chemical equilibrium abundances corresponding to a 
temperature $T_{\rm had}\eq\Tc$.\footnote{I am not stating the same for the
slope of their momentum distribution because the hadronizing quark-gluon
matter already undergoes transverse flow which blueshifts the 
momentum spectrum. The abundances and the ``chemical temperature'' 
extracted from them are not affected by flow \cite{He98}.} Before
dicussing what happens to this ``chemical temperature'' after 
hadronization has been completed, let us first have look at the data.

%%%%%%%%%%%%%%%%%%%%%%%%%%%%%%%%%%%%%%%%%%%%%%%%%%%%%%%%%%%%%%%%%%%%%%%%%%%%
\subsection{Hadrosynthesis at SPS and RHIC and pre-established hadronic 
chemical equilibrium}
\label{sec5b}
%%%%%%%%%%%%%%%%%%%%%%%%%%%%%%%%%%%%%%%%%%%%%%%%%%%%%%%%%%%%%%%%%%%%%%%%%%%%%

\begin{figure}[ht] 
\begin{center}
  \epsfig{file=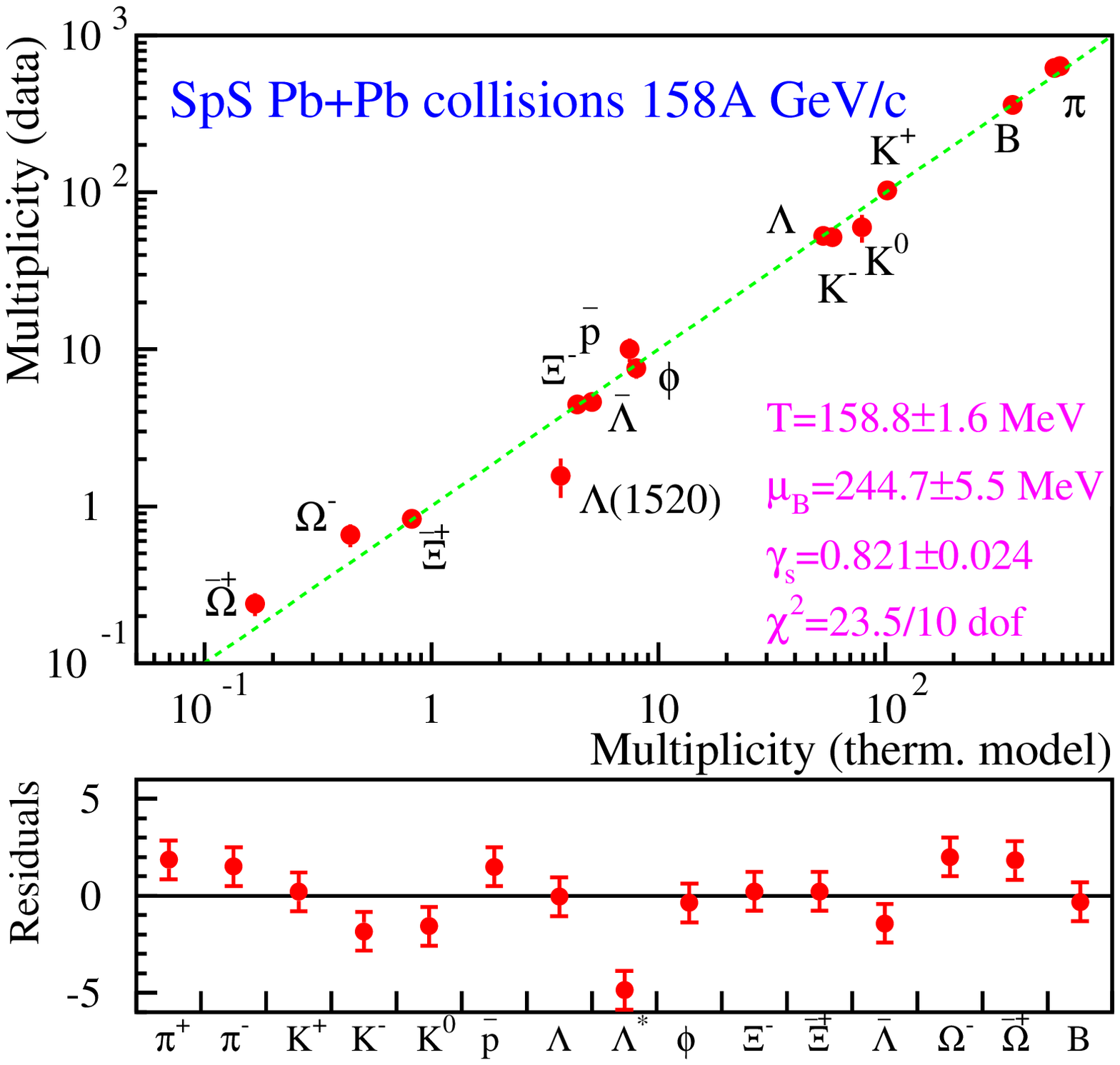,width=7.1cm}
\hspace*{5mm}
  \epsfig{file=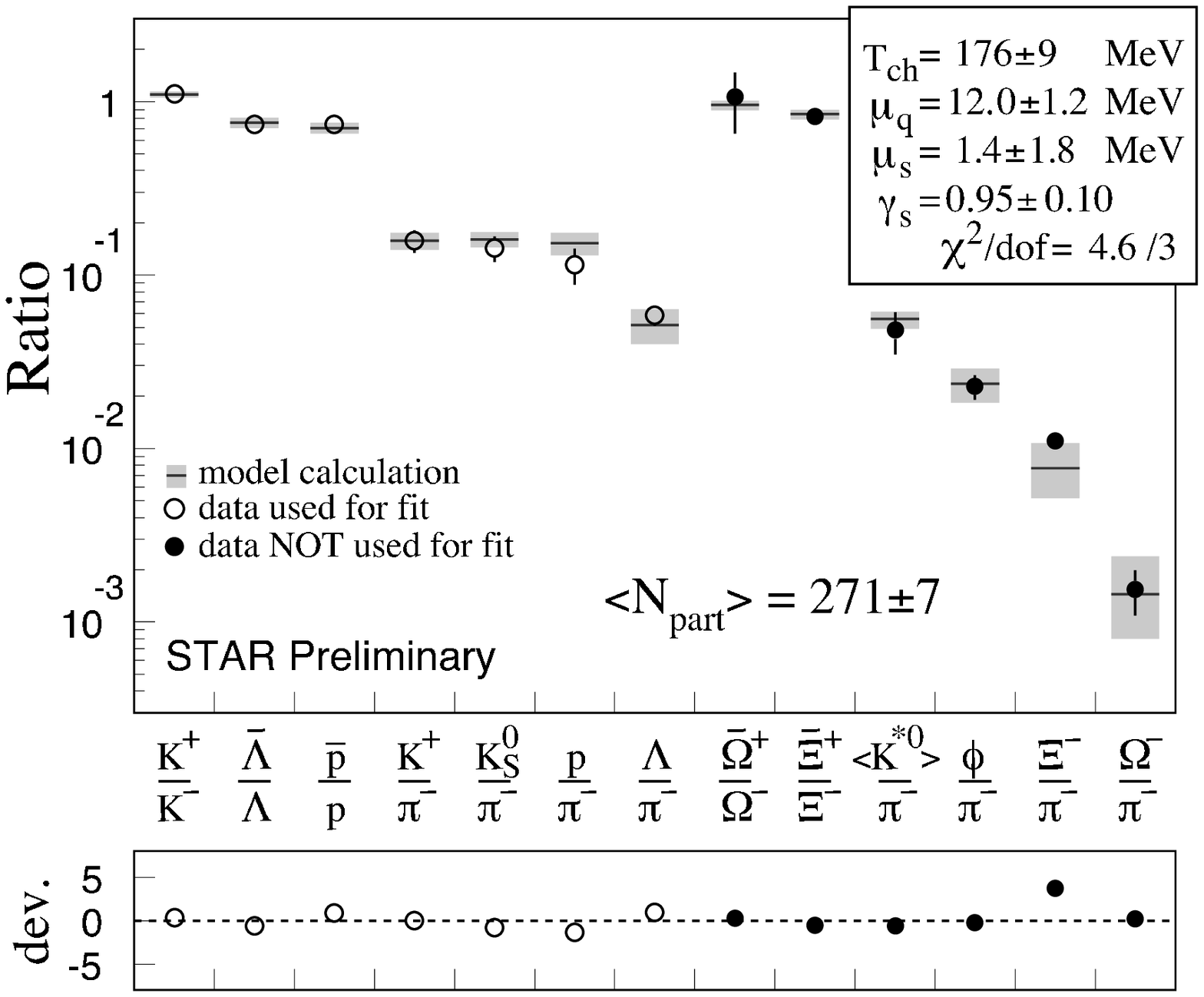,width=7cm,height=6.67cm}
\end{center}
\vspace*{-7mm}
\caption{Chemical equilibrium fits to the hadron abundances from Pb+Pb 
collisions at $\scm\eq17$\,GeV \cite{vL02,BGS98} (left) and from Au+Au 
collisions at  $\scm\eq17$\,GeV \cite{BMMRS01,vB02}. The baryon chemical 
potential $\mu_B$ used on the left and the quark chemical potential
$\mu_q$ used on the right are related by $\mu_B\eq3\mu_q$. The strangeness
chemical potential $\mu_s$ is not an independent fit parameter, but follows 
from overall strangeness neutrality. The strangeness saturation factor 
$\gamma_s$ \cite{Let92} is approximately 1 in both cases.
\label{F17} 
} 
\vspace*{-2mm}
\end{figure}

In Figure~\ref{F17} I show the particle multiplicity ratios for stable
(with respect to strong interactions) hadron emitted in Pb+Pb collisions
at the SPS (left) and in Au+Au collisions at RHIC (right). In both cases
the data have been fit to an ``apparent chemical equilibrium distribution'' 
(\ref{mep}); the resulting fit parameters are given in the 
Figure. The left panel presents the quality of the fit in a form 
%
%%%%%%%%%%%%%%%%%%%%%%%%% Fig. 17 %%%%%%%%%%%%%%%%%%%%%%%%%%%%%%%%%%%%%%%%%
%%%%%%%%%%%%%%%%%%%%%%%%%%%%%%%%%%%%%%%%%%%%%%%%%%%%%%%%%%%%%%%%%%%%%%%
%
introduced by Becattini \cite{Bec96} where the measured multiplicities
on the vertical axis are plotted against the multiplicities from the
thermal model after optimization of the thermal parameters. If 
Eq.~(\ref{mep}) gave a perfect description of the data, all points would
lie exactly on the diagonal. For clarity the differences between data
and optimized thermal model are plotted (in terms of standard deviations)
on a linear scale at the bottom. In the right panel the data are compared
with bands of predicted ranges whose width is controlled by the statistical
and systematic error of the thermal fit parameters \cite{PBM94}. Note 
that in the fit of the RHIC data only a fraction of the data was used 
for the fit while the rest (including ratios involving the rare doubly
and triply strange $\Xi$ and $\Omega$ hyperons) were successfully predicted
by this fit.

These fits tell us several things. First, not surprisingly, the baryon 
chemical potential at RHIC is a lot smaller than at the SPS, due to 
increased transparency of the target nucleus for the incoming projectile
baryons. Second, the value for the strangeness saturation factor $\gamma_s$
is almost equal to 1 in both cases; I will return to this observation in 
the following subsection. Third, and most importantly, the ``chemical
decoupling temperature'' $T_{\rm chem}$ is in both cases much higher
than the kinetic freeze-out temperature $\Tdec$ extracted from the shape
of the momentum spectra. At the SPS $T_{\rm chem}{\,\approx\,}160$\,MeV
and at RHIC $T_{\rm chem}{\,\approx\,}175$\,MeV. Interpreting $T_{\rm chem}$
in the Maximum Entropy spirit as a Lagrange multiplier which tells us
at which energy density statistical hadronization happened we see that
the data corrspond to energy densities which are very, very close to
the critical energy density for deconfinement predicted by lattice QCD.
According to lattice QCD, hadrons in thermodynamic equilibrium cannot 
exist at temperatures higher than about 175\,MeV if $\mu_B\eq0$, and this
limit goes down if $\mu_B{\,\ne\,}0$. So these chemical decoupling 
temperatures are about as high as they could possibly be for a system of 
hadrons!

Furthermore, we know from the spectra and elliptic flow that the fireball
is collectively expanding and therefore undergoes rapid cooling. So we
face the following conundrum: hadron formation is impossible at energy 
densities $e$ corresponding to $T{\,>\,}T_{\rm chem}$, but once the
hadrons are formed at $\ec$ the system does not remain at this energy
density but keeps diluting very rapidly. So if the hadrons first appeared 
with arbitrary abundances they would have had no time at all to adjust
their abundances to chemical equilibrium values at 
$T_{\rm chem}{\,\approx\,}\Tc$ by inelastic rescattering! The possibility
that the measured chemical freeze-out temperature 
$T_{\rm chem}{\,\simeq\,}160-170$\,MeV reflects a kinetically established
chemical equilibrium is therefore logically excluded. The apparent chemical
equilibrium at $T_{\rm chem}{\,\approx\,}\Tc$ seen in the hadron abundances
must thus be a ``pre-established equilibrium'' in the sense of statistical
hadronization according to the Maximum Entropy Principle \cite{BH97,Heinz99}.

%
%%%%%%%%%%%%%%%%%%%%%%%%% Fig. 17a %%%%%%%%%%%%%%%%%%%%%%%%%%%%%%%%%%%%%%%%%
\begin{figure}[ht] 
\begin{center}
  \epsfig{file=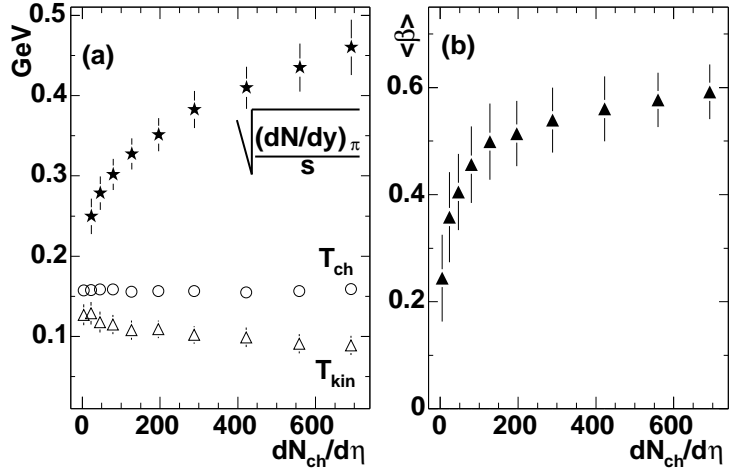,width=10cm}
\end{center}
\vspace*{-7mm}
\caption{{\bf (a)} Pion multiplicity $dN_\pi/dy$ per unit overlap area $S$
(stars), chemical decoupling temperature $T_{\rm ch}$ (circles), thermal 
(kinetic) decoupling temperature $T_{\rm kin}$ (open triangles), and 
{\bf (b)} the average transverse flow velocity $\langle \beta_{\rm T}\rangle$ 
at kinetic decoupling as a function of collision centrality 
(measured by the total charged multiplicity pseudorapidity density 
$dN_{\rm ch}d\eta$ at midrapidity) from $200\,A$\,GeV Au+Au collisions 
at RHIC \protect\cite{Tkin_chem}.
\label{F17a} 
} 
\vspace*{-2mm}
\end{figure} 
%%%%%%%%%%%%%%%%%%%%%%%%%%%%%%%%%%%%%%%%%%%%%%%%%%%%%%%%%%%%%%%%%%%%%%%
%

In addition to being {\em logically} excluded, there is also strong 
{\em empirical} evidence against a kinetic interpretation of the observed 
chemical decoupling temperature: First, it takes the same universal value
of about 170\,MeV in high energy $e^+e^-$ \cite{Bec96}, $pp$ and 
$\bar p p$ \cite{BH97}, and nucleus-nucleus collisions \cite{Beca98}.
Second, in Au+Au collisions at RHIC, $T_{\rm chem}$ is found to be 
independent of collision centrality (open circles in the left panel of 
Fig.\,\ref{F17a}), in contrast to the observed significant centrality 
dependence of the thermal decoupling temperature and average radial
flow velocity extracted from the momentum spectra (triangles in 
Fig.\,\ref{F17a} -- see Fig.\,\ref{F8} for comparison and the discussion 
in Section~\ref{sec4a8}). As discussed earlier, any kinetic decoupling 
process is a competition between local
thermalization and global expansion rates. The latter depends on the
size and lifetime on the collision fireball, as reflected by the clear
dependence of the radial flow velocity on the impact parameter of the 
collision system in the right panel of Fig.~\ref{F17a}. The larger
fireballs formed in central collisions live longer, develop more radial
flow and cool down to lower kinetic freeze-out temperatures than the 
smaller fireballs from more peripheral collisions. The chemical 
decoupling temperature $T_{\rm chem}$, on the other hand, shows no 
dependence on the collision centrality (size of the fireball) at all;
it can therefore not be controlled by a similar competition between
scattering and expansion as seen in the kinetic decoupling temperature.
{\bf $\bm{T_{\rm chem}}$ and $\bm{T_{\rm kin}}$ rest conceptually on different 
footings.} $T_{\rm chem}$ is not a kinetic decoupling temperature, but
a Lagrange multiplier in the Maximum Entropy framework whose value 
indicates the critical energy density at which hadrons are formed.

So if, as we concluded from the momentum spectra, after hadronization 
the system continues to expand and cool to lower temperatures, why doesn't 
the chemical decoupling temperature $T_{\rm chem}$ follow suit? The answer 
must be that, after being formed, the hadrons are unable to change their 
abundance ratios any further by inelastic rescattering. If this were not 
the correct interpretation, it would be very difficult to see how about 
a dozen different hadron species with abundances varying over 3 orders 
of magnitude could conspire to look like chemical equilibrium at 
$T{\,\approx\,}\Tc$. This would be an unlikely accident indeed. In 
Sec.~\ref{sec5d} we will see theoretical arguments why chemical 
decoupling has to happen more or less directly at $\Tc$.

%%%%%%%%%%%%%%%%%%%%%%%%%%%%%%%%%%%%%%%%%%%%%%%%%%%%%%%%%%%%%%%%%%%%%%%%%%%%
\subsection{Strangeness enhancement}
\label{sec5c}
%%%%%%%%%%%%%%%%%%%%%%%%%%%%%%%%%%%%%%%%%%%%%%%%%%%%%%%%%%%%%%%%%%%%%%%%%%%%%

\begin{figure}[ht] 
\begin{center}
  \epsfig{file=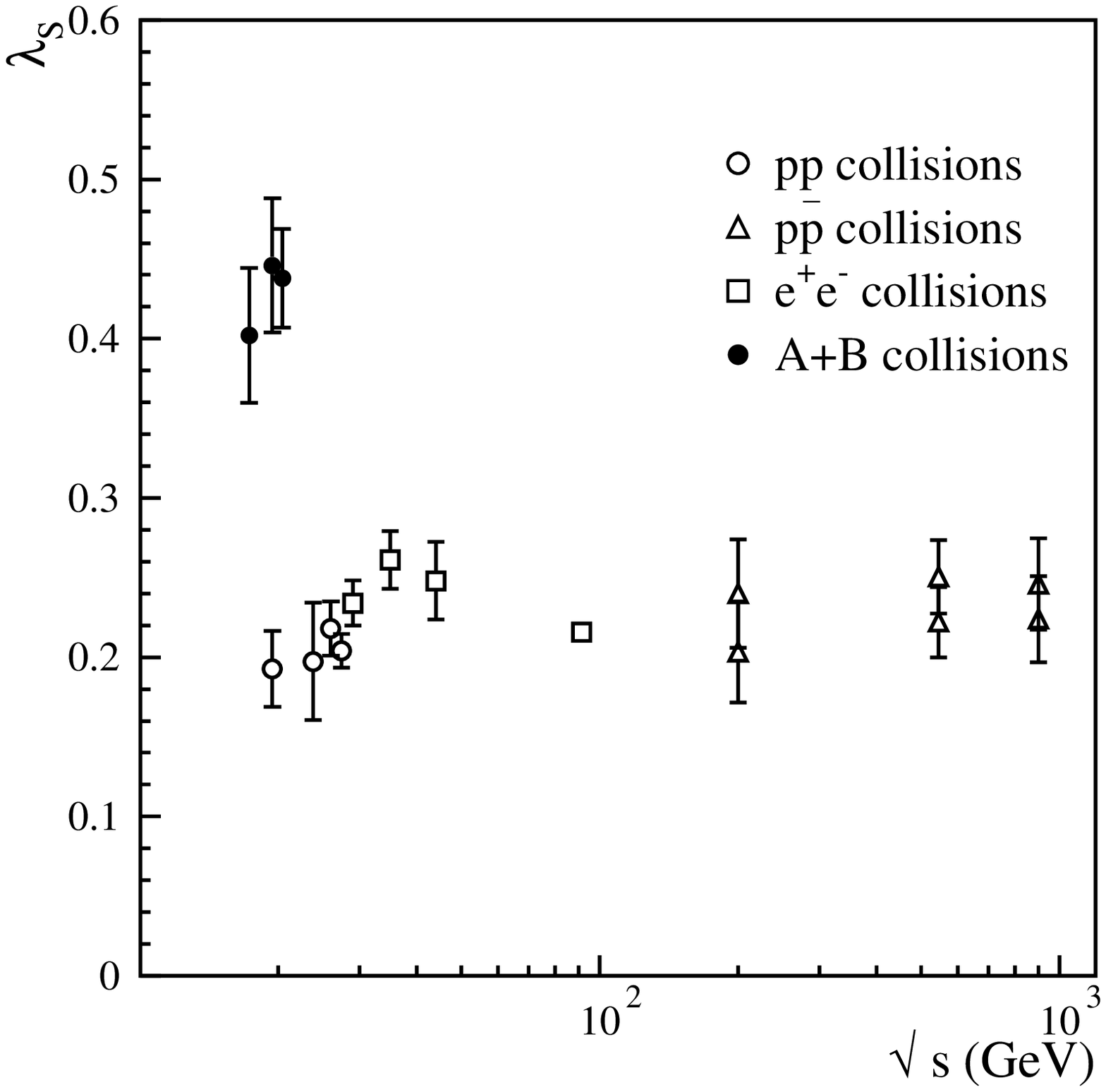,width=7cm,height=5.5cm}
\hspace*{5mm}
  \epsfig{file=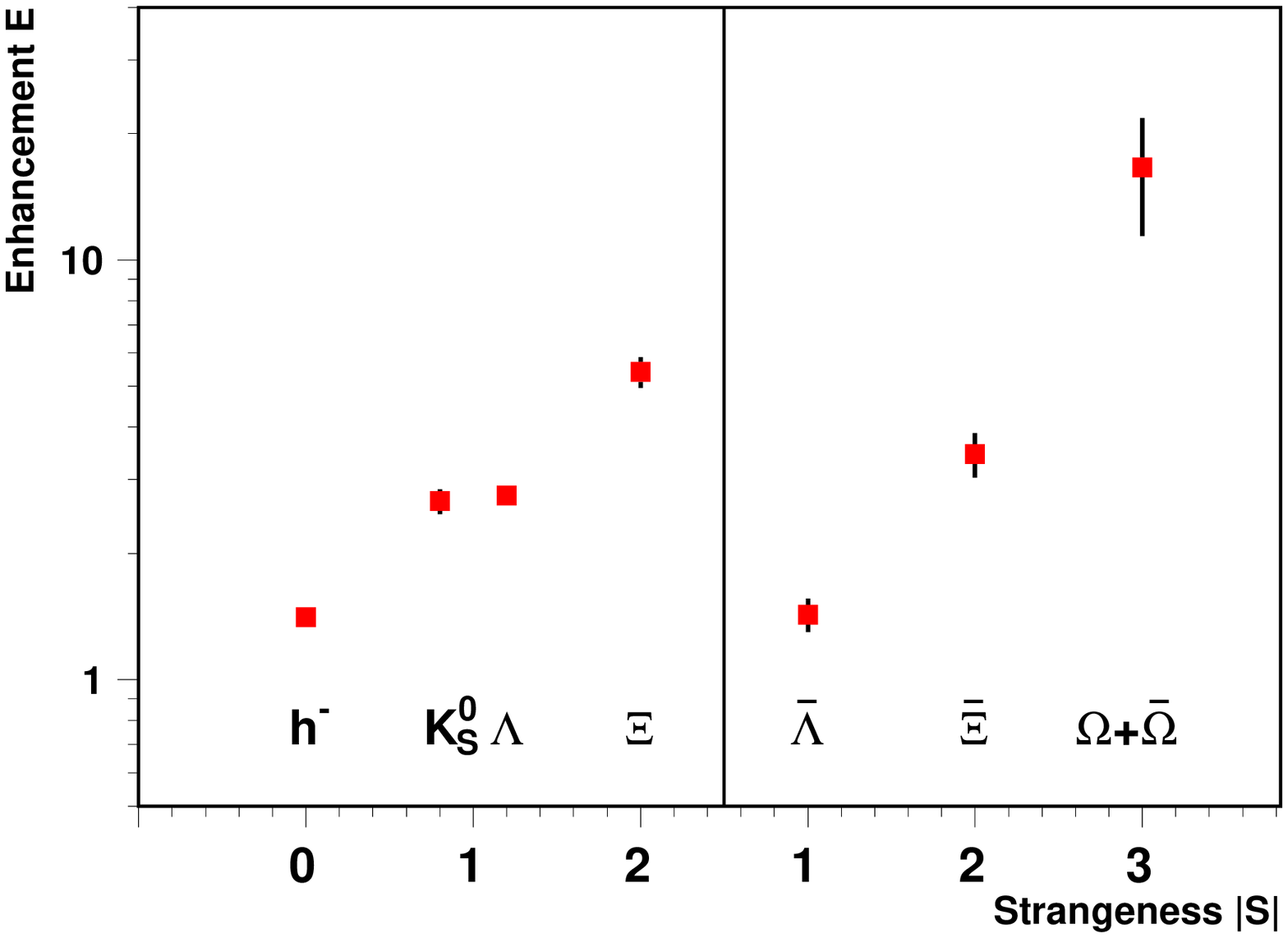,width=8cm}
\end{center}
\vspace*{-7mm}
\caption{Left: The strangeness suppression factor of produced strange 
    vs. nonstrange quarks, $\lambda_s\eq2\la\bar s s\ra/\la\bar u 
    u{+}\bar d d\ra\vert_{\rm produced}$, in elementary particle and heavy-ion 
    collisions as a function of $\sqrt{s}$ \cite{BGS98}. The two points 
    each for $p\bar p$ collisions reflect the inclusion (exclusion) 
    of the initial valence quarks. 
    Right:  Enhancement factor for the midrapidity yields per 
    participating nucleon in 158\,$A$\,GeV/$c$ Pb+Pb relative to
    p+Pb collisions for various strange and non-strange hadron 
    species \cite{Lietava}. 
\label{F18} 
} 
\vspace*{-2mm}
\end{figure}

The fact that the chemical decoupling temperature has an uncanny 
resemblance with the predicted phase transition temperature is,
in fact, only half the story. A similar picture of statistical 
hadronization at the critical energy density $\ec$ arises 
even from an analysis of $e^+e^-$, $pp$ and $p\bar p$ collisions 
\cite{BH97}. What is really dramatically different in heavy-ion 
collisions is the level of strangeness saturation reflected in the
apparent chemical equilibrium state: The left panel of Fig.~\ref{F18} 
shows that the overall fraction of strange particles is about twice 
as high in heavy-ion collisions as in elementary particle collisions! 
In other words, strangeness is suppressed relative to its chemical
equilibrium saturation value in elementary particle collisions,
but this strangeness suppression has disappeared in $A{+}A$ collisions. 
According to the preceding discussion this extra strangeness cannot 
have been produced by final state hadronic rescattering; 
it thus reflects the properties of the prehadronic state.
This points to a new, fast strangeness production mechanism, either 
before or during hadronization. Rapid strangeness production
from thermal gluons was predicted as one of the key characteristics of a 
QGP \cite{RM82}.

%
%%%%%%%%%%%%%%%%%%%%%%%%% Fig. 18 %%%%%%%%%%%%%%%%%%%%%%%%%%%%%%%%%%%%%%%%%
%%%%%%%%%%%%%%%%%%%%%%%%%%%%%%%%%%%%%%%%%%%%%%%%%%%%%%%%%%%%%%%%%%%%%%%
%

A large fraction of the strangeness suppression seen in elementary 
particle collisions can actually be understood within the statistical
approach. Strong interactions conserve net strangeness {\em exactly},
i.e. $S\eq0$ without any fluctuations. This means that in principle 
strangeness conservation should not be treated in the grand canonical 
ensemble as in Eqs.~(\ref{rand}) and (\ref{mep}), where it is conserved 
only on average, but in the canonical ensemble were $S$ is fixed
exactly. In the infinite volume limit this makes no difference,
but for the small ``fireball'' volumes created in elementary particle
collisions it is a major correction. Strange hadron get suppressed in
this way since, together with the strange hadron, always a second hadron
with balancing strangeness has to be created inside the same small 
volume at the same time, and this requires more energy. However,
even if this canonical effect were the complete explanation of the 
suppression in $e^+e^-$ and $pp$ collisions (which seems quantitatively
unlikely \cite{Soll97}), the disappearance of the effect in heavy-ion 
collisions would still be most interesting: The absence of canonical
strangeness suppression in nuclear collisions tells us that creation
of a strange hadron at a given position in the fireball does {\rm not}
require the production of a particle with balancing strangeness nearby
(as in $pp$), but strangeness can be balanced by production of an 
anti-strange hadron on the other side of the nuclear fireball! 
Now we know that all microscopic QCD processes which create $s\bar s$
pairs are local processes, i.e. the pairs are created at the same
point. If the final strange hadron abundaces don't know about this 
anymore, but behave as if the strange and antistrange hadrons were 
created independently and statistically distributed over the entire 
nuclear fireball, the necessary implication is that they have lost
their memory of the locality of the primary QCD process from which they
came and have been able to communicate over macroscopic fireball
distances. I cannot imagine how this is possible without a significant
amount of strangeness diffusion {\em before} hadronization.

A very striking way of plotting the ``strangeness unsuppression''
is shown in the right panel of Fig.~\ref{F18} \cite{Lietava}: relative 
to $p$+Pb collisions, the number of produced strange hadrons per 
participating nucleon is the more strongly enhanced the more strange 
(anti)quarks for its formation are required. For $\Omega$ and 
$\bar\Omega$ this enhancement factor is about 15! The tendency 
shown in Fig.~\ref{F18} is completely counterintuitive for hadronic 
rescattering mechanisms, where multistrange (anti)baryons are 
suppressed by higher thresholds than kaons and $\Lambda$'s; but 
it is perfectly consistent with a statistical hadronization picture 
\cite{Bialas} where multi-strange particles profit more from the 
global strangeness enhancement than singly strange hadrons.  

%%%%%%%%%%%%%%%%%%%%%%%%%%%%%%%%%%%%%%%%%%%%%%%%%%%%%%%%%%%%%%%%%%%%%%%%%%%%
\subsection{Chemical kinetics after hadronization}
\label{sec5d}
%%%%%%%%%%%%%%%%%%%%%%%%%%%%%%%%%%%%%%%%%%%%%%%%%%%%%%%%%%%%%%%%%%%%%%%%%%%%%

The kinetic freeze-out temperatures of about 120\,MeV extracted 
from he hadron momentum spectra tell us that after hadronization 
at $\Tc{\,\approx\,}170$\,MeV the hadrons continue to rescatter for 
quite a while. This rescattering can be simulated in a hadronic
cascade such as URQMD \cite{Bass99}, using everything that is known 
about hadron masses and cross sections from the Particle Data Tables.
%
%%%%%%%%%%%%%%%%%%%%%%%%% Fig. 19 %%%%%%%%%%%%%%%%%%%%%%%%%%%%%%%%%%%%%%%%%
\begin{figure}[ht] 
\begin{center}
\begin{minipage}[h]{8.5cm}
  \epsfig{file=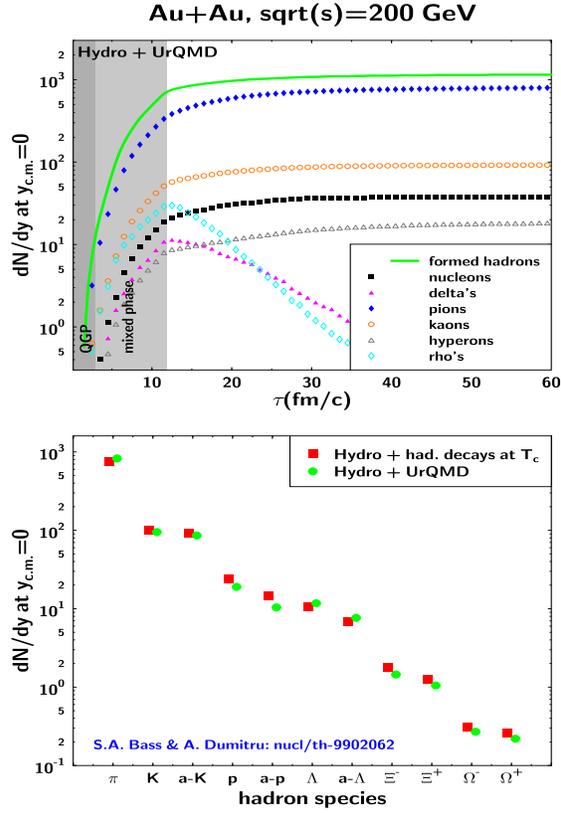,width=8cm,height=11cm}
\end{minipage}
\begin{minipage}[h]{6.5cm}
\vspace*{-7mm}
\caption{Time-dependence of the chemical composition of an expanding
hadronic fireball which was initialized at $\Tc$ with chemical
equilibrium abundances \cite{Bass99}. Unstable hadrons are seen to
decay as stable hadrons approach their asymptotic abundances (top).
In the bottom panel, the squares indicate the initial chemical
equilibrium yields for stable hadrons, after taking into account
feeddown from unstable resonance decays, whereas the circles indicate
the final stable hadron yields including hadronic rescattering.
\label{F19} 
} 
\end{minipage}
\end{center}
\vspace*{-7mm}
\end{figure} 
%%%%%%%%%%%%%%%%%%%%%%%%%%%%%%%%%%%%%%%%%%%%%%%%%%%%%%%%%%%%%%%%%%%%%%%
%
These tables tell us that hadron cross sections are strongly dominated
by resonances at well-defined collision energies, for example
$\pi{+}N{\,\to\,}\Delta{\,\to\,}\pi{+}N$, 
$\pi{+}\pi{\,\to\,}\rho{\,\to\,}\pi{+}\pi$, 
$\pi{+}K{\,\to\,}K^*{\,\to\,}\pi{+}K$, etc. These resonances have large
cross sections, hence these types of collisions happen frequently
and are able to keep up with the collective expansion of the fireball 
by continually re-equilibrating the hadron momentum distributions
to the falling temperature. However, almost all of these resonances
have the tendency to decay again into exactly the same hadrons from
which they were created (up to charge exchange). The cross sections
for strangeness exchange processes are considerably smaller, and
even smaller are inelastic hadronic processes which create or annihilate
strange quark-antiquark pairs. The resonant collisions thus do not
modify the total measured yields of pions, kaons, and nucleons. For
example the total pion number
\beq{total}
   N_\pi^{\rm total} = N_\pi + 2 N_\rho + N_\Delta +
   N_{K^*} + \dots
\eeq
is frozen in at $\Tc$, and similarly for other stable hadrons.
This is seen in the bottom part of Fig.~\ref{F19} which shows that,
if URQMD is initialized with thermodynamic equilibrium abundances 
and a hydrodynmiccal generated transverse flow profile similar
to the discussion in Sec.~\ref{sec4}, the stable particle yields 
hardly change if they are calculated directly at $\Tc$ by simply 
letting all unstable resonances decay, but switching off all collisions,
or if they are calculated at the end of the entire hadronic cascade.

Looking a bit more carefully at the bottom part of Fig.~\ref{F19} we
see that URQMD tends to lose predominantly (anti)baryons 
(and thus a fraction of the initial enhancement of multi-strange 
(anti)baryons) by baryon-antibaryon annihilation during the 
rescattering stage. It was recently shown \cite{RS00} that this is to
a large extent a manifestation of the lack of detailed balance in the 
codes which include processes like $\bar p p\to n\pi$ (with $n=5-6$)
but not their inverse. Rapp and Shuryak \cite{RS00} argue that, as the 
system cools below $T_{\rm chem}$, pions and kaons don't annihilate but 
instead build up a positive chemical potential which enhances the 
probability for the inverse reaction and strongly reduces the net 
annihilation of antibaryons. This is really fortunate, because it
is this lack of abundance-changing processes during the hadronic 
expansion stage which allows us to glimpse the hadronization process
itself through the final hadronic abundances, in spite of intense, 
resonance-mediated {\em elastic} rescattering among the hadrons 
between hadronization at $T_{\rm chem}\approx 170$\,MeV and kinetic 
freeze-out at $T_{\rm f}\approx 120$\,MeV.

Of course, the abundances of unstable resonances are {\em not} 
frozen in at $\Tc$ (see top panel of Fig.~\ref{F19}): Due to their strong 
coupling to the cooling pion fluid (to which cooling they actively
contribute), their abundances readjust to the decreasing temperature.
Note, however, that their abundances don't fall with the usual 
free-space exponential decay law, but more slowly since they keep 
being recreated by resonant hadronic interactions. But as the 
temperature drops, their abundances decrease. Detailed balance 
dictates that, the larger their pionic decay width, the later the 
resonance yields decouple and the smaller a temperature their
decoupling abundances reflect. If we can reconstruct the resonances
from their decay products we therefore expect resonance yields
which are smaller than predicted from the thermal fit of the stable
hadron yields:
\beq{res}
   \left(\frac{K^*}{K}\right)_{\rm meas} < 
   \left(\frac{K^*}{K}\right)_{\rm T_{rm chem}}\qquad {\rm etc.}
\eeq
Such studies are presently under way.

%%%%%%%%%%%%%%%%%%%%%%%%%%%%%%%%%%%%%%%%%%%%%%%%%%%%%%%%%%%%%%%%%%%%%%%%%%%%
\subsection{Primordial hadrosynthesis: measuring the critical temperature
for deconfinement}
\label{sec5e}
%%%%%%%%%%%%%%%%%%%%%%%%%%%%%%%%%%%%%%%%%%%%%%%%%%%%%%%%%%%%%%%%%%%%%%%%%%%%%

We can summarize our insights gained in this chapter by stating as
our {\bf Fourth Lesson:} The hadron abundances ``freeze out'' directly
at hadronization. What we see in the SPS and RHIC data are hadron 
abundances established during {\em primordial hadrosynthesis} via
a statistical hadronization process. The measured abundance ratios
give for the hadronization temperature 
$T_{\rm had}{\,\approx\,}\Tc{\,\approx\,}170$\,MeV, confirming 
predictions from lattice QCD. The QCD phase transition temperature
has thus been measured. This measurement was possible because at SPS
and RHIC the fireball expands so rapidly that in the late hadronic 
rescattering stage no inelastic chemical processes happen anymore.
In this way the hadron abundances open a window onto the hadronization
phase transition even though their momentum distributions continue to 
change and cool for several more fm/$c$.     

%
%%%%%%%%%%%%%%%%%%%%%%%%% Fig. 20 %%%%%%%%%%%%%%%%%%%%%%%%%%%%%%%%%%%%%%%%%
\begin{figure}[ht] 
\vspace*{-2mm}
\begin{center}
\begin{minipage}[h]{8.5cm}
  \epsfig{file=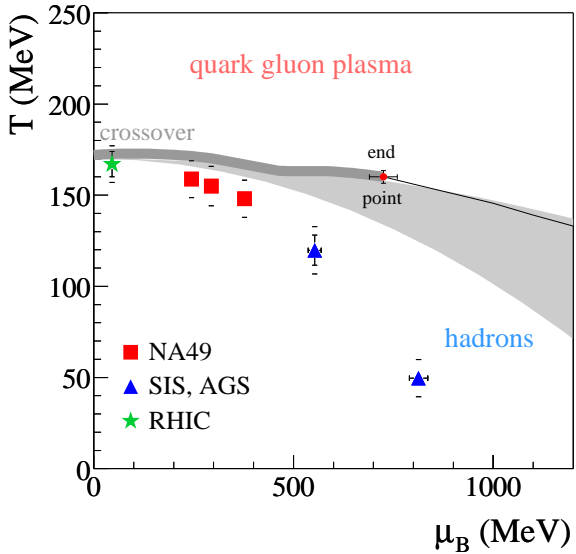,width=8.5cm}
\end{minipage}
\begin{minipage}[h]{6cm}
\vspace*{-7mm}
\caption{QCD phase diagram with chemical freeze-out points extracted from 
hadron abundance ratios from heavy ion ($A{\,\simeq\,}200$) collisions 
at center of mass energies $\scm$ ranging from 2.4 to 200\,GeV \cite{vL02}.
The curve and shaded region indicate estimates for the deconfinement
phase transition at finite baryon chemical potential $\mu_B$ from recent
lattice QCD calculations \cite{FK02,Allton02}. The critical endpoint
where the transition changes from first order (high $\mu_B$) to continuous
crossover (low $\mu_B$) is also indicated (although for
unrealistically large quark masses \cite{FK02}). Figure taken from 
Ref.~\cite{vL02}.
\label{F20} 
} 
\end{minipage}
\end{center}
\vspace*{-7mm}
\end{figure} 
%%%%%%%%%%%%%%%%%%%%%%%%%%%%%%%%%%%%%%%%%%%%%%%%%%%%%%%%%%%%%%%%%%%%%%%
%
Figure~\ref{F20} shows the chemical freeze-out points extracted from 
hadron yields in $A{+}A$ collisions with $A{\,\simeq\,}200$, compiled
for heavy-ion fixed-target and collider experiments from 
$\scm{\,\approx\,}2.4$\,GeV (SIS) to $\scm{\,\approx\,}200$\,GeV (RHIC),
and compares them to the latest estimates for the hadronization
phase transition from lattice QCD \cite{FK02,Allton02}. While for
the SPS and RHIC the chemical decoupling temperatures are consistent
with the predicted hadronization temperatures, the values extracted 
at the Brookhaven AGS and in particular at the SIS at GSI lie
considerably below the phase transition. In these cases statistical
hadronization can not explain the values for $T_{\rm chem}$. It is
likely that, due to the larger net baryon densities in lower-energy 
heavy-ion collisions, inelastic hadronic rescattering processes happen
faster than at higher energies and are able to lead to kinetic 
readjustment of the chemical temperatue below $\Tc$. One should also note,
however, that the ``quality'' of the apparent chemical equilibrium 
seen at the AGS and SIS is not as good as at the higher energies
since many of the heavier hadrons (in particular the strange and
multistrange antibaryons) are too rare to be measured reliably. The
data points labelled by ``SIS,AGS'' in Figure~\ref{F20} thus comprise
a much smaller number of measured hadron yields, in particular
in the strange sector where only $K$, $\bar K$, $\Lambda$ and
$\phi$ are available at these energies. 

%%%%%%%%%%%%%%%%%%%%%%%%%%%%%%%%%%%%%%%%%%%%%%%%%%%%%%%%%%%%%%%%%%%%%%%%%%%%%%
\section{PARTON ENERGY LOSS AND JET QUENCHING}
\label{sec6}
%%%%%%%%%%%%%%%%%%%%%%%%%%%%%%%%%%%%%%%%%%%%%%%%%%%%%%%%%%%%%%%%%%%%%%%%%%%%%%

In Section~\ref{sec4} we saw that in Au+Au collisions at RHIC the soft 
momentum particles with $\pperp\lapp2$\,GeV/$c$ behave hydrodynamically.
For larger $\pperp$ the hydrodynamic description gradually breaks down.
We saw this in Sec.~\ref{sec4b} when discussing the elliptic flow
and how it breaks away from the hydrodynamically predicted increase
with $\pperp$ above $\pperp{\,\simeq\,}2$\,GeV/$c$. In the single
%
%%%%%%%%%%%%%%%%%%%%%%%%% Fig. 22 %%%%%%%%%%%%%%%%%%%%%%%%%%%%%%%%%%%%%%
\begin{figure}[ht] 
\begin{center}
\begin{minipage}[h]{9.5cm}
  \epsfig{file=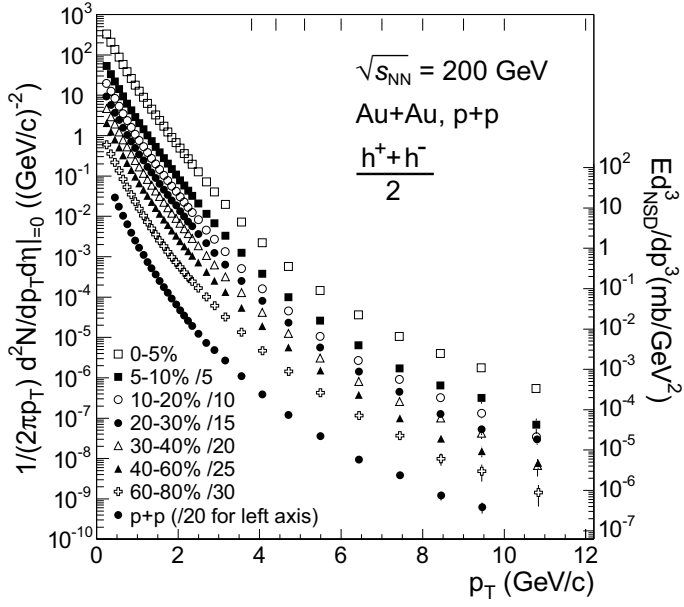,width=9cm}
\end{minipage}
\begin{minipage}[h]{5cm}
\vspace*{-2mm}
\caption{Charged hadron transverse momentum spectra in p+p and 
Au+Au collisions at $\scm\eq200$\,GeV \protect\cite{highpt_STAR}. 
The Au+Au data are binned in different centrality classes as 
indicated. For central and semicentral Au+Au collisions, note the 
transition from an exponential behaviour at $\pperp\protect\lapp2$\,GeV/$c$ 
to a power law for $\pperp{\,>\,}4$\,GeV/$c$.
\label{F22} 
}
\end{minipage}
\end{center}
\vspace*{-3mm}
\end{figure} 
%%%%%%%%%%%%%%%%%%%%%%%%%%%%%%%%%%%%%%%%%%%%%%%%%%%%%%%%%%%%%%%%%%%%%%%%
%
particle spectra the transition from collective hydrodynamic behaviour 
to hard scattering manifests itself by a change of shape: above 
$\pperp\gapp3-4$\,GeV/$c$ the spectra change from a thermal exponential 
shape to a power law as predicted by perturbative QCD. Figure~\ref{F22}
shows this for Au+Au collisions at RHIC, together with a comparison 
spectrum from p+p collisions. This power law affects fewer than 0.1\% 
of all produced hadrons, but it is these rare high-$\pperp$ hadrons on
which we will focus our attention in this section.

These high-$\pperp$ hadrons stem from the fragmentation of 
even-higher-$\pperp$ partons which, as noted in Sec.~\ref{sec3a}, are 
created very early in the collision, at 
$\tau_{\rm form}{\,\simeq\,}1/\pperp$.
On their way out of the evolving collision fireball they interact with
the hot matter and probe its properties (density, opacity, etc.). In 
this respect they are very similar to positrons in PET (= Positron 
Emission Tomography) where one injects a positron-emitting source into 
some living organ and then explores properties of that organ by detecting
these positrons and their energies outside the body. Measuring the
properties of hard hadrons emitted from a relativistic heavy-ion
collisions thus corresponds to Parton Emission Tomography, but 
since this would have the same acronym which might lead to all kinds of
confusion, and since we don't really detect partons but rather their 
hadronic fragments which form a jet, I follow M. Gyulassy \cite{tomography}
and call this method {\bf JET} (for {\bf Jet Emission Tomography}).

%%%%%%%%%%%%%%%%%%%%%%%%%%%%%%%%%%%%%%%%%%%%%%%%%%%%%%%%%%%%%%%%%%%%%%%%%%%%
\subsection{Radiative energy loss of a fast parton}
\label{sec6a}
%%%%%%%%%%%%%%%%%%%%%%%%%%%%%%%%%%%%%%%%%%%%%%%%%%%%%%%%%%%%%%%%%%%%%%%%%%%%%

The idea of jet quenching by parton energy loss goes back to an 
unpublished preprint by J.D.~Bjorken in 1982 \cite{Bj82}, but he 
incorrectly identified elastic parton scattering as the dominant 
medium interaction by which the fast parton loses energy. The correct 
mechanism, namely induced gluon bremsstrahlung, was identified by 
Gyulassy and collaborators \cite{G_jet} and first fully evaluated 
by Baier, Dokshitzer, Mueller, Peign\'e, and Schiff \cite{BDMPS}. 
Its quantitative effect on the energy loss has now been calculated 
by several groups under various approximations (see \cite{B03} for 
a recent review).

%
%%%%%%%%%%%%%%%%%%%%%%%%% Fig. 23 %%%%%%%%%%%%%%%%%%%%%%%%%%%%%%%%%%%%%%
\begin{figure}[ht] 
\begin{center}
  \epsfig{file=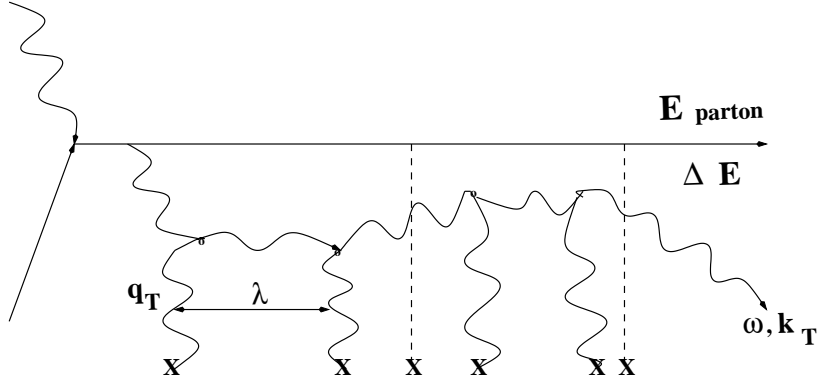,bb=163 88 449 705,width=5cm,angle=270}
\caption{Typical gluon radiation diagram (from \protect\cite{B03}).
\label{F23} 
} 
\end{center}
\end{figure} 
%%%%%%%%%%%%%%%%%%%%%%%%%%%%%%%%%%%%%%%%%%%%%%%%%%%%%%%%%%%%%%%%%%%%%%%%
%

In QCD the dominant process by which a fast parton with 
$E_{\rm parton}{\,\gg\,}1$\,GeV loses energy is by induced
gluon radiation, see Fig.~\ref{F23}. The color charge of the
fast parton (which could be a quark or a gluon) interacts with
the color charges of the medium (which can be modeled as
static external charges since their motion can be neglected
relative to that of the fast particle) and, as a result, emits
a bremsstrahlung gluon. What makes things different from and 
more complicated than the analogous process in QED is the fact
that the emitted gluon itself carries color charge and again 
interacts with the color charges in the medium (see Fig.~\ref{F23}), 
and that all these medium interactions must be added up coherently. 
In contrast to the fast particle, whose trajectory can be taken as
a straight line, the reinteractions of the much softer emitted
gluon with the medium induce a random walk in its transverse
momentum $k_{\rm T}$. The reinteractions of the emitted gluon with
the medium are characterized by a mean free path $\lambda_{\rm g}$,
also indicated in the Figure. 

In the limit $E_{\rm parton}{\,\to\,}\infty$ and for a thick target,
$L{\,\gg\,}\lambda_{\rm g}$, BDMS \cite{BDMPS} found for the energy loss
\beq{BDMS}
   \Delta E_{_{\rm BDMS}} = \frac{C_R \alpha_s}{4}\,
   \frac{\mu^2}{\lambda_{\rm g}}\, L^2\, \tilde v
\eeq
where $C_R$ is the color Casimir eigenvalue of the fast parton (${=\,}N_c$
for gluons), $\mu$ is the color Debye screening length of the medium
(controlled by its density), and $\tilde v{\,\sim\,}1{-}3$ is a factor
which depends logarithmically on the thickness $L$ of the medium and
the gluon mean free path, $\tilde v{\,\sim\,}\ln(L/\lambda_{\rm g})$.
The ratio
\beq{tc}
   q \simeq \frac{\mu^2}{\lambda_{\rm g}} \simeq \rho 
   \int d^2q_\perp\,q_\perp^2\,\frac{d\sigma}{d^2q_\perp} \propto 
   \alpha_s^2\,\rho
\eeq
plays the role of a momentum transport coefficient of the medium and
exhibits the explicit dependence of the energy loss on the density $\rho$
of the medium. The non-Abelian nature of QCD reflects itself in the
characteristic {\em non-linear} dependence of the energy loss 
(\ref{BDMS}) on the thickness $L$ of the medium. The extra factor of 
$L$ compared to the naive expectation arises from the random walk
in $k_{\rm T}$ of the emitted gluon which increases its probability
to separate itself (decohere) from the fast parton linearly with $L$.

The asymptotic expression (\ref{BDMS}) was improved for ``thin'' plasmas
and finite energy kinematic effects by Gyulassy, Levai and Vitev (GLV)
\cite{GLV} by use of an expansion in powers of the ``opacity'' of the 
medium (i.e. the integral over density times cross section along the 
path of the fast parton, $\bar n\eq{L}/\lambda_{\rm g}$). Their leading 
order contribution, which strongly dominates the energy loss, is
\beq{glv}
   \Delta E_{_{\rm GLV}}^{(1)} = \frac{C_R \alpha_s}{N(E)}\,
   \frac{\mu^2}{\lambda_{\rm g}}\, L^2\, \ln\frac{E}{\mu}.
\eeq
While the characteristic quadratic $L$ dependence of Eq.~(\ref{BDMS}) 
is not altered by keeping only the first term in the opacity expansion,
the dependence of the energy loss on the parton energy 
$E\eq{E}_{\rm parton}$ is quite different, due to the factor $N(E)$
in the denominator. For $E{\,\to\,}\infty$ this factor approaches
the BDMS value, $N(E){\,\to\,}4$, but as the parton energy $E$ 
decreases $N(E)$ increases rapidly, suppressing energy loss for
low-$E$ partons. In the range 2\,GeV${\,\leq\,}E{\,\leq\,}10$\,GeV,
for fixed $L$ the fractional energy loss $\Delta E_{_{\rm GLV}}^{(1)}/E$
is almost constant \cite{levai}.

%%%%%%%%%%%%%%%%%%%%%%%%%%%%%%%%%%%%%%%%%%%%%%%%%%%%%%%%%%%%%%%%%%%%%%%%%%%%
\subsection{Soft vs. hard particle production}
\label{sec6b}
%%%%%%%%%%%%%%%%%%%%%%%%%%%%%%%%%%%%%%%%%%%%%%%%%%%%%%%%%%%%%%%%%%%%%%%%%%%%%

According to QCD and asymptotic freedom, the production of high-$\pperp$
partons can be calculated perturbatively and is proportional to the number
of binary nucleon-nucleon collisions which for a nucleus-nucleus collision
$A{+}B$ at impact parameter $b$ is given by
\beq{bin}
    N_{\rm coll}(b) = 
    \sigma_0 \int dx\,dy\,T_A(x{+}b/2,y)\cdot T_B(x{-}b/2,y).
\eeq
Here $\sigma_0$ is the total inelastic nucleon-nucleon cross section,
the integral is over the transverse plane, and 
$T_A(x,y)\eq\int\rho_A(x,y,z)\,dz$ is the nuclear thickness function
of a nucleus of mass $A$ with density profile $\rho_A$. The impact 
parameter $b$ can be estimated from the total charge multiplicity,
by equating equal fractions of the total multiplicity divided by the 
maximum multiplicity measured in the most central collisions with 
the corresponding fractions of the total cross section, calculated
geometrically as a function of $b$ \cite{KN01}. Due to the fact that 
for fixed impact parameter $B$ the number of produced hadrons fluctuates,
this determination of $b$ from $dN_{\rm ch}/dy$ is uncertain by a
fraction of 1\,fm, which induces a corresponding uncertainty in the
relation between $N_{\rm coll}$ and $dN_{\rm ch}/dy$ \cite{KN01}.

The argument that hard particle production should scale 
${\,\sim\,}N_{\rm coll}$ exploits the fact that hard particles
are produced on short time scales $\tau{\,\sim\,}1/\pperp$,
and that hard particle production on successive nucleons in the
nucleus therefore happens incoherently. The same is not true
for soft hadron production which involves the coherent scattering
of a projectile nucleon with several target nucleons, due to
the Landau-Pomeranchuk-Migdal (LPM) effect of a finite formation (or 
``decoherence'') time (see the last paper in \cite{G_jet} for an
explanation of the LPM effect). The net result of the LPM effect 
is a reduction of soft particle production by destructive interference,
resulting in its phenomenologically observed approximate scaling with 
the number $N_{\rm part}$ of participating (or ``wounded'') nucleons: 
each struck nucleon contributes only once to soft particle production, 
and suffering more than one collision in sequence does not increase 
the soft particle yield. The number $N_{\rm part}$ of wounded nucleons 
can again be calculated geometrically, by a formula similar to 
Eq.~(\ref{bin}) but involving the sum of the nuclear thickness 
functions instead of their product (see e.g. Ref.~\cite{KHHET01}). 

%%%%%%%%%%%%%%%%%%%%%%%%%%%%%%%%%%%%%%%%%%%%%%%%%%%%%%%%%%%%%%%%%%%%%%%%%%%%
\subsection{Suppression of high-$\bm{\pperp}$ hadrons in central Au+Au 
            collisions at RHIC}
\label{sec6c}
%%%%%%%%%%%%%%%%%%%%%%%%%%%%%%%%%%%%%%%%%%%%%%%%%%%%%%%%%%%%%%%%%%%%%%%%%%%%%

In the 5\% most central Au+Au collisions at $\scm\eq200$\,GeV at RHIC,
the average number of participating nucleons is 
$\la N_{\rm part}\ra\eq344$ (i.e. 172 times higher than in a p+p collision)
whereas the number of binary nucleon-nucleon collisions is 
$\la N_{\rm coll}\ra\eq1074$ (i.e.1074 times higher than in a p+p collision)
\cite{KN01}. The difference between ``participant scaling'' (expected 
for soft particles) and ``binary collision scaling'' (expected for hard 
processes) is in this case a factor of 1074/172\,=\,6.2. The ``nuclear 
modification factor''
\beq{raa}
  R_{AA} = \frac{1}{N^{AA}_{\rm coll}}\, 
  \frac{dN^{AA}/dyd\pperp}{dN^{pp}/dyd\pperp}
\eeq
for $A\eq197$ (gold) should therefore be $R_{AA}\eq1$ for particles
whose production scales with $N_{\rm coll}$, and $R_{AA}{\,\simeq\,}0.16$ 
for particles whose production scales with $N_{\rm part}$. Hence, 
as a function of increasing $\pperp$, we should expect $R_{AA}$ to 
rise from a value near 1/6 at low $\pperp$ to about 1 at high $\pperp$.

However, it has been known for over 25 years from minimum bias p+$A$ 
collisions at Fermilab \cite{Cronin} that $R_{AA}$ reaches values even 
larger than 1 at $\pperp\gapp2$\,GeV/$c$. The explanation of this 
``Cronin effect'' goes as follows: At high transverse momenta the 
$\pperp$-spectrum of hadrons produced in p+$A$ is not only shifted 
{\em up} in normalization by a factor $N_{\rm coll}{\,\simeq\,}A$ as 
appropriate for a hard scattering process, but on top of this also 
{\em horizontally} towards higher $\pperp$ (which at fixed $\pperp$,
of course, manifests itself as an additional gain, due to the falling
nature of the $\pperp$-spectrum). The horizontal shift towards larger
$\pperp$ arises from the fact that the partons inside the projectile 
nucleon, before making the hard collision which ends up producing the
measured high-$\pperp$ hadron, already suffer multiple elastic 
collisions with other target nucleons which they encounter first, 
thereby acquiring transverse momentum $k_\perp$ which grows in a 
random walk with the square root of the number of these elastic 
collisions. Once the hard inelastic collision happens, the projectile 
parton already brings in this ``initial $k_\perp$'', thereby giving 
and extra $k_\perp$ kick to the produced hard parton. Since this 
extra $k_\perp$ is a decreasing fraction of the observed $\pperp$ 
as $\pperp$ becomes larger, the Cronin enhancement should disappear 
as $\pperp{\,\to\,}\infty$. (For the same reason the Cronin effect
should become weaker as $\scm$ increases.) Hence, for large $\pperp$, 
$R_{AA}$ should indeed approach the value 1, but not from {\em below} 
due to soft scaling at low $\pperp$, but rather from {\em above} due 
to Cronin enhancement at intermediate $\pperp$ \cite{wang}. This behavior 
is seen in the two top curves (labelled ``no $dE/dx$'') in the right part
%
%%%%%%%%%%%%%%%%%%%%%%%%% Fig. 24 %%%%%%%%%%%%%%%%%%%%%%%%%%%%%%%%%%%%%%
\begin{figure}[ht] 
\begin{center}
  \epsfig{file=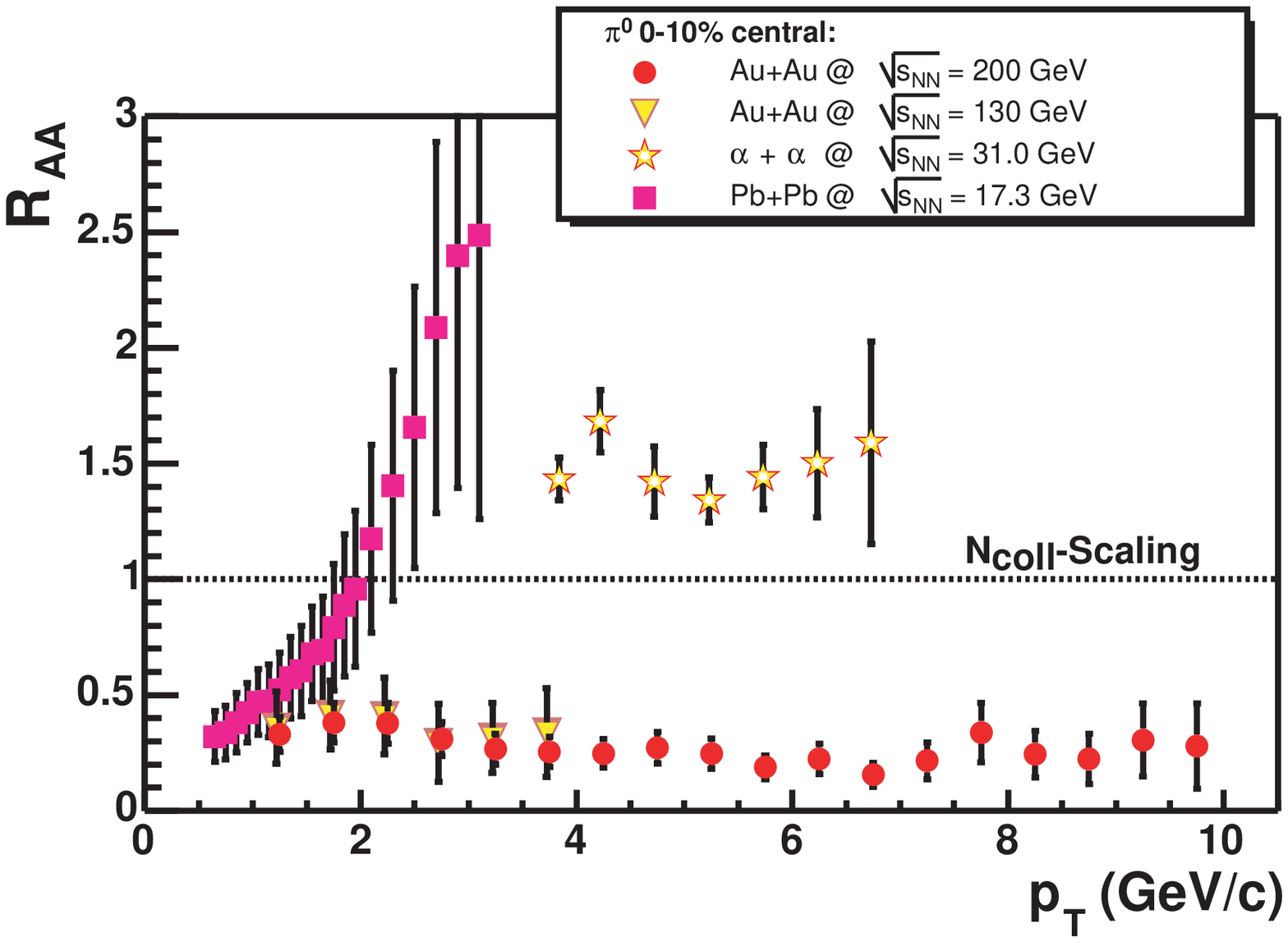,width=7.5cm,height=6cm}
  \epsfig{bbllx= 67pt, bblly=315pt, bburx=295pt, bbury=535pt,
          file=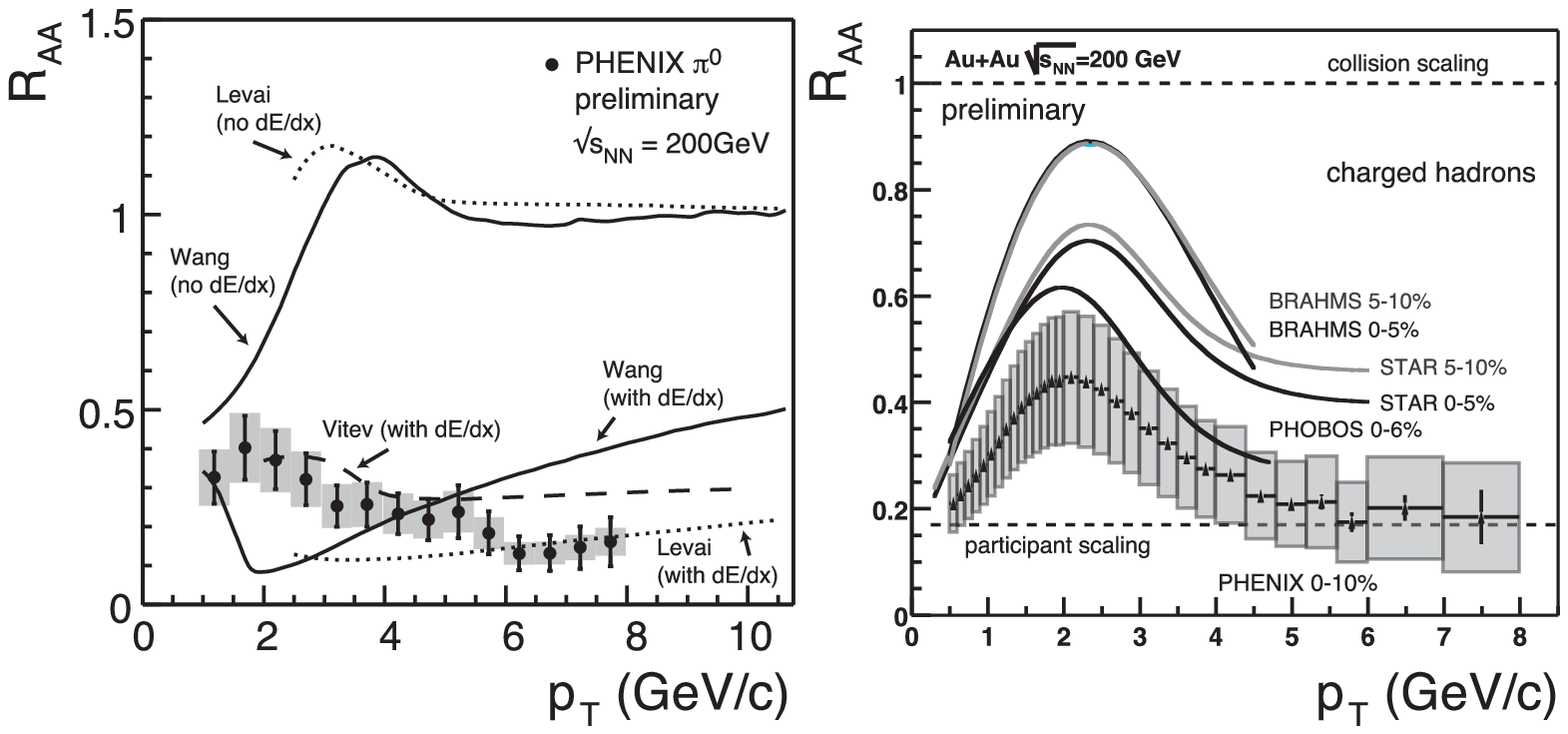, width=6.5cm, height=5.7cm,clip=}
\vspace*{-7mm}
\caption{Left: The nuclear modification ratio $R_{AA}$ defined 
in Eq.~(\protect\ref{raa}) for neutral pions measured in $\alpha{+}\alpha$
collisions at the ISR \protect\cite{ISR}, in Pb+Pb collisions
at the SPS \protect\cite{WA98_pi0}, and in Au+Au collisions at RHIC
\protect\cite{RAA_PH,RAA_PH_highpt}. (Figure taken from 
Ref.~\protect\cite{Reygers}.)
Right: $R_{AA}$ for $\pi^0$ in central Au+Au collisions at RHIC
\protect\cite{RAA_PH} together with theoretical calculations 
\protect\cite{levai,wang,vitev}. (Figure taken from Ref.~\cite{PeitzQM}.) 
\label{F24} 
} 
\end{center}
\vspace*{-5mm}
\end{figure} 
%%%%%%%%%%%%%%%%%%%%%%%%%%%%%%%%%%%%%%%%%%%%%%%%%%%%%%%%%%%%%%%%%%%%%%%%
%
of Fig.~\ref{F24}.

The left panel of Figure~\ref{F24} shows the measured nuclear 
modification factor for neutral pion production in Pb+Pb collisions 
at the SPS and in Au+Au collisions at RHIC. For comparison also 
results from $\alpha{+}\alpha$ collisions at the ISR
in the range 4\,GeV/$c{\,\leq\,}\pperp{\,\leq\,}7$\,GeV/$c$ are included
which exhibit the Cronin effect in this $\pperp$ region. The Pb+Pb data
at $\scm\eq17$\,GeV show a dramatic growth of $R_{AA}$ with increasing 
$\pperp$, starting from participant scaling at low $\pperp$ but then
exceeding the binary collision limit and exhibiting a strong Cronin
enhancement above $\pperp\gapp2$\,GeV/$c$. (This Cronin effect is larger
than in the ISR data, due to both the larger collision system, implying
a larger initial $k_\perp$, and the lower center of mass eneergy.)

\begin{figure}[ht] 
\begin{center}
  \epsfig{file=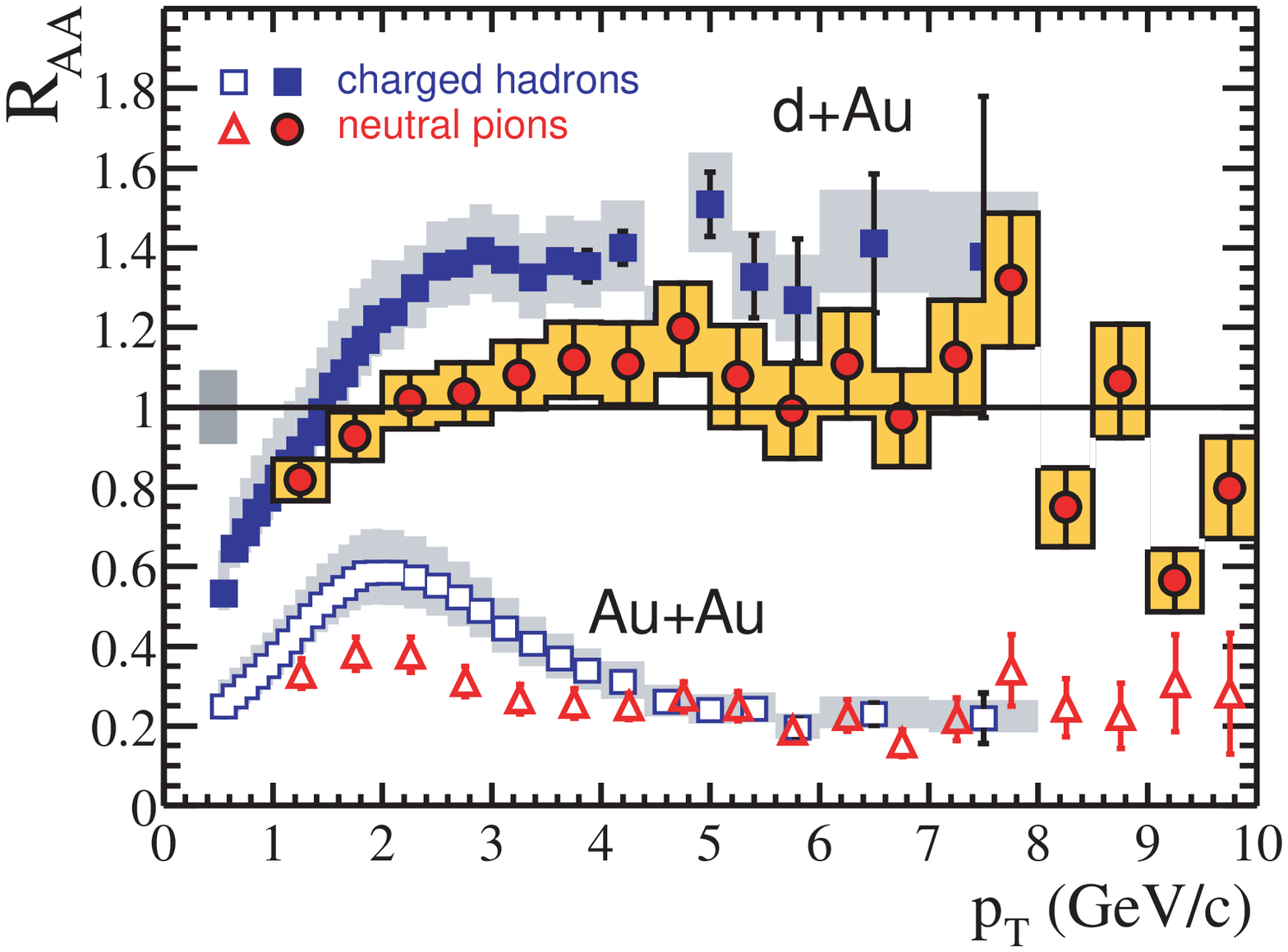,width=8cm}
  \epsfig{file=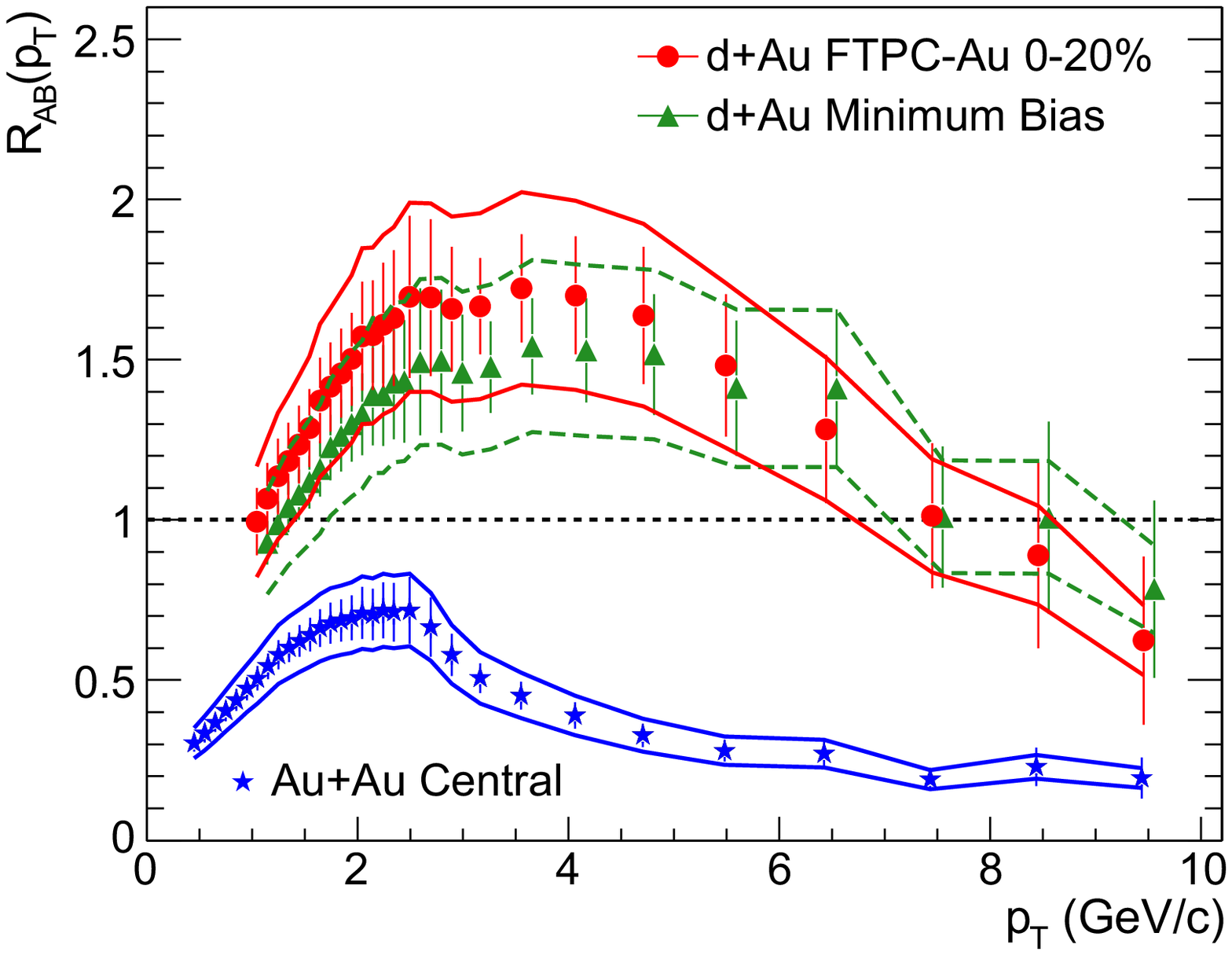,width=7.5cm,height=5.67cm}
\vspace*{-2mm}
\caption{The nuclear modification ratio $R_{AA}$ for neutral pions (left)
and charged hadrons (left and right), measured in d+Au and central Au+Au 
collisions at $\scm\eq200$\,GeV, by PHENIX \cite{Raa_dAu_PH} (left) and 
by STAR \cite{Raa_dAu_STAR} (right). At intermediate $\pperp$, 
$1.5$\,GeV/$c{\,<\,}\pperp{\,<\,}4$\,GeV/$c$, the $R_{AA}$ is 
systematically larger for charged particles than for $\pi^0$, indicating 
a larger $R_{AA}$ for kaons and protons than for pions in this 
$\pperp$-range.
\label{F25} 
} 
\end{center}
\vspace*{-7mm}
\end{figure}

In stark contrast, the RHIC Au+Au data at both 130 and 200\,GeV c.m. 
energy hardly rise at all above the participant scaling level. With 
some good will $R^{\pi^0}_{AA}$ is seen to feature a small peak
around $\pperp{\,simeq\,}2$\,GeV $R_{AA}$, but at a level of less than
50\% of the binary collision limit. At higher $\pperp$, instead of
increasing towards 1, $R_{AA}$ decreases again towards a value 
of about 0.2. No sign of the Cronin effect and of binary collision 
scaling anywhere! If we jump ahead to Fig.~\ref{F25} which shows in 
the lower parts of both panels (labelled by ``Au+Au'') the 
$\pperp$-dependence of $R_{AA}$ for all charged particles, 
and if we assume that charged and neutral pions (triangles in
the left panel) have the same nuclear modification factor, we see 
that $R_{AA}$ for heavier charged hadrons must be larger than for
pions in the region 1\,GeV/$c\lapp\pperp\lapp4$\,GeV/$c$, but also
drop back to well below 1 for $\pperp\gapp4{-}5$\,GeV/$c$.
 
The conclusion we must draw from these data is that in central Au+Au
collisions at RHIC high-$\pperp$ hadron production is suppressed 
by at least a factor 5 compared to the perturbative QCD prediction of 
binary collision scaling plus Cronin effect (top curves in the right panel
of Fig.~\ref{F24}). In fact, looking at various collision centralities
in Au+Au collisions, it was shown explicitly by the PHOBOS Collaboration
\cite{Back:2003qr} that above $\pperp\gapp4$\,GeV the charged hadron
spectra scale again linearly with $N_{\rm part}$ (i.e. like a soft 
process), instead of scaling with $N_{\rm coll}$ as expected for a hard
process. Furthermore, a crucial control experiment has recently been
completed where the nuclear modification factor was measured in 
d+Au collisions (top parts of Fig.~\ref{F25}).
%
%%%%%%%%%%%%%%%%%%%%%%%%% Fig. 25 %%%%%%%%%%%%%%%%%%%%%%%%%%%%%%%%%%%%%%%%%
%%%%%%%%%%%%%%%%%%%%%%%%%%%%%%%%%%%%%%%%%%%%%%%%%%%%%%%%%%%%%%%%%%%%%%%
%
One sees that in deuteron-gold collisions $R_{AA}\eq{R}_{\rm dAu}$ 
increases from participant scaling at low $\pperp$ to above 1 at
$\pperp{\,\simeq\,}1.5$\,GeV/$c$ and shows the expected Cronin 
enhancement (rather than a suppression) which persists at least up
to $\pperp\eq8$\,GeV/$c$. Again all charged particles taken together
(which includes heavier hadrons such as kaons and protons) show a larger 
Cronin enhancement than pions alone (left panel in Fig.~\ref{F25}), and in 
central d+Au collisions the Cronin effect is a bit larger than in minimum
bias collisions (right panel in Fig.~\ref{F25}), as expected from the 
random walk argument for the initial $k_\perp$. 

So the suppression of high-$\pperp$ hadrons in central Au+Au collisions 
must be a final state effect which involves the dense matter created in 
the heavy-ion collision. It can not be explained as an initial state 
effect, i.e. by some property of the internal structure of the gold 
nucleus which shows up when probing it at high energies (such as the 
existence of a ``color glass condensate'', see the short discussion in 
Sec.~\ref{sec3d} and the recent review \cite{ILM02}): such an
effect should still be visible (i.e. some suppression should persist) 
if only one of the colliding nuclei is gold \cite{KLM03}.

The right panel in Figure~\ref{F24} shows that only theories which 
include an energy loss $dE/dx$ for the fast partons that produce
the high-$\pperp$ hadrons can account for the $R_{AA}$ data at RHIC.
Although the different theoretical predictions shown in the graph, 
which use different approximations for the implementation of the 
radiative energy loss, give somewhat different $\pperp$-dependences 
of $R_{AA}$ and do not agree perfectly with the data, it is obvious
that a significant amount of energy loss is required to account
for the observed suppression by more than a factor 5 relative to 
the expected binary collision scaling plus Cronin (upper curves).

The observed scaling with $N_{\rm part}$ instead of $N_{\rm coll}$
suggests a very interesting geometric interpretation: Due to the
approximately constant density inside an atomic nucleus, the 
number of participating nucleons is proportional to the volume of 
nuclear matter affected by the collision, 
$N_{\rm part}{\,\propto\,}V_{\rm fireball}^{\rm init}$. The number
of binary collisions scales very accurately as 
$N_{\rm coll}{\,\propto\,}N_{\rm part}^{4/3}$. So the difference
between participant and binary collision scaling is a factor 
$N_{\rm part}^{-1/3}{\,\propto\,}V_{\rm 
fireball}^{-1/3}{\,\propto\,}1/R_{\rm fireball}$, i.e. a 
surface/volume ratio: the high-$\pperp$ hadron production data
in central Au+Au collisions at RHIC are consistent with the simple
hypothesis that high-$\pperp$ hadrons are only emitted from the
surface, but not from the interior of the fireball! Now why should this
be the case? 

%%%%%%%%%%%%%%%%%%%%%%%%%%%%%%%%%%%%%%%%%%%%%%%%%%%%%%%%%%%%%%%%%%%%%%%%%%%%
\subsection{Jet quenching in central Au+Au collisions at RHIC --- ``JET 
of the QGP''}
\label{sec6d}
%%%%%%%%%%%%%%%%%%%%%%%%%%%%%%%%%%%%%%%%%%%%%%%%%%%%%%%%%%%%%%%%%%%%%%%%%%%%%

\begin{figure}[ht] 
\vspace*{-2mm}
\begin{center}
  \epsfig{file=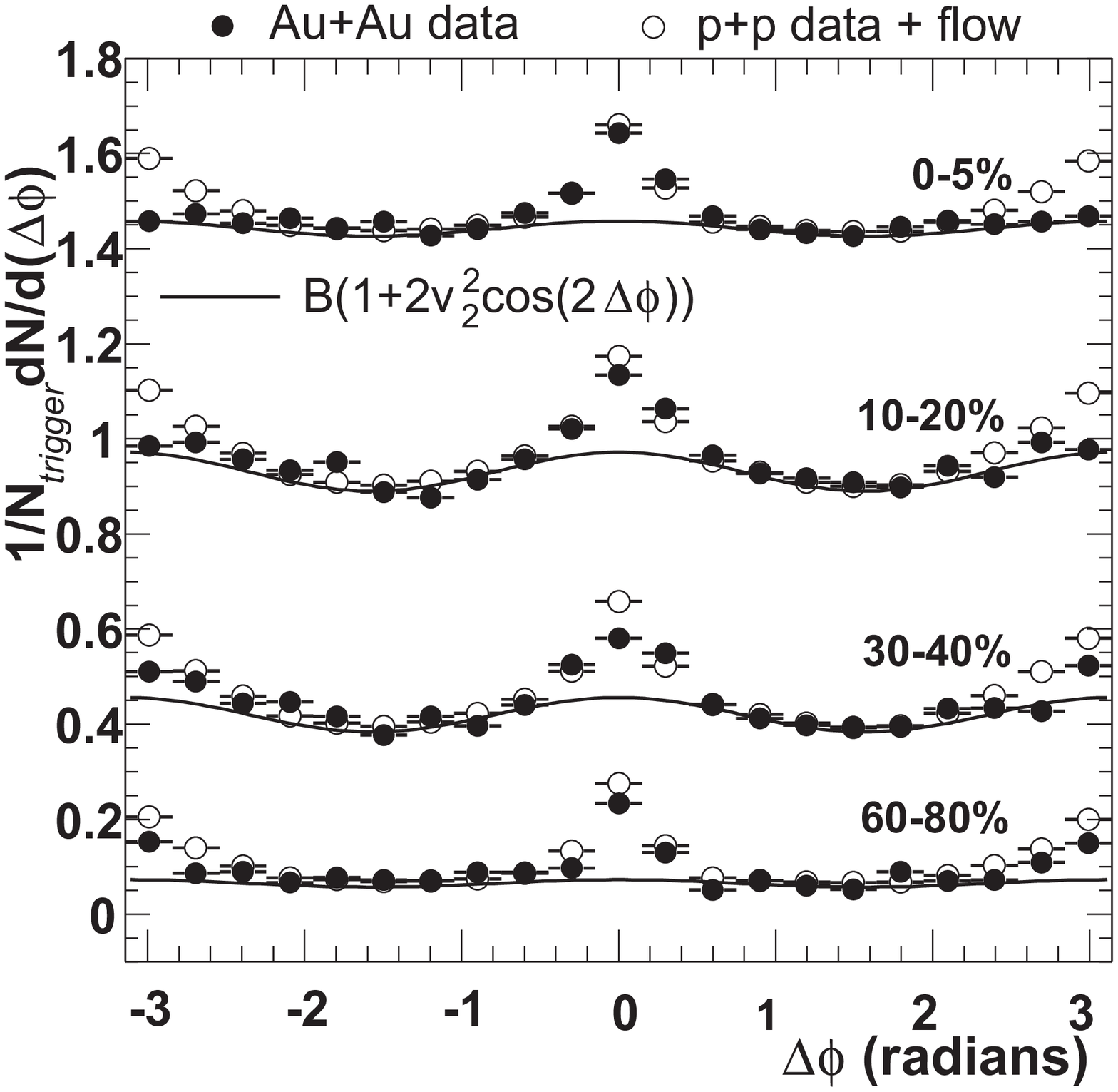,width=7.5cm}
  \epsfig{file=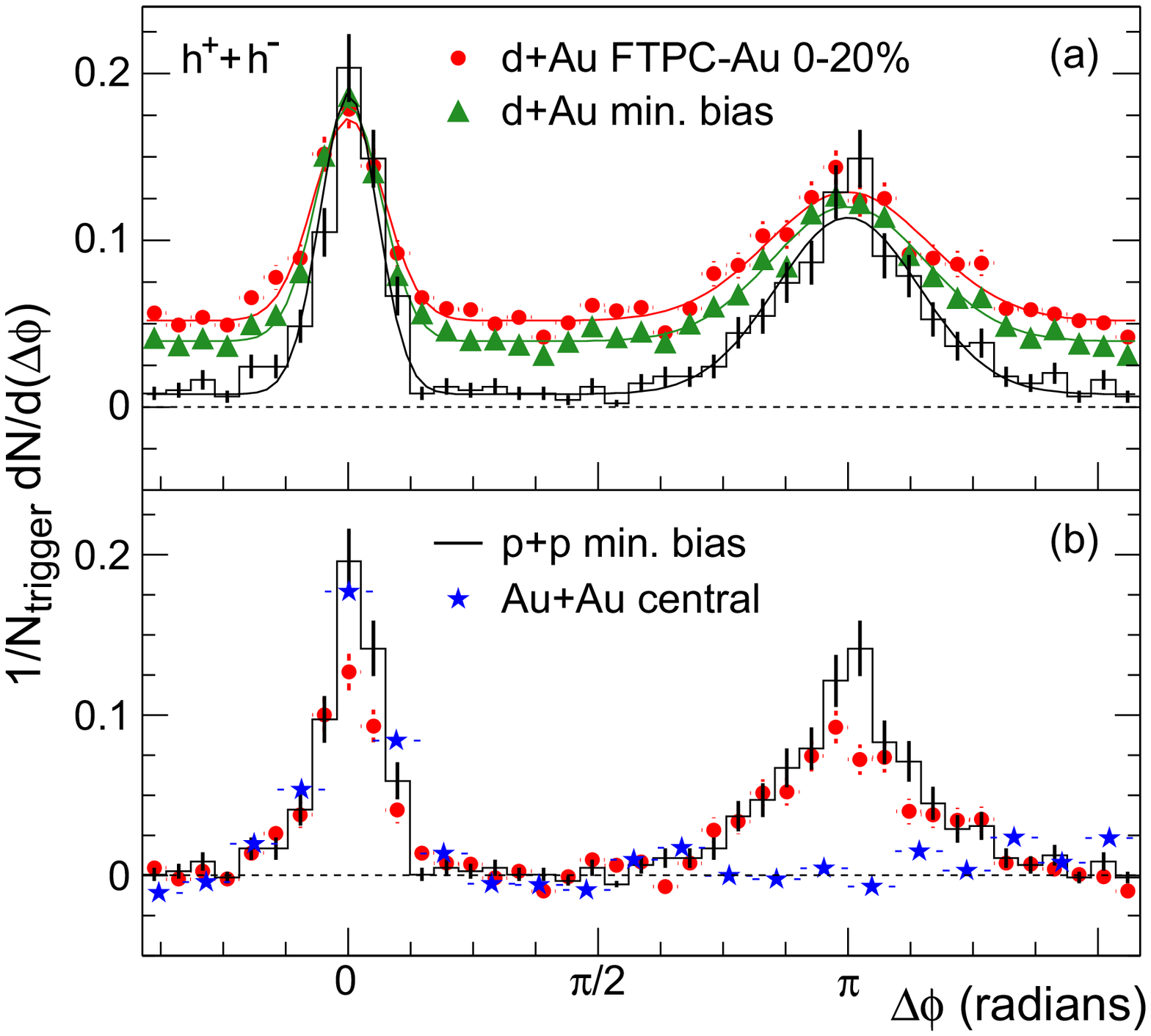,width=7.5cm,height=6.7cm}
\vspace*{-2mm}
\caption{Jet suppression in central Au+Au collisions at $\scm\eq200$\,GeV
at RHIC. Left: Azimuthal angular correlations in Au+Au collisions, in 
four centrality bins, compared with properly scaled p+p data to which
the effect of elliptic flow $v_2$ measured in the same $\pperp$- and 
centrality window has been added. (For details see Ref.~\cite{Jetsupp}.)
Whereas such a superposition of p+p data and elliptic flow can
reproduce the peripheral (60-80\%) Au+Au data, it overpredicts
the far-side correlations at $\Delta\phi\eq\pi$ for semicentral
and central collisions. Right: Azimuthal angular correlations in p+p
as well as min. bias and central d+Au collisions (upper diagram), and 
in p+p, central d+Au and central Au+Au collisions (lower diagram) 
\cite{Raa_dAu_STAR}. For details about the subtraction procedure
in the bottom part see the original paper \cite{Raa_dAu_STAR}. 
\label{F26} 
} 
\end{center}
\vspace*{-7mm}
\end{figure}

Experiments provide a unique answer also to this question 
\cite{Jetsupp}: Jets from partons which have to travel a significant 
distance through the dense fireball matter formed in the collision
lose so much energy that they are no longer recognizable as jets and
become part of the hot matter of soft particles. To show this 
experimentally one must first find jets. This is not easy due to
the huge number of soft hadrons with $\pperp{\,<\,}2$\,GeV which 
contribute so much transverse energy that the usual method of finding 
a spike in $E_{\rm T}$ in a certain small bin of solid angle fails. 
Another method for finding jets does work, however: One triggers on a
fast particle (say, with $4{\,<\,}\pperp{\,<\,}6$\,GeV/$c$ \cite{Jetsupp}) 
and then looks for azimuthal angular correlations with other, not too 
soft hadrons (say, with $\pperp{\,>\,}2$\,GeV/$c$, in order to remove
most of the background from uncorrelated soft hadrons). Since fast partons
fragment into a jet of hadrons pointing within a relatively narrow 
angular cone (``jet cone''), jets manifest themselves by positive
angular correlations at small angles relative to the fast trigger 
particle. This is shown in Fig.~\ref{F26} where it gives rise to the
peak at $\Delta\phi\eq0$. 

%
%%%%%%%%%%%%%%%%%%%%%%%%% Fig. 26 %%%%%%%%%%%%%%%%%%%%%%%%%%%%%%%%%%%%%%%%%
%%%%%%%%%%%%%%%%%%%%%%%%%%%%%%%%%%%%%%%%%%%%%%%%%%%%%%%%%%%%%%%%%%%%%%%
%
In perturbative QCD, hard partons are produced in pairs with
$180^\circ$ opening angle in the pair center of mass frame. This 
leads to a second azimuthal angular correlation peak at 
$\Delta\phi\eq\pi$. The solid histogram in the right panel
of Fig.~\ref{F26} shows this for p+p collisions at $\scm\eq200$\,GeV
\cite{Raa_dAu_STAR}. The ``far-side'' peak is flatter and wider than
the ``near-side'' peak, due to well-understood trigger effects and
a non-zero probability for semihard gluon emission by one of the two
fast partons. Except for a somewhat higher background from uncorrelated  
pairs due to increased soft particle production (``pedestal effect''), 
the same pattern is seen in minimum bias and central d+Au collisions
(Fig.~\ref{F26}, upper part of the right panel). 

However, as one
goes to Au+Au collisions and moves from peripheral to central collisions
(bottom to top in the left panel of Fig.~\ref{F26}), the away-side peak
at $\Delta\phi\eq\pi$ disappears. Whatever enhanced angular correlation
survives in almost central Au+Au collisions at $\Delta\phi\eq\pi$ is
perfectly compatible with the measured elliptic collective flow $v_2$
in the $\pperp$-range where the angular correlations are measured
\cite{Jetsupp}. Whereas the far-side jet correlations should be 
localized in a narrow window of relative pseudorapidity $\Delta\eta$ 
between the correlated particles, elliptic flow is a global collective 
effect which is correlated with the reaction plane and therefore present
at both small and large relative pseudorapidity. By subtracting the 
correlations for pairs with $\Delta\eta{\,>\,}0.5$ from those for
pairs with $\Delta\eta{\,<\,}0.5$ one in fact completely removes the
elliptic flow contribution \cite{Jetsupp}. For semicentral Au+Au
collisions the result of this procedure is a strongly reduced far-side
angular correlation peak, and in central Au+Au collisions it is 
{\em completely absent} (lower diagram in the right panel of Fig.~\ref{F26})!

%
%%%%%%%%%%%%%%%%%%%%%%%%% Fig. 27 %%%%%%%%%%%%%%%%%%%%%%%%%%%%%%%%%%%%%%%%%
\begin{figure}[ht] 
\vspace*{-2mm}
\begin{center}
  \epsfig{file=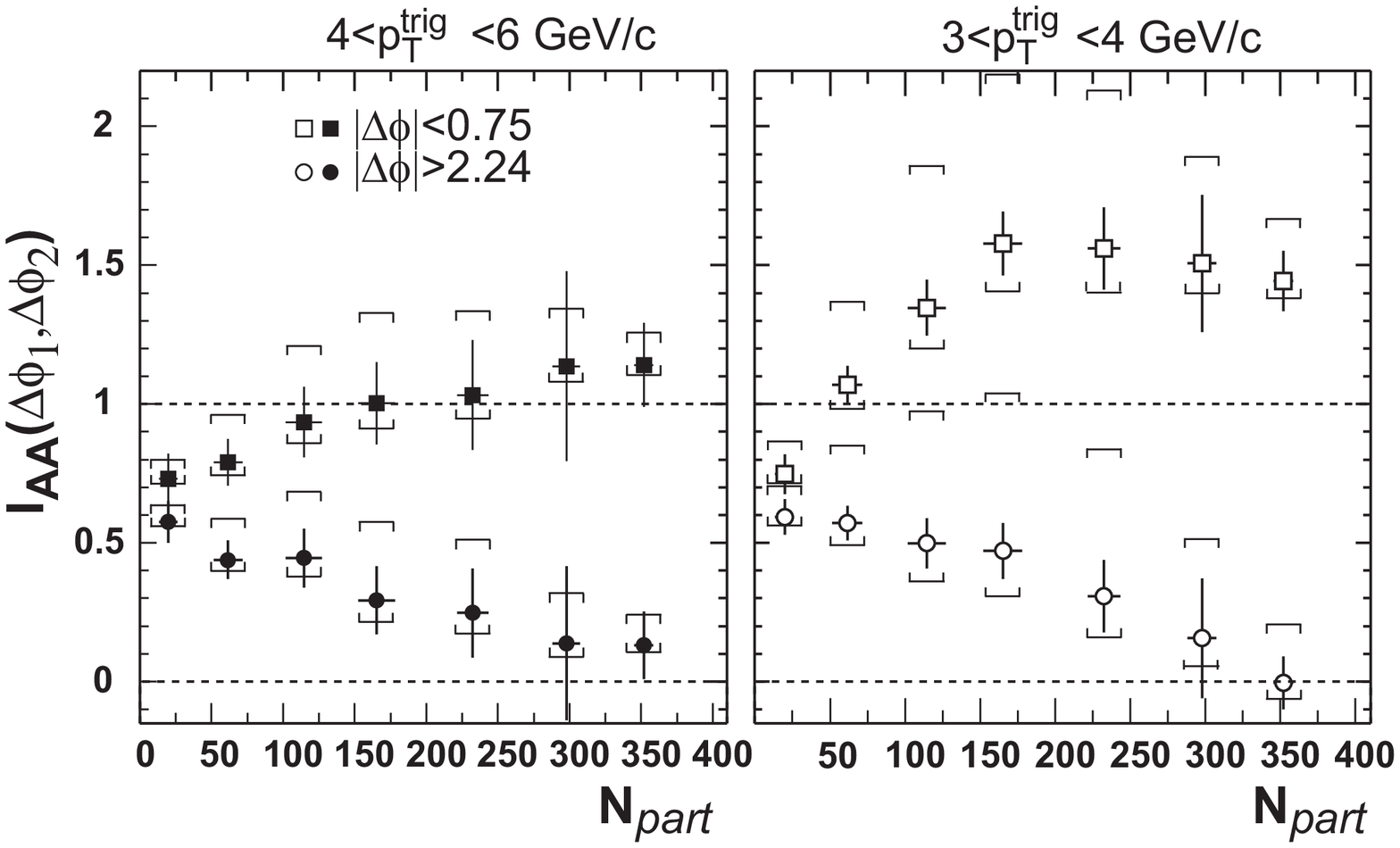,width=10cm}
  \hspace*{8mm}
  \epsfig{file=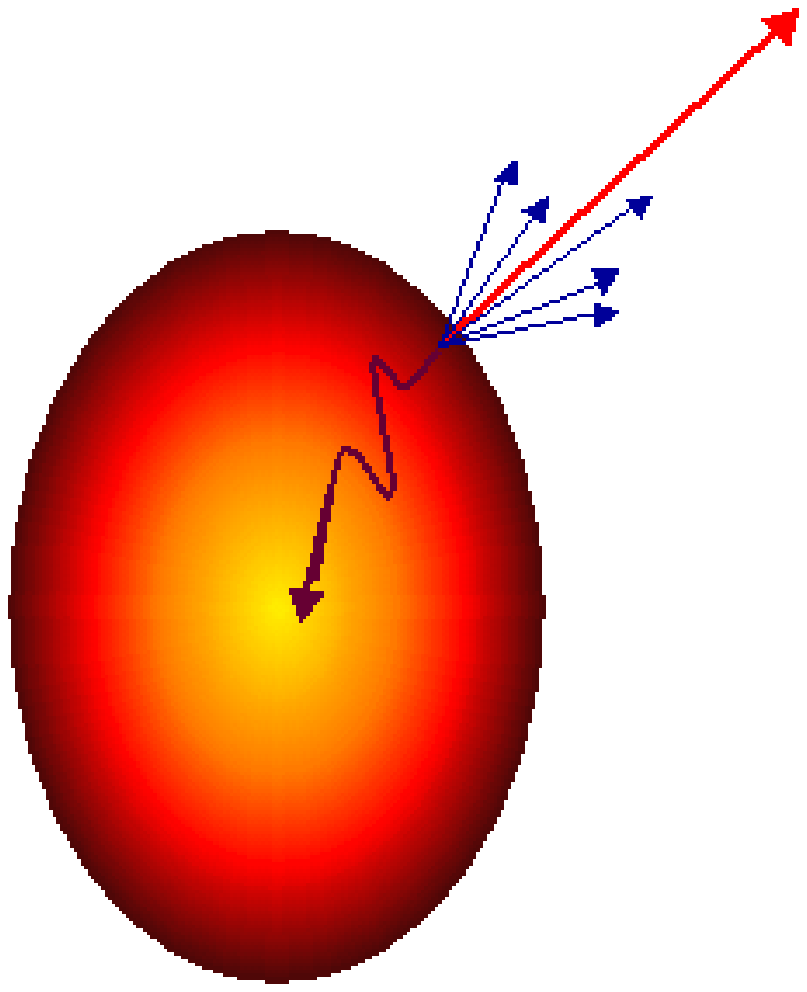,bb=186 200 426 547,width=4cm}
\vspace*{-5mm}
\caption{Left: Suppression of near-side (squares) and far-side angular
correlations (circles) in Au+Au collisions at RHIC, for two different lower 
$\pperp$ cuts for the trigger particle \protect\cite{Jetsupp}.
Note the disappearance of the far-side correlations in central collisions
(i.e. for large numbers $N_{\rm part}$ of participating nucleons).
Right: Cartoon illustrating outward jet emission from the fireball 
surface and quenching of its inward-moving partner by the dense interior.
\label{F27} 
} 
\end{center}
\vspace*{-3mm}
\end{figure} 
%%%%%%%%%%%%%%%%%%%%%%%%%%%%%%%%%%%%%%%%%%%%%%%%%%%%%%%%%%%%%%%%%%%%%%%
%
One can further quantify this by comparing the area under both 
near- and far-side peaks of the azimuthal correlation function
(after subtracting elliptic flow) with the corresponding areas in p+p
collisions. Figure~\ref{F27} shows the ratio of these areas
for the near-side peak as squares and for the far-side peak as circles.
For very peripheral Au+Au collisions both ratios are close to,
but slightly less than 1. The reduction below 1 can be understood as
a trigger effect \cite{Hardtke}. As the Au+Au collisions become
more central (i.e. $N_{\rm part}$ increases), the near-side
correlations seem to increase, in particular for lower values
of the trigger-$\pperp$, which may indicate an incomplete
subtraction of elliptic flow effects. The away-side correlations,
on the other hand, get weaker in less peripheral collisions and completely
vanish in central collisions. If the elliptic flow were larger than
assumed in the subtraction, this would bring down both the near- and
far-side correlation strength, rendering the disappearance of the 
far-side jets even more dramatic.

The interpretation suggested by these data is indicated by the cartoon
in the right part of Figure~\ref{F27}: The high-$\pperp$ trigger hadron
identifies a jet moving outward from the fireball. The yield of these
high-$\pperp$ hadrons scales by a factor 
$N_{\rm part}^{-1/3}{\,\propto\,}1/R_{\rm fireball}$ more weakly than
expected from perturbation theory, so these outward-moving jets come 
only from a thin surface layer. Fast partons created in the interior 
and trying to move out, as well as the inward-moving partners of the 
observed outward-moving jets, lose so much energy in the dense medium 
that they no longer make hadrons with $\pperp{\,>\,}2$\,GeV which
are included in the angular correlation function.
(If they still produced hadrons with $\pperp{\,>\,}2$\,GeV and 
had lost no energy but only their angular correlation with the fast 
trigger particle, the total yield at high $\pperp$ should still scale
with $N_{\rm coll}$, i.e. $R_{AA}$ should be 1.) 

This amounts to ``complete jet suppression'': The hot and dense medium
(which we know from Sec.~\ref{sec4} must be a thermalized QGP) suppresses
all jets from fast partons except for a small fraction being produced
near the surface and moving outward. Since in jet fragmentation the
leading hadron carries on average about half the $\pperp$ of the hard
parton, a trigger hadron with a $\pperp$ between 4 and 6 GeV/$c$ 
corresponds to $\pperp{\,\simeq\,}10$ for both primary partons. The
complete disappearance of the far-side jet in central Au+Au collisions
requires the far-side parton to lose at least 6 GeV of momentum such
that its fragments are all below the 2\,GeV/$c$ cut imposed on the
angular correlation function. The energy lost by these partons thus 
becomes part of the thermalized and hydrodynamically expanding
medium at $\pperp{\,<\,}2$\,GeV/$c$.

So here is our {\bf Fifth Lesson}: The quark-gluon plasma strongly
suppresses jets. Inward-moving partons with $\pperp\lapp10$\,GeV 
don't make it to the opposite edge and become part of the low-$\pt$
``thermal soup''. Only outward-moving jets formed near the fireball 
surface survive. The era of JET (Jet Emission Tomography) of the 
quark-gluon plasma has begun:
by investigating the $A$-, $b$- and emission-angle dependence of the
far-side jet suppression and combining it with information extracted
from soft particle emission we should be able to accurately determine
the fireball conditions (density and temperature) in the early 
stage.

%%%%%%%%%%%%%%%%%%%%%%%%%%%%%%%%%%%%%%%%%%%%%%%%%%%%%%%%%%%%%%%%%%%%%%%
\section{CONCLUSIONS AND DISCLAIMERS}
\label{sec7}
%%%%%%%%%%%%%%%%%%%%%%%%%%%%%%%%%%%%%%%%%%%%%%%%%%%%%%%%%%%%%%%%%%%%%%% 

The title of these lectures has been ``Concepts in Heavy-Ion Physics'',
but these lecture notes do not give a comprehensive overview of all
the relevant concepts. In $3\times75$ minutes one simply can't cover 
everything. Notable omissions include (I give references to good recent
reviews) $J/\psi$ suppression \cite{Satz} as well as direct photon and 
thermal dilepton emission \cite{R92,Gale}.
% , and 2-particle Hanbury Brown-Twiss (HBT) interferometry 
% \cite{HJ99,TW02}. 
While RHIC has not yet produced sufficient data on the first two of 
these items, $J/\psi$ suppression has been a hot subject at the SPS, 
and some very interesting signals were also seen in the photon and 
dilepton spectra \cite{HJ00,Heinz:2000ba}. I also omitted a more
detailed discussion of the very interesting quark-coalescence approach 
to the production of hadrons at intermediate transverse momenta 
($1.5\,{\rm GeV}/c{\,<\,}\pt{\,<\,}6$\,GeV/$c$) where some really 
nice systematics has been discovered which strongly indicates
that in this kinematic region quark counting rules, causing a 
meson-baryon splitting of the elliptic flow and hadron yields instead
of the hydrodynamic mass splitting observed at lower $\pt$  
\cite{Muller:2004kk}. I may include a more detailed discussion of 
these topics in some future lecture notes. 

Relativistic heavy-ion physics is making progress in huge strides.
These lectures have taught us a few important lessons which, I think,
are now solidly established, but with the steady stream of RHIC data 
new, smaller lessons are being taught to us by Nature every week and
keep complementing the picture literally as I sit here writing
these notes. (Just compare these notes and the many new data I present 
here with the copied transparencies distributed at the summer school! 
And this write-up doesn't even mention everything that happened during 
the last year \dots) Nevertheless, let me repeat once more what I think 
are the important and firm conclusions one can draw at this point in 
time (note that, in spite of the new evidence and significantly improved 
data, these still agree with what I said at the school over a year ago):

\begin{itemize}
\item
  The heavy-ion reaction zone undergoes violent explosion; this proves
  the existence of strong thermal pressure and intense rescattering
  of the quanta produced in the collision. The bulk of the hadron data
  at $p_\perp{\,<\,}2$\,GeV is well described by hydrodynamics, and 
  decoupling occurs at $T_f{\,\approx\,}100-120$ MeV with 
  average transverse flow 
  $\langle \beta_{\rm T}\rangle{\,\approx\,}0.5{-}0.6$ 
  ($\Longrightarrow$ ``The Little Bang'').
\item
  The measured elliptic flow in non-central collisions almost exhausts
  the hydrodynamic upper limit. This provides stringent limits on the
  time scale for thermalization, $\tau_{\rm therm}{\,<\,}1$\,fm/$c$,
  and shows that the quark-gluon plasma is a strongly coupled fluid and
  thermalization in the QGP is controlled by non-perturbative processes. 
\item
  At $\tau_{\rm therm}$ the energy density of the thermalized system 
  exceeds the critical value for quark deconfinement by more than an 
  order of magnitude. Hence, this thermalized state must be a Quark-Gluon 
  Plasma. At RHIC energies, the QGP lives for about $5{-}7$\,fm/$c$ 
  before hadronizing. 
  At the time of hadron formation the elliptic flow has already saturated.
\item
  Soft hadronization at $T_{\rm crit}{\,\approx\,}170$\,MeV is 
  a statistical process which generates a chemical equilibrium 
  distribution. This equilibrium is not due to kinetic processes 
  involving inelastic scattering between hadrons, but the result
  of a statistical formation process following the principle of 
  maximum entropy. Due to rapid expansion and dearth of inelastic 
  processes among hadrons, the hadronic chemical composition 
  decouples immediately. Therefore the critical temperature for 
  hadronization $\Tc\eq{T}_{\rm chem}$ can be extracted from the 
  final hadron yields. One finds $\Tc{\,\approx\,}170$\,MeV, 
  confirming lattice QCD predictions.
\item
  The QGP induces strong energy loss for high-$p_\perp$ partons. 
  Partons with $p_\perp\lapp10$\,GeV which travel inward through the 
  fireball lose enough energy to become indistinguishable from the soft
  hadron background (``complete'' jet quenching). Only outward-moving 
  surface jets are emitted, explaining why high-$p_\perp$ hadron 
  production ($p_\perp{\,\leq\,}8$\,GeV) is found to scale 
  ${\,\propto\,}N_{\rm part}$ instead of 
  ${\,\propto\,}N_{\rm coll}{\,\propto\,}N_{\rm part}^{4/3}$ as 
  perturbatively expected.
\end{itemize}

%%%%%%%%%%%%%% acknowledgements %%%%%%%%%%%%%%%%%%%%%%%%%%%%%%%%%%%%%%%%%%%
\noindent
\section*{ACKNOWLEDGEMENTS}

I would like to thank the summer school organizers for their warm 
hospitality and for creating a wonderful summer school spirit. I am 
grateful to M.~van Leeuwen, Ziwei Lin, K.~Reygers and Th.~Ullrich for 
help with some of the figures. I would also like to thank Nick Ellis for 
his constant prodding and his patience with my slow writing. I hope 
it was worth it! This work was supported in part by the U.S. Department 
of Energy under Grant DE-FG02-01ER41190.

%%%%%%%%%%%%%%%%%%%%%%% references %%%%%%%%%%%%%%%%%%%%%%%%%%%%%%%%%%%%%%%%

\end{document}